\RequirePackage[english=usenglishmax]{hyphsubst}

\documentclass[
		twoside,openright,titlepage,numbers=noenddot,headinclude,
	 	footinclude=true,cleardoublepage=empty,
		dottedtoc, 
		BCOR=5mm,paper=a4,fontsize=11pt, 
		english,american, 
		]{scrreprt} 
                
\PassOptionsToPackage{utf8}{inputenc} 
\usepackage{inputenc}

\PassOptionsToPackage{eulerchapternumbers,listings,
 pdfspacing, subfig,beramono,
parts}{classicthesis}

\newcounter{dummy} 
\providecommand{\mLyX}{L\kern-.1667em\lower.25em\hbox{Y}\kern-.125emX\@}

\usepackage{lipsum} 

\usepackage{csquotes}
\PassOptionsToPackage{%
backend=bibtex8,bibencoding=ascii,%
language=auto,%
style=numeric-comp,%
sorting=none,
maxbibnames=10, 
natbib=true 
}{biblatex}
\usepackage{biblatex}
 
 \usepackage{amsmath}

\PassOptionsToPackage{T2A,T1}{fontenc} 
\usepackage{fontenc}

\PassOptionsToPackage{russian,english}{babel}  
\usepackage{babel}

\usepackage{textcomp} 

\usepackage{scrhack} 

\usepackage{xspace} 

\usepackage{mparhack} 

\usepackage{fixltx2e} 

\PassOptionsToPackage{smaller}{acronym} 
\usepackage{acronym} 

\PassOptionsToPackage{pdftex}{graphicx}
\usepackage{graphicx} 

\usepackage{tabularx} 
\usepackage{caption}
\usepackage{subfig}  

\usepackage{listings} 
\PassOptionsToPackage{pdftex,hyperfootnotes=false,pdfpagelabels}{hyperref}
\usepackage{hyperref}  
\hypersetup{
colorlinks=true, linktocpage=true, pdfstartpage=3, pdfstartview=FitV,
breaklinks=true, pdfpagemode=UseNone, pageanchor=true, pdfpagemode=UseOutlines,%
plainpages=false, bookmarksnumbered, bookmarksopen=true, bookmarksopenlevel=1,%
hypertexnames=true, pdfhighlight=/O,
urlcolor=webbrown, linkcolor=RoyalBlue, citecolor=webgreen, 
}

\makeatletter
\@ifpackageloaded{babel}
{
\addto\extrasamerican{

}
\addto\extrasngerman{

}
}{\relax}
\makeatother

\usepackage{classicthesis}

\usepackage{amsmath}
\usepackage{amssymb}
\usepackage{cancel}
\usepackage{empheq}
\usepackage{float}
\usepackage{multicol}
\usepackage{multirow}
\usepackage{verbatim}
\usepackage{chngcntr}
\usepackage{epigraph}

\usepackage{minitoc}

\usepackage{changepage}

\counterwithout{footnote}{chapter}

\makeatletter
\renewcommand{\@seccntformat}[1]{%
  \csname the#1\endcsname
  \csname suffix@#1\endcsname 
  \quad
}
\renewcommand{\thesubsubsection}{\normalfont\alph{subsubsection})}
\renewcommand{\p@subsubsection}{\thesubsection.}
\newcommand{\suffix@subsubsection}{.}
\makeatother

\makeatletter
\AtBeginDocument{%
  \renewcommand*{\AC@hyperlink}[2]{%
    \begingroup
      \hypersetup{hidelinks}%
      \hyperlink{#1}{#2}%
    \endgroup
  }%
}
\makeatother
\AtBeginEnvironment{acronym}{%
 }

\numberwithin{equation}{section}
\usepackage{geometry}
\usepackage{frcursive}
\usepackage{etoc}
\etocsettocstyle{\normalfont\chapter*{Contents}}{}

\colorlet{mdtRed}{red!50!black}

\newcommand{\HRule}{\rule{.9\linewidth}{.6pt}}
\newcommand{\be}{\begin{equation}}
\newcommand{\ee}{\end{equation}}
\newcommand{\bi}{\begin{itemize}}
\newcommand{\ei}{\end{itemize}}
\newcommand{\bea}{\begin{eqnarray}}
\newcommand{\eea}{\end{eqnarray}}
\newcommand{\ba}{\begin{array}}
\newcommand{\ea}{\end{array}}
\newcommand{\TR}{\mathcal T}
\newcommand{\Par}{\mathcal P}
\newcommand{\mL}{\mathcal{L}}
\newcommand{\mM}{\mathcal{M}}
\newcommand{\mO}{\mathcal{O}}
\newcommand{\Tr}{\mbox{Tr}}

\usepackage{calligra}
\newcommand{\setfont}[2]{{\fontfamily{#1}\selectfont #2}}

\DeclareMathAlphabet{\mathcalligra}{T1}{calligra}{m}{n}
\DeclareFontShape{T1}{calligra}{m}{n}{<->s*[2.2]callig15}{}
\newcommand{\brs}{\mbox{\setfont{frc}{s}}}
\renewbibmacro{in:}{}

\preto\fullcite{\AtNextCite{\defcounter{maxnames}{99}}}
\AtBeginDocument{\renewcommand{\thepart}{\scshape\roman{part}}}

\begin{document}

\dominitoc[n]

\frenchspacing 

\raggedbottom 

\selectlanguage{english} 

\pagenumbering{roman} 

\pagestyle{plain} 

\newlength\tocrulewidth
\setlength{\tocrulewidth}{0.5pt}


\begin{titlepage}

\newcommand{\ttitle}{Notes from the Bulk} 
\newcommand{\supname}{Prof. Nicola \textsc{Maggiore}} 
\newcommand{\examname}{} 
\newcommand{\degreename}{Doctor of Philosophy} 
\newcommand{\addressname}{} 

\newcommand{\Subjectname}{Theoretical Physics} 
\newcommand{\keywordnames}{} 
\newcommand{\univname}{\href{https://unige.it/}{University of Genova}} 
\newcommand{\deptname}{\href{https://www.difi.unige.it/it}{Physics Department}} 
\newcommand{\doctoraldegreein}{\href{https://www.difi.unige.it/it/dottorato}{Physics and Nanoscience - XXXVI cycle}} 
\newcommand{\facname}{\href{https://scienze.unige.it/}{School of Mathematical, Physical and Natural Sciences}} 

\begin{center}

\vspace*{.05\textheight}
{\scshape\LARGE \univname\\{\Large\deptname\par}}
\begin{figure} [H]
\centering
\includegraphics[height=5cm]{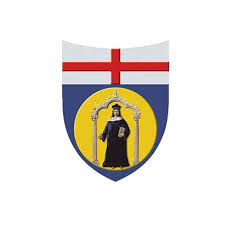} 
 \end{figure}
\textsc{\Large Doctoral Thesis}\\[0.5cm] 

\HRule \\[0.6cm] 
\spacedallcaps{\huge 
\ttitle
}\vspace{0.25cm} 
\HRule \\[1.5cm] 
 
\begin{minipage}[t]{0.4\textwidth}
\begin{flushleft} \large
\emph{Author :}\\
\textcolor{mdtRed}{Erica \textsc{Bertolini}}
\end{flushleft}
\end{minipage}
\begin{minipage}[t]{0.4\textwidth}
\begin{flushright} \large
\emph{Advisor :} \\
\textcolor{mdtRed}{\supname} 
\end{flushright}
\end{minipage}\\[3cm]
 
\vfill

\large \textit{A thesis submitted in fulfillment of the requirements\\[0.1cm] for the Doctoral degree}\\[0.3cm] 
\textit{in}\\[0.4cm]
\doctoraldegreein
\\[2cm] 
 
\vfill

{\large March 2024}\\[2cm] 
\vfill
\end{center}
\end{titlepage}


\thispagestyle{empty}

\hfill

\vfill


\cleardoublepage

\thispagestyle{empty}
\refstepcounter{dummy}

\pdfbookmark[1]{Dedication}{Dedication} 

\vspace*{3cm}
\textit{\`A ma grand-m\`ere Gis\`ele}
\begin{center}

\end{center}

\medskip

\begin{center}

\end{center} 

\cleardoublepage

\pdfbookmark[1]{Abstract}{Abstract} 

\begingroup
\let\clearpage\relax
\let\cleardoublepage\relax
\let\cleardoublepage\relax

\chapter*{Abstract}
In most cases Quantum Field Theories (QFTs) are considered without boundaries and have been successful in providing descriptions of fundamental interactions, including gravity and cosmology. This is because one is generally interested in bulk effects, where the boundary can be neglected. Nevertheless, boundaries do exist, and in some cases, their effects are self-evident and dominant. Important phenomena pertaining to condensed matter physics, like the Fractional Quantum Hall Effect  and the behavior of Topological Insulators have been explained in terms of topological QFTs with boundaries. This is rather counterintuitive : topological QFTs, when considered without boundaries, have a vanishing Hamiltonian and no energy-momentum tensor. They might appear as the least physical theories one can imagine. Despite this, when a boundary is introduced, an extremely rich physics emerges, which can be observed experimentally. The scope of this Ph.D thesis is to study the effects of the presence of a boundary from a Quantum Field Theoretical perspective, searching for new physics and explanations of observed phenomena. In particular, thanks to the formal QFT setting, the issue of the existence of local, accelerated, edge modes in Hall systems is analyzed and understood in terms of the bulk-to-boundary approach as related to a curved background in topological QFTs with boundary. Within this formalism the induced metric on the boundary can be associated to the ad hoc potential introduced in the phenomenological models in order to obtain such non-constant edge velocities. This also leads to the prediction of local modes for Topological Insulators, and Quantum Spin Hall systems in general. The paradigm for which only topological QFTs have a physical content on the boundary is broken, and also non-Topological Quantum Field Theories such as fracton models and Linearized Gravity are shown to have non-trivial boundary dynamics. Indeed due to the breaking of their defining symmetry both models have a current algebra of the Kac-Moody type on the boundary. In the case of fractons this algebra is in a generalized form, which also appears in some kinds of higher order Topological Insulators, a sign of a possible relation between these materials and edge states of fracton quasiparticles. Concerning the theory of Linearized Gravity, instead, the algebra is a standard Kac-Moody one, whose presence was suspected, but never proved before. Physical results on the boundary range between condensed matter, elasticity and (massive) gravity models. A collateral result, which enrich this Thesis, is the building of a new covariant QFT for fractons with a peculiar gauge structure. This new model better highlight the properties of these quasiparticles.

\endgroup			

\vfill 

\cleardoublepage

\pdfbookmark[1]{Publications}{Publications} 

\chapter*{Publications} 

The original content of this Thesis is mainly based on

\bi
\item[\cite{Bertolini:2020hgr}]\fullcite{Bertolini:2020hgr} (Chapter \ref{ch nonTFT})\\
\item[\cite{Bertolini:2021iku}]\fullcite{Bertolini:2021iku}  (Section \ref{sec CSinCS})\\
\item[\cite{Bertolini:2022sao}]\fullcite{Bertolini:2022sao} (Section \ref{sec BFinCS})\\
\item[\cite{Bertolini:2022ijb}]\fullcite{Bertolini:2022ijb} (Section \ref{sec MaxThFract})\\
\item[\cite{Bertolini:2023juh}]\fullcite{Bertolini:2023juh} (Section \ref{sec gauge})\\
\item[\cite{Bertolini:2023sqa}]\fullcite{Bertolini:2023sqa} (Section \ref{sec Frac+bd})\\
\item[\cite{Bertolini:2023wie}]\fullcite{Bertolini:2023wie} (Section \ref{sec LG+bd})\\
\ei

\cleardoublepage

\setlength{\epigraphwidth}{0.5\textwidth}

\pdfbookmark[1]{Acknowledgements}{Acknowledgements} 
\epigraph{\begin{otherlanguage*}{russian}
Простите, не поверю, ― ответил Воланд, ― этого быть не может. \textbf{Рукописи не горят.}
\end{otherlanguage*}}{Il Maestro e Margherita - M.A.Bulgakov}

\begingroup

\let\clearpage\relax
\let\cleardoublepage\relax
\let\cleardoublepage\relax

\chapter*{Acknowledgements}

First of all I want to thank people I had the pleasure to work with : Alberto Blasi, Andrea Damonte, Filippo Fecit, Giulio Gambuti and Giandomenico Palumbo. I also want to thank Dario Ferraro, Paolo Meda, Maura Sassetti and Niccol\`o Traverso Ziani for enlightening discussions  and information about some phenomenological and experimental scenarios. Special thanks to all those people I encountered during my last year of traveling, especially the international group I met in Vietnam, in particular Samuel Valach and Osama Khlaif, I hope to see you soon guys ! Last but not least (on the contrary) the most important thanks goes to Nicola Maggiore, there is too much I would say, but in the end, here, a "thank you" summarize it perfectly.\\
\\
I also thank the Galileo Galilei Institute (GGI) in Firenze, Italy, the International Center for Interdisciplinary Science and Education (ICISE) in Quy Nhon, Vietnam, and the Nordita Institute in Stockholm, Sweden, for hospitality during part of these works.\\
\\
These works have been partially supported by  INFN Scientific Initiative GSS~: ``Gauge Theory, Strings and Supergravity'' and by MIUR grant ``Dipartimenti di Eccellenza” (100020-2018-SD-DIP-ECC\_001).


\endgroup 

\pagestyle{scrheadings} 

\cleardoublepage

\refstepcounter{dummy}

\pdfbookmark[1]{\contentsname}{tableofcontents} 

\setcounter{tocdepth}{1} 

\setcounter{secnumdepth}{3} 

\manualmark
\markboth{\spacedlowsmallcaps{\contentsname}}{\spacedlowsmallcaps{\contentsname}}
\tableofcontents 
\automark[section]{chapter}
\renewcommand{\chaptermark}[1]{\markboth{\spacedlowsmallcaps{#1}}{\spacedlowsmallcaps{#1}}}
\renewcommand{\sectionmark}[1]{\markright{\thesection\enspace\spacedlowsmallcaps{#1}}}

\clearpage

\begingroup 
\let\clearpage\relax
\let\cleardoublepage\relax
\let\cleardoublepage\relax


\refstepcounter{dummy}
\pdfbookmark[1]{\listtablename}{lot} 

\listoftables
        
\vspace{8ex}
\newpage


\refstepcounter{dummy}
\pdfbookmark[1]{Acronyms}{acronyms} 

\markboth{\spacedlowsmallcaps{Acronyms}}{\spacedlowsmallcaps{Acronyms}}

\chapter*{Acronyms}

\acrodefplural{TQFT}{Topological Quantum Field Theories}
\acrodefplural{CFT}{Conformal Field Theories}
\acrodefplural{QFT}{Quantum Field Theories}
\begin{acronym}[FQHEEE]
\acro{BC}{Boundary Conditions}
\acro{CFT}{Conformal Field Theory}
\acro{DoF}{Degrees of Freedom}
\acro{EoM}{Equations of Motion}
\acro{FQHE}{Fractional Quantum Hall Effect}
\acro{GNC}{Gaussian Normal Coordinates}
\acro{KM}{Kac-Moody}
\acro{LG}{Linearized Gravity}
\acro{QAH}{Quantum Anomalous Hall}
\acro{QFT}{Quantum Field Theory}
\acro{QSH}{Quantum Spin Hall}
\acro{TI}{Topological Insulators}
\acro{TQFT}{Topological Quantum Field Theory}
\acro{$n$D}{$n$-dimensional}
\end{acronym}  
\textcolor{white}{\ac{$n$D}}

\endgroup 

\cleardoublepage

\pagenumbering{arabic} 

\cleardoublepage 


\ctparttext{\sloppy Boundaries exist in Nature. Their presence is usually swept under the rug, when teaching classes, except then saying, rather vaguely indeed, that ``boundary effects'' should be taken into account, which substantially affect the idealized bulk-only theories. Think, for instance, to the inexistent \textit{infinitely long} solenoids, or to the ideal \textit{infinitely extended} parallel plates of a capacitor. But boundaries have a long history of surprising successes.
}\fussy
\part{State of the Art - the Physics is in the Boundary}\label{partI}


\chapter{Boundaries in time} 

\label{ch intro} 


Boundaries are everywhere in real life. What would it even mean for an everyday object to be without boundaries? We also set conceptual boundaries, limits to our lives, properties, even countries. But not in physics. We want it to be ideal, thus everything is infinite and perfect. But physics is the language of Nature, and Nature does not care for what we want. Indeed the presence of boundaries in a model may have strong consequences, even leading to unexpected phenomena, or a whole new physical interpretation, and the aim of this Thesis is to dwell on these aspects, searching for new physics through the powerful tools of \acf{QFT}. In the long history of physics, the first boundary effect has been conceptualized in the '40s by Casimir \cite{Casimir:1948dh}, where he related the existence of vacuum pressure as to the presence two conducting parallel plates. This was later experimentally verified \cite{Lamoreaux:1996wh}. In field theory, the pioneering work which must be referred to is \cite{Symanzik:1981wd}, where Symanzik, using the Casimir effect as a playground,  defined the boundary in field theory as the surface, or the line, which separates propagators, $i.e.$ two-points Green functions, in the sense that propagators computed between points lying on opposite sides of the boundary must vanish. This approach relies on very general principles of field theory, like locality and power counting, and not much space is left to arbitrariness. For instance, the conditions which must by fulfilled by the quantum fields on the boundary  (of the Dirichlet, Neumann or Robin type), are not put by hand in the theory, but are those which naturally come out from the request of separability of propagators. This approach has been very fruitful also in the study of \acp{TQFT} with boundary \cite{Blasi:2008gt,Amoretti:2013xya,Amoretti:2013nv}. \acp{TQFT} are characterized by the absence of physical local observables, the only observables being global properties of the manifold they are built on, like the genus, or the numbers of holes and handles \cite{Birmingham:1991ty}. In other words, \acp{TQFT} have vanishing Hamiltonian and energy-momentum tensor, thus no physical content of their own. Things change when a boundary is taken into account. One of the first, and still more important, example has been the classification of all known rational \acp{CFT}, obtained by the introduction of a boundary in three-dimensional (3D) topological Chern-Simons theory \cite{Moore:1989yh}. In that case, the boundary is a circle on a flat bulk 3D manifold. Later, the introduction of a boundary in \acp{TQFT} gave a field theoretical predictive description of the edge modes of the \ac{FQHE} \cite{Wen:1992vi,Stone:1990iw,Frohlich:1994vq,Bieri:2010za,Froehlich:2018cdi} and of another kind of topological state of matter, the \ac{TI}, both in 3D and 4D \cite{Hasan:2010xy,Qi:2010qag,Hasan:2010hm,moorenature}. In particular, the Chern-Simons coupling constant has been found to be tightly related to the filling factor of \ac{FQHE} \cite{Zhang:1992eu} and to the central charge of the \ac{KM} algebras \cite{Kac:1967jr,Moody:1966gf} formed by conserved currents on the edge of Chern-Simons theory \cite{Blasi:1990jq,Emery:1991tf,Balachandran:1991dw}. Similarly, another class of \acp{TQFT}, the BF models \cite{Horowitz:1989ng,Karlhede:1989hz,Birmingham:1991ty,Blasi:2005vf},  which can be defined on any spacetime dimensions, have been found to describe the bulk theory of the \ac{TI} \cite{Cho:2010rk,Balachandran:1992qg}. Hence, the introduction of a boundary in \acp{TQFT} is the only way to give local, physical and measurable observables to theories which otherwise show ``only'' formal interest \cite{Chen:2015gma,Geiller:2019bti,Tiwari:2017wqf}, and major results have been achieved in condensed matter theory, mainly. Thus what is striking is that when dealing with a boundary in a \ac{TQFT}, local \ac{DoF} arise, and set the appropriate conditions for uncovering new physics. However 
 much less investigated is the fact that boundary physics is not a peculiarity  of \acp{TQFT} only. For instance, to a completely different framework belongs the  there is the so called AdS/CFT correspondence, also known as \textit{gauge-gravity duality}, where a higher dimensional gravity theory in Anti-de-Sitter space can be mapped into an induced, lower dimensional, \ac{CFT} through a well established dictionary \cite{Witten:1998qj,Klebanov:2000me,Polchinski:2010hw}. Interestingly, it started in the '90s in the context of string theory, and it was later  shown to  have deep consequences in condensed matter theory (again!), even leading to promising phenomenological results concerning, for instance,  the theory of superconductivity and of strange metals \cite{Hartnoll:2009sz,Herzog:2009xv,Zaanen:2015oix}. This duality is sometimes called AdS/CMT \cite{Sachdev:2010ch}.
 This Thesis lies on the solid ground of first principles of \ac{QFT} and takes inspiration from the work of Symanzik to treat a kind of holography without gravity \cite{McGreevy,Amoretti:2014kba}. The aim is to draw insights from phenomenology, in order to apply the formal tools of \ac{QFT} to physical problems. For instance we will update well established facts involving \acp{TQFT} with boundary and quantum Hall phenomena to explain new effects which cannot be justified within the standard context of a \ac{QFT} with boundary.
In particular we will extend the flat bulk-to-boundary paradigm to a curved one and we shall see that this will allow us to give a theoretical description of accelerated edge modes recently observed in some Hall systems \cite{Bocquillon} without any heuristic introduction, which is the usual approach. The induced metric on the bulk will then be interpreted as a confining/interacting potential in the sample, in the same way as the coupling constant of the Chern-Simons theory, for instance, is related to the filling factor of the \ac{FQHE}. This approach also leads to new predictions for non-constant edge velocities in \ac{QSH} systems. Concerning standard \acp{QFT}, recently it has been shown \cite{Maggiore:2019wie,Bertolini:2020hgr} that also non-\acp{TQFT} has nontrivial boundary dynamics. We will also consider boundary effects in new physical theories. That is the case for instance of fracton models, which describe quasiparticles with restricted mobility that are widely drawing the attention of the community, both theoretical, and experimental, from many areas of physics and mathematical physics \cite{Chamon:2004lew,Haah:2011drr,Vijay:2016phm,Vijay:2015mka,Nandkishore:2018sel,Pretko:2020cko,Seiberg:2020wsg}. These models have the peculiarity of being related to new phases of matter, but also, quite surprisingly to  the theory of \ac{LG}, thus with gravitational waves~: from microscopic to ``cosmological'' sizes. This makes such models quite interesting from the point of view of \ac{QFT} in general (with and without boundary).
 We shall indeed see that both the theory defined by the covariant fracton symmetry and \ac{LG} share a rich boundary dynamics, in the context of condensed matter and 3D theories of gravity.\\[10px]
The Thesis is organized as follows. Part \ref{partI} is dedicated to the introduction~: Chapter \ref{ch intro} gives the general setting and motivations, presenting some historically relevant examples of the role played by boundaries in \ac{QFT}. We will briefly review the Casimir effect (Section \ref{sec casimir}), Symanzik's paper on the Casimir effect \cite{Symanzik:1981wd} (Section \ref{sec symanzik}), and we finally give an example of AdS/CFT and the corresponding gauge/gravity dictionary (Section \ref{sec ads/cft}). In Chapter \ref{QFTapproach} we present the approach and techniques applied in this Thesis to study \acp{QFT} with boundary. Part \ref{partII} concerns \acp{TQFT}, which are briefly introduced in Chapter \ref{ch TFT}. In Chapter \ref{ch CSandBF} the 3D Chern-Simons and BF abelian theories with boundary are presented, in flat spacetime, with their physical consequences and phenomenology. This will serve as an introduction to Chapter \ref{ch CSandBFinCS}, where the 3D Chern-Simons and BF models are studied with boundary in curved spacetime, in order to find an explanation of observations for which the standard flat bulk-to-boundary approach is not enough. The study of 4D Maxwell theory with boundary \cite{Bertolini:2020hgr}, showed that also non-\acp{TQFT} display nontrivial physics on the boundary, and this leads to Part \ref{partIII}, which focuses on boundary effects on non-\acp{TQFT}. After a brief overview of the 4D Maxwell theory with boundary in Chapter \ref{ch nonTFT}, Chapter \ref{ch fractons} is dedicated to fracton models, where a new covariant \ac{QFT} for fractons is built (Section \ref{sec MaxThFract}), which agrees with the Literature (Section \ref{sec literature}). The peculiar gauge structure of the fracton theory is analyzed (Section \ref{sec gauge}) and finally the theory with boundary is studied (Section \ref{sec Frac+bd}). The strong relation between fractons and \ac{LG} gives a starting point for considering \ac{LG} with boundary in Chapter \ref{ch LG}. Part \ref{partIV} concludes the Thesis with some final remarks. \\[10px]The original contributions that are collected in this Thesis concern in particular Chapter \ref{ch CSandBFinCS}, Chapter \ref{ch fractons} (specifically Sections \ref{sec MaxThFract}, \ref{sec gauge} and \ref{sec Frac+bd}), and Chapter \ref{ch LG} (Section \ref{sec LG+bd}).  \\[20px]
In this introductory Part it can be of interest to present some well known cases where boundaries had a leading role. The scope of what follows is to both present the general physical idea  which motivated the papers, and to describe the technique applied in each situation, in order to highlight differences and similarities which inspired this Thesis.

\section{1948 : Casimir Effect}\label{sec casimir}
The first time boundary effects were investigated was in 1948 in \cite{Casimir:1948dh}. Inspired by a previous work on the interaction between atoms and a plate \cite{Casimir:1947kzi}, in \cite{Casimir:1948dh}, in the context of classical electrodynamics, H.B.G. Casimir studied  the interaction between two perfectly conducting plates, focusing on the change of electromagnetic vacuum energy. The idea was to look at if and how two parallel flat boundaries (plates) of conducting material could alter the vacuum energy just as a consequence of their existence. The system is composed of a perfectly conducting cubic cavity of volume $L^3$ with an additional conducting square plate of side $L$ placed parallel to the $xy$ face of the cube. This will play the role of the boundary, first placed at a small distance $a\ll L$ from the $xy$ face and then at a very large distance $a\sim L/2$. The interaction between the plate and the $xy$ face is given by the difference in energy between the two situations, which has a finite value given by
	\be\label{dE}
	\delta E=\frac{1}{2}\sum \hbar\omega|_a-\frac{1}{2}\sum \hbar\omega|_{L/2}\ ,
	\ee
where the sums are over all the possible resonance frequencies. Given the wave numbers $k_x=\frac{\pi}{L}n_x\ ,\ k_y=\frac{\pi}{L}n_y\ ,\ k_z=\frac{\pi}{a}n_z$, where $n_x,\ n_y$ and $n_z$ are positive integers, for a big enough cavity, $i.e.$ for large $L$, one can treat the $x,\ y$ components of the wave number (those parallel to the conductive boundary) as continuous variables for which the sum can be replaced by an integral. In polar coordinates, with $x^2\equiv k_x^2+k_y^2$, the first term at the right hand side (r.h.s.) of \eqref{dE} is
	\be
	\frac{1}{2}\sum\hbar\omega|_a=\hbar c\frac{L^2}{2\pi}\sum^\infty_{0}\theta_n\int^\infty_0x\,dx\sqrt{x^2+n_z^2\tfrac{\pi^2}{a^2}}
	\ee
where $\theta_n$ is meant to indicate that the term with $n _z= 0$ has to be multiplied by $1/2$ since to every $k_i$ correspond two standing waves unless one of the $n_i$ is zero. This can be ignored in the case of the continuous variables. For very large $a$, corrisponding to the second term at the r.h.s. of \eqref{dE}, also this last summation can be replaced by an integral and it is therefore easily seen that the interaction energy \eqref{dE} is given by
	\be
	\delta E=\hbar c\frac{\pi^2L^2}{4a^3}\left[\sum^\infty_{0}\theta_n\int^\infty_0du\sqrt{n_z^2+u}-\int^\infty_0\!\int^\infty_0du\,dn_z\sqrt{n_z^2+u}\right]\ ,
	\ee
with $u\equiv \frac{a^2}{\pi^2}x^2$. At this point the integral can still diverge for big enough $k$. This physically means that high frequency waves are not affected by the cavity, nor by the position of the plates: the zero pt energy would diverge as if unaffected. The solution is to introduce a function which cuts away the divergent contribution by going sufficiently rapidly to zero in $k$. This function and the cutting frequency depend on the material of the cavity, but will not intervene in the approximation that will be taken into account. Indeed by following this approach, and using Euler-Maclaurin formula to estimate the difference between the integrals, one finds
	\be
	\frac{\delta E}{L^2}=-\hbar c\frac{\pi^2}{720}\frac{1}{a^3}\ .
	\ee
There is thus a force given by
	\be\label{Cas for}
	F=-\frac{1}{L^2}\frac{\partial \delta E}{\partial a}=-\hbar c\frac{\pi^2}{240}\frac{1}{a^4}\ .
	\ee
This means that two uncharged parallel conductive plates (boundaries), in vacuum, experience an attractive force between them. The  existence of such an attractive force between the two plates is the consequence of vacuum energy pressure of electromagnetic waves. In evaluating this force one can see that it is observable only for very small values of $a$, $i.e.$ the distance between the plates. Indeed it took many years before a first direct measure, in 1997, was made possible and verified the effect \cite{Lamoreaux:1996wh}.

\section{1981 : Symanzik's method}\label{sec symanzik}
Many years later Symanzik, in \cite{Symanzik:1981wd}, applied the ``high flexibility and perfection of the QFT formalism'' ($sic$) to recover the above result of the casimir force \eqref{Cas for} as a \ac{QFT} with boundary for the first time. The Casimir effect was indeed the perfect playground to investigate the meaning of a boundary in \acp{QFT}. Indeed how could \ac{BC} be implemented without spoiling the renormalizability of the theory? What are the consequences on the observables? These are all legit and fundamental questions questions when dealing with a \ac{QFT} with boundary.
\subsection*{Boundary conditions}
The model considered by Symanzik in \cite{Symanzik:1981wd} is that of a free massive scalar field in euclidean four-dimensional spacetime \footnote{the original papar \cite{Symanzik:1981wd} works in generic $\nu$-dimensional spacetime, but for the introductory scope of this Thesis it is sufficient to work within the standard four-dimensional spacetime ($i.e.\ \nu=4$), in order to better highlight crucial points and compare the results and procedure with those of the previous Section.}, described by the lagrangian density
	\be
	\mathcal L_0=-\frac{1}{2}\partial_\mu\phi\partial^\mu\phi-\frac{1}{2}m^2\phi^2\ .
	\ee
The model is considered to have a 3D boundary $\partial\mM$ (the third dimension corresponding to euclidean time) described by the equation $f(x)=0$, which separates the space into two regions: 
$\mM$ with  $f(x)>0$, and its complementary $\mM'$ with $f(x)<0$.This is therefore a two-sided boundary, and the massive scalar theory lives on both sides of it. As we shall see this is different to what will be considered in the original part of this Thesis, which is a one-sided boundary. While this last would be more of a ``confined'' situation, where the theory is in an interiior region delimited by the boundary, the present one is more similar to a defect or interface \cite{Saleur:1998hq,Saleur:2000gp,Fradkin:2006mb,Mintchev:2007qt,}. Due to the presence of the boundary, an augmented lagrangian density needs to be considered
	\be\label{INTRO-Lbd}
	\mL=\mL_0-\delta(f(x))\phi(x-\epsilon)\partial_\mu\phi(x)\partial_\mu f(x)\ ,
	\ee
where $\epsilon$ is a vector normal to the boundary $\partial\mM$. Since on the boundary the field $\phi(x)$ and its derivative with respect to the normal of the boundary ($\partial_n\phi(x)$) are independent, this additional contribution to the lagrangian implements Dirichlet and Neuman \ac{BC} in  the two sides of the boundary respectively, as follows
	\begin{align}
	\phi(x)\to0&\qquad x\to\partial\mM\mbox{ from }\mM\\
	\partial_n\phi(x)\to0&\qquad x\to\partial\mM\mbox{ from }\mM'\ .
	\end{align}
From \eqref{INTRO-Lbd} the generating functional of Green function can be computed to be
	\be\label{Z}
	Z[J]=\int[d\phi] e^{\int\mL+\int J\phi}\propto e^{\frac{1}{2}\int_\mM J(x)G_D(x,y)J(y)+\frac{1}{2}\int_{\mM'} J(x)G'_N(x,y)J(y)}\ ,
	\ee
where $G_D$ is the Dirichlet Green function in $\mM$ given by
	\begin{equation}\label{SymPropD}
		\begin{array}{rclcl}
		(m^2-\partial_x^2)G_D(x,y)&=&\delta(x-y)&&\mbox{if }x,y\in\mM\\
		G_D(x,y)&=&0&&\mbox{if }x\in\partial\mM,\ y\in\mM\ ,
		\end{array}
	\ee
and $G'_N$ the Neumann Green function in $\mM'$
	\begin{equation}\label{SymPropN}
		\begin{array}{rclcl}
		(m^2-\partial_x^2)G'_N(x,y)&=&\delta(x-y)&&\mbox{if }x,y\in\mM'\\
		\partial_nG'_N(x,y)&=&0&&\mbox{if }x\in\partial\mM,\ y\in\mM'\ .
		\end{array}
	\ee
Therefore the presence of the boundary affects the two-pt functions (propagators) of the theory as shown by \eqref{SymPropD} and \eqref{SymPropN}, as a consequence of the surface term introduced in \eqref{INTRO-Lbd}. Indeed the main consequence is that in \eqref{Z} there are no correlations between points in $\mM$ and in $\mM'$, have thus been decoupled from each other by the boundary, and propagators of points separated by the boundary have to vanish.
\subsection*{Casimir effect}
We have seen in Section \ref{sec casimir} that the presence of two parallel boundaries (plates) affects the vacuum energy through their presence, leading to a Casimir force \eqref{Cas for} which attract one-another. In the context of a model described by the lagrangian \eqref{INTRO-Lbd}, the vacuum energy for the scalar field is
	\be
	E_{\bar\mM}=-\lim_{T\to\infty}\frac{1}{T}\ln\int[d\phi] e^{\int_0^Tdt\int_{\bar\mM}dx\mL(\phi,\partial\phi)}\ ,
	\ee
where $\bar\mM$ denotes the spatial part of the boundary $\mM$, $i.e.$ $\mM=\bar\mM\times[0,T]\ ,\ \partial\mM=\partial\bar\mM\times[0,T]$.The relevant quantity is the total energy, which is the ground-state energy in $\bar\mM$, for instance with Dirichlet \ac{BC}, plus the one in the complementary region $\bar\mM'$ with Neumann \ac{BC}. The boundary-independent contribution can be omitted, leaving surface terms only. Therefore in the context of the Casimir effect, considering Dirichlet conditions on the inner sides of two parallel plates, at distance $a$, the total energy is the same as if the two Neumann regions were absent. Thus for a \ac{QFT} computation of the Casimir effect the surface graphs have merely the $\phi\partial_n\phi$ vertices on $\partial\mM$ appearing in \eqref{INTRO-Lbd}, for which the expansion is
	\be\label{INTRO-casE}
	E^{(D)}_{\bar\mM}+E^{(N)}_{\bar\mM}-const=-\lim_{T\to\infty}\frac{1}{2T}\left(2\Tr\overline{\partial_nG}+\frac{1}{2}2^2\Tr \,\overline{\partial_nG}\cdot\overline{\partial_nG}
+...\right)
	\ee
where $\Tr$ is the trace on the surface $\partial\mM$, and $\overline{\partial_nG}$ is
	\be
	\overline{\partial_nG}=-\frac{1}{2}\pi^{-2}\Gamma(2)\left[(X-X')^2+a^2\right]^{-2}a\ ,
	\ee
with $X$ and $X'$  the $2+1$-dimensional coordinates on the plates. While on the r.h.s. 
 of \eqref{INTRO-casE} the odd terms all vanish, the even ones can be summed, giving
	\be
	\frac{1}{area}\left(E^{(D)}_{\bar\mM}+E^{(N)}_{\bar\mM}-const\right)=-2^{-4}\pi^{-2}\Gamma(2)\zeta(4)a^{-3}=-\frac{1}{1440}\frac{\pi^2}{a^3}\ ,
	\ee
where $\zeta(4)=\sum_{n=1}^{\infty}n^{-\nu}|_{\nu=4}=\pi^4/90$, and $\Gamma(2)=1$ is the Gamma function. 
 Eq.\eqref{INTRO-casE} is analogous to the Casimir potential obtained in Section \ref{sec casimir}, keeping in mind that here $\hbar=c=1$.

\section{1997 : Holography and AdS/CMT }\label{sec ads/cft}
The AdS/CFT correspondence, conjectured for the first time by Maldacena in 1997 \cite{Maldacena:1997re}, is another example of the successful approaches involving boundaries. It shows dualities between $d+1$-dimensional gravity bulk theories and their holographic counterparts on their $d$-dimensional boundaries, where the extra ``energy'' dimension run from zero to infinity  \cite{Polchinski:2010hw,Klebanov:2000me}. The AdS/CFT holographic correspondence, a.k.a. gauge/gravity duality, was originally conjectured in string theory and captured the attention of many physicists.
For instance a typical paper on the topic had abstract like this \cite{Witten:1998qj}\\[-3px]
\begin{adjustwidth}{15px}{15px}
{\small Recently, it has been proposed by Maldacena that large $N$ limits of certain conformal field theories in $d$ dimensions can be described in terms of supergravity (and string theory) on the product of $d+1$-dimensional AdS space with a compact manifold. Here we elaborate on this idea and propose a precise correspondence between conformal field theory observables and those of supergravity: correlation functions in conformal field theory are given by the dependence of the supergravity action on the asymptotic behavior at infinity. In particular, dimensions of operators in conformal field theory are given by masses of particles in supergravity. As quantitative confirmation of this correspondence, we note that the Kaluza-Klein modes of Type IIB supergravity on $AdS_5\times S^5$ match with the chiral operators of $\mathcal N= 4$ super-Yang-Mills theory in four dimensions. With some further assumptions, one can deduce a
Hamiltonian version of the correspondence and show that the $\mathcal N= 4$ theory has a large $N$ phase transition related to the thermodynamics
of AdS black holes.}
\end{adjustwidth}
Later on the duality received much attention in condensed matter theory, enough to introduce for that case a new acronym (AdS/CMT). The bulk/boundary correspondence turned out to be a powerful new technique to study strongly coupled systems, reviewed for instance in \cite{Hartnoll:2009sz,Herzog:2009xv,Sachdev:2010ch,Zaanen:2015oix,Amoretti:2017xto}. 
Despite the completely different context, the dictionary  still involve a $d+1$-dimensional gravity theory in anti-de-Sitter (AdS) spacetime, $i.e.$ with negative curvature and metric of the form
	\be\label{ads metric}
	ds^2=\tfrac{L^2}{r^2}\left(dr^2+\eta_{\mu\nu}dx^\mu dx^\nu\right)\ ,
	\ee
whose $d$-dimensional boundary \ac{CFT} is related to phenomena in condensed matter theory \cite{Zaanen:2015oix}. Starting from the ``quantum gravity'' $d+1$-dimensional bulk action $S_{QG}$ and bulk field $\phi$, the dictionary prescribes \cite{Gubser:1998bc,Witten:1998qj} that the boundary $d$-dimensional \ac{QFT} has source $\phi_0$ related to the boundary value of the bulk field $\phi$, and is obtainable as a boundary limit of the on-shell ($i.e.$ saddle point) gravitational path integral as
	\be\label{Zads}
		Z_{QFT}[\phi_0]=\int[dA]e^{iS_{QFT}+i\int\phi_0\mathcal O(A)}=Z_{QG}[\phi\to\phi_0]\sim e^{iS_{QG}}|_{\phi\to\phi_0}\ .
	\ee
Thus the bulk field $\phi$ is related to the boundary operator $\mathcal O$ through its boundary value $\phi_0$ becoming a source for such operator
	\be\label{duals}
	\phi|_{bd}=\phi_0\leftrightarrow\mathcal O\ .
	\ee
This is the core of the duality and the key to computing fundamental quantities (n-pt functions) in the boundary theory through
	\be
	\langle\mathcal O(x_1)...\mathcal O(x_n)\rangle=\frac{\delta^{(n)}\log Z_{QFT}[\phi_0]}{\delta{\phi_0(x_1)}...\delta{\phi_0(x_n)}}\sim\frac{\delta^{(n)} Z_{QG}^{os}[\phi_0]}{\delta{\phi_0(x_1)}...\delta{\phi_0(x_n)}}\ .
	\ee

\subsection*{An example : holographic conductivity}
The easyest way to quickly show the peculiar features of this duality is to try and compute conductivity applying this the dictionary. This means that we want to compute Ohm's law through the AdS/CFT correspondence, thus starting from a theory general relativity with boundary. In a 3+1-dimensional asymptotically AdS theory of gravity, electromagnetism is described by the following Einstein-Maxwell action
\be
S_{bulk}=\int d^4x\sqrt{-g}\left[\frac{1}{2\kappa^2}\left(R+\frac{6}{L^2}\right)-\frac{1}{4e^2}F_{\mu\nu}F^{\mu\nu}\right]\ ,
\ee
where $L$is the AdS radius, and $\mu\nu...$ are the bulk $d+1$ indices ($m,n...$ will be those of the boundary). We have thus a bulk theory with a charged black hole, which we say to be at a chemical potential $\mu$ \cite{Chamblin:1999tk}. The Reissner-N\"ordstrom black hole solution, which comes from solving Einstein's \ac{EoM}, is
	\be
	ds^2=\frac{L^2}{r^2}\left[-f(r)dt^2+\frac{dr^2}{f(r)}+\eta_{\mu\nu}dx^\mu dx^\nu\right]\ ,
	\ee
which is of the form \eqref{ads metric} and with
	\be
	f(r)=1-\left(1+\tfrac{r_h^2\mu^2}{\gamma^2}\right)\left(\frac{r}{r_h}\right)^3+\tfrac{r_h^2\mu^2}{\gamma^2}\left(\frac{r}{r_h}\right)^4\quad;\quad\gamma\equiv\frac{2e^2L^2}{\kappa^2}\ ,
	\ee
where $r_h$ is the radius of the outer horizon of the black hole. It is also known that one can identify a temperature at the horizon of black holes, which is the Hawking temperature \cite{Hawking:1982dh}. In the case of this Reissner-N\"ordstrom black hole it is
	\be
	T=\frac{1}{4\pi r_h}\left(3-\frac{r_h^2\mu^2}{\gamma^2}\right)\quad\Rightarrow\quad r_h=r_h(\mu,T)\ .
	\ee
This means that black holes in the bulk correspond to placing the boundary field theory at some finite temperature, and the bulk spacetime corresponds to real time dynamics in the boundary thermal field theory. The duality bulk fields-boundary sources \eqref{duals} coming from the prescription \eqref{Zads} here says
	\be
	g_{mn}\leftrightarrow T_{mn}\mbox{ (stress)}\quad;\quad A_m\leftrightarrow J_m\mbox{ (current)}\ .
	\ee
Concerning the gauge field $A_m $, from solving its \ac{EoM} with proper conditions at the horizon such that it vanishes, one finds that the temporal component takes the form
	\be
	A_0=\phi_0(x)+\phi_1(x)r=\mu\left(1-\tfrac{r}{r_h}\right)\ .
	\ee
Then the dictionary tells us that the $leading$ $\phi_0$ and $subleading$ $\phi_1$ terms of this solution can be identified respectively with the source for the current $J^0$ and the expectation value $\langle J^0\rangle$ \cite{Hartnoll:2009sz,Herzog:2009xv,Sachdev:2010ch,Zaanen:2015oix}. Which means that the source for $J^0$ is the chemical potential $\mu$, while the expectation value $\langle J^0\rangle$ is the charge density. The presence of an electric field with frequency $\omega$ is related to the $x$-component of the gauge field through $\partial_tA_x = Ee^{i\omega t}$, for which $A_x = \frac{E}{i\omega}e^{i\omega t}$ on the boundary. In the bulk, the leading order terms in $A_x$ take the form
\be
A_x = \frac{E}{i\omega}e^{i\omega t}+\langle J_x\rangle r+...\ ,
\ee
from which the subleading term $\langle J_x\rangle$ can be determined through the \ac{EoM} of the bulk. Considering the fact that this implies the presence of a $g_{tx}$ component of the metric \cite{Hartnoll:2009sz}, from Maxwell and Einstein equations, together with ingoing \ac{BC} at the
horizon, we get
\be\label{Ax'}
(fA_x')'+\frac{\omega^2}{f}A_x=\tfrac{4\mu^2}{\gamma^2r_h^2}A_x\ .
\ee
Although this cannot be solved analytically, it can be done numerically and from the Ohm's relation $J=\sigma E$, we can thus determine the response $\langle J_x\rangle$ in terms of the source, from which the ratio is the conductivity
\be\label{sigma}
\sigma(\omega)=\frac{1}{e^2}\frac{A'_x}{i\omega}\ .
\ee
What is surprizing of this result, which highlight the power of this duality, is that starting from general relativity with a charged black hole and a boundary, one obtain the ``conductivity'' \eqref{sigma}, which plotted with the numerical solution recovered from \eqref{Ax'}, reproduces a behaviour seen in graphene and typical of any \ac{CFT} in  2 + 1 dimensions \cite{Sachdev:2011fcc,Hartnoll:2009sz}\footnote{Notice that a delta function appear in the analysis, which can however be eliminated when considering the theory on a lattice (so to break translational invariance). This discussion anyway lies outside the scope of this introductory Section and Thesis.}.

\chapter{The QFT Approach in this thesis}\label{QFTapproach}
We have seen that it is possible to treat theories with boundary in the context of different formulations~: through standard electrodynamics as in the Casimir effect, by means of the formalism of \ac{QFT}, in the case of Symanzik's work, or looking at holography as a theory of gravity displaying microscopic condensed matter phenomena on its asymptotic conformal boundary. All of this with fruitful results. In this Section an additional case is considered, representing the technique and attitude  in terms of which the original works contained in this Thesis is built. The principles around which the Thesis develops are those of \ac{QFT} : symmetry, covariance, locality and power-counting, the only external input  is given by the introduction of the boundary itself. No additional requests are made to recover the induced physics, it will be the theory itself that will give all the information needed.

\section{The symmetry, the action and the boundary}\label{sec symm}
The main ingredient from which we start building the theory is the symmetry, described by a transformation $\delta_\alpha$ acting on the fields, which here are collectively represented as $\phi_A(x)$,  as follows
	\be\label{symm}
	\delta_\alpha\phi_A=\mathcal W_{\alpha A}\ ,
	\ee
and leaving the most general, power-counting-compatible, action invariant, $i.e.$
	\be\label{dS=0}
	\delta_\alpha S_{inv}[\phi_A]=0\ .
	\ee
Here the index $\alpha$ in $\delta_\alpha$ is a generic one, meaning that the transformation might change the nature of the involved field (like the case of supersymmetry \cite{Wess:1992cp}). In the following we will only deal with transformation where no $\alpha$-index appear. This is the case for instance of the standard gauge transformation
	\be\label{dgauge}
	\delta_{gauge} A_\mu=\partial_\mu\Lambda
	\ee
 whose invariant 4D action describes Maxwell theory
	\be\label{Max}
	S_{Max}=-\frac 1 4 \int d^4x F_{\mu\nu}F^{\mu\nu}
	\ee
with
	\be
	\quad F_{\mu\nu}\equiv\partial_\mu A_\nu-\partial_\nu A_\mu\ .
	\ee
In 3D the same symmetry give rise to the topological Chern-Simons action \cite{Birmingham:1988ap}, or the Maxwell-Chern-Simons one \cite{Deser:1981wh} depending on the mass dimension of the gauge field. From just one symmetry and one field, three completely different models come out~: we have the theory of electromagnetism, and by lowering the dimensions one can  either have a theory of mostly mathematical interest, as is Chern-Simons, or one which gives the 3D photon a topological mass \cite{Deser:1981wh}. Not to mention non-abelian extensions. Infinitesimal  diffeomorphisms
	\be
	\delta_{diff}h_{\mu\nu}=\partial_\mu\xi_\nu+\partial_\nu\xi_\mu\ ,
	\ee
for a rank-2 symmetric tensor field define the theory of \ac{LG}, where the gauge parameter $\xi_\mu(x)$ is a vector. There are also more ``exotic'' cases that we will encounter, and the discussion will be postponed to the relevant Chapter. In Path Integral formalism the fundamental objects (observables) are related to the generating functional $Z[J]$ and in particular to the generating functional of the connected Green's functions $Z_c[J]$, defined by
	\be\label{Z1}
	Z[J]=e^{iZ_c[J]}=\int [d\phi_A]e^{iS_{inv}+S_J}\ .
	\ee
In \eqref{Z1} a source $J^A(x)$ is coupled to the field $\phi_A(x)$ 
	\be\label{SJ}
	S_J=\int d^dx J^A\phi_A\ ,
	\ee
which allows to compute the Green functions $G^{(n)}(x_1,...,x_n)$ as
	\be
	\left.\frac{\delta^nZ_c[J]}{\delta J^{A_1}(x_1)...\delta J^{A_n}(x_n)}\right|_{J=0}=i^{n-1}\langle T(\phi_{A_1}(x_1)...\phi_{A_n}(x_n))\rangle=G_{A_1...A_n}^{(n)}(x_1,...x_n)\ ,
	\ee
where $T$ represents the time-ordered product. Of particular interest are the 1- and 2-pt functions, $i.e.$ mean value of the field and propagators
	\begin{align}
	\left.\frac{\delta Z_c[J]}{\delta J^A(x)}\right|_{J=0}&=\langle\phi_A\rangle\label{G1}\\
	\left.\frac{\delta^2Z_c[J]}{\delta J^A(x)\delta J^B(y)}\right|_{J=0}&=i\langle T(\phi_A(x)\phi_B(y))\rangle=G^{(2)}_{AB}(x-y)\ .\label{G2}
	\end{align}
If the transformation \eqref{symm} is linear, which is the case for an abelian theory, the symmetry of \eqref{dS=0} can be written in a functional way as
	\be\label{WI2}
	\left[\int d^dx\sum_{\phi_A}\mathcal W_{\alpha A}\frac{\delta}{\delta\phi_A}\right]S_{inv}=0\ ,
	\ee
known as Ward identity. The symmetry implies a conserved current equation for the source $J^A(x)$ in \eqref{SJ}. This can be immediately visualized if we consider for instance the gauge symmetry \eqref{dgauge} and the Maxwell action \eqref{Max} with a source term \eqref{SJ} where the current is $J^\mu(x)$. In that case the identity \eqref{WI2}, for every value of the gauge parameter $\Lambda(x)$ \eqref{dgauge}, gives
	\be
	\partial_\mu J^\mu=0\ .
	\ee
A gauge theory described by the action \eqref{dS=0}, in order to be well defined, needs a gauge-fixing so to render the path integral finite by identifying one representative for each gauge orbit \cite{Faddeev:1967fc}, and allowing for the computation of observables (propagators for instance). Physical results should not depend on the gauge choice. In the case of theories with boundary that choice, usually, but not necessarily,  is the axial one, where a component normal to the boundary is set to be zero through a proper gauge fixing term $S_{gf}$. This term breaks covariance \cite{Bassetto:1991ue}, as does the presence of the boundary itself. The boundary is introduced in the theory by means of a Heaviside theta distribution  $\theta(f(x))$ in the action, where $f(x)=0$ describes the boundary $\partial M$ of the manifold $\mathcal M$. In this case the theory has a one-sided boundary. This differs from the approach of Symanzik \cite{Symanzik:1981wd}, where the theory was defined on both sides of the boundary. Thus in our case we will not have to require any separability condition as \eqref{SymPropD} on the propagators. However still the question naturally arises of which \ac{BC} should be imposed on the bulk fields or, rather, which are the most general \ac{BC} which naturally emerge as a consequence of the presence of a boundary itself, without introducing them by hand. Indeed a statement of the procedure we want to build in this Thesis is that we want to get rid of the dependence on any particular choice. To do so we consider an additonal action term $S_{bd}$ analogous to Symanzik's term in his ``augmented lagrangian'' \eqref{INTRO-Lbd}, compatible with power-counting and locality, which could have a form like the following
	\be\label{SbdINTRO}
	S_{bd}=\int d^dx\delta(f(x))\phi_AT^{AB}\phi_B\ ,
	\ee
where $T^{AB}$ is a generic operator that may contain spacetime derivatives. While Sy\-man\-zik imposes \ac{BC} and then sees which is the boundary term in the action which implements them, we generalize that approach not imposing any \ac{BC} \textit{a priori}, but 
writing the most general boundary term, compatible with locality and power counting, to find out which are the most general \ac{BC} compatible with it. The question as whether to require covariance or not in the boundary term will depend on the model. Clearly one could for instance start from a non-covariant term and recover covariance as a particular case, however this could introduce an unnecessary high number of free parameters, which may render the analysis more complicated, hiding physical results in a proliferation of coefficients. On the other hand the request of covariance could prevent from reaching any result at all, as we shall see for instance in the case of Chern-Simons theory in Section \ref{sec CS}.  The presence of the boundary spoils the gauge invariance of the bulk action. This boundary term $S_{bd}$ \eqref{SbdINTRO} allows to obtain the most general \ac{BC} for the theory as a variational principle for the full action
	\be\label{StotINTRO}
	S_{tot}=S_{inv}+S_{bd}+S_{gf}+S_J\ .
	\ee
\section{The breaking of the symmetry and current algebra}\label{sec brokensymm}
We started this Chapter by saying that the symmetry is the main ingredient of a \ac{QFT}, the first step from which a theory can be built. But it is even more important in the present case, since the symmetry, or better, its breaking by the introduction of the boundary, is also the first step  to recover the induced lower-dimensional theory. As a consequence, the content of this Section depends on the specific symmetry considered, and the scope here is to give an idea of the procedure that we will follow in this Thesis. Indeed the presence of the boundary translates in a breaking of the symmetry which thus results in a breaking of the Ward identity \eqref{WI2}, which acquires a nonzero value at its r.h.s. Such Ward identity is known \cite{Mack:1988nf,Becchi:1988nh} to imply conservation laws,  in the sense that, taken on-shell ($i.e.\ J=0$), it can be written as an equation of the form 
		\be\label{INTROcc}
		\partial\cdot(...)|_{\partial\mM}=0\ ,
		\ee
associated to the breaking, where ``$(...)$'' is a linear function of the bulk fields $\phi_A(x)$. This will be one of the ingredients thanks to which one should be able to build the boundary theory, as we shall see.

\subsection*{Commutator algebra and Kac-Moody}
Through the relations \eqref{G1} and \eqref{G2}, by taking functional derivatives with respect to the source of the broken Ward identity, it is possible to obtain an algebraic structure on the boundary $\partial\mM$
	\be\label{WI->alg}
	\frac{\delta}{\delta J(x')}\mbox{(Broken Ward id.)}\quad\to\quad\mbox{boundary algebra}\ .
	\ee
Since the on-shell breaking of the identity represents a conserved current on the boundary \eqref{INTROcc}, this means that the commutation relations coming from \eqref{WI->alg} represents an algebra of currents. In fact typically this algebraic structure turns out to be of the \ac{KM} type \cite{Kac:1967jr,Moody:1966gf,Goddard:1986bp}, which is a recurrent one in the context of theories with boundary \cite{Blasi:1990bk,Emery:1991tf,Blasi:2008gt,Blasi:2011pf}. For instance one can think of the two-dimensional case, for which the current algebra takes the following form
		\begin{equation}\label{[j,j]}
		[J^a(x),J^b(x')]=i\hbar f^{abc}J^c(x)\delta(x-x')+\frac{i\hbar^2}{2\pi}k\delta^{ab}\partial^{(x)}\delta(x-x')\ ,
		\end{equation}
where $f^{abc}$ are the structure constants of an algebra $g$, and $k$ is known as the central charge. When the theory is abelian the structure constants disappear, and the r.h.s. is proportional to a derivative of the Dirac delta function with the central charge as coefficient. This is also called the Schwinger term \cite{Schwinger:1959xd} and is a second order quantum effect. To relate this current algebra to the proper \ac{KM} one, which is associated to a compact and finite-dimensional Lie algebra $g$ defined by the commutation relation
		\begin{equation}\label{eq:KM-gen-T}
		[T_m^a,T_n^b]=if^{abc}T^c_{m+n}+km\delta^{ab}\delta_{m,-n}\ ,
		\end{equation}
where $m,n\in\mathbb{Z}$, and $a,b,c\in\{1,...,\mathrm{dim}g\}$, it is necessary to introduce a \ac{KM} field 
		\begin{equation}
		T^a(z)=\sum_nT^a_{-n}z^n\ ,
		\end{equation}
related to the current in \eqref{[j,j]} through
		\begin{equation}\label{eq:J-km}
		J^a(x)=\frac{\hbar}{L}T^ae^{2\pi i\frac{x}{L}}\ ,
		\end{equation}
where the definition is taken on a unitary circle of period $L$ which can then be extended to infinity. Following these definitions the current algebra \eqref{[j,j]} can be written in terms of the \ac{KM} one \eqref{eq:KM-gen-T}, where one can see that the central charge $k$ is the element that commutes with all the other elements of the algebra $T_m^a$. Algebraic structures of this kind are frequent in physics, indeed, \ac{KM} algebras are tightly related to \acp{QFT} with boundary. Starting from the pioneering paper \cite{Moore:1989yh}, where all rational 2D conformal field theories were derived from the 3D Chern-Simons theory with a boundary, it is now almost paradigmatic that conserved currents exist on the boundary of topological field theories, forming \ac{KM}-type algebras with a central charge proportional to the inverse of the coupling constant of the bulk theory.
This will be particularly relevant in this Thesis to better interpret the boundary physics and identify physical observables.

\section{The induced lower-dimensional theory} \label{holographic123}
Section \ref{sec symm} was the building of the theory starting from the symmetry, and the introduction of the boundary. That procedure is always valid~: given a symmetry, this defines an invariant action and, with the addition of the boundary, proper gauge-fixing and boundary term are considered to deal with the model. On the other hand Section \ref{sec brokensymm} deals with the consequences of the aforementioned introduction of boundary~: the breaking of the symmetry/Ward identity, the emergence of a conserved current on the boundary, and of an algebraic structure. Thus the results are highly model/symmetry dependent. Therefore that part should only be seen as a guideline on how to proceed and what to possibly expect. Even though in all the cases presented in this Thesis everything works fine, this does not mean, a priori, that it always should. Indeed results may depend on how much ``exotic'' the symmetry is, or the general setting that is chosen. For instance the theory could be defined on a flat or curved background spacetime, and in the second case one should expect computations to be more tricky, if possible at all. One has to keep in mind that it is not necessarily always true that an induced boundary physics can be recovered from any theory just by adding a boundary. Indeed, until recently, it was believed that this only worked for a class of theories, $i.e.$ \acp{TQFT}, and not in general. What we shall see in this Thesis is that apparently it is more universal than expected.\\[5px]
If after the building of the theory as Section \ref{sec symm} prescribes, results of the kind mentioned in Section \ref{sec brokensymm} are recovered, the scope of the present Section is to show how to interpret the obtained results, and translate them into the induced theory, always keeping in mind that the spirit is to not impose anything by hand. All the results obtained up to this level should be enough to determine the boundary theory, indeed~: 
	\begin{enumerate}
	\item we can identify the \ac{DoF} of the induced model by solving the on-shell conserved current equation \eqref{INTROcc}. This will give a relation between the bulk fields $\phi(x)$ evaluated on the boundary, and new boundary fields $\varphi (X)$, obtained as the \textbf{general} solution of \eqref{INTROcc}. This relation must be preserved, which implies that the boundary fields $\varphi(X)$ must transform in a way that does not affect it. That is what defines the transformation on the fields $\delta_{bd}\varphi$ under which the lower dimensional action $S_{d-1}[\varphi]$ must be invariant, $i.e.$ 
		\be\label{dSd-1=0}
		\delta_{bd}S_{d-1}=0\ .
		\ee
	\item The commutation relations obtained through \eqref{WI->alg}, when written in terms of the new boundary fields $\varphi(X)$, could be interpreted as an equal time canonical commutation relation
		\be
		\left[\ q(X)\ ,\ p(X')\ \right]=i\delta^{(d-2)}(X-X')\ ,
		\ee
from which canonical variables can be identified which constrain the free parameters of the action $S_{d-1}$ through
		\be\label{p,qINTRO}
		p=\frac{\delta S_{d-1}}{\delta\dot q}\ .
		\ee
	\item To consolidate the relation between the bulk, represented by the action $S_{tot}[\phi]$ \eqref{StotINTRO}, and the boundary, whose action is $S_{d-1}[\varphi]$ satisfying \eqref{dSd-1=0} and \eqref{p,qINTRO}, the remaining step is to match the \ac{BC} of the bulk theory, written in terms of the boundary fields $\varphi(X)$, with the \ac{EoM} of the $d-1$-dimensional theory, $i.e.$
		\be
		\mbox{bulk \ac{BC}}\ \leftrightarrow\ \mbox{boundary \ac{EoM}}\ .
		\ee
This, which we call ``holographic contact'' makes it possible to relate (some of) the remaining free parameters of $S_{d-1}$ with those of the bulk, which may also involve the ``coupling'' constant.
	\end{enumerate}
Again it is important to remark that up to this point nothing has been imposed besides the presence of the boundary, and everything that is obtained simply comes from applying first principles of \ac{QFT}. Once we have properly collected the information that the theory is giving us by following these steps, nothing is left if not to physically interpret the theory that has just been recovered. This can be done for instance by computing propagators, analyzing the \ac{EoM}, or looking at the energy momentum tensor.
\section*{notations}

If not otherwise stated, the following notations  concerning indices and coordinates will be used throughout the Thesis
	\begin{empheq}{align}
	\alpha,\beta,...,\mu,\nu,...&=\{0,...,d-1\}\\
	a,b,c,...&=\{0,...,d-2\}\\
	\textsc{a,b,c},...&=\{1,...,d-2\}\ .
	\end{empheq}
In the case $d=3$ the last line (that corresponds to the index referring to the single spatial coordinate on the boundary) we will just use the corresponding number (or letter). Moreover bulk and boundary coordinates are respectively
\begin{empheq}{align}
x^\mu=&(x^0,...,x^{d-1})\\
X^m=&(x^0,...,x^{d-2})\ .
\end{empheq}
For instance in Chapter \ref{ch CSandBFinCS} we will have radial coordinates as follows
\begin{equation}
	\begin{split}
	\mu,\nu,\rho,...=& \{0,1,2\}=\{t,r,\theta\}\\
	i,j,k,...=& \{0,2\}=\{t,\theta\}\ ,
	\end{split}\label{1.1}
\end{equation}
\begin{empheq}{align}
x=&(x^0,x^1,x^2)=(t,r,\theta)\quad\mbox{on the 3d bulk}\label{1.2}\\
X=&(x^0,x^2)=(t,\theta)\quad\mbox{on the 2d boundary }r=R\ .\label{1.3}
\end{empheq}
Exceptions to this notation will be in Section \ref{sec MaxThFract}, \ref{sec gauge}, and \ref{sec GW}, where no boundary is present, and thus the distinction will be between space and time indices as
	\begin{empheq}{align}
	\alpha,\beta,...,\mu,\nu,...&=\{0,...,d-1\}\\
	a,b,c,...&=\{1,...,d-1\}\ .
	\end{empheq}
Mostly plus Minkowskian metric will be used $\eta_{\mu\nu}=\mbox{diag}(-1,1,...,1)$, and when dealing with curved spacetime we remark that the Levi-Civita tensor $\epsilon^{\mu\nu\rho}(x)$ is written in terms of the corresponding symbol $\tilde\epsilon^{\mu\nu\rho}$ as follows
\begin{equation}
\epsilon^{\mu\nu\rho}(x)=\frac{\tilde\epsilon^{\mu\nu\rho}}{\sqrt{-g(x)}}\ ,
\label{epsilon}\end{equation}
where $g(x)$ is the determinant of the bulk metric tensor $g_{\mu\nu}(x)$, with Lorentzian signature. Finally, concerning the constants, we will adopt the standard \ac{QFT} convention where $\hbar=c=1$ dimensionless, thus everything will be expressed in terms of mass dimensions, keeping in mind that $[\partial_x]=1$ and $[x]=-1$.

\cleardoublepage 


\ctparttext{\aclp{TQFT} represent a paradigmatic example of how  boundaries may be relevant in quantum field theory. In fact \acp{TQFT} are characterized by the defining property of having $global$ observables only, and not $local$. This means that the word ``observable'' in this context is a kind of oxymoron, being 
of geometrical, rather than physical, nature. These results are quite important, and motivated the great interest in the community of theoretical physicists which arose on \acp{TQFT} after a couple of seminal papers at the end of the eighties of the past century \cite{Witten:1988ze,Witten:1988hf}, where problems typical of mathematics and mathematical physics were for the first time successfully faced by quantum field theory methods. On the other hand, the fact that \acp{TQFT} have vanishing energy-momentum tensor, hence vanishing energy density and, above all, vanishing Hamiltonian, justified as well the colder attitude of the larger part of less formal physicists toward \acp{TQFT}.  The introduction of boundaries in \acp{TQFT} drastically changed this scenario.

}

\part{Topological Field Theories} \label{partII}

\chapter{An overview of TQFTs (1988)} 
\numberwithin{equation}{chapter}

\label{ch TFT} 

\setcounter{minitocdepth}{1}
\minitoc

\aclp{TQFT} are characterized by the defining property that their observables do not depend on the metric of the background spacetime. As a consequence, \acp{TQFT} do not display $physical$ observables, which are local, but ``only'' $geometrical$ ones, $i.e.$ global properties of the manifolds, like handles, knots and so on \cite{Witten:1988ze,Birmingham:1991ty}. An easy way to highlight this peculiar property is to compute the energy-momentum tensor of a \ac{TQFT} : the result is zero or, more precisely, the only contribution comes from the unphysical gauge-fixing sector. The situation drastically changes if a boundary is introduced in the background spacetime. The presence of a boundary breaks everything which can be broken, starting from translations and rotations, hence Lorentz invariance. In \acp{TQFT} the boundary also breaks gauge invariance, thus, if a boundary is present, integration by parts gives a nonvanishing contribution, and on the lower dimensional boundary unbroken residual symmetries survive.This will allow for local \ac{DoF} to emerge on the boundary, and promote the theory to a physical one. However, before discussing the introduction of a boundary in a \ac{TQFT}, it is interesting to briefly display the properties that defines a theory a topological one.  In  \cite{Birmingham:1991ty} it is stated that four are the characteristics that makes a theory a \ac{TQFT}. Indeed a topological quantum field theory consists of
	\begin{enumerate}	
	\item a collection of fields $\Phi(x)$ (which are Grassmann graded) defined on a Riemannian manifold $(\mM,g)$ ;
	\item a nilpotent operator $\brs$, which is odd with respect to the Grassmann grading ;
	\item physical states defined to be $\brs$-cohomology classes ;
	\item an energy-momentum tensor which is $\brs$-exact, $i.e.$
		\be\label{TTFT}
		T_{\alpha\beta} =\brs V_{\alpha\beta}(\Phi,g) 
		\ee
	for some functional $V_{\alpha\beta}$ of the fields and the metric.
	\end{enumerate}
The definition comes from BRS formulation \cite{Becchi:1974md,Becchi:1975nq}, where $\Phi(x)$ includes the gauge field, ghosts, and multipliers, and $\brs$ is the nilpotent, $i.e.$ $\brs^2 = 0$, BRS operator. This is the content of 1. and 2., while point 3. says that the physical Hilbert space is defined by the condition 
	\be
	\brs|phys\rangle= 0\ ,
	\ee
and  has thus states of the form
	\be
	|phys\rangle’ = |phys\rangle + \brs|x\rangle\sim|phys\rangle\ ,
	\ee
due to nilpotency, where ``$\sim$'' implies an equivalence between the states. This equivalence means that the physical Hilbert space is composed of states which are $\brs$-closed ($\brs|x\rangle$=0) modulo $\brs$-exact ($|x\rangle=\brs|y\rangle$) states, $i.e.$ are $\brs$-cohomology classes. The variation of any functional $\mO$ of the fields $\Phi(x)$ is a BRS transformation, therefore from BRS invariance of the vacuum
\be\label{<QO>=0}
\langle \brs\mO\rangle=0\ ,
\ee
where ``$\langle...\rangle$'' represents the vaccum expectation value. The fourth prerequisite has a very important consequence on the generating functional
	\be\label{ZTFT}
	Z=\int[d\Phi]e^{-S_q}\ ,
	\ee
where $S_q$ si the complete quantum action, which includes the classical action $S_c$ together with the necessary gauge fixing and ghost terms and is BRS-invariant by construction $i.e.\ \brs S_q=0$. Considering the theory on a manifold $\mM$ with metric $g_{\alpha\beta}(x)$, the energy momentum tensor given in pt.4. by \eqref{TTFT} is such that the generating functional $Z$ \eqref{ZTFT} is zero under an infinitesimal change in the metric\footnote{Metric independence of the functional measure is assumed. To show that this assumption is in fact realized, one needs to check for metric anomalies \cite{Birmingham:1991ty}.}, in fact
	\be
		\begin{split}
		\delta_gZ=&\int[d\Phi]e^{-S_q}\left(-\tfrac 1 2\int_\mathcal{M}d^nx\sqrt{-g}\,T_{\alpha\beta}\,\delta g^{\alpha\beta}\right)\\
			=&\int[d\Phi]e^{-S_q} \brs\chi=\langle\brs\chi\rangle=0\ ,
		\end{split}
	\ee
where we used the BRS invariance of the vacuum \eqref{<QO>=0}, the definition of energy-mo\-men\-tum tensor as the variation of the action with respect to the metric $i.e.$
\be
\delta_g S_q=\frac 1 2\int_\mathcal{M}d^nx\sqrt{-g}\,T_{\alpha\beta}\,\delta g^{\alpha\beta} \ ,
\ee
and where we defined
	\be
	\chi\equiv-\tfrac 1 2\int_\mathcal{M}d^nx\sqrt{-g}\, V_{\alpha\beta}\,\delta g^{\alpha\beta}\ .
	\ee
The term ``topological'' can thus be interpreted as this \textbf{metric independence}. The generating functional $Z$ \eqref{ZTFT} does not depend on the local structure of the manifold, but only on global properties~: $Z$ is a topological invariant. However, it is not the only invariant. The vacuum expectation value of an observable
	\be
	\langle\mO\rangle=\int[d\Phi]e^{-S_q}\mO(\Phi)\ \label{(2.6)}
	\ee
is another topological invariant, $i.e.$ $\delta_g\langle\mO\rangle=0$ if \cite{Witten:1988ze}
	\be
	\delta_g\mO=\brs R\quad;\quad \brs\mO=0\ ,
	\ee
which means that $\mO$ has to be in the $\brs$-cohomology class  ($i.e.$ it is a BRS invariant operators which is not $\brs$-exact), and which satisfy $\delta_g\mO = \brs R$ for some $R$. Then we have that
	\be
	\delta_g\langle\mO\rangle=\int[d\Phi]e^{-S_q}(\delta_g\mO-\delta_gS_q\cdot\mO) = \langle\brs(R + \chi\mO)\rangle = 0\ ,
	\ee
again due to \eqref{<QO>=0}. Another property of topological field theories is the absence of dynamical excitations, $i.e.$ there are no propagating degrees of freedom. Indeed, from 1.-4., we have
	\be
	\langle phys'|H|phys\rangle=\langle phys'|\int T_{00}|phys\rangle=\langle phys'|\int \brs V_{00}|phys\rangle=0\ ,\label{(2.15)}
	\ee
where $H$ is the Hamiltonian, which means that the energy of any physical state is zero, and hence there are no physical excitations.
\subsection*{The classification}
The above definition of a \ac{TQFT} allows for a classification of such theories : one can have Witten type or Schwarz type \acp{TQFT} \cite{Birmingham:1988ap}. The discriminant comes from the shape of the quantum action, for which the two models acquire different features. For instance this difference also reflects on the names  \cite{Witten:1989ig,Witten:1990bs} : Witten type theories are also called $cohomological$ due to the structure of its observables, while the Schwarz type models are said to be $quantum$ as a consequence of its non-triviality (classical action).
\bi
\item Concerning \textbf{Witten type} theories \cite{Witten:1988ze,Witten:1988xj}, the defining feature is that the complete quantum action $S_q$ can be written as a BRS variation, $i.e.$ 
	\be
	S_q = \brs V\ , \label{(2.10)}
	\ee
$i.e.$ it is exact for some functional $V(\Phi, g)$ of the fields. This immediately implies
	\be
	T_{\alpha\beta} = \brs\left(\tfrac{2}{\sqrt{-g}}\tfrac{\delta V}{\delta g^{\alpha\beta}}\right) \ ,
	\ee 
which through 4. ensures of the topological nature of the model. 
 It is also possible to show \cite{Birmingham:1991ty} that the main property of Witten type theories \eqref{(2.10)}, implies exactness of the functional $Z$ at the semiclassical level ($i.e.$ small $\hbar$). Examples of Witten theories are topological Yang-Mills and topological $\sigma$-model.
\item Concerning  \textbf{Schwarz type} theories \cite{Schwarz:1978cn,Witten:1988hf}, the quantum action is of the form
	\be
	S_q(\Phi,g) = S_c(\Phi) + \brs V(\Phi,g)\ , \label{(2.13)}
	\ee
where $S_c(\Phi)$ is the metric-independent classical action, while the rest comes from gauge-fixing procedure. Metric independence of the classical action ensures that its energy momentum tensor vanishes, and the complete one thus depends only on gauge-fixing and gost sector as
	\be
	T_{\alpha\beta} = \brs\left(\tfrac{2}{\sqrt{-g}}\tfrac{\delta V}{\delta g^{\alpha\beta}}\right) \ , \label{(2.14)}
	\ee
which, again, satisfies feature 4. of a \ac{TQFT}.
\ei
To the second class, $i.e.$ Schwarz theories, belong both Chern-Simons and BF theories\footnote{the higher dimensional $n > 3$ class of non-Abelian BF theories possess some “non-standard” properties \cite{Birmingham:1991ty}.}, which will be the subject of study in the next Chapters, for the case of the abelian 3D models, described by the following classical actions
	\begin{align}
	S_{CS}&=\int d^3x\epsilon^{\mu\nu\rho}A_\mu\partial_\nu A_\rho\label{CSTFT}\\
	S_{BF}&=\int d^3x\epsilon^{\mu\nu\rho}B_\mu\partial_\nu A_\rho\ ,\label{BFTFT}
	\end{align}
where metric independence is ensured by the presence of the Levi-Civita symbol, which in curved spacetime is a density and comes with a metric contribution at the denominator of its tensorially-extended quantity \eqref{epsilon} $\epsilon^{\mu\nu\rho}(x)=\epsilon^{\mu\nu\rho}/\sqrt{-g(x)}$, which compensate the integration measure $d^3x\,\sqrt{-g}$. As a curiosity, notice that the name of the BF theory is not an acronym, but directly comes from the aspect of the 3D theory. From the definition of the Maxwell field strength 
	\be
	F_{\mu\nu}\equiv \partial_\mu A_\nu-\partial_\nu A_\mu\ ,
	\ee
the abelian classical action \eqref{BFTFT} is indeed written as ``BF''. From \eqref{CSTFT} and \eqref{BFTFT} one can see that both Schwarz-type Lagrangians  transform under gauge the transformations 
	\be
	\delta A_\mu=\partial_\mu\lambda\quad;\quad\delta' B_\mu=\partial_\mu\lambda'\ ,
	\ee
 as total derivatives
	\begin{align}
	\delta S_{CS}&=-\int d^3x\partial_\mu\left(\epsilon^{\mu\nu\rho}\lambda\partial_\nu A_\rho\right)\\
	\delta' S_{BF}&=-\int d^3x\partial_\mu\left(\epsilon^{\mu\nu\rho}\lambda'\partial_\nu A_\rho\right)\quad;\quad\delta S_{BF}=0\ ,
	\end{align}
which is a sign of the existence of a lower dimensional theory in the presence of a boundary, as we shall see in the following Chapter.

\chapter{Chern-Simons and BF with boundary} 
\label{ch CSandBF} 
\numberwithin{equation}{section}
The scope of this Chapter is to briefly review the physics that emerges when one introduces a boundary in a \ac{TQFT} in flat spacetime, and in particular to the 3D abelian Schwarz theories discussed above : Chern-Simons and BF. The first Section \ref{sec CS} concerns the abelian Chern-Simons theory with boundary and its relation to the edge states of \ac{FQHE}. This will be the first practical example of the results that one can obtain by applying the \ac{QFT} procedure presented in Chapter \ref{QFTapproach} to a specific model, as was done in \cite{Maggiore:2017vjf}. The theoretical observation of physical edge states on the boundary of Chern-Simons theory was done for the first time in 1992 with a different approach \cite{Wen:1992vi}, and we shall see that both analysis are in agreement. The Section will thus also serve as a testing ground for the general procedure displayed in Chapter \ref{QFTapproach} by means of a practical example. After looking at the Chern-Simons case with boundary,  in Section \ref{sec BFflat} we will apply the same steps to the abelian 3D BF model with boundary, following \cite{Amoretti:2014iza}, and analyzing its relation with \ac{TI}, discovered in 2010.


\section{Chern-Simons and the edge states of FQHE (1992)}\label{sec CS}
In 1990 the low energy effective field theory of the edge states of \ac{FQHE} was derived, and shown to describe properties of chiral Luttinger liquids \cite{Wen:1990se}. Under this regard, it is known \cite{Dunne:1998qy,Tong:2016kpv} that the abelian Chern-Simons theory, when coupled to matter sources as
	\be
	S=S_{CS}+S_J=\kappa\int d^3x\epsilon^{\mu\nu\rho}A_\mu\partial_\nu A_\rho+\int d^3x\,J^\mu A_\mu\ ,
	\ee
where $\kappa$ is the ``coupling'' constant. This is relevant in the context of planar electrodynamics, indeed the spatial component of its \ac{EoM}
	\be
	\frac{\delta S}{\delta A_a}=-\kappa\epsilon^{0ab}F_{0b}+J^a=0
	\ee
 describes Hall conductivity
	\be\label{J=sigmaE}
	J^a=\sigma^{ab}E_b=\kappa\epsilon^{0ab}E_b
	\ee
$i.e.$ the presence of a current which is perpendicular to the usual electric field $E_b=F_{0b}$. Inspired by this relation, two years after the discovery of the edge states of \ac{FQHE}, in 1992, it was shown \cite{Wen:1992vi} that these Hall edge states, represented by chiral Luttinger liquids, could be obtained as  boundary modes of the abelian Chern-Simons model. Following \cite{Maggiore:2017vjf} and the steps of Chapter \ref{QFTapproach}, we will see how such a result can be recovered in the context of \acp{QFT} with boundary. The abelian Chern-Simons theory is described by the action \eqref{CSTFT}, where a flat planar boundary on $x^2=0$ can be introduced by means of  Heaviside step distribution. Thus the action is
	\be
	S_{CS}=\frac{\kappa}{2}\int d^3x\theta(x^2)\epsilon^{\mu\nu\rho}A_\mu\partial_\nu A_\rho\label{CS4}\ .
	\ee
Notice that the coupling constant $\kappa$ could be reabsorbed by a redefinition of the gauge field $A_{\mu}(x)$, however when dealing with boundaries, it can be useful to maintain it explicit in order to keep track of the contribution of the boundary. Indeed we will see that this will be identified with a physical quantity, namely the filling fraction $\nu$ of the quantum Hall effect \cite{Dunne:1998qy,Tong:2016kpv}.  The full action is given by \eqref{StotINTRO}, $i.e.$
	\be
	S_{tot}=S_{CS}+S_{gf}+S_J+S_{bd}\ ,
	\ee
where in this case the gauge-fixing, source and boundary terms are of the following form
	\begin{align}
	S_{gf}&=\int d^3x\theta(x^2)bA_2\label{SgfCS}\\
	S_J&=\int d^3x\theta(x^2)J^iA_i\label{SJCS}\\
	S_{bd}&=\int d^3 \delta(x^2)\left(\tfrac{a_1}{2}A_0^2+a_2A_0A_1+\tfrac{a_3}{2}A_1^2\right)\ .\label{SbdCS}
	\end{align}
The axial gauge fixing $A_2=0$ has been chosen and implemented through the \ac{EoM} of the multiplier field $b(x)$, and where the matrix $T^{AB}$ in \eqref{SbdINTRO} has the form 
	\be
	T^{ij}\equiv\frac{1}{2}\left(\begin{array}{cc}
	a_1&a_2\\
	a_2&a_3
	\end{array}\right)\ ,
	\ee
depending on the three constants $a_1,\ a_2,\ a_3$. No derivatives in the boundary term appears due to power-counting, since the gauge field has mass dimensions $[A_\mu]=1$. From the \ac{EoM} of the full action, through the variational principle $\lim_{\epsilon\to0}\int^\epsilon_0 dx^2\, \mbox{\ac{EoM}}$, one recovers the \ac{BC} of the theory, which, neglecting those that require for the gauge field to vanish, are of the form
	\be\label{BCCS}
	A_0-vA_1|_{x^2=0}=0\ ,
	\ee
where $v$ is a constant coefficient that depends on the bulk coupling $\kappa$ and on the boundary coefficient $a_i$, $i.e.$ $v=v(\kappa,a_i)$. This \ac{BC} will serve as a trigger to identify the physical theory once we have built the 2D induced action. It is interesting to notice that this same fundamental equation \eqref{BCCS}, that here comes naturally as a general \ac{BC} for the theory through Symanzik's principle, in \cite{Wen:1992vi} was imposed as a gauge-fixing condition. This would however raise the ambiguity of whether the results would just be peculiar of that specific gauge choice. But physical results should not depend on the gauge choice~: what would happen if, for instance, a covariant gauge-fixing, which is of a completely different nature, was to be considered? In \cite{Wen:1992vi} the gauge-fixing directly intervene on the physical results through its equation \eqref{BCCS}, while in terms of the \ac{QFT} procedure outlined in Chapter \ref{QFTapproach} that equation naturally comes out as a \ac{BC} for the theory, while the gauge choice only simplifies the computations, without affecting the final results. For instance, the \ac{KM} algebra on the boundary of Chern-Simons theory has been shown to exist, for flat spacetime, both in covariant \cite{Blasi:1990jq} and axial \cite{Emery:1991tf} gauge. The presence of the boundary at $x^2=0$ breaks the gauge invariance, which also implies a breaking of the Ward identity, as discussed in Section \ref{sec brokensymm}. By means of the \ac{EoM} of the theory one can thus recover the following integrated broken Ward identity
	\be\label{BrokenWICS}
	\int_0^\infty dx^2\partial_iJ^i=-\kappa\epsilon^{0ij}\partial_iA_j|_{x^2=0}\ ,
	\ee
where $\epsilon^{0ij}$ is the 2D Levi-Civita symbol. By means of functional derivatives with respect to the current, this broken identity gives an equal-time algebraic structure of the \ac{KM} type \cite{Kac:1967jr,Moody:1966gf}
	\be\label{KMCS}
	\left[A_1(X)\ ,\ A_1(X')\right]=-\frac{i}{\kappa}\partial_1\delta(x^1-x'^1)\ ,
	\ee
where we notice that the central charge is proportional to the inverse of the coupling. Finally, the identity \eqref{BrokenWICS}, taken on-shell, gives a conserved current equation, which can be solved as
	\be\label{dofCS}
	\epsilon^{0ij}\partial_iA_j|_{x^2=0}=0\quad\Rightarrow\quad A_i|_{x^2=0}=\partial_i\phi\ .
	\ee
With this equation we now have all the information needed to build the 2D model at the boundary of Chern-Simons, and we can do this by following the steps of Section \ref{holographic123} :
	\begin{enumerate}
	\item The boundary \ac{DoF} are identified as the scalar field $\phi(X)$ in \eqref{dofCS}, and the symmetry preserving its definition is the shift symmetry
		\be
		\delta_s\phi=\eta\ ,
		\ee
with $\eta$ a constant.
	\item Through the \ac{KM} algebra, canonical variables are identified to be
		\be
		q=\phi\quad;\quad p=-\kappa\partial_1\phi\ .
		\ee
	\item From 1. and 2., together with power-counting and locality, we get the following shift-invariant action
		\be
		S_{2D}=\int d^2X\left[-\kappa\partial_0\phi\partial_1\phi+c\left(\partial_1\phi\right)^2\right]\ ,
		\ee
	for which the matching between the \ac{BC} \eqref{BCCS} and its EoM as $\partial_1\ac{BC}=\ac{EoM}$ gives a constraint on the free parameter $c$ to be
		\be
		c=\kappa v\ ,
		\ee
where $v$ depends on the parameters $a_i$ of the boundary action $S_{bd}$ \eqref{SbdCS}.
	\end{enumerate}
The 2D induced theory is thus described by the action
	\be\label{CSS2D}
	S_{2D}=\kappa\int d^2X\left[-\partial_0\phi\partial_1\phi+v\left(\partial_1\phi\right)^2\right]\ ,
	\ee
where the \ac{EoM} is
	\be\label{CSEOM}
	\frac{\delta S_{2D}}{\delta\phi}=\partial_0\partial_1\phi-v\partial_1\partial_1\phi=0\ .
	\ee

\subsection*{The phenomenology}
What's the physics described by this boundary theory? One can notice that the \ac{BC} \eqref{BCCS}, written in terms of the boundary scalar field $\phi(x)$ through the solution \eqref{dofCS} is 
	\be
	\partial_0\phi+v\partial_1\phi=0\ ,
	\ee 
which describe a scalar field (boson), that is chiral, $i.e.$ satisfy 
	\be\label{CSchiral}
	\phi(x^0,x^1)=\phi(x^1-vx^0)\ .
	\ee
The \ac{BC} \eqref{dofCS}, which is not imposed, as we said, plays a very important role in the identification of the nature of the edge \ac{DoF}. Experimentally indeed, the edge states of the \ac{FQHE} and of \ac{TI} are fermionic, while the corresponding bulk theories are completely bosonic, being described in terms of gauge fields. But the \ac{BC} have been recognized to be the conditions for the fermionization of bosonic \ac{DoF} \cite{Aratyn:1983qjz, Aratyn:1983bg, Amoretti:2013xya}. The scalar bosonic degree of freedom \eqref{CSchiral} found on the 2D boundary represents a chiral Weyl fermion. Indeed in Weyl fermion theory all local operators are of this form \eqref{CSchiral}, and according to the fermion-boson correspondence \cite{Aratyn:1983qjz,Kane} one can associate a density operator $\rho(X)$ to the Weyl fermions $\psi(X)$ as
	\be
	\rho=:\psi^\dag\psi:\ ,
	\ee
which can be written as a chiral boson as
	\be\label{rho=phiCS}
	\rho=\frac{1}{2\pi}\partial_1\phi\ .
	\ee
In these terms the boundary \ac{EoM} \eqref{CSEOM} appears as a wave equation
	\be\label{(2.3)}
	\partial_0\rho-v\partial_1\rho=0\ ,
	\ee
where the density $\rho(x)$ propagates along the boundary with constant velocity $v$. This corresponds to Tomonaga-Luttinger theory, whose action coincides to the 2D one we found \eqref{CSS2D} induced by the bulk Chern-Simons theory. The density $\rho(x)$ is known to satisfy \cite{Wen:1992vi} a commutation relation of the \ac{KM} type
	\be
	\left[\rho(x)\ ,\ \rho(x')\right]=\frac{\nu}{4\pi}\partial_1\delta(x-x')\ ,
	\ee
where $\nu$ is the filling factor of the \ac{FQHE}. This algebra is exactly the one recovered in \eqref{KMCS} from the broken Ward identity \eqref{BrokenWICS}, since  from \eqref{dofCS} and \eqref{rho=phiCS} we have that $\rho\propto\partial_1\phi\propto A_1|_{x^2=0}$. This finally allows to identify the coupling $\kappa$ of Chern-Simons theory with a physical quantity, $i.e.$ the filling factor $\nu$, as
	\be
	\kappa=\frac{1}{4\pi\nu}\ .
	\ee

\section{ The 3D BF model and Topological Insulators (2010)}\label{sec BFflat}

Many years after the first studies on the quantum Hall effect and its edge states, a new kind of topological phase of matter  was observed~: \acl{TI} \cite{Fu:2006djh,Hasan:2010xy,moorenature,Qi:2010qag}. These new materials belong to the category of Hall systems, and are distinguished by the fact that their bulk is invariant under time-reversal ($\TR$) transformation, and so are its edge modes, which are characterized by spin currents. This gave them the name of \ac{QSH} systems. Indeed, while for standard Hall systems the presence of a time-reversal breaking magnetic field is necessary, in the case of the \ac{QSH} the edge states are protected by that symmetry, and a manifestation of a spin-orbit interaction \cite{Qi:2010qag}. Following this discovery, a natural question raised as to whether edge states of \ac{QSH} systems could emerge from a \ac{TQFT} with boundary, as it happened for the \ac{FQHE} and Chern-Simons. In 2010 the answer arrived, and it was shown \cite{Cho:2010rk} that the edge states of three- and four-dimensional topological insulators could be related to the three- and four-dimensional BF models with boundary respectively. We remind that this is the other \ac{TQFT} of Schwarz type, which in the abelian 3D case is represented by the classical action \eqref{BFTFT}. Indeed one can see that the bulk BF theory, compared to the Chern-Simons one, has the additional property of being invariant under time-reversal. In fact defining the Time-Reversal transformation $\TR$ to act on coordinates as 
	\be\label{TxBF}
	\TR x^0=-x^0\quad;\quad \TR x^1=x^1\quad;\quad \TR x^2=x^2\ ,
	\ee
the 3D BF action \eqref{BFTFT} is invariant for fields that transform as follows
	\be\label{TABBF}
	\TR A_0=A_0\quad;\quad \TR A_{1,2}=-A_{1,2}\quad;\quad \TR B_0=-B_0\quad;\quad \TR B_{1,2}=B_{1,2}\ .
	\ee
According to this the gauge field $A_\mu(x)$ transforms as an electromagnetic potential, while $B_\mu(x)$ as a spin current, remarking the physical relation with \ac{QSH}. This invariance can also be transposed to the boundary states, thus mirroring the characteristic of topological insulators. In this Section we will review the 3D case that was studied as a \ac{QFT} with boundary in \cite{Amoretti:2014iza}. This will serve as a starting point for the original work presented in the next Chapter.\\[5px]
The full action of the model with boundary is
	\be
	S_{tot}=S_{BF}+S_{gf}+S_J+S_{bd}\ ,
	\ee
composed of the following bulk, gauge-fixing, source and boundary terms 
	\begin{align}
	S_{BF}&=\kappa\int d^3x\theta(x^2)\epsilon^{\mu\nu\rho}B_\mu\partial_\nu A_\rho\\
	S_{gf}&=\int d^3x\theta(x^2)\left(bA_2+d\,B_2\right)\label{SgfBF}\\
	S_J&=\int d^3x\theta(x^2)\left(J^iA_i+K^iB_i\right)\label{SJBF}\\
	S_{bd}&=\int d^3 \delta(x^2)\left(\tfrac{a_1}{2}A_iA^i+\tfrac{a_2}{2}B_iB^i+a_3A_iB^i\right)\ ,\label{SbdBF}
	\end{align}
where again a flat boundary  at $x^2=0$ has been introduced through a theta term $\theta(x^2)$, an axial gauge fixing has been chosen for both $A_\mu(x)$ and $B_\mu(x)$ gauge fields, and a covariant boundary term has been defined in \eqref{SbdBF}. Notice that here a Lorentz-invariant boundary term has been chosen, and a particular combination of the coefficients ($a_3=0$) allows for it to be invariant under the time-reversal $\TR$ transformation defined in \eqref{TxBF} and \eqref{TABBF}, thus mimicking the property of the bulk. This reflects on the boundary and the \ac{BC}, which, excluding Dirichlet solutions, give
	\be\label{BFBC}
	a_1A^i-\epsilon^{0ij}B_j=0\ ,
	\ee
for $a_1a_2=\kappa-1$, and $a_3=0$. The presence of the boundary implies a breaking of the Ward identity, one for each field, giving the following integrated quantities
	\begin{align}
	\int_0^\infty dx^2\partial_iJ^i&=-\kappa\epsilon^{0ij}\partial_iB_j|_{x^2=0}\label{BrokenWIBF1}\\
	\int_0^\infty dx^2\partial_iK^i&=-\kappa\epsilon^{0ij}\partial_iA_j|_{x^2=0}\ .\label{BrokenWIBF2}
	\end{align}
Going on-shell ($i.e.$ at vanishing external sources $K=J=0$) in \eqref{BrokenWIBF1} and \eqref{BrokenWIBF2} one can also recover the boundary \ac{DoF} to be two scalar fields
	\begin{align}
	\epsilon^{0ij}\partial_iB_j|_{x^2=0}&=0\quad\Rightarrow\quad B_j|_{x^2=0}=\partial_j\zeta\\
	\epsilon^{0ij}\partial_iA_j|_{x^2=0}&=0\quad\Rightarrow\quad  A_j|_{x^2=0}=\partial_j\Lambda\ .
	\end{align}
A semidirect sum of \ac{KM} algebras can be finally derived by applying functional derivatives with respect to the currents in the broken identities \eqref{BrokenWIBF1} and \eqref{BrokenWIBF2}. For instance differentiating \eqref{BrokenWIBF1} with respect to $J^j(x')$ one gets, at equal time,
	\be
	\left[B_1(X)\ ,\ A_1(X')\right]=\frac{i}{\kappa}\delta(x^1-x'^1)\ ,
	\ee
which can be interpreted as a canonical commutation relation for the boundary scalar fields, thus identifying the canonical variables to be
	\be\label{BFCC}
	q=\kappa\zeta\quad;\quad p=\partial_1\Lambda\ .
	\ee
The induced 2D shift-invariant action compatible with the canonical variables  identified in \eqref{BFCC} and whose \ac{EoM} matches the Time-Reversal-invariant \ac{BC} \eqref{BFBC} of the bulk is the following
	\be
	S_{2D}=\kappa\int d^2X\left[\partial_0\zeta\partial_1\Lambda+\frac{1}{2a_1}(\partial_1\zeta)^2+\frac{a_1}{2}(\partial_1\Lambda)^2\right]\ .
	\ee
Notice that it seems that starting with a covariant boundary action $S_{bd}$ \eqref{SbdBF} leads to a 2D action that shows a strong-weak coupling duality $i.e.\ \zeta\leftrightarrow\Lambda\ , \ a_1\leftrightarrow\frac{1}{a_1}$, and the \ac{EoM}  can be  understood as the continuity equations for fluids, 
	\be\label{()}
	\partial_0\rho_i+v_i\partial_1\rho_i=0\ ,
	\ee
with $i=1,2$, $v_1\equiv\frac{1}{a_1}\ ,\ v_2\equiv a_1$, and densities $\rho_1\equiv\partial_1 \zeta,\ \rho_2\equiv \partial_1\Lambda$. This does not represent a time-reversal invariant current, which is described by chiral bosons moving at equal and opposite velocities. This property is however recovered if $a_1=1$, for which
	\be\label{(51)}
	\partial_0\rho_\pm\pm\partial_1\rho_\pm=0\ ,
	\ee
where $\rho_\pm\equiv \partial_1\left(\zeta\pm\Lambda\right)$, which describes bosons moving at the speed of light in opposite directions. We will see in the next Chapter, and in particular in Section \ref{sec BFinCS}, that by considering a non-covariant boundary action $S_{bd}$ \eqref{SbdBF}, we will be able to recover finite values for the velocities.


\chapter{metric dependence of the edge states of TQFTs} 

\label{ch CSandBFinCS} 
\setcounter{minitocdepth}{1}

We have just seen that, all measurable quantities related to edge states of Hall systems, like for instance filling factors $\nu$ and chiral velocities of the edge modes $v$, have been obtained for $flat$ bulk manifolds, with planar, or radial (in the case of Chern-Simons on a disk \cite{Dunne:1998qy} for instance) boundary. However experiments are pushing theoretical investigations, since recently  $accelerated$ chiral bosons have been observed on the edge of some particular Hall systems \cite{Bocquillon}, which cannot be explained by the usual flat Chern-Simons with boundary paradigm  (\ac{TQFT} on flat spacetime with planar boundary), which unavoidably yields $constant$ edge velocities, as we have shown in Chapter \ref{ch CSandBF}. The standard approach to deal with these cases is phenomenological, and basically consists in adding a suitable potential to the 2D Luttinger action, in order to reproduce time-dependent velocities \cite{Wen:1989mw,Wen:1990qp,Kane95,Hashi18,Wen:1991ty}. The price of this way of solving the problem is that the whole ``holographic'' construction of finding the right lower dimensional action without any $ad\ hoc$ extension fails.  Similar experimental evidence is currently sought in the other relevant topological state of matter, $i.e.$ the topological insulators. In all the cases we mentioned above, the lower dimensional edge dynamics depends on the bulk only through its parameters, and the details of the bulk manifold are somehow hidden by the particular flat choice. But is there any hidden dependence of the boundary physics on the bulk manifold? It appears interesting and reasonable to ask if, how and where the details of the metric reveal themselves in the edge observables, to see which are the quantities depending on the bulk metric and, maybe even more interestingly, which are the physical quantities really universal, or topologically protected. Thus the aim of this Chapter  is to investigate an alternative approach to reproduce accelerated edge modes in Hall systems without the need of adding any empirical potential, while keeping the holographic construction intact. It consists in considering the bulk theory on a generic, rather that flat, background manifold. While the topological invariant action of course does not depend on the particular spacetime metric, Symanzik's boundary term and the boundary itself certainly do, and it is an interesting issue to find out if, how and where this metric dependence reflects on the holographic 2D theory and, more interestingly, on physical observables. The holographic 2D theory induced on the boundary of the abelian Chern-Simons theory is the Floreanini-Jackiw action \cite{Floreanini:1987as}, describing edge modes moving with constant chiral velocities, which are indeed observed \cite{kane87}. The velocity of the edge excitations is the main observable of both Hall systems and \ac{TI}. We will see that a remnant of the bulk metric remains on these most relevant physical observables, $i.e.$ the chiral velocity of the edge modes, which will become $local$. In other words, the edge modes of the Hall systems, when described by a \ac{TQFT} built on a generic manifold, are accelerated and, moreover, the velocities also depend on the position of the quasiparticle on the boundary, not only on time. The dependence on the bulk metric manifests only through the determinant of the induced metric on the boundary, hence it is mild, as one might expect due to the topological character of the bulk theory. Still, the local effects are reproduced, without any empirical modification of the 2D holographic Luttinger theory. 
\\[5px]
The Chapter is composed of two main Sections, each concerning the effect of a curved spacetime in one of the two abelian Schwarz \ac{TQFT} with boundary~:  Chern-Simons theory in Section \ref{sec CSinCS}, and the 3D BF model in Section \ref{sec BFinCS}.

\section{Edge states of  Chern-Simons theory}\label{sec CSinCS}
The present Section, which concenrns the Chern-Simons model in curved spacetime with boundary, is organized as follows
\etocsetnexttocdepth{2}
\begingroup
\parindent=0em
\etocsettocstyle{\rule{\linewidth}{\tocrulewidth}\vskip1.25\baselineskip}{\vskip-0.75\baselineskip\rule{\linewidth}{\tocrulewidth}\vskip1\baselineskip}
\makeatletter
  \edef\scr@tso@subsection@indent
    {\the\dimexpr\scr@tso@subsection@indent-\scr@tso@section@indent}
  \def\scr@tso@section@indent{0pt}
\makeatother
\localtableofcontents 
\endgroup
\noindent
In particular in Section \ref{sec2} the Chern-Simons theory on a cylindrical spacetime $\mathbb{R} \times D$, in a not necessarily flat, Lorentzian background is introduced, and the \ac{EoM}, together with the most general \ac{BC} are derived. We find the Ward identity, broken by the boundary, and we study the existence of a \ac{KM} algebra, with constant and positive central charge. In Section \ref{sec3} we derive the 2D theory, holographically induced on the boundary of the 3D Chern-Simons theory. The holographic contact is imposed, which relates the parameters of the edge theory to the bulk ones. The dependence on the bulk through the induced metric on the boundary is discussed. Our concluding remarks are summarized in Section \ref{sec4}.

\subsection{Chern-Simons with boundary at $r=R$}\label{sec2}
It is convenient, in this first approach, to work in \ac{GNC}, where the line element takes the form
\begin{equation}
ds^2=g_{\mu\nu}(x)dx^\mu dx^\nu =\  dr^2+\gamma_{ij}(x)dX^idX^j\ .
\label{1.9}
\end{equation}
The adoption of the \ac{GNC} is particularly useful for calculations in which one is given a hypersurface $\partial\mM$, $i.e.$ a $(d-1)$-dimensional embedded submanifold of the d-dimensional manifold $\mM$ \cite{Wald:1984rg}. In the case treated here, being $\mM$ diffeomorphic to a cylinder $\mM \simeq \mathbb{R} \times D$, this choice is the most natural one, since it immediately provides the induced metric on the boundary. The \ac{GNC} are characterized by the presence of one coordinate (the one normal to the hypersurface, in our case $r$) such that $g_{rr} =1$ and the off-diagonal terms vanish~: $g_{rt}=g_{r\theta}=0$ \cite{Weinberg:1972kfs}. These are three extra conditions which might be seen as  ``gauge conditions''  on the metric \cite{Carroll:2004st,dInverno:1992gxs}, corresponding to the threefold coordinate freedom $x^\mu\rightarrow x^{\prime\;\mu}= x^{\prime\;\mu}(x)$. Simple examples of \ac{GNC} are the Cartesian coordinates on Minkowski space, or polar coordinates in Euclidean 2 and 3-space, or the Robertson-Walker coordinates used in cosmology \cite{Carroll:2004st}.  
Using these coordinates, the determinant of the induced metric on the boundary
\begin{equation}
\gamma_{ij}(X)=\frac{\partial x^\mu}{\partial X^i}\frac{\partial x^\nu}{\partial X^j}g_{\mu\nu}(x)\ ,
\label{1.10}
\end{equation}
is \cite{Carroll:2004st}
\begin{equation}
\sqrt{-\gamma}=\sqrt{-g}\ ,
\label{1.11}
\end{equation}
which, in particular, holds on the boundary $r=R$. 
Lastly,  in curved spacetime the Heaviside step distribution is a scalar, and its derivative, as in flat spacetime, is (see Appendix \ref{appA})
\begin{equation}
\nabla_\mu\theta(R-r)=
-\delta_\mu^r\delta(R-r)\ .
\label{1.12}
\end{equation}

\subsubsection*{The action}\label{sec2.1}
We consider the abelian 3D Chern-Simons theory on a cylindrical spacetime $\mathbb{R} \times D$, where the model is confined to the closed subspace $0\leq r\leq R$.  This is achieved by introducing a Heaviside step distribution $\theta(R-r)$ in the action. The Chern-Simons bulk term is then
\begin{equation}
S_{bulk}
=
\frac{\kappa}{2}\int d^3x\;  \theta(R-r)\; \tilde\epsilon^{\mu\nu\rho}\;A_\mu\partial_\nu A_\rho\ .
\label{2.1}\end{equation}
We remark that although the coupling constant $\kappa$ could be reabsorbed by a redefinition of the fields, it is useful to leave it explicit, in order to keep track of the contributions of the bulk theory to the physics of the boundary. Moreover, since the Chern-Simons theory, being topological,  has vanishing energy momentum tensor, there is no constraint on the sign of $\kappa$ from requiring a positive energy density. Still, $\kappa$ should be a positive constant, as we shall see later. 
The gauge fixing term is chosen to be
\begin{equation}
S_{gf }
=\int d^3x\; \sqrt{-g}\;  \theta(R-r)\; b\;n^\mu A_\mu\ ,
\label{2.2}\end{equation}
where $n^\mu=(0,1,0)$ is a unit vector. The field $b(x)$ is the Nakanishi-Lautrup  Lagrange multiplier \cite{Nakanishi:1966zz,Lautrup:1967zz} which implements
 the radial gauge choice
\begin{equation}
\frac{\delta S}{\delta b}
=n^\mu A_\mu
=A_r
=0\ .
\label{x2.3}\end{equation}
The choice of the radial gauge-fixing $A_r=0$ is analogous to the choice of the \ac{GNC} \eqref{1.9}. Both are the most convenient choices in presence of the boundary $r=R$~: the latter on the metric $g_{\mu\nu}(x)$ and corresponds to the reparametrization invariance, the former on the gauge field $A_\mu(x)$ and comes from gauge invariance. Both transformations (ordinary gauge symmetry and diffeomorphism invariance) are broken by the boundary, and physical results should not depend on them \cite{Blasi:1990jq,Emery:1991tf}.
External sources are coupled  to the $A_\mu(x)$ field, through the term 
\begin{equation}
S_{ext}=\int d^3x\; \sqrt{-g}\;   \theta(R-r)\; J^\mu A_\mu\ .
\label{2.3}\end{equation}
In addition, the presence of a boundary induces an extra term in the action, as the most general boundary term compatible with power counting:
\begin{equation}
S_{bd}=\int d^3x\, \sqrt{-g}\,  \delta(r-R)\,\frac{1}{2}\; T^{ij}A_iA_j\ ,
\label{2.4}
\end{equation}
where $T^{ij}=T^{ji}$ is a symmetric matrix. 
Notice that in a curved spacetime all the coefficients appearing in the action may depend on the metric, which is dimensionless, and hence on the coordinates, but only through the metric. In particular, $T^{ij}$ in $S_{bd}$ might depend implicitly on the coordinates $T^{ij}=T^{ij}(\gamma(X))$, where $\gamma(X)$ is the induced metric. An explicit dependence of $T^{ij}$ on the coordinates ``outside'' the induced metric $(T^{ij}=T^{ij}(\gamma(X);X))$ is forbidden, since in the flat limit $T^{ij}$ should reduce to a constant symmetric matrix~: $\left.T^{ij}(\gamma(X))\right|_{\gamma^{ij}=\eta^{ij}}=T^{ij}$. A similar boundary term appears also in \cite{Elitzur:1989nr} for Chern-Simons theory. In both cases, the main reason is to provide, by means of the modified equations of motion, the most general \ac{BC} which need to be fixed. The introduction of such a local functional allows to introduce the \ac{BC} in a systematic way by means of a variational principle, and not by hand, the boundary term must only satisfy the general requirements of power counting and locality. Gauge invariance and/or residual 2D covariance must not be required on it, otherwise, one would not recover the boundary dynamics which characterizes \ac{TQFT} (see for instance \cite{Elitzur:1989nr}, where the prescription of introducing a non covariant boundary term $\propto \int_{\partial Y} Tr A_1A_2$ is adopted in order to get consistent \ac{BC}).
Finally, the total action of the theory, considering the bulk, gauge-fixing, external source and boundary terms, is
\begin{equation}
S=S_{bulk} +S_{gf} + S_{ext} + S_{bd}\ .
\label{2.6}\end{equation}

\subsubsection*{Equations of motion and boundary conditions}\label{sec2.2}

From the action $S$ \eqref{2.6} we get the \ac{EoM}
\begin{equation}\label{2.7}
\frac{\delta S}{\delta A_\lambda}=\theta(R-r)\left(\kappa\epsilon^{\lambda\mu\nu}\partial_\mu A_\nu+b\;n^\lambda+J^\lambda\right)+\delta(R-r)\delta^\lambda_j\left(\frac{\kappa}{2}\epsilon^{i1j}+T^{ij}\right)A_i=0\ .
\end{equation}
Applying the operator 
$\lim_{\epsilon\to R}\int_{\epsilon}^{R}dr$ to the \ac{EoM} \eqref{2.7}, the following  \ac{BC} can be derived
\begin{equation}
\left. \left(\frac{\kappa}{2}\frac{\tilde\epsilon^{1ij}}{\sqrt{-g}}+T^{ij}\right)A_j \right|_{r=R}= 0\ .
\label{2.9}\end{equation}
In a more compact way, the \ac{BC} can be written in matricial form
\begin{equation}
M^{ij}A_j=0\ ,
\label{2.12}\end{equation}
where
\begin{equation}
M=\left(
\begin{array}{cc}
c_1 & c_2-\tilde\kappa \\
c_2+\tilde\kappa & c_3
\end{array}
\right)\ ,
\label{2.13}\end{equation}
having defined
\begin{equation}
c_1\equiv T^{00}\ ;\ c_2\equiv T^{02}=T^{20} \ ; \ c_3\equiv T^{22}
\label{2.14}
\end{equation}
and 
\begin{equation}
\tilde\kappa\equiv\frac{\kappa}{2}\frac{\tilde\epsilon^{012}}{\sqrt{-g}}\ .
\label{2.15}
\end{equation}
Even though in \eqref{2.15} the value of the 012-component of the Levi-Civita symbol is $\tilde\epsilon^{012}=1$, we choose to keep it explicit, to enhance the 
fact that $\tilde\kappa(X)$ is a scalar, and not a scalar density,  as it would appear by hiding $\tilde\epsilon^{012}$.
The most general solution of the \ac{BC} \eqref{2.12} which does not involve vanishing components of the gauge field is  
\begin{equation}
A_t + v A_\theta=0\ ,
\label{2.16}\end{equation}
where the coefficient $v(X)$  in \eqref{2.16} must be different from zero and is not constant, since it depends on the parameters appearing in the action $S$ \eqref{2.6} according to the following relations
	\bi
	\item if $c_1= 0 \ ;\ c_2 = \tilde\kappa\ ;\  c_3\neq 0$
		\begin{equation}
		2\tilde\kappa A_t+c_3A_\theta=0\label{2.17}\quad \Rightarrow\quad v=\frac{c_3}{2\tilde\kappa}\ ;
		\end{equation}
	\item if $c_1\neq 0\ ;\ c_2=-\tilde\kappa\ ;\ c_3=0$
		\be
		c_1A_t-2\tilde\kappa A_\theta=0\label{2.18}\quad \Rightarrow\quad v=-\frac{2\tilde\kappa}{c_1}\ ;
		\ee
	\item if $c_1\neq 0\ ;\ c_2\neq\pm\tilde\kappa\ ;\ c_3\neq0\ ;\ c_1c_3-c_2^2+\tilde\kappa^2=0$
		\be
		\left\{
		\begin{array}{ll}
		c_1A_t+(c_2-\tilde\kappa)A_\theta=0\ &\Rightarrow\ v=\frac{c_2-\tilde\kappa}{c_1}\\
		\mbox{or}&\\
		(c_2+\tilde\kappa)A_t+c_3A_\theta=0\ &\Rightarrow\ v=\frac{c_3}{c_2+\tilde\kappa}\ .
		\end{array}
		\right.
		\ee
	\ei
The parameter $v$ appearing in the \ac{BC} \eqref{2.16} depends on $c_i$ \eqref{2.14}, which are the components of the boundary parameter $T^{ij}$ appearing in $S_{bd}$ \eqref{2.4}, and on $\tilde\kappa$ \eqref{2.15}. Hence, $v$ may depend on the coordinates through the induced metric $\gamma^{ij}$  and on its determinant $\gamma$ (which in \ac{GNC} is equal to $g$). No explicit dependence on the coordinates $X=\{t,\theta\}$ is possible because, as we said, in the flat limit $T^{ij}$ should be constant. The \ac{BC} \eqref{2.16} is of the same type of that derived in \cite{Geiller:2017xad} by an action principle, in analogy to our approach. The difference is that in \cite{Geiller:2017xad}
the analogue of the matrix $T^{ij}$ is constant, which can be obtained after a rescaling by $\sqrt{-g}$. As a consequence, the parameter $v$ appearing in the \ac{BC} in \cite{Geiller:2017xad}  is constant. In other words, it is always possible to rescale $v$ to a constant by means of a coordinate choice. The main goal of this Section, as we shall see, is that we will be able to relate the parameter $v$ to a measurable quantity, which is the velocity of the chiral edge modes. For this reason we do not rescale $v$ to an arbitrary constant value, but we let $v$ to be determined by a phenomenological input. It is important not to rescale $v$ to a constant value in order to be able to account for $accelerated$ chiral bosons on the edge of certain Hall systems, which are not explained by a constant $v$. 
This is an important point, which concerns two different perspectives. Our approach is the same to the abelian Chern-Simons description of the Hall systems. From a pure field theoretical point of view, abelian Chern-Simons model is a free theory, with no coupling constant, for the simple reason that it can be reabsorbed by a rescaling of the gauge field. It is only the nonabelian extension of Chern-Simons theory which displays a true coupling constant. In condensed matter theory the perspective is different, somehow opposite. The coupling constants are not rescaled by field redefinitions, but are fixed by external phenomenological inputs. This reflects even in a different terminology. For instance, the abelian Chern-Simons ``coupling constant'', which is oxymoric in field theory, is often called ``Chern-Simons level'' by a condensed matter-oriented reader, because it is fixed by its relation to the filling factor of Landau levels~: $\kappa=\frac{1}{2\pi\nu}$. In curved spacetime the metric plays the role of a dynamical field, which we treat exactly in the same way~: we do not rescale the metric by choosing a particular coordinates set, but we let the parameter $v$ to be constrained by the observed chiral velocities of the edge modes. We have seen that $v$ depends on the induced metric and its determinant, and hence we relate the induced metric (which depends on the bulk metric of Chern-Simons theory) to an observable quantity. And this is the only way, so far, to take into account the observed accelerated chiral velocities, which are not explained by a flat background of Chern-Simons theory alone. If we were not interested in a phenomenological interpretation of our model, we would rescale $\kappa$ to one and $v$ to whatever value, including zero, which would correspond to a Dirichlet condition on one component of the gauge field. 

\subsubsection*{Ward identity}\label{sec2.3}

The covariant derivative of the \ac{EoM} \eqref{2.7} is
\begin{equation}\label{2.20}
	\begin{split}
\nabla_\lambda\frac{\delta S}{\delta A_\lambda}&=\theta(R-r)\left[\kappa\epsilon^{\lambda\mu\nu}\nabla_\lambda\nabla_\mu A_\nu+\nabla_\lambda\left(b\;n^\lambda\right)+\nabla_\lambda J^\lambda\right]-\\
	&-\delta(R-r)\left(\kappa\epsilon^{1kj}\partial_kA_j+b+J^r\right)=0\ ,
	\end{split}
\end{equation}
where the \ac{BC} \eqref{2.9} have been used to cancel the $\delta(R-r)$ term on the r.h.s. of \eqref{2.7}.
Noting that
\begin{equation}
\epsilon^{\mu\nu\rho}\nabla_\mu\nabla_\nu A_\rho=0\;,
\end{equation}
we find
\begin{equation}\label{2.23}
\theta(R-r)\nabla_\lambda\left( b\;n^\lambda+J^\lambda\right)-\delta(R-r)\left(\kappa\epsilon^{1kj}\partial_kA_j+b+J^r\right)=0\ .
\end{equation}
Keeping in mind that the covariant divergence is
\begin{equation}\label{2.23'}
\nabla_\mu V^\mu=\frac{1}{\sqrt{-g}}\partial_\mu\left(V^\mu\sqrt{-g}\right)\quad;\quad\nabla_k V^k=\frac{1}{\sqrt{-g}}\partial_k\left(V^k\sqrt{-g}\right)\ ,
\end{equation}
where we used \eqref{1.11} in the second identity, multiplying \eqref{2.23} by $\sqrt{-g}$ and integrating
along the coordinate normal to the boundary, we get\footnote{Notice that in general for an integration along $r$ we should use the invariant measure $\sqrt{g_{rr}}\; dr$ in order to preserve the transformation properties under diffeomorphisms,  but $g_{rr} = 1$ in GNG \eqref{1.9}.}
\begin{align}
0&=\int^\infty_0dr\;\left\{\theta(R-r)\left[\partial_r (b\sqrt{-g})+\partial_\lambda\left(J^\lambda\sqrt{-g}\right)\right]-\delta(R-r)\left(\kappa\tilde\epsilon^{1kj}\partial_kA_j\right)\right\}-\nonumber\\
&-\int_0^\infty dr\;\delta(R-r)\left[\left(b+J^r\right)\sqrt{-g}\right]\label{2.24}\nonumber\\
&=\int^\infty_0dr\;\left[\theta(R-r)\partial_k\left(J^k\sqrt{-g}\right)-\delta(R-r)\kappa\tilde\epsilon^{1kj}\partial_kA_j\right]-\left[\left(b+J^r\right)\sqrt{-g}\right]_{r=0}\nonumber
\\
&=\int^\infty_0dr\;\sqrt{-g}\left[\theta(R-r)\nabla_kJ^k-\delta(R-r)\kappa\epsilon^{1kj}\partial_kA_j\right]\ ,
\end{align}
where we integrated by parts and used the fact that, evaluating the \ac{EoM} \eqref{2.7} for $\lambda=r$ and then going at $r=0$,  we have $\left[b(x)+J^r(x)\right]_{r=0}=0$. 
In fact, \eqref{2.7} at $r=0$ ($\lambda=r$) reads
\begin{equation}\label{b=0}
\left[b+J^r - 2\tilde\kappa \,  \Big( \partial_\theta A_t  - \partial_t A_\theta \Big)\right]_{r=0} = 0\ ,
\end{equation}
and we notice that $A_\theta(x)$ necessarily vanishes at $r=0$, together with its time derivatives, while $A_t(x)$, which in principle might not vanish at $r=0$, at the origin must have vanishing angular derivatives, in order to be well defined,  hence the result. 
Therefore,  the contribution involving the Lagrange multiplier and $J^r$ cancel 
 out, leaving only
\begin{equation}
\int_0^Rdr\; \sqrt{-g}\nabla_kJ^k =
\left.\kappa\tilde\epsilon^{1kj}\partial_kA_j
\right|_{r=R}\ .
\label{2.25}
\end{equation}
Eq.\eqref{2.25} is the Ward identity of the theory, broken at its r.h.s. by the presence of the boundary, which is crucial for the determination of the boundary algebra and of the 2D theory holographically induced on the boundary. Notice that it holds for any bulk metric, and it is simply the curved extension of its flat counterpart \eqref{BrokenWICS} \cite{Maggiore:2017vjf}. From \eqref{2.25}, at vanishing external sources $J^k(x)=0$ ($i.e.$ going on-shell), we find
\begin{equation}
\left.\epsilon^{1kj}\partial_kA_j\right|_{r=R;J=0}=0\ ,
\label{2.26}\end{equation}
which describes a conserved current on the closed boundary $r=R$, whose most general solution is \cite{Nash:1983cq,Warner}
\begin{equation}
A_i(X)=\partial_i\Phi (X) + \delta_{i2}\; C\ ,
\label{2.27}\end{equation}
where $C$ is a constant (which we will show shortly being equal to zero), and $\Phi(X)$ is a scalar field that will play the role of \ac{DoF} of the 2D boundary theory. The components of the gauge field on the boundary are then
\begin{equation}
A_t (t,\theta)=\partial_t\Phi(t,\theta)\quad;\quad A_\theta(t,\theta)=\partial_\theta\Phi(t,\theta)+C\ .
\label{2.28}\end{equation}
Since we are considering a closed boundary, we also have to impose periodicity conditions on the fields~:
\begin{equation}
A_i(t,\theta)=A_i(t,\theta+2\pi )\quad \Rightarrow\quad \Phi(t,\theta)=\Phi(t,\theta+2\pi )\ .
\label{2.29}\end{equation}
The value of the constant $C$ in \eqref{2.27} is found by applying the mean value theorem for holomorphic functions, which states that if $f$ is analytic in a region $D$, and $a\in D$, then $f(a)=\frac{1}{2\pi}\oint_{{\cal C}(a)} f$, where ${\cal C}(a)$ is a circle centered in $a$. 
In our case (3D) taking  for $\cal C$ the circular boundary $r=R$ centered at $r=0$ allows us to write
\begin{equation}
\oint_{ring\; R} A_\theta(x) 
=\cancel{\left.\Phi(t,\theta)\right|_{\theta=0}^{\theta=2\pi }} +
 2\pi \; C\ ,
\label{2.30}\end{equation}
where we used the periodicity condition of the boundary field $\Phi(X)$ \eqref{2.29}. This is the value of the bulk field at the center of the ring, and using the fact that  $A_{\theta}(t,r=0,\theta)=0$, we finally get
\begin{equation}
C=0\ .
\label{2.31}\end{equation}
\subsubsection*{Algebra }\label{sec2.4}

The generating functional of the connected Green functions $Z_c[J]$ is defined, as usual, in the following way
\begin{equation}
e^{iZ_c[J]}=\int [dA][db]\;e^{iS[A,b;J]}\  ,
\label{2.32}
\end{equation}
where $S$ is the total action \eqref{2.6}. From \eqref{2.32} we get the 1- and 2-points Green functions
\begin{empheq}{align}
\left.\frac{\delta Z_c[J]}{\delta J^i(x)}\right|_{J=0}=&\;\langle A_i(x)\rangle\label{2.33}\\
\left.\frac{\delta^{(2)} Z_c[J]}{\delta J^i(x)\delta J^j(x')}\right|_{J=0}\equiv&\Delta_{ij}(x,x')=i\langle T(A_i(x)A_j(x'))\rangle\ ,\label{2.34}
\end{empheq}
where the time-ordered product is defined as
\begin{equation}
\langle T(A_j(x)A_l(x'))\rangle\equiv\theta(t-t')\langle A_j(x)A_l(x') \rangle +
\theta(t'-t)\langle A_l(x') A_j(x) \rangle\ .
\label{2.37}
\end{equation}
The algebra is obtained by making the functional derivative with respect to $J^l(x')$ of the Ward identity \eqref{2.25}, and then going on-shell, $i.e$ putting $J=0$~:
\begin{equation}
\frac{ \delta}{\delta J^l(x')} \int_0^R dr\; \sqrt{-g}\;\nabla_k\,J^k(x)=
\kappa\; \tilde\epsilon^{1kj}\; \partial_k 
\left. 
\frac{\delta^{(2)} Z_c}{\delta J^l(x') \delta J^j(x) }
\right |_{r=R;J=0}\ .
\label{2.35}\end{equation}
Therefore the first step is to compute
	\begin{equation}\label{2.35'}
		\frac{\delta}{\delta J^l(x')}\nabla_k J^k(x)=\frac{\delta}{\delta J^l(x')}\left[\frac{1}{\sqrt{-g}}\partial_k\left(J^k\sqrt{-g}\right)\right]
		=\frac{1}{\sqrt{-g}}\partial_l\tilde\delta^{(3)}(x-x')\ ,
	\end{equation}
where we used the relation \eqref{1.5} betweeen the scalar and density delta function and the fact of working in \ac{GNC} \eqref{1.11}.
Using \eqref{2.35'} on the l.h.s.  of \eqref{2.35}, we obtain 
\begin{equation}
\partial_l\tilde\delta^{(2)}(X-X') = 
i\kappa \left.\tilde\epsilon^{1kj}\partial_k\langle T(A_j(x)A_l(x'))\rangle\right |_{r=R}\ .
\label{2.36}\end{equation}
To write this we used the fact that for any $r'\leq R$ we have $\int_0^Rdr\,\tilde\delta(r-r')=1$, in fact, since by definition $\tilde\delta(r-r')\equiv-\partial_r\theta(r'-r)$:
	\begin{align}
		\int_0^Rdr\,\partial_r\theta(r'-r)f(r)&=\int_0^R\left\{\partial_r\left[\theta(r'-r)f(r)\right]-\theta(r'-r)\partial_rf(r)\right\}\nonumber\\
		&=\left.\theta(r'-r)f(r)\right|_0^R-\int_0^Rdr\,\theta(r'-r)\partial_rf(r)\\
		&=	\left\{\begin{array}{cccccc}
			f(R)-f(0)-\int^R_0dr\,\partial_rf(r)&=&0&\ & \mbox{if}\ r'>R&\\
			0-f(0)-\int^{r'}_0dr\,\partial_rf(r)&=&-f(r')&\ & \mbox{if}\ r'\leq R&\ .
			\end{array}\right.\nonumber
	\end{align}
Choosing $l=\theta$ in \eqref{2.36} we have~:
\begin{align}
\partial_\theta\tilde\delta^{(2)}(X-X')
&=
i \kappa \tilde\epsilon^{1kj} \partial_k 
 \left[
 \theta(t-t') \langle A_j(X)A_\theta(X')\rangle +
 \theta(t'-t) \langle A_\theta(X')A_j(X)\rangle
 \right]\nonumber\\
 &=
 i\kappa \tilde\epsilon^{102} \left(\partial_t\theta(t-t')\right)
 \left(
 \langle A_{\theta}(X)A_\theta(X')\rangle -
 \langle A_\theta(X')A_{\theta}(X)\rangle
 \right) \label{2.38}
 \\
 & +\kappa \tilde\epsilon^{1kj} \left[
i \theta(t-t') \langle \bcancel{\partial_kA_j(X)}A_\theta(X')\rangle +
 i\theta(t'-t) \langle A_\theta(X')\bcancel{\partial_kA_j(X)}\rangle
 \right] \nonumber\ ,
\end{align}
where we used the on-shell condition \eqref{2.26}. By defining
\begin{equation}
\left[A_j(X),A_\theta(X')\right]
\equiv
\langle A_j(X)A_\theta(X')\rangle -
 \langle A_\theta(X')A_j(X)\rangle\ ,
\label{2.39}\end{equation}
we get
\begin{equation}
\partial_\theta\tilde\delta^{(2)}(X-X')
=
-i\kappa\tilde\epsilon^{012} \partial_t\theta(t-t') \left[A_\theta(X),A_\theta(X')\right] \ .
\label{2.40}\end{equation}
Finally, integrating over time we find the equal time commutator
\begin{equation}
\tilde\epsilon^{012}\left.\left[A_\theta(X),A_\theta(X')\right]\right|_{t=t'} =
\frac{i}{\kappa}\,
\partial_\theta\tilde\delta(\theta-\theta')\ ,
\label{2.43}\end{equation}
By applying the same procedure to the case $l=t$ of \eqref{2.36}, we find the equal time commutator
\begin{equation}
\left.\left[A_\theta(X),A_t(X')\right]\right|_{t=t'}=0\ .
\label{2.45}\end{equation}
Eq.\eqref{2.43} represents an abelian \ac{KM} algebra
identical to the one found in the case of Chern-Simons theory with planar boundary in flat space \eqref{KMCS}. We see that the existence of a \ac{KM} algebra and its central charge depends neither on the bulk metric nor on the details of the boundary. The central charge 
\begin{equation}
c=\frac{1}{\kappa}
\label{2.49}\end{equation}
must be positive (for the unitarity of the associated \ac{CFT} {\cite{Mack:1988nf,Becchi:1988nh}), and for this reason the coupling constant of the Chern-Simons theory has to be positive as well
\begin{equation}
\kappa>0\ .
\label{2.50}\end{equation}
Notice that this algebraic method is the only way to determine the sign of the Chern-Simons coupling constant, since this theory, being topological, has vanishing stress-energy tensor, and hence the usual argument based on the positivity of the energy density cannot be applied here.

\subsection{2D boundary theory}\label{sec3}

\subsubsection*{The 2D action}

In Section \ref{sec2.3}, we identified the solution of the conserved current equation \eqref{2.26} as the boundary \ac{DoF}. This will allow us to find the 2D dynamics on the boundary, in fact we can express the equal time commutation relation \eqref{2.43} in terms of the boundary field $\Phi(X)$, by using \eqref{2.28}
\begin{equation}\label{3.1}
[\partial_\theta\Phi(X),\tilde\epsilon^{012}\partial_{\theta'}\Phi(X')]=i\frac{1}{\kappa}\;\partial_\theta\tilde\delta(\theta-\theta')\ ,
\end{equation}
which, rewritten as
\begin{equation}\label{3.2}
[\Phi(X), \kappa\tilde\epsilon^{012}\partial_{\theta'}\Phi(X')]=i\;\tilde\delta(\theta-\theta')\ ,
\end{equation}
can be interpreted as a canonical commutation relation in curved spacetime
\begin{equation}\label{3.3}
[q(t,x),p(t,x')]=i\;\tilde\delta(x-x')\ ,
\end{equation}
with $p(t,x)$ a density \cite{Carroll:2004st}, provided that we match the canonical variables of the 2D theory as follows
\begin{equation}\label{3.4}
q(X)\equiv\Phi(X)\quad;\quad p(X)\equiv \tilde\epsilon^{012}\kappa\partial_\theta\Phi(X)\ .
\end{equation}
Here again, the presence of the $\tilde\epsilon^{012}$ factor in \eqref{3.4} is of great importance, since it makes $p(X)$ a scalar density, according to the standard definition {\cite{Basler:1991st}. Once identified the canonical variables, we can begin the search for the induced boundary theory, by  writing the most general  Lagrangian compatible with power counting. Noting that the scalar field has vanishing mass dimension, we find
\begin{equation}
\mathcal{L}_{2D}=\sqrt{-\gamma}\;
(a^{ij}\partial_i\Phi\partial_j\Phi+b^i\partial_i\Phi+c)\ ,
\label{3.5}
\end{equation}
where the coefficients $a^{ij},\ b^i,\ c$
\begin{enumerate}
\item must be tensor quantities, $i.e.$ symmetric tensor of rank (2,0), contravariant vector and scalar respectively, in order that the Lagrangian \eqref{3.5} is a scalar density\; ;
\item must have mass dimension 0, 1 and 2, respectively\; ;
\item may depend on the scalar field $\Phi(X)$, the metric $\gamma_{ij}(X)$ and/or its determinant $\gamma(X)$ (but not on their derivatives), since in 2D the scalar field $\Phi(X)$, like the metric, is dimensionless, and thus the power counting is preserved. In other terms, the 2D action \eqref{3.5} is written as a derivative expansion.
\end{enumerate}
We also notice that the scalar field $\Phi(X)$ is defined by means of \eqref{2.28} up to a shift transformation
\begin{equation}
\delta_{shift}\Phi=\alpha\ ,
\label{3.6}
\end{equation}
with $\alpha$ constant, which implies that the action 
\begin{equation}
S_{2D}=\int d^2X\; {\cal L}_{2D}
\label{3.7}\end{equation}
describing the boundary theory should possess the same symmetry as well, $i.e.$
\begin{equation}
 \delta_{shift}S_{2D}=0\;.
 \label{3.8}\end{equation}
An immediate consequence of the shift symmetry \eqref{3.8} is that $a^{ij}$ and $c$ must be constant with respect to $\Phi$. For what concerns $b^i$, it is easily seen that it may admit a linear dependence on the scalar field. In fact, if $b^i=b_1^i+b_2^i\Phi$, where $b_1^i$ and $b_2^i$ do not depend on $\Phi$, but may depend on the induced metric $\gamma_{ij}$ and on its determinant $\gamma$, we have that 
\begin{equation}
\delta_{shift}\int d^2X\; \sqrt{-\gamma} \;
b^i\partial_i\Phi=
\int d^2X\; \sqrt{-\gamma} \;
\frac{\partial b^i}{\partial\Phi}\alpha\partial_i\Phi=
\alpha\int d^2X\; \sqrt{-\gamma} \;\nabla_i\left(\frac{\partial b^i}{\partial\Phi}\Phi\right)\ ,
\label{linearb}\end{equation}
which is a vanishing boundary term. 
Therefore, the corresponding term in the 2D Lagrangian \eqref{3.5} does not contribute, being a boundary term as well.
\begin{equation}
\int d^2X\; \sqrt{-\gamma} \;
b^i\partial_i\Phi=
\int d^2X\; \sqrt{-\gamma} \;
(b_1^i\partial_i\Phi+b_2^i\Phi\partial_i\Phi)=
\int d^2X\; \sqrt{-\gamma} \;\nabla_i
(b_1^i\Phi+\tfrac{1}{2}b_2^i\Phi^2)\ .
\end{equation}
The most general 2D action satisfying the shift symmetry  is 
 \begin{equation}
	\begin{split}
S_{2D} &= \int d^2X\; \sqrt{-\gamma} \;
	(a^{ij}\partial_i\Phi\partial_j\Phi+c) \\
&=\int d^2X\; \sqrt{-\gamma}\;\left(a_0\partial_t\Phi\partial_t\Phi+2a_1\partial_t\Phi\partial_\theta\Phi+a_2\partial_\theta\Phi\partial_\theta\Phi +c \right),
\label{3.10}
	\end{split}
\end{equation}
where $a_0\equiv a^{00}$, $a_1\equiv a^{02}$ and $a_2\equiv a^{22}$. The last term does not contribute to the equations of motion of the scalar field $\Phi$ and will be omitted. The most general dependence of the dimensionless $a^{ij}$ on the induced metric $\gamma^{ij}$ is
\begin{equation}\label{3.10'}
a^{ij}=\alpha^{ij}(\gamma)+\beta(\gamma)\;\gamma^{ij}\ ,
\end{equation}
where $\alpha^{ij}(\gamma)$ and $\beta(\gamma)$ may depend at most on the metric determinant. What is left now is to verify under which conditions the Lagrangian ${\cal L}_{2D}$ is compatible with the canonical variables identified from the \ac{KM} algebra through \eqref{3.4}. This is reached by requiring
\begin{equation}\label{3.11}
\frac{\partial\mathcal{L}_{2D}}{\partial\dot q}=p=\tilde\epsilon^{012}\kappa\partial_\theta\Phi\ .
\end{equation}
The l.h.s.  of \eqref{3.11} is
\begin{equation}\label{3.12}
\frac{\partial\mathcal{L}_{2D}}{\partial\dot q}=\frac{\partial\mathcal{L}_{2D}}{\partial(\partial_t\Phi)}=\sqrt{-\gamma}(2a_0\partial_t\Phi+2a_1\partial_\theta\Phi)\ ,
\end{equation}
therefore we must ask
\begin{equation}
a_0=0\quad;\quad a_1=\tilde\kappa 
\quad\quad
\; (a_2\ \mbox{free})\ ,\label{3.13}
\end{equation}
where $\tilde\kappa$ is the scalar function related to the Chern-Simons coupling constant through \eqref{2.15}. Eq.\eqref{3.13} represents a constraint on the metric dependence of $a^{ij}$ in \eqref{3.10'}, in fact it must be
\begin{empheq}{align}
0&=a_0=a^{00}=\alpha^{00}+\beta\;\gamma^{00}\\
\tilde\kappa&=a_1=a^{02}=\alpha^{02}+\beta\;\gamma^{02}\ .
\end{empheq}
Since we do not want to impose any unnecessary condition on the components of the induced metric $\gamma^{ij}$, that leads to the request
\begin{equation}
\beta=0\ ,
\end{equation}
therefore
\begin{equation}
a_2=a^{22}=\alpha^{22}(\gamma)\ ,
\label{a2}\end{equation}
which means that $a_2$ is a free parameter depending at most on the metric determinant~: $a_2=a_2(\gamma)$. Hence, up to terms which do not contribute to the equation of motion of the scalar field $\Phi(X)$, the action $S_{2D}$ \eqref{3.7} is 
\begin{equation}\label{3.14}
S_{2D}=\int d^2X\sqrt{-\gamma}\;\left(
\tilde\kappa\partial_t\Phi
+a_2\partial_\theta\Phi
\right)\partial_\theta\Phi \ .
\end{equation}

\subsection{Holographic contact}

From the action $S_{2D}$ \eqref{3.14} we get the following 	\ac{EoM}\\
\begin{equation}\label{3.15}
\partial_\theta\left(
\partial_t\Phi+\frac{a_2}{\tilde\kappa}\;\partial_\theta\Phi\right)=0\;,
\end{equation}
where we divided by the Chern-Simons coupling constant $\kappa$. Our task is now to find under which conditions the \ac{BC} \eqref{2.16}, written in terms of the boundary field $\Phi(X)$ by means of \eqref{2.28} (with $C=0$ \eqref{2.31})
\begin{equation}
\partial_t\Phi+v\,\partial_\theta\Phi=0\ ,
\label{3.16}
\end{equation}
can be related to the \ac{EoM} \eqref{3.15}. 
The \ac{BC} \eqref{3.16}
is the equation of a chiral boson whose velocity $v$, defined by \eqref{2.16}, may depend on the induced metric $\gamma^{ij}$ through the coefficients $T^{ij}$ appearing in $S_{bd}$ \eqref{2.4}. We recover here the physical interpretation of the parameter $v$ appearing in the \ac{BC} \eqref{2.16} as the chiral velocity of the edge modes living on the boundary of Chern-Simons theory, which are known, in the framework of \ac{FQHE}, to be measurable quantities. In most Hall systems, the observed chiral velocity is constant, and this is achieved by a Chern-Simons theory built on a flat 3D spacetime with planar boundary \cite{Maggiore:2017vjf}. The novelty we are finding here, is that $v$ is now a $local$ quantity, in particular depending on time, which is a consequence of considering the Chern-Simons theory on a curved, instead of flat, spacetime. This corresponds to $accelerated$ edge chiral bosons, which have been indeed recently observed, and which cannot be explained by putting a boundary on a flat background. The fact that $v$ is a phenomenological parameter requires that it should be determined by experimental inputs, hence, as stressed at the beginning of the Section, a choice of coordinates, which would set $v$ to an arbitrary, possibly constant, or even vanishing, value, should not be done $a\ priori$. We might rather say that the $right$ choice of coordinates is that of the laboratory frame where the chiral local velocity is measured. 
Compatibility between the \ac{BC} \eqref{3.16} and the \ac{EoM} \eqref{3.15} is reached if we require that \ac{EoM}$\equiv\partial_\theta$\ac{BC}, or introducing, as in Section \ref{sec CS}, the field $\rho(t,\theta)\equiv\frac{1}{2\pi}\partial_\theta\Phi(t,\theta)$. In any case, the following condition must hold
\begin{equation}\label{3.17}
v=\frac{a_2}{\tilde\kappa}\ .
\end{equation}
Eq.\eqref{3.17} relates the 
parameters of the two theories, thus establishing a (holographic) link between the bulk ($v,\tilde\kappa$) and the induced boundary ($a_2$). 
From \eqref{3.17} we see that, because of \eqref{a2}, the holographic contact makes $v$ depend only on the determinant $\gamma$ of the induced metric $\gamma_{ij}$, and not on its components. This constitutes a restriction on all possible $v$ appearing in the \ac{BC} \eqref{2.16}, which, instead, might depend on the determinant of the induced metric $and$ on its components as well. This constraint, however, does not spoil the possible local character of $v$.
Eq. \eqref{3.17} is a condition on the parameters which appears in $S_{bd}$ for the holographic bulk-boundary contact to be possible. Therefore the edge velocity depends on the \ac{BC} for the Chern-Simons gauge field.
The issue of the determination of the chiral velocities has been discussed in details by Wen (in \cite{Wen:1989mw} and \cite{Wen:1990qp} for instance), who remarked that it cannot be determined from the bulk action (nor be rescaled to an arbitrary value), and that it is thus appropriate to take the velocity $v$ as a phenomenological parameter. This is the phenomenological counterpart of the considerations made in \cite{Frohlich:1990xz}~: Chern-Simons theory provides merely an effective large-distance description which captures the general symmetries of the manybody theory and has, as such, intrinsic limitations. For instance, it should not be expected to capture certain details such as the velocity of the edge chiral bosons. In \cite{Frohlich:1990xz} it is correctly argued that
the  determination of the chiral velocity should come from  gauge-breaking terms present in the full Lagrangian, which, however cannot be identified with the gauge fixing term, since physical results should not depend on the gauge choice, hopefully. In our approach, it is clear that gauge-breaking term present in the full Lagrangian thanks to which the \ac{TQFT} acquires local degrees of freedom is $S_{bd}$ (2.5). 
 It is the breaking of the gauge symmetry which starts the game: the breaking of the gauge symmetry due to the boundary manifests itself in the breaking of the Ward identity \eqref{2.25}, which yields the irrotationality condition \eqref{2.26}, which identifies the local bosonic boundary degrees of freedom. It is from the broken Ward identity that the algebra \eqref{2.43} is derived, and then all the rest follows, as described. None of these results depend on the particular gauge choice, as we already remarked.
The 2D action \eqref{3.14} can now be entirely written  in terms of the parameters appearing in the 3D bulk theory \eqref{2.6} as follows
 \begin{equation}
S_{2D}=\frac{\kappa}{2}\int d^2X\; \tilde\epsilon^{012}\;
\left(\partial_t\Phi+v\,\partial_\theta\Phi\right)\partial_\theta\Phi\ ,
\label{3.18}
\end{equation}
where $\kappa$ is the Chern-Simons coupling constant and $v(t,\theta)$, appearing in the \ac{BC} \eqref{2.16}, depends on the parameters of the boundary action $S_{bd}$ \eqref{2.4} and, through $\tilde\kappa$, on the determinant of the induced metric $\gamma(t,\theta)$.
We observe that the only dependence of the action $S_{2D}$ on the bulk metric is concealed in $v(t,\theta)$. 
We might say that, in this sense, the metric form of \eqref{3.18} is protected. Physically, the 2D action \eqref{3.18} can be immediately identified with the Luttinger theory \cite{Haldane:1981zza}, relevant example of the bosonization phenomenon, for which the density operator $n(t,\theta)$, written in terms of chiral fermions, is
\begin{equation}
n(t,\theta)=:\psi^\dagger(t,\theta)\,\psi(t,\theta):\ .
\label{3.19}
\end{equation}
We have seen for the flat case discussed in Section \ref{sec CS} that bosonization, $i.e.$ fermion/boson correspondence, is achieved through the identification
\begin{equation}
n(t,\theta)=\frac{1}{2\pi}\partial_\theta\Phi(t,\theta)\ ,
\label{3.20}
\end{equation}
where $\Phi(X)$ is just the chiral boson \eqref{3.16} found as the edge state of the Chern-Simons theory. The density operator \eqref{3.19} satisfies the following commutation relation
\begin{equation}
[n(t,\theta),n(t,\theta')]=i\frac{\nu}{2\pi}\partial_\theta\delta(\theta-\theta')\ ,
\label{3.21}
\end{equation}
where $\nu$ is the filling factor of the \ac{FQHE}. Eq.\eqref{3.21}, with the bosonisation relation \eqref{3.20},
is exactly the \ac{KM} algebra we found in \eqref{3.1}, by means of which
we can physically interpret the field $A_\theta(t,\theta)$ as the density operator $n(t,\theta)$ \eqref{3.19}. 
Here we also notice how the boundary theory makes sense for $any$ dependence of $v$ on $X$. Indeed, from the perspective of \eqref{3.16}, the \ac{EoM} derived from the action \eqref{3.14} reads
\begin{equation}
\partial_tn+\partial_\theta(vn)=0\ ,
\end{equation}
which represents the continuity equation of a 1D fluid of velocity $v$ and density $n$, as already remarked in the case of flat bulk metric in Section \ref{sec CS} and in \cite{Blasi:2008gt}. This, again, confirms that $v$, from the perspective of the Chern-Simons theory, is simply a free phenomenological parameter.
As a consequence, we can associate the parameters of the action $S_{2D}$ \eqref{3.18} to physical quantities. In particular, by matching through \eqref{3.20} the algebra \eqref{3.1}, written in terms of the boundary field $\Phi(X)$, with the algebra \eqref{3.21} written in terms of the density operator $n(t,\theta)$, we obtain the well known relation between the filling factor $\nu$ and the coupling constant of the Chern-Simons theory $\kappa$  \cite{Blasi:2008gt,Maggiore:2017vjf}
\begin{equation}
\kappa=\frac{1}{2\pi\nu}\;.
\label{3.22}\end{equation}
Hence, from \eqref{3.16}, we can identify $v(t,\theta)$ as the spacetime-dependent velocity of the chiral boson $\Phi(X)$.
Notice that, because of \eqref{3.17}, the chiral boson turn left or right depending on the sign of $a_2$ in \eqref{3.18}, being $\tilde\kappa$ positive due to the positivity of the central charge. This is in agreement with \cite{Wen:1990se}, where the edge excitations were studied 
on a disc and on a cylinder, $i.e.$ on curved boundaries in flat spacetime, which belong to the cases studied in this Section. It is interesting to remark that we recover those results, in particular the algebra and the Luttinger theory, in a quite different way. 

\subsection{Summary of results and discussion}\label{sec4}

The aim of the original work presented in this Section was to understand if the geometry of the bulk spacetime affected in some way the boundary physics of Chern-Simons theory, which, in the flat case, we have seen to reproduce, on the boundary, the theory of edge states of the \acl{FQHE}.  Naively, starting from a \ac{TQFT} one would expect a mild dependence on the bulk metric. Instead, our analysis shows that the velocity of the edge chiral boson indirectly depends on the bulk metric, through the determinant of the induced metric. Our work might therefore be the first step towards a theoretical framework for the recently observed accelerated chiral boson on the edge of particular Hall systems, which cannot be explained by the usual description in terms of flat Chern-Simons theory with boundary. On the other hand, on the boundary of Chern-Simons theory we find the usual \ac{KM} algebraic structure of the flat background case, with the same central charge proportional to the inverse of the Chern-Simons coupling. Therefore, the \ac{KM} algebra appears to be insensitive to both the bulk metric and the type of boundary (at least, it is the same for planar and radial boundary). What is certainly true is that the Chern-Simons bulk theory in flat spacetime, when considered with a boundary, generates local \ac{DoF} on the 2D induced theory,  corresponding to the edge states of \ac{FQHE}, with $constant$ chiral velocities. The possibility of local velocities is not captured by the flat approach alone. 
\\

A few remarks are in order.
\begin{itemize}
\item \textbf{Technical remarks :} the determinant of the metric depends on the coordinates that are chosen, and locally one might always find coordinates such that the determinant is a constant. Moreover, in 2+1 dimensions the boundary of Chern-Simons theory is well known to be a \ac{CFT}, hence depends only upon the conformal class of the boundary metric, and in every conformal class there is a metric of constant determinant. This would correspond to choosing a reference frame where the velocity of the chiral boson is normalized at a given constant value. This represents a good, physical reason for not choosing a metric with constant determinant. The boundary conformal structure is ensured by the existence of an algebraic structure of the \ac{KM} type. Had we finished here, we would not have a good reason for not choosing, amongst the metrics in \ac{GNC}, those with constant determinant. This would have been a without loss of generality choice, and we would have achieved our result as far as only the conformal structure described by the boundary algebra is concerned. But this is only half of our results~: on the $planar$ boundary of Chern-Simons theory with \textit{flat} bulk metric it is known to exist the Luttinger theory of the chiral boson. We derived the corresponding action 
following the procedure described in Section \ref{sec3}. The action \eqref{3.18} is not invariant under change of coordinates, which hence is not allowed at this stage. Nonetheless, the Luttinger theory which we found on the boundary of the Chern-Simons theory with non-flat metric is not 
independent from its bulk counterpart. We would not have found this result if, in the previous step, we had frozen the bulk metric by choosing coordinates with constant determinant (or a reference frame where the chiral velocity is constant). Instead, a  memory of the bulk metric survives, in the velocity of the chiral boson, which depends only on the determinant of the bulk metric. Not having made the choice of constant determinant leaves therefore the possibility of highlighting the residual dependence on the bulk metric, and also of selecting the bulk metric by means of a physical input, which is the velocity of the edge chiral bosons, which is a physical observable. This is not so different from what is usually done with the Chern-Simons coupling constant, which is fixed by means of the physical requirements of the incompressible Hall fluid, by normalizing the external current coupled to the electromagnetic gauge field to the known relationship between Chern-Simons coupling constant and filling factor $\kappa=\frac{1}{2\pi\nu}$ as in \eqref{J=sigmaE}. Yet, speaking of a coupling constant for the $abelian$ Chern-Simons theory is rather inappropriate, since it is always possible to reabsorb the coupling constant by means of a redefinition of the gauge fields. Only in the non-abelian case a true coupling constant for Chern-Simons theory exists. Nevertheless, normalizing to one the Chern-Simons coupling constant would be equivalent to loosing the entire structure of Landau levels of {\textsc{(F)QHE}}, and for that reason we keep it alive. We see an analogy between the two situations: as we do not normalize to one the Chern-Simons coupling constant in the abelian case in order to describe the filling factors of the \ac{FQHE}, for the same reason we do not choose coordinates such that the determinant is constant, in order to keep the possibility of fixing the bulk metric by measuring the (possibly time dependent) velocity of the edge chiral boson.

\item \textbf{The phenomenological approach : }it is useful to make a comparison with the approach usually adopted in the Literature to describe non-constant velocities of the edge excitations in the Hall systems. It is known that the fermions in a quantum Hall droplet are constrained by a confining potential whose gradient determines the velocity of the chiral edge modes \cite{Kane95}. For instance, the confinement may involve potentials that depend on space and/or time due to deformations or interactions in the sample \cite{Hashi18} and typically, in the phenomenological framework, one can invoke these considerations in order to take into account non-constant chiral velocities, which therefore, in the microscopic model, depend on the different possible potentials in the sample. 
In general, the paradigmatic way to describe the edge excitations has been illustrated by Wen \cite{Wen:1989mw,Wen:1990qp,Wen:1991ty}, and well summarized in \cite{Kane95}. One starts with a 3D Chern-Simons action in $flat$ spacetime, from which the bulk \ac{DoF} are eliminated by integrating out the time-component of the gauge field $A_0(x)$. Then, the irrotational constraint on the spatial components of the gauge field is \textit{imposed}~: $\vec\nabla\times \vec A = 0$. The boundary \ac{DoF} is then identified by the scalar field characterizing the solution $\vec A = \vec\nabla\Phi$. Finally, the ``appropriate effective action'' at the edge is written as the sum of two terms:  the Luttinger action, describing a chiral boson with $constant$ velocity, to which is possibly added an interaction term, which depends on the form of the edge confining potential and on the details of the electron-electron interactions. If this latter interaction is assumed to be local, this obviously implies non-constant velocities. Usually, to simplify the description, the interactions are taken as piecewise functions, in order to take into account, for instance, the screening effects which are present in the leads. More details can be found in \cite{safi} and in \cite{perfetto}. This latter paper is related to an interesting experiment \cite{Kamata}. Other experiments are described in \cite{brasseur,lin}.
To make it short, the total interaction is split into two parts~: a constant velocity piece $v$ contained in the Luttinger action and an additional local velocity term in the interaction. This is the framework for the description of edge excitations with non-constant velocities. Alternatively, according to our approach, there is no need of $imposing$ the irrotational condition which gives rise to the scalar boundary degree of freedom. Rather, it emerges as the breaking term of the Ward identity \eqref{2.25}, evaluated on shell. Similarly, the 2D action \eqref{3.18} is not $chosen$ as an effective action, but is recovered by means of a general procedure which we described as an ``holografic contact'' between the \ac{BC} on the bulk gauge field and the \ac{EoM} of the 2D scalar field. Finally, we showed that a bulk Chern-Simons model on curved spacetime yields a possible $non$-$constant$ chiral velocity, where the locality property is inherited solely from the induced metric determinant from a bulk-boundary correspondence.
An intriguing way to interpret this fact, in relation to the microscopic models, is to consider it as an equivalent description~: a change of potential can be effectively encoded in the Chern-Simons theory by a bulk metric. As the Chern-Simons coupling constant $\kappa$ encodes all the possible values of the filling factor $\nu$, similarly the Chern-Simons bulk metric encodes the spacetime properties of the (confining) potential, in terms of the induced metric determinant.

\item \textbf{On current experiments :} one of the main results of this Thesis is that the measurable chiral velocity $v$ of the edge modes is related not only to the Chern-Simons level $\kappa$, as it is known already for a flat background, but also to the determinant of the metric, through which $v$ can acquire a time dependence, which can be measured.
It is indeed possible to detect time dependent edge chiral velocity, as well as to investigate non-standard geometries (for what concerns the quantum Hall effect). In \cite{Bocquillon} the boundary of a Hall system with $\nu=2$ is considered. The two channels (left and right) interact, and the problem diagonalizes into one fast and one slow mode, and the velocity is not constant and is measured. For what concerns non conventional bulk geometries, recently, in \cite{kumar},
the \ac{FQHE} has been observed in graphene. Here, the electrons live on a two-dimensional ``suspended'' membrane made by carbon atoms. There are some differences with respect to the standard case (Dirac-type linear dispersion relation rather than quadratic, presence of an additional degree of freedom like a pseudo-spin, for instance), which makes the phenomenology similar to that of \ac{FQHE}, but with different values of the plateaux. The nice thing is that graphene technically allows to make samples with various shapes like Corbino discs,  and observations seem sensitive to the bulk geometry. Moreover, always using graphene, there are attempts to realize quantum Hall systems in curved space, like described in 
\cite{Wagner:2019nyo} and in \cite{Can:2014ota}. Now, we do not know if the results presented in this Section can immediately be applied to these particular experimental sets, but the remarkable fact is that a recent experimental research activity exists, which concerns the main topic of this Chapter~: the possibility of having boundary chiral modes whose time dependent velocities are sensitive to the bulk geometry. 
\end{itemize}


\newpage
\section{BF description of 2D accelerated chiral edge modes} \label{sec BFinCS}

This Section, which concerns the study of the 3D BF model in curved space as a natural continuation of the results obtained in the previous Section for Chern-Simons, is organized as follows
\etocsetnexttocdepth{2}

\begingroup
\parindent=0em
\etocsettocstyle{\rule{\linewidth}{\tocrulewidth}\vskip1.25\baselineskip}{\vskip-0.75\baselineskip\rule{\linewidth}{\tocrulewidth}\vskip1\baselineskip}
\makeatletter
  \edef\scr@tso@subsection@indent
    {\the\dimexpr\scr@tso@subsection@indent-\scr@tso@section@indent}
  \def\scr@tso@section@indent{0pt}
\makeatother
\localtableofcontents 
\endgroup
\noindent
In particular in Section \ref{sec modelBFinCS} we prepare the tools to face the problem~: we introduce a radial boundary in the gauge-fixed 3D BF action, and derive \`a la Symanzik the most general \ac{BC} on the two gauge fields of the theory. As we said, the boundary breaks gauge invariance, and this reflects in the breakings of the two Ward identities describing the broken gauge symmetry. The breaking are particularly fruitful, as they lead us to identify the 2D scalar degrees of freedom and the \ac{KM} algebra formed by the edge conserved currents. Requiring the positivity of the \ac{KM} central charge constrains the BF coupling to be positive. In Section \ref{sec bdBFinCS} the 2D theory is derived as the holographic projection of the 3D bulk theory. The contact is realized in Section \ref{sec HCBFinCS} by matching the \ac{BC} on the bulk side and the \ac{EoM} on the boundary side. The resulting action involves two scalar fields and is more complicated than the simple Luttinger theory found on the edge of Chern-Simons model. We give a physical interpretation of the 2D theory, and we find three possibilities for the motion of the edge quasiparticles: same directions, opposite directions and a single-moving mode. But, requiring that the Hamiltonian of the 2D theory is bounded by below, { the case of edge modes moving in the same direction is ruled out}. We are therefore left with physical situations characterized by edge excitations moving with opposite velocities, (examples are \ac{FQHE} with $\nu=1-1/n$, with $n$ positive integer \cite{Wen:1995qn}, and Helical Luttinger Liquids phenomena  \cite{Wu2006HelicalLA}) 
 or a single-moving mode (Quantum Anomalous Hall \cite{Qi:2010qag,Liu:2008xej,Yu:2010hth}). In Section \ref{sec TRBFinCS} a strong restriction is obtained by requiring Time Reversal symmetry, which uniquely identifies modes with equal and opposite  velocities, and we know that this is the case of Topological Insulators. The novelty, with respect to the flat bulk background, is that the modes have local velocities, which corresponds to \acl{TI} with $accelerated$ edge modes. In Section \ref{sec concBFinCS} we summarize and discuss our results.

\subsection{The model : bulk and boundary}\label{sec modelBFinCS}
\subsubsection*{The action}
We consider the abelian 3D BF model on a manifold diffeomorphic to a cylinder of radius $R$. The boundary is introduced by means of a Heaviside step function in the bulk action, constraining the radial coordinate to $r\leq R$. The BF bulk action is 
\begin{equation}\label{BF-2.1}
	S_{BF}=\kappa\int d^3x\,\theta(R-r)\,\tilde\epsilon^{\mu\nu\rho}\partial_\mu A_\nu B_{\rho}\ ,
	\end{equation}
where $A_\mu(x)$ and $B_\mu(x)$ are two gauge fields with mass dimensions $[A]=[B]=1$ and $\kappa$ is a constant which will be determined by physical inputs, as we shall see later. Differently from the previous case, in this Section we decide not to fix \ac{GNC} (we will see that indeed this does not affect the final results). According to this the properties of distributions in general curved spacetimes (such as the derivative of the Heaviside step function) are given in Appendix \ref{appA}. We choose the radial gauge, implemented by the following gauge-fixing term
	\begin{equation}
	S_{gf}=\int d^3x\sqrt{-g}\,\theta(R-r)\left[\left(b A_\mu+d\; B_{\mu}\right) n^\mu\right]\ ,
	\end{equation}
where $n^\mu=(0,1,0)$ is a vector and $b(x),\ d(x)$ are scalar Nakanishi-Lautrup Lagrange multipliers \cite{Nakanishi:1966zz,Lautrup:1967zz}:
	\begin{equation}
	\frac{\delta S}{\delta b}=n^\mu A_\mu=A_r=0\quad;\quad	\frac{\delta S}{\delta d}=n^\mu B_\mu=B_r=0\ .
	\end{equation}
The external source term is
	\begin{equation}
	S_{ext}=\int d^3x\sqrt{-g}\,\theta(R-r)\,\left(J^\mu A_\mu+\hat J^\mu B_\mu\right)\ ,
	\end{equation}
where $J^\mu(x)$ and $\hat J^\mu(x)$ are vectors. The presence of a boundary induces, as an additional contribution, the most general boundary term compatible with power-counting and locality \cite{Symanzik:1981wd}
	\begin{equation}\label{BF-2.5}
	S_{bd}=\int d^3x\,\sqrt{-\gamma}\,\delta(r-R)\left(\frac{\alpha^{ij}}{2}A_i A_j+\frac{\beta^{ij}}{2}B_iB_j+\zeta^{ij}A_i B_j\right)\ ,
	\end{equation}
where $\alpha^{ij}=\alpha^{ji},\  \beta^{ij}=\beta^{ji}$ and $\zeta^{ij}$ are dimensionless tensors which depend, at most, on the induced metric (components $\gamma^{ij}$ and/or determinant $\gamma$) in the following way
	\begin{eqnarray}
	\alpha^{ij}=&\hat\alpha^{ij}({\gamma})+\hat\alpha({\gamma})\gamma^{ij}&=\hat\alpha^{ij}+\hat\alpha\gamma^{ij}\label{BF-coeff1}\\
	\beta^{ij}=&\hat\beta^{ij}({\gamma})+\hat\beta({\gamma})\gamma^{ij}&=\hat\beta^{ij}+\hat\beta\gamma^{ij}\label{BF-coeff2}\\
	\zeta^{ij}=&\hat\zeta^{ij}({\gamma})+\hat\zeta({\gamma})\gamma^{ij}&=\hat\zeta^{ij}+\hat\zeta\gamma^{ij}\ .\label{BF-coeff3}
	\end{eqnarray}
Thus, as in the previously discussed Chern-Simons case, a non-covariant boundary term has been chosen. Notice that covariance could be recovered by taking $\hat\alpha^{ij}=\hat\beta^{ij}=\hat\zeta^{ij}=0$. In the flat limit the parameters $\alpha^{ij}$, $\beta^{ij}$ and $\zeta^{ij}$ in $S_{bd}$ are constant. Finally, the total action, containing BF bulk, gauge fixing, external sources and boundary terms, is
	\begin{equation}\label{BF-2.8}
	S=S_{BF}+S_{gf}+S_{ext}+S_{bd}\ .
	\end{equation}
	
\subsubsection*{Equations of motion and boundary conditions}\label{BF-secBC}

From the total action $S$ \eqref{BF-2.8}, the \ac{EoM} of the theory follow
	\begin{align}
	\frac{\delta S}{\delta A_\lambda}&=\theta(R-r)\left[-\kappa\epsilon^{\mu\lambda\rho}\partial_\mu B_\rho+b\, n^\lambda+J^\lambda\right]+\delta^\lambda_i\tfrac{\sqrt{-\gamma}}{\sqrt{-g}}\delta(r-R)\left[-\kappa\frac{\tilde\epsilon^{i1j}}{\sqrt{-\gamma}}B_j+\alpha^{ij}A_j+\,\zeta^{ij}B_j\right]=0\label{BF-2.9}\\[5px]
	\frac{\delta S}{\delta B_\lambda}&=\theta(R-r)\left[\kappa\epsilon^{\mu\nu\lambda}\partial_\mu A_\nu+d\, n^\lambda+\hat J^{\lambda}\right]+\delta^\lambda_i\tfrac{\sqrt{-\gamma}}{\sqrt{-g}}\delta(r-R)\left[\beta^{ij}B_j+\zeta^{ji}A_j\right]=0\ .\label{BF-2.12}
	\end{align}
From the \ac{EoM} \eqref{BF-2.9}, \eqref{BF-2.12} we get the \ac{BC} of the theory by applying the operator $\lim_{\epsilon\to R}\int^R_\epsilon$, $i.e.$\\
	\begin{empheq}{align}
	\lim_{\epsilon\to R}\int^R_\epsilon dr\eqref{BF-2.9} :\qquad&\left.\alpha^{ij}A_j+\left(-\kappa\frac{\tilde\epsilon^{i1j}}{\sqrt{-\gamma}}+\,\zeta^{ij}\right)B_j\right|_{r=R}=0\label{BF-BC1}\\
	\lim_{\epsilon\to R}\int^R_\epsilon dr\eqref{BF-2.12} :\qquad&\left.\zeta^{ji}A_j+\beta^{ij}B_j\right|_{ r=R}=0\ .\label{BF-BC2}
	\end{empheq}
Eq.\eqref{BF-BC1} and \eqref{BF-BC2} can be written in a compact form as follows
	\begin{equation}\label{BF-B.16}
	\left.\Lambda^{IJ}X_J\right|_{r=R}=0\ ,
	\end{equation}
where $I,J=\{i;j\}=\{0,2;0,2\}$,
\begin{equation}
\Lambda^{IJ}\equiv\left(\begin{array}{cc}
\alpha^{ij} & \zeta^{i j}-\kappa\epsilon^{i1 j} \\
\zeta^{j i} & \beta^{ i j} \\
\end{array}\right)=\left(
\begin{array}{cccc}
\alpha^{00} & \alpha^{02} & \zeta^{00} & \zeta^{02}-\hat\kappa\\
 \alpha^{20} & \alpha^{22} & \zeta^{20}+\hat\kappa & \zeta^{22}\\
 \zeta^{00} & \zeta^{20} & \beta^{00} & \beta^{02}\\
 \zeta^{02} & \zeta^{22}& \beta^{02} & \beta^{22}
\end{array}
\right)\ , \label{BF-Lambda}
\end{equation}
and
\begin{equation}
X_J\equiv\left(\begin{array}{c}
A_j\\
B_{j}
\end{array}\right)\ ,
\label{BF-B.18}
\end{equation}
where we defined
\begin{equation}\label{BF-B.19}
\hat\kappa\equiv\frac{\kappa\tilde\epsilon^{012}}{\sqrt{-\gamma}}\ .
\end{equation}
Here again we leave $\tilde\epsilon^{012}$ explicit (instead of simply putting it equal to 1) to keep explicit the tensor nature of all quantities. For instance, in this way, it is immediate to see that $\hat\kappa$ is a scalar function. The \ac{BC} \eqref{BF-B.16} defines a linear, homogeneous system of four equations and four variables $A_i|_{r=R}$ and $B_i|_{r=R}$, for which, requiring $\det\Lambda=0$, it is possible to write three of the fields in terms of one:
	\begin{empheq}{align}
	B_\theta(X)&= -l_1B_t(X)\label{BF-bcs2'}\\
	A_\theta(X)&= -l_2B_t(X)\label{BF-bcs4'}\\
	A_t(X)&= -l_3B_t(X)\label{BF-bcs6'}\ ,
	\end{empheq}
where $l_{1,2,3}$ depend on the coefficients of the total action \eqref{BF-2.8} and therefore, in general, are local functions of the induced metric on the boundary $\gamma_{ij}(X)$. Their explicit form can be found in Appendix \ref{BF-appl} (\eqref{BF-l1}, \eqref{BF-l2} and \eqref{BF-l3}). Notice that to exclude Dirichlet-like solutions ($i.e.\ A_i|_{r=R}=B_i|_{r=R}=0$), which would trivialize the boundary 2D physics,  we must require  $l_i\neq0$ and $l_i^{-1}\neq0$.

\subsubsection*{Ward identities}

The covariant divergence of the \ac{EoM} \eqref{BF-2.9} is
\begin{align}
		\nabla_\lambda\frac{\delta S}{\delta A_\lambda}=&\nabla_\lambda\left[\theta(R-r)\left(-\kappa\epsilon^{\mu\lambda\rho}\partial_\mu B_\rho+ n^\lambda\,b+J^\lambda\right)\right]\label{BF-2.17}\\
		=&\tfrac{1}{\sqrt{-g}}\delta(r-R)\,\kappa\,\tilde\epsilon^{i1j}\partial_i B_j+\tfrac{1}{\sqrt{-g}}\partial_\lambda\left[\theta(R-r) n^\lambda\,b\sqrt{-g}\right]+\nabla_\lambda\left[\theta(R-r)J^\lambda\right]=0\ ,\nonumber
	\end{align}
where we used the \ac{BC} \eqref{BF-BC1}, the fact that  $\epsilon^{\mu\nu\rho}\nabla_\mu\partial_\nu B_{\rho}=0$, the definition of covariant derivative of the step function \eqref{BF-1.9}  and the formula for the covariant divergence \eqref{2.23'}
	\begin{equation}\label{BF-div-cov'}
	\nabla_\mu V^\mu=\frac{1}{\sqrt{-g}}\partial_\mu\left(V^\mu\sqrt{-g}\right)\ .
	\end{equation} 
By multiplying \eqref{BF-2.17} by $\sqrt{-g}$ and integrating over the coordinate normal to the boundary $r=R$, we get
	\begin{align}
		0=&\int_0^{+\infty}dr\,\left\{\delta(r-R)\,\kappa\,\tilde\epsilon^{i1j}\partial_i B_j+\partial_\lambda\left[\theta(R-r) n^\lambda\,b\sqrt{-g}\right]+\sqrt{-g}\,\nabla_\lambda\left[\theta(R-r)J^\lambda\right]\right\}\nonumber\\
		=&\left.\kappa\tilde\epsilon^{i1j}\partial_iB_j\right|_{r=R}-\cancel{b\,\sqrt{-g}|_{r=0}}+\int_0^{+\infty}dr\,\sqrt{-g}\,\nabla_\lambda\left[\theta(R-r)J^\lambda\right]\ ,\label{BF-int ward}
	\end{align}
where we adopted the convention according to which the $\theta$ components of the fields, their $\theta$-derivatives and the Lagrange multipliers vanish at $r=0$ :
	\begin{equation}
	\left.A_\theta=B_\theta=\partial_\theta A_t=\partial_\theta B_t=b=d\right|_{r=0}=0\ .
	\end{equation}
A comment here may be useful~: the invariant measure for integrating along $r$ can be identified as induced from the bulk, as follows
	\begin{equation}
	\int d^3x\sqrt{-g}\delta^{(2)}(X-X')=\int d^3x\sqrt{-g}\frac{\tilde\delta^{(2)}(X-X')}{\sqrt{-\gamma}}=\int dr\frac{\sqrt{-g}}{\sqrt{-\gamma}}=\int dr\,\sqrt{g_{rr}}\ ,
	\end{equation}
 where we used \eqref{BF-medamath}, and $X=(t,\theta)$ are the boundary coordinates. However, in the specific case of \eqref{BF-int ward}, the $\sqrt{g_{rr}}$ factor in the integration can be omitted, simply dividing \eqref{BF-2.17} by $\sqrt{g_{rr}}$. Using \eqref{BF-1.9} in \eqref{BF-int ward}, we finally find
	\begin{equation}\label{BF-ward1}
	\int_0^Rdr\,\sqrt{-g}\nabla_\lambda J^\lambda=\left(-\kappa\tilde\epsilon^{i1j}\partial_iB_j+\sqrt{-g}J^r\right)_{r=R}\ ,
	\end{equation}
which is the Ward identity corresponding to the gauge transformation of the gauge field $A_\mu(x)$, broken at its r.h.s. by the presence of the boundary. By applying the same procedure to the \ac{EoM} \eqref{BF-2.12}, with $d|_{r=0}=0$, we obtain a second broken Ward identity~:
	\begin{equation}\label{BF-ward2}
	\int_0^Rdr\,\sqrt{-g}\nabla_\lambda \hat J^\lambda=\left(-\kappa\tilde\epsilon^{i1j}\partial_iA_j+\sqrt{-g}\hat J^r\right)_{r=R}\ .
	\end{equation}
From \eqref{BF-ward1} and \eqref{BF-ward2}, going on-shell ($i.e.\  J=\hat J=0$), we find
	\begin{empheq}{align}
	\left.\tilde\epsilon^{i1j}\partial_i B_{j}\right|_{r=R}=&0\label{BF-2.26}\\
	\left.\tilde\epsilon^{i1j}\partial_i A_j\right|_{r=R}=&0\ ,\label{BF-2.27}
	\end{empheq}
which describe two conserved currents on the boundary $r=R$. The most general solutions to these equations are \cite{Nash:1983cq,Warner}
	\begin{empheq}{align}
	B_i(X)=&\partial_i\psi(X)+\delta_{i2}\hat C\label{BF-2.28}\\
	A_i(X)=&\partial_i\varphi(X)+\delta_{i2}C\ ,\label{BF-2.29}
	\end{empheq}
where $C$ and $\hat C$ are two constants and $\varphi(X)$ and $\psi(X)$ are scalar boundary fields  with zero mass dimensions, $i.e.\ [\varphi]=[\psi]=0$, which should be identified with the boundary \ac{DoF}. Being on a closed, periodic boundary, we need to specify periodicity conditions
	\begin{empheq}{align}
	A_i(t,\theta)=A_i(t,\theta+2\pi ) &\Rightarrow \varphi(t,\theta)=\varphi(t,\theta+2\pi )\label{BF-period1'}\\
	B_i(t,\theta)=B_i(t,\theta+2\pi ) &\Rightarrow \psi(t,\theta)=\psi(t,\theta+2\pi )\ .\label{BF-period2'}
	\end{empheq}
The values of the constants $C$ and $\hat C$ in \eqref{BF-2.28}, \eqref{BF-2.29} are found by applying the mean value theorem for holomorphic functions, as in the previous Section \ref{sec CSinCS}, which states that if $f$ is analytic in a region $D$, and $a\in D$, then $f(a)=\frac{1}{2\pi}\oint_{{\cal C}(a)} f$, where ${\cal C}(a)$ is a circle centered in $a$. In our case taking  for $\cal C$ the circular boundary $r=R$ centered at $r=0$ allows us to write
	\begin{equation}
	A_{\theta}(t,r=0,\theta)=\oint_{ring\; R} A_\theta(x) 
	=\cancel{\left.\varphi(t,\theta)\right|_{\theta=0}^{\theta=2\pi }} +
	 2\pi \; C\ .
	\label{BF-holo A}
	\end{equation}
From the requirement $A_{\theta}(t,r=0,\theta)=0$, it follows that
	\begin{equation}
	C=0\ .
	\label{BF-C=0}
	\end{equation}
Analogously, we also get
	\begin{equation}
	\hat C=0\ .
	\label{BF-hatC=0}
	\end{equation}

\subsubsection*{Algebra}

The generating functional of the connected Green functions $Z_c[J,\hat J]$ is defined
\begin{equation}
e^{iZ_c[J,\hat J]}=\int [dA][dB][db][dd]
e^{iS[A,B,b,d;J,\hat J]}\  ,
\label{BF-xA.36}
\end{equation}
from which the 1- and 2-points Green functions are derived
\begin{empheq}{align}
\left.\frac{\delta Z_c[J]}{\delta J^i(x)}\right|_{J=0}=&\;\langle A_i(x)\rangle\label{BF-xA.37}\\
\left.\frac{\delta^{(2)} Z_c[J]}{\delta J^i(x)\delta J^j(x')}\right|_{J=0}\equiv&\Delta_{ij}(x,x')=i\langle T(A_i(x)A_j(x'))\rangle\ ,\label{BF-xA.38}
\end{empheq}
where $T$ is the time-ordered product
\begin{equation}
\langle T(A_l(x)A_j(x'))\rangle\equiv\theta(t-t')\langle A_l(x)A_j(x') \rangle +
\theta(t'-t)\langle A_j(x') A_l(x) \rangle\ .
\label{BF-xA.39}
\end{equation}
In order to compute the propagator \eqref{BF-xA.38}, we need the following result
	\begin{equation}\label{BF-funcJ}
		\frac{\delta}{\delta J^i(x')}\nabla_\lambda J^\lambda(x)=\frac{\delta}{\delta J^i(x')}\left[\frac{1}{\sqrt{-g}}\partial_\lambda\left(J^\lambda\sqrt{-g}\right)\right]
		=\frac{1}{\sqrt{-g}}\partial_i\tilde\delta^{(3)}(x-x')\ ,
	\end{equation}
where we used \eqref{BF-div-cov'} and the relation between scalar and density Dirac delta distributions \eqref{1.5}. Taking now the functional derivative with respect to the external sources $J,\hat J$ of the  broken Ward identities \eqref{BF-ward1} and \eqref{BF-ward2}, we get\\[10px]
$\frac{\delta}{\delta J^k(x')}\eqref{BF-ward1}\;:$
	\begin{equation}\label{BF-2.30}
		\begin{split}
		\int^R_0dr \;\partial_k\tilde\delta^{(3)}(x-x')&=\left.- \kappa\tilde\epsilon^{i1j}\partial_i\frac{\delta^{(2)}Z_c}{\delta J^k(X')\delta\hat J^{j}(X)}\right|_{J=\hat J=0}\\
		\partial_k\tilde\delta^{(2)}(X-X')&=-i\kappa\tilde\epsilon^{i1j}\partial_i\langle T\left(A_k(X')B_{j}(X)\right)\rangle\\
		&=-i\kappa\tilde\epsilon^{012}\left[B_{\theta}(X),A_k(X')\right]\partial_t\theta(t-t')\ ,
		\end{split}
	\end{equation}
where we used \eqref{BF-xA.38},  \eqref{BF-xA.39} and \eqref{BF-funcJ}. Going on-shell, $i.e.$ using \eqref{BF-2.26}, we have
	\begin{equation}\label{BF-2.31}
	\tilde\epsilon^{012}\left[B_{\theta}(X),A_k(X')\right]\partial_t\theta(t-t')=\frac{i}{\kappa}\partial_k\tilde\delta^{(2)}(X-X')\ .
	\end{equation}
Setting $k=t$ in \eqref{BF-2.31} and integrating over time, we get the equal time commutator
	\begin{equation}\label{BF-2.32}
	\left[ B_\theta(X),A_t(X')\right]=0\ .
	\end{equation}
Analogously, choosing $k=\theta$ we get 
	\begin{equation}\label{BF-3.41}
	\left.\tilde\epsilon^{012}\left[B_\theta(X),A_\theta(X')\right]\right|_{t=t'}=\frac{i}{\kappa}\partial_\theta\tilde\delta(\theta-\theta')\ .
	\end{equation}
We can repeat this to find the whole current algebra\\
$\frac{\delta}{\delta \hat J^{k}(x')}\eqref{BF-ward1}\;:$\\
	\begin{equation}\label{BF-2.34}
		\begin{split}
		0=\left.\tilde\epsilon^{i1j}\partial_i\frac{\delta^{(2)}Z_c}{\delta\hat J^{k}(X')\delta\hat J^{j}(X)}\right|_{J=0}&=\tilde\epsilon^{i1j}\partial_i\langle T\left(B_{k}(X')B_{j}(X)\right)\rangle\\
		&=\tilde\epsilon^{012}\left[B_{\theta}(X),B_{k}(X')\right]\partial_t\theta(t-t')\ ,
		\end{split}
	\end{equation}
which leads to
	\begin{equation}\label{BF-2.35}
	\left.\left[ B_\theta(X),B_{k}(X')\right]\right|_{t=t'}=0\ .
	\end{equation}
In the same way, from $\frac{\delta}{\delta J^k(x')}\eqref{BF-ward2}$ we get
	\begin{equation}\label{BF-2.37}
	\left[A_\theta(X),A_k(X')\right]_{t=t'}=0\ .
	\end{equation}
Finally, from $\frac{\delta}{\delta\hat J^{k}(x')}\eqref{BF-ward2}$ we have
	\begin{empheq}{align}
	\left[A_\theta(X),B_{t}(X')\right]_{t=t'}=&0\\
	\tilde\epsilon^{012}\left[A_\theta(X),B_{\theta}(X')\right]_{t=t'}=&\frac{i}{\kappa}\partial_\theta\tilde\delta(\theta-\theta')\ .\label{BF-3.55}
	\end{empheq}
Summarizing, the equal time commutators are
	\begin{empheq}{align}
	\left[ B_\theta(X),A_t(X')\right]=&0\label{BF-2.41}\\
	\tilde\epsilon^{012}\left[B_\theta(X),A_\theta(X')\right]=&\frac{i}{\kappa}\partial_\theta\tilde\delta(\theta-\theta')\label{BF-2.42}\\
	\left[ B_\theta(X),B_{k}(X')\right]=&0\label{BF-2.43}\\
	\left[A_\theta(X),A_k(X')\right]=&0\label{BF-2.44}\\
	\left[A_\theta(X),B_{t}(X')\right]=&0\label{BF-2.45}\\
	\tilde\epsilon^{012}\left[A_\theta(X),B_{\theta}(X')\right]=&\frac{i}{\kappa}\partial_\theta\tilde\delta(\theta-\theta')\ .\label{BF-2.46}
	\end{empheq}
One can observe that by using the property of the delta function $\delta'(x)=-\delta'(-x)$, the commutators \eqref{BF-2.42} and \eqref{BF-2.46} represent the same relation. 
 Eq.\eqref{BF-2.41}-\eqref{BF-2.46} describe  a semidirect sum of \ac{KM} algebras \cite{Kac:1967jr,Moody:1966gf} with central charge
	\begin{equation}\label{BF-2.66}
	c=\frac{1}{\kappa}\ ,
	\end{equation}
which, as one can expect, inherit the topological nature of the bulk theory, being independent from the metric. Again we remark that from the positivity of the central charge, being necessary for the unitarity of the \ac{CFT} \cite{Mack:1988nf,Becchi:1988nh}, we get a constraint on the coupling constant of the BF bulk model
	\begin{equation}\label{BF-2.67}
	{\kappa}>0\ .
	\end{equation}
Typically, the constraint on the coupling constants of \acp{QFT} is derived by asking the positivity of the energy density, which is the 00-component of the energy momentum tensor $T^{00}$. This cannot be achieved in \acp{TQFT}, which have vanishing Hamiltonian, as it is well known. The constraint in that case is obtained by asking the positivity of the central charge of the edge current algebra, as in the present case, and in the previous (Section \ref{sec CSinCS}). 

\subsection{2D boundary theory}\label{sec bdBFinCS}

We now focus on the construction of the boundary theory, whose \ac{DoF} are  the boundary scalar fields $\varphi(X)$ and $\psi(X)$ defined by 
the solutions \eqref{BF-2.28} and \eqref{BF-2.29} of the conserved currents equations \eqref{BF-2.26} and \eqref{BF-2.27}. To build the 2D induced theory we follow three steps :
	\begin{enumerate}
	\item identification of the 2D canonical variables in terms of boundary fields ;
	\item derivation of the most general 2D action ;
	\item  bulk-boundary correspondence (holographic contact).
	\end{enumerate}
	
\subsubsection*{2D canonical variables}\label{BF-sec3.1}

The first step is to write the commutator \eqref{BF-2.46} in terms of the boundary fields $\varphi(X)$, $\psi(X)$, using \eqref{BF-2.28} and \eqref{BF-2.29} (with $C=\hat C=0$) :
	\begin{equation}\label{BF-A.61}
	\tilde\epsilon^{012}\left[\partial_\theta\varphi(X),\partial_{\theta'}\psi(X')\right]=\frac{i}{\kappa}\partial_\theta\tilde\delta(\theta-\theta')\quad\Rightarrow\quad\tilde\epsilon^{012}\left[\varphi(X),\partial_{\theta'}\psi(X')\right]=\frac{i}{\kappa}\tilde\delta(\theta-\theta')\ .
	\end{equation}
Analogously, from \eqref{BF-2.42} we get
	\begin{equation}\label{BF-comm2}
	\tilde\epsilon^{012}\left[\psi(X),\partial_{\theta'}\varphi(X')\right]=\frac{i}{\kappa}\tilde\delta(\theta-\theta')\ .
	\end{equation}
Both the relations \eqref{BF-A.61} and \eqref{BF-comm2} can be interpreted as canonical commutation relations of the type
	\begin{equation}\label{BF-}
	\left[q(X),p(X')\right]=i\tilde\delta(\theta-\theta')\ ,
	\end{equation}
once the following identifications are done :
	\begin{empheq}{align}
	q_1\equiv\varphi\quad;&\quad p_1\equiv\kappa\tilde\epsilon^{012}\partial_\theta\psi\label{BF-A.64}\\
	\color{black}q_2\equiv\psi\quad;&\quad\color{black}p_2\equiv\kappa\tilde\epsilon^{012}\partial_\theta\varphi\ .\normalcolor\label{BF-CV2}
	\end{empheq}
Therefore the \ac{DoF} of the 2D theory are equivalently described by either of the sets of canonical variables \eqref{BF-A.64} or \eqref{BF-CV2}.

\subsubsection*{The 2D action}\label{BF-sec2Daction}

To find the most general 2D action, we make a derivative expansion in the boundary fields $\varphi(X)$ and $\psi(X)$, compatible with power-counting (we remind that $[\varphi]=[\psi]=0$)
	\begin{align}
	S_{2D}[\varphi,\psi]&=\int d^2X\,\mathcal L_{2D}\label{BF-A.65}\\
	&=\int d^2X\,\sqrt{-\gamma}\left(a^{ij}\partial_i\varphi\partial_j\varphi+b^{ij}\partial_i\psi\partial_j\psi+c^{ij}\partial_i\varphi\partial_j\psi+d^i\partial_i\varphi+f^i\partial_i\psi+h\right)\ ,\nonumber
	\end{align}
where $a^{ij}=a^{ji},$ $b^{ij}=b^{ji}$ and $c^{ij}$ are tensors, $d^i$ and $f^i$ are vectors and $h$ is a scalar, with mass dimensions
	\begin{equation}\label{BF-m dim}
	[a]=[b]=[c]=0\quad;\quad[d]=[f]=1\quad;\quad[h]=2\ .
	\end{equation}
The coefficients appearing in \eqref{BF-A.65} may depend on the induced metric $\gamma_{ij}(X)$ and/or on the boundary fields, but not on their derivatives. The definitions of the scalar fields $\varphi(X)$ and $\psi(X)$ \eqref{BF-2.28} and \eqref{BF-2.29} are invariant under the shift transformations $\delta_s,\delta'_s$ defined as
	\begin{empheq}{align}
	\delta_{s}\varphi=&\eta\label{BF-sh-phi}\\
	\delta'_{s}\psi=&\eta'\ .\label{BF-sh-vphi}
	\end{empheq}
Consequently, the 2D lagrangian $\mathcal L_{2D}$ in \eqref{BF-A.65} must be shift-invariant as well
	\begin{equation}\label{BF-sh-S}
	\delta_s\mathcal L_{2D}=\delta'_s\mathcal L_{2D}=0\ ,
	\end{equation}
which implies
	\begin{equation}\label{BF-A.72}
	S_{2D}[\varphi,\psi]=\int d^2X\,\sqrt{-\gamma}\left(a^{ij}\partial_i\varphi\partial_j\varphi+b^{ij}\partial_i\psi\partial_j\psi+c^{ij}\partial_i\varphi\partial_j\psi+
d^i\partial_i\varphi+f^i\partial_i\psi\right)\ ,
	\end{equation}
with
\begin{equation}
	\begin{split}
\frac{\partial a^{ij}}{\partial\varphi}=\frac{\partial b^{ij}}{\partial\varphi}=\frac{\partial c^{ij}}{\partial\varphi}=0&\\
\frac{\partial a^{ij}}{\partial\psi}=\frac{\partial b^{ij}}{\partial\psi}=\frac{\partial c^{ij}}{\partial\psi}=0&\ ,
	\end{split}
\end{equation}
and 
\begin{equation}\label{BF-d,f}
	\begin{split}
&\frac{\partial d^{i}}{\partial\varphi}\partial_i\varphi+\frac{\partial f^{i}}{\partial\varphi}\partial_i\psi=0\\
&\frac{\partial d^{i}}{\partial\psi}\partial_i\varphi+\frac{\partial f^{i}}{\partial\psi}\partial_i\psi=0\ .
	\end{split}
\end{equation}
Hence, the coefficients $d^i$ and $f^i$ may still depend on the boundary fields $\varphi(X)$ and $\psi(X)$, provided that the constraints \eqref{BF-d,f} hold.
    In $S_{2D}[\varphi,\psi]$ \eqref{BF-A.72} we omitted the scalar term $h$ because, being metric-dependent only, it does not contribute to the \ac{EoM} of the boundary theory.
As we did for $S_{bd}$ \eqref{BF-2.5}, we parametrize the metric dependence of the rank-2 tensors as follows~:
	\begin{empheq}{align}
	a^{ij}&=\hat a^{ij}+\hat a\,\gamma^{ij}\label{BF-coeff-a}\\
	b^{ij}&=\hat b^{ij}+\hat b\,\gamma^{ij}\label{BF-coeff-b}\\
	c^{ij}&=\hat c^{ij}+\hat c\,\gamma^{ij}\ ,\label{BF-coeff-c}
	\end{empheq}
where the hat means dependence on the metric determinant at most. As observed for the coefficients of the boundary action $S_{bd}$ \eqref{BF-2.5}, in the flat limit the tensors appearing in the action \eqref{BF-A.72} must reduce to constant matrices, and, in particular, $c^{ij}$ to a constant symmetric matrix. The compatibility of the 2D Lagrangian in \eqref{BF-A.72} with the canonical boundary structure is ensured if the relation
	\begin{equation}\label{BF-A.73}
	\frac{\partial\mathcal L_{2D}}{\partial \dot q}=p\ ,
	\end{equation}
holds for both $q_1,p_1$ in \eqref{BF-A.64} and $q_2,p_2$ in \eqref{BF-CV2}, being equivalent descriptions of the boundary \ac{DoF}. From \eqref{BF-A.72} we have
	\begin{empheq}{align}\label{BF-}
	\frac{\partial\mathcal L_{2D}}{\partial \dot \varphi}&=\sqrt{-\gamma}\left(2a^{0i}\partial_i\varphi+c^{0i}\partial_i\psi+
d^0\right)\\
	\color{black}\frac{\partial\mathcal L_{2D}}{\partial \dot \psi}&\color{black}=\sqrt{-\gamma}\left(2b^{0i}\partial_i\psi+c^{i0}\partial_i\varphi
+f^0\right)\normalcolor\ ,
	\end{empheq}
and  the request \eqref{BF-A.73} is fulfilled if
	\begin{equation}\label{BF-A.75}
	a^{0i}=\textcolor{black}{b^{0i}=}c^{00}=d^0\textcolor{black}{=f^0}=0\quad;\quad c^{02}\textcolor{black}{=c^{20}}=
		\hat\kappa\quad;\quad (a^{22}
	,\ b^{22},\ c^{22}
	\ \mbox{free})\ ,
	\end{equation}
where $\hat\kappa(\gamma)$ has been defined in \eqref{BF-B.19} and $d^2,\ f^2$ are constrained by \eqref{BF-d,f}. Eq.\eqref{BF-A.75} represents a constraint on the metric dependence of the coefficients \eqref{BF-coeff-a}-\eqref{BF-coeff-c}, for which we have
\be
\begin{split}
a^{0i} =\hat a^{0i}+\hat a\,\gamma^{0i}&=0\quad ;\quad c^{00}=\hat c^{00}+\hat c\,\gamma^{00}=0\\
b^{0i}=\hat b^{0i}+\hat b\,\gamma^{0i}&=0\quad ;\quad c^{02}=\hat c^{02}+\hat c\,\gamma^{02}=\hat\kappa\ . \label{BF-A.78}
\end{split}
\ee
Since we do not want to impose unnecessary conditions on the induced metric $\gamma^{ij}$ (we are interested in determining if and how the 2D theory keeps memory of the bulk through the induced metric on the boundary), we must ask
	\begin{equation}\label{BF-A.79}
	\hat a=\hat b=\hat c=0\quad\Rightarrow\quad a^{ij}=\hat a^{ij}(\gamma)\quad;\quad b^{ij}=\hat b^{ij}(\gamma)\quad;\quad c^{ij}=\hat c^{ij}(\gamma)\ .
	\end{equation}
From \eqref{BF-A.75} and \eqref{BF-A.79}, we also get
\begin{equation}
\hat a^{0i}=\hat b^{0i}=\hat c^{00}=0\quad;\quad\hat c^{02}=\hat c^{20}=\hat\kappa\ .
\end{equation}
Applying \eqref{BF-A.75} and \eqref{BF-A.79} to the action $S_{2D}$ \eqref{BF-A.72}, we obtain
	\begin{equation}\label{BF-A.80}
		\begin{split}
		S_{2D}[\varphi,\psi]=\int d^2X\sqrt{-\gamma}&\left[\hat a^{22}(\partial_\theta\varphi)^2+\hat b^{22}(\partial_\theta\psi)^2+2\hat\kappa\partial_t\varphi\partial_\theta\psi+\hat c^{22}\partial_\theta\varphi\partial_\theta\psi+d^2\partial_\theta\varphi+ f^2\partial_\theta\psi\right]\ ,
		\end{split}
	\end{equation}	
where all the coefficients (but $d^2$ and $f^2$) may depend on the determinant $\gamma(X)$ of the induced metric $\gamma_{ij}(X)$, but not on its components. 
The \ac{EoM} of the action $S_{2D}[\varphi,\psi]$ are
	\begin{empheq}{align}
	\frac{\delta S_{2D}[\varphi,\psi]}{\delta\varphi}&=-\frac{1}{\sqrt{-\gamma}}\partial_\theta\left[\sqrt{-\gamma}\left(2\hat a^{22}\partial_\theta\varphi+2\hat \kappa\partial_t\psi+\hat c^{22}\partial_\theta\psi+
d^2\right)\right]=0\label{BF-eom1NC}\\
	\frac{\delta S_{2D}[\varphi,\psi]}{\delta\psi}&=-\frac{1}{\sqrt{-\gamma}}\partial_\theta\left[\sqrt{-\gamma}\left(2\hat b^{22}\partial_\theta\psi+2\hat \kappa\partial_t\varphi+\hat c^{22}\partial_\theta\varphi
	+f^2\right)\right]=0\ ,\label{BF-eom2NC}
	\end{empheq}
where we used \eqref{BF-d,f} and the fact that  $\sqrt{-\gamma}\hat\kappa$ is constant, in order to write these equations as $\frac{1}{\sqrt{-\gamma}}\partial_\theta[...]$, $i.e.$ as a $\theta$-derivative.

\subsection{Holographic contact}\label{sec HCBFinCS}

We now consider the generic solutions of the \ac{BC} \eqref{BF-bcs2'}-\eqref{BF-bcs6'}, where the bulk gauge fields $A_i(X)$ and $B_i(X)$ are now replaced by their boundary values $\partial_i\varphi(X)$ and $\partial_i\psi(X)$, defined in \eqref{BF-2.28}-\eqref{BF-2.29} and with $C=\hat C=0$ \eqref{BF-C=0}-\eqref{BF-hatC=0} :
	\begin{empheq}{align}
	\partial_\theta\psi&=-l_1\partial_t\psi\label{BF-bcs2}\\
	\partial_\theta\varphi&=-l_2\partial_t\psi\label{BF-bcs4}\\
	\partial_t\varphi&=-l_3\partial_t\psi\label{BF-bcs6}\ .
	\end{empheq}
Eq.\eqref{BF-bcs2} describes a chiral boson $\psi(X)$ moving at the 2D edge of the bulk with velocity  $v_\psi=\frac{1}{l_1}$. In the same way, by using \eqref{BF-bcs4} in \eqref{BF-bcs6}, we find that $\varphi(X)$ is a chiral boson as well, satisfying
	\begin{equation}\label{BF-bcs8}
	\partial_t\varphi-\frac{l_3}{l_2}\partial_\theta\varphi=0\ ,
	\end{equation}
moving with velocity $v_\varphi=-\frac{l_3}{l_2}$. What is important to remark now is that, differently from what happens in flat spacetime (Chapter \ref{ch CSandBF} and \cite{Blasi:2008gt,Maggiore:2017vjf,Amoretti:2014iza}) and in analogy to the case of Chern-Simons theory in curved spacetime with radial boundary (Section \ref{sec CSinCS} and \cite{Bertolini:2021iku}), on the edge of a generic bulk manifold we find two chiral bosons moving with $local$, rather than constant, velocities. In fact, both velocities, at this stage, can depend on the determinant and/or on the components of $\gamma_{ij}$ which, in general, are local quantities.
To establish the holographic contact, we consider the \ac{EoM} \eqref{BF-eom1NC}, \eqref{BF-eom2NC} with $d^2=f^2=0$ (since the \ac{BC} are homogeneously linear in the derivatives)
	\begin{empheq}{align}
	&\partial_\theta\left(2\frac{\hat a^{22}}{\hat\kappa}\partial_\theta\varphi+2\partial_t\psi+\frac{\hat c^{22}}{\hat\kappa}\partial_\theta\psi\right)=0\label{BF-eom1NC'}\\
	&\partial_\theta\left(2\frac{\hat b^{22}}{\hat\kappa}\partial_\theta\psi+2\partial_t\varphi+\frac{\hat c^{22}}{\hat\kappa}\partial_\theta\varphi\right)=0\ .\label{BF-eom2NC'}
	\end{empheq}
The holographic contact is realized by inserting into the \ac{EoM} \eqref{BF-eom1NC'}, \eqref{BF-eom2NC'} the \ac{BC} solutions \eqref{BF-bcs2}-\eqref{BF-bcs6}, which gives
\begin{empheq}{align}
-2\frac{l_2}{\hat\kappa}\hat a^{22}+2-\frac{l_1}{\hat\kappa}\hat c^{22}&=0\\
-2\frac{l_1}{\hat\kappa}\hat b^{22}-2l_3-\frac{l_2}{\hat\kappa}\hat c^{22}&=0\ .
\end{empheq}
We can write two of the three boundary parameters ($e.g.\ \hat a^{22},\hat b^{22}$) in terms of the remaining one ($\hat c^{22}$) and of the bulk coefficients ($\hat\kappa,l_i$):
	\begin{empheq}{align}
	\hat a^{22}&=+\hat\kappa\frac{1}{l_2}\left(1-\frac{l_1}{2\hat\kappa}\hat c^{22}\right)\label{BF-aHC}\\
	\hat b^{22}&=-\hat\kappa\frac{l_3}{l_1}\left(1+\frac{l_2}{2\hat\kappa l_3}\hat c^{22}\right)\ .\label{BF-bHC}
	\end{empheq}
Remember that $\hat{a}^{22}$ and $\hat{b}^{22}$, defined in \eqref{BF-coeff-a} and \eqref{BF-coeff-b}, must depend on the determinant of the induced metric only, and not on its components. On the other hand, the coefficients $l_i$ \eqref{BF-l1}-\eqref{BF-l3} may depend on both the determinant and the components of $\gamma_{ij}$. Therefore, we have to tune the parameters of the boundary action \eqref{BF-2.5} in order that $\hat{a}^{22}$ and $\hat{b}^{22}$ have the right dependence. For instance, one easy way to achieve this is to set $\hat\alpha=\hat\beta=\hat\zeta=0$ in \eqref{BF-coeff1}-\eqref{BF-coeff3}.
These two equations are consequence of the bulk (\ac{BC})-boundary (\ac{EoM}) correspondence, from which we can find out the physics of the 2D induced theory. We do this by inserting them back into the 2D action \eqref{BF-A.80} (with $d^2=f^2=0$):
	\begin{equation}\label{BF-S2DHC}
		S_{2D}[\varphi,\psi]=
		\kappa\int d^2X\tilde\epsilon^{012}\left\{\left[\tfrac{1}{l_2}(\partial_\theta\varphi)^2- \tfrac{l_3}{l_1}(\partial_\theta\psi)^2\right]-\tfrac{\hat c^{22}}{2\hat\kappa}\tfrac{l_1}{l_2}\left(\partial_\theta\varphi-\tfrac{l_2}{l_1}\partial_\theta\psi\right)^2+2\partial_t\varphi\partial_\theta\psi\right\}\ .
	\end{equation}

\subsubsection*{Physical interpretation}\label{BF-PhysInt}

The holographic contact has been imposed by inserting into the \ac{EoM} the \ac{BC} which represent two chiral bosons. Therefore we expect that the 2D theory should describe two chiral bosons as well. This fact appears evident by considering the following linear combination
	\begin{equation}\label{BF-phi+-} 
	\Phi^\pm\equiv\varphi\pm \psi\ .
	\end{equation}
In terms of these new fields the action $S_{2D}[\varphi,\psi]$ \eqref{BF-S2DHC} writes
\be
S_{2D}[\Phi^+,\Phi^-]=S_{2D}[\Phi^+]+S_{2D}[\Phi^-]+
\kappa\int d^2X\;\tilde\epsilon^{012}\partial_\theta\Phi^+\partial_\theta\Phi^-
\left(
\frac{2\hat\kappa(l_1+l_2l_3)-\hat{c}^{22}(l_1^2-l_2^2)}{4\hat\kappa l_1l_2}
\right)\ ,
\label{BF-S+-tot}\ee
where
\be
S_{2D}[\Phi^\pm]=\frac{\kappa}{2}\int d^2X\; \tilde\epsilon^{012}
\partial_\theta\Phi^\pm(\pm\partial_t\Phi^\pm + v_\pm\partial_\theta\Phi^\pm)\ ,
\label{BF-Lutt+-}\ee
and
\be
v_\pm=\tfrac{2\hat\kappa(l_1-l_2l_3)-\hat{c}^{22}(l_1\mp l_2)^2}{4\hat\kappa l_1l_2}\ .
\label{BF-vpm}\ee
The action $S_{2D}[\Phi^+,\Phi^-]$ \eqref{BF-S+-tot} decouples into the sum of the  Luttinger actions \eqref{BF-Lutt+-}
\be
S_{2D}[\Phi^+,\Phi^-]=S_{2D}[\Phi^+]+S_{2D}[\Phi^-]\ ,
\label{BF-S2D}\ee
provided that the following condition on the parameters of the theory holds
	\begin{equation}\label{BF-decoup}
2\hat\kappa(l_1+l_2l_3)-\hat{c}^{22}(l_1^2-l_2^2)=0\ .
	\end{equation}
Once decoupled, we may identify the fields $\Phi^+(X)$ and $\Phi^-(X)$ as Right (R) and Left (L) modes moving at the radial edge of the 3D bulk theory with velocities $\pm v_\pm$ respectively \cite{Kane}, where
\be
v_\pm=\frac{1\mp l_3}{l_2\pm l_1}\qquad\mbox{if}\ \ l_1^2-l_2^2\neq 0\ ,
\label{BF-}\ee
and
\be
\begin{array}{lcllcl}
v_+=\dfrac{1}{l_1}&;& v_-=\dfrac{1}{l_1}-\dfrac{\hat{c}^{22}}{\hat\kappa}&\quad\mbox{if}\ \ l_1=l_2&;& l_3=-1\label{BF-}\\
v_+=-\dfrac{1}{l_1}+\dfrac{\hat{c}^{22}}{\hat\kappa}&;&v_-=-\dfrac{1}{l_1}&\quad\mbox{if}\ \ l_1=-l_2&;&l_3=1\ .
\end{array}
\ee
As a consequence of the holographic contact, the metric dependence of the boundary parameters  \eqref{BF-A.79} is transferred to the $l_i$ coefficients through \eqref{BF-aHC} and \eqref{BF-bHC}. This makes $v_\pm$ depend on the determinant of the induced metric $v_\pm=v_\pm(\gamma)$. We therefore remark the crucial point that the fact of dealing with a curved bulk spacetime has the primary consequence that the velocities of the edge modes depend on both time and space $v_\pm=v_\pm(t,\theta)$, differently to what happens for flat backgrounds.
Hence, from $v_\pm$ we see that the edge action $S_{2D}$ \eqref{BF-S2D} may describe three classes of physical situations, tuned by the $local$ bulk parameters $l_i$ and by $\hat{c}^{22}$~:
\begin{enumerate}
\item $\pmb{v_+v_->0}$: LR movers with opposite velocities.\\
It is realized if 
\be
\frac{1-l^2_3}{l_2^2-l_1^2}>0\qquad\mbox{if}\ \ l_2^2-l_1^2\neq 0\ ,
\label{BF-}\ee
or 
\be
\hat\kappa-\hat{c}^{22}l_1>0\qquad\mbox{if}\  \ l_1=\pm l_2\ \ ,\ \ l_3=\mp 1\ .
\label{BF-}\ee
This situation describes generic chiral Luttinger liquids \cite{Wen:2004ym}, but also helical ones \cite{Wu2006HelicalLA}.
In fact
ordinary \ac{TI} \cite{moorenature,Hasan:2010xy,Hasan:2010hm,Qi:2010qag,Cho:2010rk}, characterized by edge modes moving in opposite directions with equal velocities
\be
v_+=v_-\qquad\mbox{(Topological Insulators)}\ ,
\label{BF-v+=v-}\ee
fall into this category. It is easy to see that the condition \eqref{BF-v+=v-} is satisfied provided that
\bea
l_1+l_2l_3&=&0 \label{BF-l1+l2l3=0}\\
\hat{c}^{22}&=&0\ .
\label{BF-c22=0}
\eea
The equal and opposite edge velocities therefore are
\be
v_+=v_-=\frac{1}{l_2}\ ,
\label{BF-}\ee
which still, for a generic bulk manifold, may have a spacetime dependence.
\item $\pmb{v_+v_-<0}$: LR movers in the same direction.\\
It is realized if 
\be
\frac{1-l^2_3}{l_2^2-l_1^2}<0\qquad\mbox{if}\  l_2^2-l_1^2\neq 0
\label{BF-3.59}\ee
or 
\be
\hat\kappa-\hat{c}^{22}l_1<0\qquad\mbox{if}\   l_1=\pm l_2\ \ ,\ \ l_3=\mp 1\ .
\label{BF-3.60}\ee
Also in this case we can recover the particular case of a pair of Hall systems \cite{Wen:1995qn}, with edge excitations moving in the same direction with the same velocity
\be
v_+=-v_-\qquad\mbox{(pair of Hall systems)}\ ,
\label{BF-v+=-v-}\ee
realized if
\bea
l_2+l_1l_3 &=& 0 \label{BF-}\\
\hat{c}^{22}&=& \frac{2\hat\kappa}{l_1}\ . \label{BF-}
\eea
The velocities of the edge modes in this case are
\be
v_+=-v_-=\frac{1}{l_1}\ .
\label{BF-}\ee
\item $\pmb{v_+v_-=0}$: L or R mover not moving, which characterizes the \ac{QAH} Insulators  \cite{Qi:2010qag}. This happens when
\be
l_3=\pm1\qquad\mbox{if}\ \ l_1\mp l_2\neq0\ ,
\label{BF-}\ee
which means
\be
v_\pm=0\quad ;\quad  
v_\mp=\frac{2}{l_2\mp l_1}\quad ;
\quad \hat{c}^{22}=\frac{2\hat\kappa}{l_1\mp l_2}\ .
\label{BF-v+-=0}\ee
\end{enumerate}
Some comments are in order. First of all we notice that, since the BF coupling constant $\kappa$ must be positive,
 $\hat{c}^{22}=0$ uniquely identifies L and R modes moving on the edge of the 3D bulk with opposite velocities ($\pmb{v_+=v_-}$). 
Hence, as for chiral velocities, $\hat{c}^{22}$ should be determined either by a phenomenological input or by a symmetry principle. Now, \ac{TI} belong to this class of edge excitations, and are topological phases of electrons which respect Time Reversal ($\TR$) symmetry \cite{Qi:2010qag,Cho:2010rk}. This suggests that $\hat{c}^{22}=0$ might be related to the conservation of $\TR$-symmetry and, conversely, $\hat{c}^{22}\neq 0$ to its violation, like it happens, for instance, in the case of the \ac{QAH} Insulators \cite{Qi:2010qag}, described by case 3.
We shall come back to this point in the next Subsection. 
A second comment comes from the fact that 
it is necessary that the Hamiltonian corresponding to the action \eqref{BF-S2D} is positive definite. This request yields constraints on the bulk parameters of the model, $i.e.$ the ``coupling'' constant $\hat\kappa$ \eqref{BF-B.19}, the parameters $\alpha^{ij}$ \eqref{BF-coeff1}, $\beta^{ij}$ \eqref{BF-coeff2} and $\zeta^{ij}$ \eqref{BF-coeff3} appearing in $S_{bd}$ \eqref{BF-2.5}, together with the parameter $\hat{c}^{22}$. We recall that the parameters $l_i$ depend on the bulk through \eqref{BF-l1}, \eqref{BF-l2} and \eqref{BF-l3} and appear in the action $S_{2D}[\Phi^+,\Phi^-]$ through $v_\pm$.
The canonical variables defined in \eqref{BF-A.64} and \eqref{BF-CV2} in terms of $\Phi^\pm$ \eqref{BF-phi+-} write
\begin{equation}
		\begin{split}
		q_1&=\frac{\Phi^++\Phi^-}{2}\quad;\quad p_1=\frac{\kappa\tilde\epsilon^{012}}{2}\partial_\theta\left(\Phi^+-\Phi^-\right)\\
		q_2&=\frac{\Phi^+-\Phi^-}{2}\quad;\quad p_2=\frac{\kappa\tilde\epsilon^{012}}{2}\partial_\theta\left(\Phi^++\Phi^-\right)\ .
		\end{split}
	\end{equation}
The Hamiltonian density of the model therefore is
	\begin{equation}\label{BF-T00}
		\begin{split}
		\mathcal H_{2D}&=p_1\dot q_1+p_2\dot q_2-\mathcal L_{2D}\\
		&=-\frac{1}{2\kappa\tilde\epsilon^{012}}\left[v_+(p_1+p_2)^2+v_-(p_1-p_2)^2\right]\\
		&=-\frac{\kappa\tilde\epsilon^{012}}{2}\left[v_+(\partial_\theta\Phi^+)^2+v_-(\partial_\theta\Phi^-)^2\right]\ .
		\end{split}
	\end{equation}
Positive energy density means $\mathcal H_{2D}>0$,  and since $\kappa>0$ \eqref{BF-2.67}, requiring the coefficients of the squared terms to be positive gives the following constraint
	\begin{equation}
	\mathcal H_{2D}>0\quad\Leftrightarrow\quad v_+\leq0,\ v_-\leq0\ .\label{BF-positiveH}
	\end{equation}
Therefore, we observe that the physical situation of edge modes moving in the same direction ($\pmb{v_+v_-<0}$) is not compatible with the positivity conditions \eqref{BF-positiveH}. The fact that the Hamiltonian is not bounded by below would lead us to discard this case, leaving us only with the cases 1 and 3. 

\subsection{The role of Time-Reversal symmetry}\label{sec TRBFinCS}

In this Section we present two alternative ways, a) and b), of introducing the Time Reversal $\TR$-symmetry in the theory with boundary. It will turn out that these two approaches, although seemingly quite different, are indeed equivalent. $\TR$-transformation is defined in the usual way as $\TR x^0=-x^0$. Due to the invariance of the line element $ds^2$, we have
	\begin{equation}\label{BF-Tgamma}
	\TR\gamma_{t\theta}=-\gamma_{t\theta}\footnote{Notice that if the metric is stationary, $i.e.\ \partial_t\gamma_{ij}=0$, then the $\TR$-invariance \eqref{BF-Tgamma} requires that $\gamma_{t\theta}=0$. }\ .
	\end{equation}
The only components of the gauge fields which change sign under $\TR$ are:
	\begin{equation}
		\begin{split}
		\TR A_r(t,r,\theta)&=-A_r(-t,r,\theta)\\
		\TR A_\theta(t,r,\theta)&=-A_\theta(-t,r,\theta)\\
		\TR B_t(t,r,\theta)&=-B_t(-t,r,\theta)\ .\label{BF-Tgf}
		\end{split}
	\end{equation}
According to this definition $A_\mu(x)$ may be associated to an electric potential and $B_\mu(x)$ to a spin current, like in the flat case \cite{Amoretti:2014iza}. 

\subsubsection{When $\TR S_{bd}=S_{bd}$}

It is immediate to see that the bulk action  $S_{BF}$ \eqref{BF-2.1} is $\TR$-invariant\footnote{Other choices of $\TR$ are possible which leave $S_{BF}$ invariant, like for instance $\TR A_t=-A_t,\ \TR B_r=-B_r,\ \TR B_\theta=-B_\theta$, which correspond to $A_\mu\leftrightarrow B_\mu$. One can show that these choices are equivalent \cite{Amoretti:2014iza}.}, and it is interesting to study which are the consequences of imposing Time-Reversal also on the boundary term  $S_{bd}$ \eqref{BF-2.5}, $i.e.$ 
	\begin{equation}\label{BF-Tsbd inv}
	\TR S_{bd}=S_{bd}\ .
	\end{equation}
Due to \eqref{BF-Tgamma}, to the fact that $\TR\gamma=\gamma$ and to the form of the coefficients \eqref{BF-coeff1}, \eqref{BF-coeff2} and \eqref{BF-coeff3}, the parameters appearing in $S_{bd}$ \eqref{BF-2.5} which transform non-trivially under $\TR$ are
	\begin{empheq}{align}
	\TR\alpha^{02}=&\hat\alpha^{02}-\hat\alpha\gamma^{02}\label{BF-xTAcoeff1}\\
	\TR\beta^{02}=&\hat\beta^{02}-\hat\beta\gamma^{02}\label{BF-xTAcoeff2}\\
	\TR\zeta^{02}=&\hat\zeta^{02}-\hat\zeta\gamma^{02}\label{BF-xTAcoeff3}\\
	\TR\zeta^{20}=&\hat\zeta^{20}-\hat\zeta\gamma^{20}\ .\label{BF-xTAcoeff4}
	\end{empheq}
Requiring the invariance \eqref{BF-Tsbd inv}, from \eqref{BF-Tgf} and \eqref{BF-xTAcoeff1}-\eqref{BF-xTAcoeff4}, we get the following constraints
	\begin{equation}\label{BF-Thp}
	\hat\alpha^{02}=\hat\alpha^{20}=0\quad;\quad\hat\beta^{02}=\hat\beta^{20}=0\quad;\quad\zeta^{00}=0\quad;\quad\zeta^{22}=0\quad;\quad\hat\zeta=0\ .
	\end{equation}
The resulting $\TR$-invariant boundary term $S_{bd}$ \eqref{BF-2.5} is
	\begin{equation}\label{BF-xB.23'}
		\begin{split}
		S_{bd}=\int d^3x\,&\sqrt{-\gamma}\,\delta(r-R)\left[\frac{\gamma^{ij}}{2}\left(\hat\alpha A_i A_j+\hat\beta B_iB_j\right)+\frac{\hat\alpha^{00}}{2}A_t A_t+\right.\\
		&\left.\ +\frac{\hat\alpha^{22}}{2}A_\theta A_\theta+\frac{\hat\beta^{00}}{2}B_tB_t+\frac{\hat\beta^{22}}{2}B_\theta B_\theta+\hat\zeta^{02}A_t B_\theta+\hat\zeta^{20}A_\theta B_t\right]\ ,
		\end{split}
	\end{equation}
and we recall that all the coefficients appearing in \eqref{BF-xB.23'} might still depend on the determinant $\gamma$ of the induced metric $\gamma_{ij}$, being therefore local quantities and not simply constants. It will be interesting to investigate which are the consequences, if any, of imposing $\TR$ on $S_{bd}$ on the holographically induced 2D theory.

\subsubsection*{Generic non-diagonal metric ${\gamma_{t\theta}\neq0}$}

 As a consequence of \eqref{BF-Thp}, in the hypothesis of $\gamma_{t\theta}\neq0$ and if $\hat\alpha$ and/or $\hat\beta$ are/is non-vanishing, the \ac{BC} \eqref{BF-BC1} and \eqref{BF-BC2} become
	\begin{empheq}{align}
	\left.\alpha^{00}A_t+\hat\alpha\gamma^{t\theta}A_\theta+(\hat\zeta^{02}-\hat\kappa)B_\theta\right|_{r=R}=&0\label{BF-BCT1}\\
	\left.\hat\alpha\gamma^{t\theta}A_t+\alpha^{22}A_\theta+(\hat\zeta^{20}+\hat\kappa)B_t\right|_{r=R}=&0\label{BF-BCT2}\\
	\left.\hat\zeta^{20}A_\theta+\beta^{00}B_t+\hat\beta\gamma^{t\theta}B_\theta\right|_{ r=R}=&0\label{BF-BCT3}\\
	\left. \hat\zeta^{02}A_t+\hat\beta\gamma^{t\theta}B_t+\beta^{22}B_\theta\right|_{ r=R}=&0\ ,\label{BF-BCT4}
	\end{empheq}
or, using the notation already adopted in \eqref{BF-B.16},
	\begin{equation}\label{BF-BCT5}
	\left.\Lambda_\TR^{IJ}X_J\right|_{r=R}=0\ ,
	\end{equation}
where $\Lambda_\TR$ is the matrix \eqref{BF-Lambda} evaluated at \eqref{BF-Thp}. Notice the explicit dependence on the off-diagonal component $\gamma_{t\theta}$ of the induced metric and on $\hat\alpha$ or $\hat\beta$. The linear system \eqref{BF-BCT5} has nontrivial solutions if $\det\Lambda_\TR=0$, $i.e.$
\begin{align}
0&=\det\Lambda_\TR\nonumber\\&=\hat\alpha\left(\gamma^{t\theta}\right)^2\left\{-\hat\alpha\det\beta+\hat\beta\left[ \hat\zeta^{20} \hat\kappa -  \hat\zeta^{02} \left(2   \hat\zeta^{20}+\hat\kappa \right)\right]\right\}+\\
&+ \alpha^{00} \left[ \alpha^{22} \det\beta-  \beta^{22}   \hat\zeta^{20} \left(  \hat\zeta^{20}+\hat\kappa \right)\right]+  \hat\zeta^{02}\left(  \hat\zeta^{02}-\hat\kappa \right) \left[  \hat\zeta^{20} (  \hat\zeta^{20}+\hat\kappa )-  \alpha^{22}   \beta^{00}\right]\ .\nonumber
\end{align}
It is easily seen that the solutions \eqref{BF-bcs2'}-\eqref{BF-bcs6'} are recovered, with
	\begin{equation}\label{BF-vi'}
	l_1\to l_1|_{\eqref{BF-Thp}}\quad;\quad l_2\to l_2|_{\eqref{BF-Thp}}\quad;\quad l_3\to l_3|_{\eqref{BF-Thp}}\ ,
	\end{equation}
and $l_{1,3}|_{\eqref{BF-Thp}}\propto\gamma^{t\theta}$, as shown in Appendix \ref{BF-appl}. Notice that for a diagonal (but not necessarily static) metric ($\gamma_{t\theta}=0$) or when $\hat\alpha=\hat\beta=0$, these solutions must be discarded, since they imply Dirichlet \ac{BC} on both fields ($l_{1,3}|_{\eqref{BF-Thp}}=0\Rightarrow B_\theta(X)=A_t(X)=0$), which, as we already remarked, would trivialize the 2D physics. The case of diagonal metric or $\hat\alpha=\hat\beta=0$ will be analyzed in the next Subsection. The procedure we followed to recover the 2D theory does not change: the bulk-boundary correspondences \eqref{BF-aHC} and \eqref{BF-bHC} still hold, with the replacements \eqref{BF-vi'}. 
From \eqref{BF-l1'} and \eqref{BF-l3'} we see that, due to the $\TR$-invariance request \eqref{BF-Thp}, the coefficients $l_1$ and $l_3$ appearing in the \ac{BC} \eqref{BF-bcs2'} and \eqref{BF-bcs6'}
explicitly depend on $\gamma^{t\theta}$. As a consequence, $\hat{a}^{22}$ \eqref{BF-aHC} and $\hat{b}^{22}$ \eqref{BF-bHC} would depend on $\gamma^{t\theta}$ as well, but we know that, due to the holographic contact, the coefficients appearing in the action $S_{2D}$ \eqref{BF-A.80} should depend on the determinant of the induced metric only, and not on its components. The only way to realize this is to set $\hat{c}^{22}=0$. In fact, in this case we have
\be
\hat{c}^{22}=0 \quad\Rightarrow \quad\frac{l_1|_{\eqref{BF-Thp}}}{l_3|_{\eqref{BF-Thp}}}=-l_2|_{\eqref{BF-Thp}}\ ,
\label{BF-c22 l1+l2l3}\ee
which does not depend on $\gamma^{t\theta}$, and, from \eqref{BF-aHC} and \eqref{BF-bHC}, 
	\begin{equation}\label{BF-HCT}
	\hat a^{22}=\frac{1}{l_2|_{\eqref{BF-Thp}}}\hat\kappa\quad;\quad\hat b^{22}=-\frac{l_3|_{\eqref{BF-Thp}}}{l_1|_{\eqref{BF-Thp}}}\hat\kappa\quad \ .
	\end{equation}
Eq. \eqref{BF-c22 l1+l2l3} coincides with Eqs.\eqref{BF-c22=0} and \eqref{BF-l1+l2l3=0}, which belong to the case 1 considered in the previous Section, where we have seen that the physical situation described by the decoupled action $S_{2D}[\Phi^+,\Phi^-]$ \eqref{BF-S2D} together with the conditions \eqref{BF-c22=0} and \eqref{BF-l1+l2l3=0}, is that of
a Luttinger model for two chiral currents with non-constant and opposite velocities. We thus established a link between the parameter  $\hat c^{22}$ and $\TR$-invariance on the boundary, which enforces the physical interpretation as edge states of \ac{TI}, as anticipated in the previous Section. 

\subsubsection*{Diagonal metric $\gamma_{t\theta}=0$}\label{BF-secTdiag}

When considering a diagonal metric or boundary coefficients \eqref{BF-coeff1}-\eqref{BF-coeff3} which depend at most on the determinant of the metric ($i.e.$ when $\gamma_{t\theta}=0$ or $\hat\alpha=\hat\beta=0$), the \ac{BC} \eqref{BF-BCT1}-\eqref{BF-BCT4} become
	\begin{empheq}{align}
	\left.\alpha^{00}A_t+(\hat\zeta^{02}-\hat\kappa)B_\theta\right|_{r=R}=&0\label{BF-BCTs1}\\
	\left.\alpha^{22}A_\theta+(\hat\zeta^{20}+\hat\kappa)B_t\right|_{r=R}=&0\label{BF-BCTs2}\\
	\left.\hat\zeta^{20}A_\theta+\beta^{00}B_t\right|_{ r=R}=&0\label{BF-BCTs3}\\
	\left. \hat\zeta^{02}A_t+\beta^{22}B_\theta\right|_{ r=R}=&0\ ,\label{BF-BCTs4}
	\end{empheq}
which represent two systems of homogeneous linear equations. 
Non-Dirichlet solutions are
	\begin{empheq}{align}
	A_t(X)=&-l_aB_\theta(X)\label{BF-xBCT1}\\
	A_\theta(X)=&-l_bB_t(X)\ ,\label{BF-xBCT2}
	\end{empheq}
where
\be
l_a=\frac{\beta^{22}}{\hat\zeta^{02}}\quad\mbox{and}\quad l_b=\frac{\beta^{00}}{\hat\zeta^{20}}\ ,
\label{BF-}\ee
provided that
\bea
\left(\hat\zeta^{02}-\hat\kappa\right)\hat\zeta^{02}-\alpha^{00}\beta^{22} &=& 0 \label{BF-}\\
\left(\hat\kappa+\hat\zeta^{20}\right)\hat\zeta^{20}-\alpha^{22}\beta^{00} &=&0\ .\label{BF-}
\eea
Following the same steps described in Section \ref{sec HCBFinCS}, we still land on the action $S_{2D}$  \eqref{BF-A.80}. The holographic contact is realized crossing the \ac{EoM} \eqref{BF-eom1NC} and \eqref{BF-eom2NC} with the \ac{BC} \eqref{BF-xBCT1} and \eqref{BF-xBCT2}, which in terms of the boundary fields $\varphi(X)$, $\psi(X)$ read~:
	\begin{empheq}{align}
	\partial_t\varphi=&-l_a\partial_\theta\psi\label{BF-bcsT1}\\
	\partial_\theta\varphi=&-l_b\partial_t\psi\ .\label{BF-bcsT2}
	\end{empheq}
The correspondence is achieved if
	\begin{equation}\label{BF-HClambT'}
	\hat a^{22}=\frac{1}{l_b}\hat\kappa\quad;\quad\hat b^{22}=l_a\hat\kappa\quad;\quad\hat c^{22}=0\ .
	\end{equation}
From the properties of the tensors $\hat a^{ij},\hat b^{ij},\hat c^{ij}$ in \eqref{BF-A.79}, it is immediate to check that $l_{a,b}$ depend only on the determinant of the induced metric $\gamma$, $i.e.\ l_{a,b}=l_{a,b}(\gamma)$. The 2D action \eqref{BF-A.80} decouples into a pair of Luttinger models, provided that the following condition holds
\begin{equation}
l_a=\frac{1}{l_b}\ ,
\end{equation}
and describes two chiral modes travelling on the edge of the 3D bulk with equal and opposite local velocities $v_\pm=l_a(\gamma)$
	\begin{empheq}{align}
	\partial_t\Phi^++l_a\partial_\theta\Phi^+&=0\label{BF-bcsT1+}\\
	\partial_t\Phi^--l_a\partial_\theta\Phi^-&=0\ ,\label{BF-bcsT2-}
	\end{empheq}
characterizing \ac{TI} \cite{Hasan:2010xy,Qi:2010qag,Cho:2010rk}. \\

\subsubsection{Inherited $\TR$-transformation}\label{BF-T2D}

It is possible to impose the $\TR$-symmetry on the theory with boundary in an alternative way with respect to what we did in the previous Subsection, reaching the same physical conclusions ($i.e.$ chiral edge modes moving with opposite velocities as the unique physical outcome of putting a boundary onto 3D BF theory). The bulk gauge fields $A_\mu(x)$ and $B_\mu(x)$ transform under $\TR$ according to \eqref{BF-Tgf}. Consequently, the boundary scalar fields $\psi(X)$ and $\varphi(X)$, being defined by \eqref{BF-2.28} and \eqref{BF-2.29}, should inherit the following $\TR$-transformations
\begin{equation}\label{BF-Tphi}
	\TR\varphi(t,\theta)=-\varphi(-t,\theta)\quad;\quad \TR\psi(t,\theta)=\psi(-t,\theta)\ ,
\end{equation}
hence, due to \eqref{BF-phi+-}, we have
\begin{equation}\label{BF-Tphi+-}
	\TR\Phi^+(t,\theta)=-\Phi^-(-t,\theta)\quad;\quad \TR\Phi^-(t,\theta)=-\Phi^+(-t,\theta)\ .
\end{equation}
The action $S_{2D}[\Phi^+,\Phi^-]$ \eqref{BF-S+-tot} is $\TR$-invariant if
\be
v_+=v_-\ ,
\label{BF-}\ee
and if the decoupling condition \eqref{BF-decoup} holds. This is precisely the situation considered in case 1 treated in the physical interpretation of Section \ref{sec HCBFinCS}, uniquely identified by the vanishing of the parameter $\hat{c}^{22}$ \eqref{BF-c22=0}. \\

We therefore checked our guess on the peculiar role played by the parameter $\hat{c}^{22}$ appearing in the action $S_{2D}[\varphi,\psi]$ \eqref{BF-A.80}, and also that the two alternative ways described in Subection \ref{sec TRBFinCS} of imposing the $\TR$-symmetry (on the boundary term $S_{bd}$ \eqref{BF-2.5} or directly on $S_{2D}[\varphi,\psi]$ through the defining relations \eqref{BF-2.28} and \eqref{BF-2.29}) are indeed physically equivalent. 
Lastly, we remark that the physical situation represented by the \ac{TI}, $i.e.$ the existence of chiral edge modes moving with equal and opposite velocities, is singled out 
by imposing $\TR$-symmetry (in either way), while asking that the Hamiltonian ${\cal H}_{2D}$ is bounded by below, as discussed in  Section \ref{sec HCBFinCS}, admits a larger class of physical situations, including $\TR$-breaking effects, but only for opposite-moving modes. The possibility of edge modes moving in the same direction is ruled out by both argumentations, and $\TR$-symmetry is more restrictive than the positive energy condition. According to this analysis, case 3 is not compatible with $\TR$-symmetry, however lower-bounded Hamiltonian does admit the possibility of a single-moving edge state. Hence it can be interpreted as a $\TR$-breaking effect associated to \ac{QAH} insulators \cite{Qi:2010qag}. In the same way the situations with $\hat c^{22}\neq0$ belonging to case 1 can be seen as other examples of symmetry breaking effects in Quantum Spin Hall systems \cite{Hasan:2010xy,Qi:2010qag,Wu2006HelicalLA}.\\

It is known \cite{Qi:2010qag,Cho:2010rk} that $\TR$-symmetry is peculiar to the  edge states of \ac{TI} in 2D, which are described by a helical Luttinger  model \cite{Wu2006HelicalLA}. This is exactly the situation we observed for $\hat c^{22}=0$ \eqref{BF-c22=0}. Therefore we may finally claim that BF model with boundary together with $\TR$-invariance is an effective description of the edge states of \ac{TI} with possibly non-constant chiral velocities. The existence of accelerated edge modes is a direct consequence of the bulk/boundary correspondence in curved spacetime~: flat spacetime analysis provides only for constant velocities. The $\hat c^{22}\neq0$ situation now can be related to $\TR$-breaking phenomena in helical Luttinger liquids \cite{Wu2006HelicalLA}. For instance (as we already remarked) we can associate the case 3 of Section \ref{BF-PhysInt} to the \acl{QAH} Effect \cite{Qi:2010qag,Liu:2008xej,Yu:2010hth}.

\subsection{Summary of results and discussion}\label{sec concBFinCS}

In this Section we considered the abelian 3D BF model on a manifold described by a generic metric, with a radial boundary which spoils the topological character of the theory. Following a method introduced by Symanzik, we added to the action a boundary term constrained only by locality and power counting, in order to find out the most general \ac{BC} on the two gauge fields involved in the theory. The boundary breaks gauge invariance, and this reflects in two broken Ward identities, from which we identified the boundary degrees of freedom, represented by two scalar fields, deriving also a semidirect sum of two \acl{KM} algebras formed by two on-shell conserved currents. As in the Chern-Simons case of Section \ref{sec CSinCS}, the central charge of the \acl{KM} algebra is proportional to the inverse of the BF coupling, and this constrains the BF coupling to be positive. Once written in terms of the scalar boundary degrees of freedom, the \acl{KM} algebra can be interpreted as the commutation relation of canonical variables, which led us to derive the corresponding 2D action. The bulk/boundary holographic correspondence is achieved by matching the \ac{BC} on the 3D gauge fields and the equations of motion of the 2D action. The resulting 2D action is a complicated functional of the pair of scalar fields but, by  means of a simple linear redefinition, the action can be written in terms of two Luttinger actions for two chiral fields, plus a mixed term, which we force to vanish by imposing a decoupling condition. At this point, the parameters which survived to the holographic contact and to the decoupling condition allow for interesting physical interpretations of the theory, which may indeed describe~:
\begin{enumerate}

\item two edge excitations moving in opposite directions. This is realized in 
Hall systems, like \acl{FQHE} with $\nu=1-1/n$ \cite{Wen:1995qn}, and edge modes of \acl{QSH} systems, like \acl{TI} (when $v_+=v_-$), possibly interacting \cite{Calzona}, or nanowires \cite{meng} with additional magnetic fields acting on the velocities up to switching one off \cite{Streda,Heedt}. In higher dimensions an effect of renormalization of chiral velocities ($i.e.\ v_+v_->0$) can be achieved by adding magnetic fields \cite{bernevig,goerbig2}, or by structural deformations \cite{goerbig} ;

\item two chiral bosons moving in the same direction. This physically corresponds to Hall systems like, for instance, Quantum Hall with $\nu=2$, or \acl{FQHE} with $\nu=2/5$ \cite{Bocquillon,Wen:1995qn,ferraro} possibly with non-constant interactions or confining potentials \cite{Wen:1989mw,Wen:1990qp,Kane95,Hashi18,Wen:1991ty} ;

\item one static and one moving mode. This is the situation of 
\acl{QAH}, where magnetic impurities break the $\TR$-symmetry of \acl{TI} \cite{Qi:2010qag,Liu:2008xej,Yu:2010hth}. This could also be explained as  an extremal effect of a magnetic field acting on the modes of the nanowires mentioned above.
\end{enumerate}
In all cases described above, the chiral velocities depend on the induced metric on the boundary, hence on time and space. Therefore, their local nature is a direct consequence of the non-flatness of the bulk metric, which thus appear as a general property of (abelian) \acp{TQFT} in curved spacetimes. Moreover, we calculated the Hamiltonian corresponding to the 2D action and we found that the request of the existence of a lower bound rules out the case 2~: Hall systems with parallel velocities, like Fractional Quantum Hall with $\nu=2/5$ \cite{Wen:1995qn} or Integer Quantum Hall with $\nu=2$ \cite{ferraro}, cannot be described by a BF theory with boundary. We then considered the role of Time Reversal symmetry, which distinguishes the two main Schwarz-type \acp{TQFT} : $\TR$-violating Chern-Simons and $\TR$-respecting BF. $\TR$-symmetry can be introduced in two different ways~: from the bulk side asking that the ``Symanzik's'' boundary term of the 3D action is $\TR$-invariant, which has consequences on the subsequent steps till the holographically induced 2D action. $\TR$-symmetry may be imposed also on the boundary side directly on the 2D action, starting from the definition of the scalar degrees of freedom. These two seemingly inequivalent ways of requiring $\TR$-invariance lead to the same outcome~: the only physical case which respects $\TR$-symmetry is a subclass of case 1 above, namely the one involving two edge excitations moving in opposite directions { with the same velocity}, which, in the general case of non-Minkowskian bulk metric, { might be local}, $i.e.$ time and space dependent. This is the case of \acl{TI}. Hence, imposing $\TR$-symmetry is much more restrictive than asking for a lower bounded Hamiltonian. Topological phases of matter displaying accelerated edge modes have been observed for instance in the Integer Quantum Hall with $\nu=2$ \cite{Bocquillon}. But, to our knowledge, generalized Topological Insulators with accelerated chiral edge modes have not been discovered yet. Here we thus predict that they should, and we presented a theoretical framework for their existence, as a direct consequence of a non-Minkowskian bulk background. In our opinion, this represents a cleaner alternative to adding an \textit{ad hoc} local potential to the pair of the decoupled Luttinger actions, which would spoil the whole holographic construction.

\cleardoublepage 

\ctparttext{\sloppy In most cases, \aclp{QFT} are considered without boundaries and have been successful in providing descriptions of fundamental interactions, including gravity and cosmology. This is because one is generally interested in bulk effects, where the boundary can be neglected. Nevertheless, boundaries do exist, and in some cases, their effects are self-evident and dominant. As we have seen in the previous Chapters, recently, important phenomena pertaining to condensed matter physics, like the \acl{FQHE} and the behaviour of \acl{TI}, have been explained in terms of topological \acp{QFT} with boundaries. This is rather counterintuitive~: topological \acp{QFT}, when considered without boundaries, have a vanishing Hamiltonian and no energy-momentum tensor. They might appear as the least physical theories one can imagine. Despite this, when a boundary is introduced, an extremely rich physics emerges, which can be observed experimentally. This is an example of the power of the boundary, and for a long time it has been believed that such power was peculiar only of these kind of theories. Until now...

}
\fussy
\part{Non-Topological Field Theories}\label{partIII}

\chapter{An overview for the non-Topological case} 

\label{ch nonTFT} 


\newcommand{\ka}{\kappa}
\newcommand{\kat}{\tilde{\kappa}}
\newcommand{\al}{\alpha}
\newcommand{\alt}{{\tilde{\alpha}}}
\newcommand{\ga}{\gamma}
\newcommand{\gat}{{\tilde\gamma}}
\newcommand{\la}{\lambda}
\newcommand{\lat}{\tilde{\lambda}}
\newcommand{\La}{\Lambda}
\newcommand{\Lat}{\tilde{\Lambda}}
\newcommand{\bet}{{\tilde{\beta}}}

\newcommand{\T}{\mathcal{T}}
\newcommand{\At}{\tilde{A}}
\newcommand{\Ft}{\tilde{F}}
\newcommand{\Gt}{\tilde{G}}
\newcommand{\xit}{\tilde{\xi}}
\newcommand{\phit}{\tilde{\phi}}
\newcommand{\Lag}{\mathcal{L}}
\newcommand{\Sa}{S^{^{(1)}}_{3D}}
\newcommand{\Sb}{S^{^{(2)}}_{3D}}
\newcommand{\Sc}{S^{^{(3)}}_{3D}}

Since the introductory part of Chapter \ref{ch intro}, we have encountered many cases of physical results induced by the presence of boundaries. The Casimir effect \cite{Casimir:1948dh} perhaps is the first highly nontrivial example of boundary effect which has been thoroughly studied in a systematic way. The role of boundaries has been largely discussed in 2D Conformal Field Theory \cite{Moore:1989yh, Cardy:2004hm}. In particular, in \cite{Moore:1989yh} the zoo of Conformal Field Theories has been tamed by means of a boundary put on the 3D topological Chern-Simons theory. In \cite{Cardy:2004hm}, instead, the role of the boundary, and in particular of the \ac{BC}, has been exploited for the study of the Virasoro algebras and their extensions (\acl{KM}, superconformal, W-algebras).  We have seen that \acp{TQFT} have vanishing Hamiltonian and energy-momentum tensor, and it is rather surprising that for such non-physical theories it has been possible to establish \cite{Blasi:1990pf,Blasi:1990bk,Emery:1991tf}
that on their lower dimensional edge, conserved currents exist, which form \ac{KM} algebras 
\cite{Kac:1967jr,Moody:1966gf}, whose central charge is inversely proportional to the coupling constant of the bulk theory, and directly related to the velocity of the boundary propagating \ac{DoF}. This property seems to be a common feature of different physical situations, like the 3D \ac{FQHE} \cite{Cappelli:2018dti, Blasi:2008gt} 
and \ac{TI} in 3D \cite{Cho:2010rk,Cappelli:2016xwp,Blasi:2011pf} and 4D \cite{Schnyder:2008tya,Fu:2006djh,Amoretti:2012hs}. 
We have also seen that in the Symanzik's approach, the boundary separates two half-spaces~: left and right hand side with respect to a plane. Single-sided boundaries can also be considered, as shown throughout this whole work, which correspond to quite different physical situations :  in these cases, the Symanzik's separation requirement on propagators is not the most natural one. It is easier, and more intuitive, as we discussed, to implement the confinement of the theory in a half-space by means of a theta Heaviside step function directly introduced in the bulk action.  The theories with singe-sided boundaries are therefore treated with a different approach, which was explained in Chapter \ref{QFTapproach}, and the physical results on the boundary do not necessarily coincide with those of the separating boundaries. This is the case, for instance, of 3D Maxwell-Chern-Simons theory, where the Maxwell term is completely transparent in case of double-sided boundary \cite{Blasi:2010gw} and algebraically active in the single-sided case \cite{Maggiore:2018bxr,Geiller:2019bti}. It is precisely the different role of the non-topological Maxwell term that motivated the study of non-\acp{TQFT}, or theories with non-topological terms, defined in half-spaces \cite{Blasi:2019wpq,Wang:2017vwr}. In fact, all the results mentioned until now were obtained for  boundaries in \acp{TQFT}~: 3D Chern-Simons and BF theories  \cite{Birmingham:1991ty, Horowitz:1989ng, Karlhede:1989hz}. What is missing is the study of physical, realistic, hence entirely non-topological, theories, defined on a half-space, and it is intriguing to investigate the role of the boundary in these cases~: which are the edge \ac{DoF}? are there conserved currents, like in the topological cases? do they form an algebra? of which type? is there a lower-dimensional holographic counterpart of the non-topological bulk theory? is this unique? what kind of physics does it describe? The first example which came to mind was of course that of the 4D Maxwell theory of electromagnetism \cite{Bertolini:2020hgr}, whose nontrivial results, which we will review shortly in this Chapter, motivated the present Part of the Thesis. We will thus dwell into more exotic cases, like fracton theories \cite{Nandkishore:2018sel,Pretko:2020cko}, or models which nowadays are of high interest, like \acl{LG}, with the aim of answering the above questions and possibly acquire some phenomenological insights.

\subsection*{The case of the 4D Maxwell theory with boundary : a new beginning}\numberwithin{equation}{chapter}
The Minkowskian 4D Maxwell theory can be confined in the half-spacetime $x^3\geq 0$ by means of the introduction in the action of the Heaviside step function $\theta(x^3)$
		\begin{equation}\label{eq:Smax}
		S_M=-\frac{\ka}{4}\int d^4x\ \theta(x_3)\; F_{\mu\nu}F^{\mu\nu}\ ,
		\end{equation}
where $F_{\mu\nu}=\partial_\mu A_\nu-\partial_\nu A_\mu$ is the electromagnetic field strength, and $A_\mu(x)$ is the gauge field, with canonical mass dimension  $[A]=1$. In \eqref{eq:Smax}  $\ka>0$ is a constant which must be positive in order to have a positive-definite energy density. Maxwell theory, being a free field theory, does not display a coupling constant, which can always be reabsorbed by redefining the gauge field $A_\mu(x)$. Nonetheless, here again, we do not normalize $\kappa$ to one, in order to be able to identify at any time the role played by the bulk action in the physics on the boundary. The gauge fixing term		
\begin{equation}\label{eq:gauge-inv}
S_{gf}=\int d^4x\ \theta(x^3)\; bA_3
\end{equation}
implements, through the Lagrange multiplier field $b(x)$ \cite{Nakanishi:1966zz,Lautrup:1967zz}, the axial gauge condition
\begin{equation}
		A_3(x)=0\ .
\label{gaugecond}\end{equation}	
On the boundary $x^3=0$, the fields and their $\partial_3$-derivatives must be treated as independent fields \cite{Karabali:2015epa, Maggiore:2019wie}. To highlight this fact, we adopt the following notation :
\begin{equation}\label{At}
		\At_a(X)\equiv\left.\partial_3A_a\right|_{x_3=0}\ ,
\end{equation}	
whose mass dimension is $[\At]=2$. Therefore we must introduce another term in the action, coupling these two independent fields, $A_\mu(x)$ and, on the boundary, $\At_a(X)$, to the external sources $J^\mu(x)$ and $\tilde J^a(X)$ respectively:
		\begin{equation}
		S_J=\int d^4x\ \left[ \theta(x^3)J^a A_a+\delta(x^3)\tilde{J}^a \At_a \right]\ .
		\end{equation}
The existence of the boundary requires an additional  contribution to the action, subject to the general constraints of locality, power counting and 3D covariance. Taking into account the presence of an additional independent gauge field $\tilde A_3(X)$ \eqref{At}, the most general boundary term has to be the following
\begin{equation}\label{eq:Sbd}
				S_{bd}=\int d^4x\ \delta(x_3)\;
				\left(a^{ab}A_a A_b+b^{abc}\partial_a A_b A_c+c^{ab}\At_a A_b\right)\ ,
\end{equation}
where 
\begin{equation}
		a^{ab}=a^{ba},\quad b^{abc}=-b^{acb},\quad c^{ab}
\label{bdpar}\end{equation}
are constant matrices, with mass dimensions $[a^{ab}]=1,\ [b^{ab}]=[c^{ab}]=0$. Notice that here again we do not require covariance, however, we shall see that in this case the theory will require it in order to recover an induced physics. The total action, consisting in bulk term, gauge fixing, external sources and boundary contribution, finally is
\begin{equation}\label{eq:Stot}
		S_{tot}=S_M+S_{gf}+S_J+S_{bd}\ .
	\end{equation}
Broken Ward identities can be recovered from both the gauge field and its $\partial_3$-derivative as follows
		\begin{align}
		\int^{+\infty}_{0}dx^3\; \partial^a J_a=&\ka\left.\partial^a\At_a\right|_{x^3=0}	\label{eq:wi1}\\
		\partial^a\tilde{J}_a|_{x^3=0}=&-\ka\left.\partial^aA_a\right|_{x^3=0}\ ,\label{eq:wi2}
		\end{align}
noticing that the Ward identity \eqref{eq:wi2}, differently from \eqref{eq:wi1}, is local and not integrated. Both identities  are broken, because of the presence of the boundary, by a linear term at their r.h.s.,  and, at vanishing external sources $\tilde J=J=0$, $i.e.$ going on shell, we find
		\begin{empheq}{align}
		\partial^a \At_a|_{x^3=0}&=0\label{subeq:cc1}\\
		\partial^a A_a|_{x^3=0}&=0\ ,\label{subeq:cc2}
		\end{empheq}
which show the existence of a couple of conserved currents on the 3D edge of 4D Maxwell theory. From the broken Ward identities \eqref{eq:wi1} and \eqref{eq:wi2}, through the procedure of Section \ref{sec brokensymm}, $i.e.$ by taking functional derivatives with respect to the currents $\tilde J^a(x),\ J^a(x)$, we also get the following equal time commutators for the conserved currents $A_a(X)$ and $\At_a(X)$, which are of \ac{KM} type
\begin{empheq}{align}			 
		[\At_0(X),A_\textsc{m}(X')]&=-\frac{i}{\ka}\partial_\textsc{m}\delta^{(2)}(X-X')\label{eq:[At0,Ai]}\\
		[A_0(X),\At_\textsc{m}(X')]&=\frac{i}{\ka}\partial_\textsc{m}\delta^{(2)}(X-X')\label{eq:[A0,Ati]}\\
		[\At_0(X),\At_a(X')]&=[A_0(X),A_a(X')]=[\At_0(X),A_0(X')]=0\ .\label{eq:[]0}
		\end{empheq}
In order to identify the correct \ac{DoF} on the 3D boundary, it is convenient to introduce a field $B_a(X)$ defined by the linear transformations
\begin{equation}
\begin{split}
B_0\equiv &\ \mu A_0+\nu\At_0 \\
B_\textsc{m}\equiv&\ \rho A_\textsc{m}+\sigma\At_\textsc{m}	\ ,		
\label{defB}\end{split}
\end{equation}
where $\mu$, $\nu$, $\rho$ and $\sigma$ are constant parameters, which can be set later at our convenience, with mass dimensions constrained by the request of the dimensional homogeneity of \eqref{defB}
\begin{equation}
			[\mu]=[\nu]+1\ ,\qquad[\rho]=[\sigma]+1\ ,\qquad[\mu]=[\rho]\quad\mathrm{and}\quad[\nu]=[\sigma]\ .
\label{dimpar}\end{equation}
In terms of $B_a(X)$, the algebra \eqref{eq:[At0,Ai]}-\eqref{eq:[]0} reduces to the only nonvanishing commutator
\begin{equation}
[B_0(X),B_\textsc{m}(X')] =\;i\;\frac{\mu\sigma-\nu\rho}{\ka}\;\partial_\textsc{m}\delta^{(2)}(X-X')\ ,
\label{eq:B0,Bi}
\end{equation}
which describes an abelian \ac{KM} algebra exactly like \eqref{[j,j]} described in Section \ref{sec brokensymm}, and whose central charge is proportional to the inverse of the Maxwell coupling $\kappa$~:
\begin{equation}
	\frac{1}{\kat}\equiv\frac{\mu\sigma-\nu\rho}{\ka}\ .\label{eq:condiz1}
\end{equation}
Therefore this shows that also for non-\ac{TQFT} an algebraic structure of \ac{KM} type might exist on the boundary. The 3D current conservation relations \eqref{subeq:cc1} and \eqref{subeq:cc2} can be solved by
\begin{empheq}{align}			 
		&\At_a(X)=\epsilon_{abc}\partial^b \tilde{\xi}^c(X)\label{eq:xit}\\
		&A_a(X)=\epsilon_{abc}\partial^b\xi^c(X)\ ,\label{eq:xi}
		\end{empheq}
where the fields $\tilde{\xi}_a(X)$ and $\xi_a(X)$ have canonical dimensions 
		\begin{equation}
		[\tilde{\xi}]=1\quad\mathrm{and}\quad[\xi]=0\ .\label{eq:dim-xi}
		\end{equation}	
Consequently, we have 
\begin{empheq}{align}
B_0=&\epsilon_{0\textsc{mn}}\partial^\textsc{m}(\mu\xi^\textsc{n}+\nu\xit^\textsc{n})=\epsilon_{0\textsc{mn}}\partial^\textsc{m}\lambda^\textsc{n}\label{eq:B0-bd}\\
		B_\textsc{m}=&\epsilon_{\textsc{m}ab}\partial^a(\rho\xi^b+\sigma\xit^b)
		=\epsilon_{\textsc{m}ab}\partial^a\tilde{\lambda}^b\label{eq:Bi-bd}\ ,			 
		\end{empheq}
where we defined
		\begin{empheq}{align}			 
			\la_a&\equiv\mu\xi_a+\nu\xit_a\label{eq:la-bd}\\
			\lat_a&\equiv\rho\xi_a+\sigma\xit_a\label{eq:lat-bd}\ .		 
		\end{empheq}
The equations \eqref{eq:la-bd} and \eqref{eq:lat-bd} define the 3D vector fields  $\la_a(X)$ and $\lat_a(X)$ which, as we shall show in what follows, are the dynamical variables in terms of which the 3D theory induced on the boundary of 4D Maxwell theory will be constructed, according to the \ac{QFT} procedure of Section \ref{holographic123}.  Notice that the defining relations \eqref{eq:B0-bd} and \eqref{eq:Bi-bd} are left invariant under the transformations 
\begin{empheq}{align}				 
				\lambda_a\quad\to&\quad\lambda_a+\partial_a\Lambda\label{eq:dla}\\
				\lat_b\quad\to&\quad\lat_b+\partial_b\tilde{\Lambda}\ ,\label{eq:dlat}				 
\end{empheq}
where $\La(X)$ and  $\Lat(X)$ are local gauge parameters. We may therefore claim that the 3D theory induced on the boundary of 4D Maxwell theory should be a gauge theory of two, possibly coupled, gauge fields, which must satisfy the three constraints of Section~\ref{holographic123}~: 
\begin{enumerate}
\item[1.] invariance under the gauge transformations
\begin{empheq}{align}
		&\delta_1\la_a(X)=\partial_a\La(X)\label{eq:gt-1}\\
		&\delta_2\lat_a(X)=\partial_a\Lat(X)\label{eq:gt-2}\ ;
\end{empheq}
\item[2.] compatibility with the equal time \ac{KM} algebra \eqref{eq:B0,Bi}\ ; 
\item[3.] compatibility with the \ac{BC} of the bulk\ .
\end{enumerate}
Following this procedure a candidate shows up to be represented by the following 3D action
\begin{equation}\label{S3DMAX}
		S_{3D}=\int d^3X\biggl(-\frac{\kat}{2}G_{ab}\Gt^{ab}+\frac{\kat}{2}\frac{\nu}{2\sigma}\Gt_{ab}\Gt^{ab}+m\epsilon^{abc}\lat_a\partial_b\lat_c\biggr)\ ,
		\end{equation}
where
	\be
	G_{ab}\equiv\partial_a\lambda_b-\partial_b\lambda_a\quad;\quad\tilde G_{ab}\equiv\partial_a\tilde\lambda_b-\partial_b\tilde\lambda_a\quad;\quad 	m\propto \frac{a^c_{\ c}}{3\rho}\ ,
	\ee
with $a^c_{\ c}$ the trace of a coefficient of the boundary term \eqref{eq:Sbd}, and whose \ac{EoM} are
		\begin{empheq}{align}
		&\partial_b\Gt^{ab}=0 \label{4.34}\\
		&\partial_b G^{ab}+\frac{2 m}{\tilde\kappa}\epsilon^{abc}\partial_b\lat_c=0  \label{4.35}\ .
		\end{empheq}
As we anticipated, notice that in order to get a contact it has been necessary to set
	\be
	a^{\alpha\beta}\propto\eta^{\alpha\beta}\quad;\quad b^{\alpha\beta\gamma}=0\quad;\quad c^{\alpha\beta}\propto\eta^{\alpha\beta}\ ,
	\ee
meaning that the boundary action term $S_{bd}$ \eqref{eq:Sbd} \textit{must} be covariant. That's exactly the opposite to what happens in \acp{TQFT} with boundary, where the request of covariance restricts the possible physical results, or even, as for the Chern-Simons case, prevents to get anything at all, and non-covariance is mandatory. \\

Until now we would have thus said that the most known and also physically relevant role played by boundaries concerned \acp{TQFT}, in particular in 3D and 4D. This fact constitutes a kind of interesting paradox~: \acp{TQFT}, indeed,  are characterized by global observables of geometrical type only, vanishing Hamiltonian, no energy-momentum tensor and lack of particle interpretation. The main point which should be stressed is therefore that 4D Maxwell theory shows a non trivial boundary dynamics, which thus is not peculiar to \ac{TQFT}, contrary to what was usually believed. There are however similarities and differences with respect to \ac{TQFT} :
\begin{itemize}
\item 
On the boundary of 4D Maxwell theory the broken Ward identities \eqref{eq:wi1} and \eqref{eq:wi2} are found, which identify two conserved currents \eqref{subeq:cc1} and \eqref{subeq:cc2}. This reminds the physics of the surface states of \acl{TI} in 3D, which suggests that  the 4D Maxwell theory  might be seen as an effective bulk theory of the 3D Topological Insulators, alternative to the 4D topological BF models \cite{Cho:2010rk}. 
\item
A \acl{KM} algebra \eqref{eq:B0,Bi} with a central charge  proportional to the inverse of the Maxwell coupling exists on the boundary, and the parameters appearing in \eqref{defB} correspond to different central charges, as represented by \eqref{eq:condiz1}, each identifying a different \ac{CFT}.
This is an important difference with respect to \acp{TQFT}, which are characterized by a one-to-one correspondence between bulk coupling constants and central charges.   The relevant boundary algebra appears to be formed by the subset \eqref{defB} of the total number of components of the bulk fields. An identical mechanism  occurs in the topological twist of N=2 Super Yang-Mills Theories \cite{Witten:1988ze}. 
\item 
The 3D action \eqref{S3DMAX} induced by the 4D bulk Maxwell theory is unique and corresponds to a new theory, which has not been studied previously. 
\item
such action describes two coupled photon-like vector fields, with a topological Chern-Simons term for one of them. We computed the propagators of the theory \cite{Bertolini:2020hgr} which show that, despite the similarity with the 3D Maxwell-Chern-Simons theory, a mechanism of topological mass generation does not take place in this case.
\item
By computing the energy-momentum tensor of the theory, a non trivial physical content is revealed \cite{Bertolini:2020hgr}. In particular,  the coefficients appearing in the 3D action can be tuned in order to have a positive definite energy density. 
\item
The holographic dictionary \cite{Zaanen:2015oix} might be improved by an additional entry involving the unitarity of the \ac{CFT} found on the boundary of 4D Maxwell theory and the positivity of the energy density of its 3D holographic counterpart, represented by the action \eqref{S3DMAX}. In fact, asking that the 00-component of the energy-momentum tensor is positive, automatically implies that the central charge of the \ac{KM} algebra \eqref{eq:B0,Bi} is positive as well, thus ensuring the unitarity of the corresponding \ac{CFT}.
\end{itemize}
Each of these points is a proof of the relevance of studying non-topological \acp{QFT} with boundary : we have at least six reasons not to stop here, but to proceed to Chapter \ref{ch fractons} and to a new non-topological theory with boundary.

\chapter{Fractons}
\label{ch fractons} 
\numberwithin{equation}{section}
Fracton phases of matter represent a novel paradigm both in condensed matter theory and high energy physics \cite{Nandkishore:2018sel,Pretko:2020cko}. Although they have originally been discovered in particular kinds of lattice models \cite{Chamon:2004lew,Haah:2011drr,Vijay:2016phm,Vijay:2015mka}, fractons have been unveiled in many different systems and frameworks, ranging from elasticity \cite{Pretko:2017kvd,Gromov:2019waa,Caddeo:2022ibe,Nguyen:2020yve}, hydrodynamics \cite{Gromov:2020yoc,Yuan:2019geh,Wang:2019mtt,Grosvenor:2021rrt,Glodkowski:2022xje} and quantum scars \cite{Pai:2019rfq,Khemani:2019vor,Sala:2019zru}, to \aclp{QFT} \cite{Seiberg:2020wsg,Pretko:2016lgv,Pretko:2016kxt,Bulmash:2018lid,Shirley:2017suz,Ma:2017aog,Ma:2018nhd,Williamson:2018ofu,Slagle:2020ugk,Slagle:2017wrc,Slagle:2018swq}, curved space \cite{Slagle:2018kqf,Bidussi:2021nmp,Jain:2021ibh,Tsaloukidis:2023bvz} and holography \cite{Yan:2018nco}. These phases are characterized by constrained dynamics, for which quasiparticle excitations are immobile or move in sub-dimensional spaces. This exotic behaviour can be encoded into conservation of multipole moment, the simplest example being the dipole. Indeed,  typically fracton models are described in terms of a non-covariant higher-rank tensor theory which shares many similarities with Maxwell theory \cite{Pretko:2016lgv,Pretko:2016kxt,Pretko:2017xar}. They are written in terms of a rank-2 symmetric tensor field $A_{ij}(x)$, whose conjugate momentum is referred to as an ``electric-like'' tensor field $E^{ij}(x)$, which plays a key role for the immobility constraint of fractons, which is recovered from a generalized Gauss law.  These novel gauge theories are intrinsically non-relativistic, and many ingredients have been introduced by hand in order to implement the main characteristics of fractons, $i.e. $ their restricted mobility. Examples of these inputs are, for instance, the Maxwell-like Hamiltonian and the Gauss law, introduced as an external constraint, and not derived from an action principle. As a consequence, terms appear with inhomogeneous numbers of derivatives, as remarked in \cite{Brauner:2022rvf}, and all these $ad\ hoc$ introductions are justified $a\ posteriori$, rather than deduced from first principles of \ac{QFT}. Despite their intrinsic non-covariance, these models share remarkable similarities both with Maxwell theory (Gauss law, Hamiltonian, electric-like field...) \cite{Pretko:2016lgv} and \ac{LG} (symmetric rank-2 field, gauge symmetry,...) \cite{Pretko:2017fbf}, which however are fully covariant theories. Motivated by these similarities, in the following Sections a covariant 4D fracton gauge theory will be built \cite{Blasi:2022mbl,Bertolini:2022ijb,Bertolini:2023juh}, taking as unique ingredients locality, power counting and the covariant fracton symmetry $\delta A_{\mu\nu}=\partial_\mu \partial_\nu \Lambda$, thus keeping strict the main spirit of this Thesis : everything must come from first principles of \ac{QFT}. This procedure will make immediately apparent the correspondence with \ac{LG}, and all the analogies with Maxwell theory will naturally come out. Embedding the ordinary non-covariant theory of fractons in a larger covariant theory will indeed  lead to recover all the known results concerning fractons \cite{Pretko:2016lgv,Pretko:2016kxt,Pretko:2017xar}. Therefore, the coherent theoretical framework of Section \ref{sec MaxThFract}, and \cite{Bertolini:2022ijb}, will allow to apply standard \ac{QFT} techniques to fractons. As a physically relevant case, in in the context of this Thesis is the introduction of a flat spatial boundary in this newly built 4D covariant fracton model. We have indeed just seen, in Chapter \ref{ch nonTFT}, that on the boundary of Maxwell theory in 4D \cite{Bertolini:2020hgr} a \ac{KM} algebra is observed, and a non-trivial theory is induced on the lower-dimensional space. Fractons seem to share many relations with the electromagnetic theory, as we shall see, but this similarity is not the only motivation for studying fractons with boundary. It is in fact known that the topological $\theta$-term \cite{tong}, which is a boundary term, generates the well known Witten effect \cite{Witten:1979ey}, relevant for instance in topological insulators \cite{Rosenberg:2010ia}. 
For what concerns fractons, boundary contributions have been introduced mainly as non-covariant Chern-Simons-like  terms coming from a generalized topological-like $\theta$-term in the bulk \cite{Pretko:2017xar,Burnell:2021reh,You:2019ciz}, inspired by the standard electromagnetic case. For instance in \cite{Pretko:2017xar}, as for the Maxwell case, a Witten-like effect is observed, for which, due to the presence of the fractonic $\theta$-term, the ``electric''  charge density at the right hand side of the Gauss law acquires an additional ``magnetic''  contribution. Furthermore on the 3D boundary of \cite{Pretko:2017xar} fractonic-like excitations seem to appear, in agreement with our results, as we shall see. 
Moreover, in \cite{Pretko:2020cko,You:2019bvu} it has been speculated that certain kinds of higher-order topological phases share some properties with boundary states of fracton quasiparticles.
Another interesting example can be found in \cite{Cappelli:2015ocj}, where a similar non-covariant Chern-Simons-like  term, built as a higher-spin generalization of the standard topological one, gives some insights in the context of dipolar behaviours of quantum Hall systems.
Therefore, boundary effects might be important also in the framework of fracton theory, and the aim of this Chapter is to study the consequences of the presence of a flat boundary in the covariant theory of fractons, following the \ac{QFT} approach pioneered by Symanzik in \cite{Symanzik:1981wd}. \\

In order to do so, in this Chapter, after giving a brief overview about fractons and their main properties in Section \ref{sec literature}, in Section \ref{sec MaxThFract} we will build the new covariant theory, which will be based on a \ac{QFT} perspective. The gauge structure of the newly found model will then be studied in Section \ref{sec gauge}, and the propagators will be computed in the vectorial gauge. Finally in Section \ref{sec Frac+bd}, having established that the theory is well defined, we will introduce a boundary and analyze its consequence as prescribed by our \ac{QFT} approach.

\section{What are fractons?}\label{sec literature}

Fractons are quasiparticles with the defining property of having restricted mobility \cite{Nandkishore:2018sel,Pretko:2020cko,Chamon:2004lew,Haah:2011drr,Vijay:2016phm,Pretko:2016lgv,Pretko:2016kxt,Pretko:2017xar,rasmussen,Pretko:2018jbi,Seiberg:2020wsg}. In particular ``true'' fractons cannot move at all, while other quasiparticles which can be traced back to fractons can move only in a subdimensional space, like ``lineons'', which move on a line  and ``planons'', which move on a plane \cite{Pretko:2020cko,Seiberg:2020wsg}. The first observations of a fracton-like behaviour appeared in lattice spin models \cite{Chamon:2004lew}, and since then many developments followed \cite{Haah:2011drr,Vijay:2016phm,Yoshida:2013sqa,Vijay:2015mka,Ma:2017aog,Shirley:2017suz}. In particular, in recent years there has been a surge of interest which have come to the forefront of modern condensed matter theory \cite{Prem:2017kxc,Pretko:2016kxt,Pretko:2016lgv,Pretko:2018jbi,Nandkishore:2018sel,Pretko:2020cko,Seiberg:2020wsg,Chamon:2004lew,Vijay:2015mka,Vijay:2016phm,Pretko:2017xar,Prem:2017qcp,Haah:2011drr,Bravyi:quantum,Shi:2017qdx,Slagle:2018kqf,Argurio:2021opr,pretkoscreening,Gromov:2018nbv, Gorantla:2022eem}, since these new kind of quasiparticles represent a new class of phases of matter, whose main property is that of being immobile in isolation, but may move by forming bound states. Fractons are found in a variety of physical settings, such as spin liquids \cite{Pretko:2016lgv} and elasticity theory \cite{Pretko:2017kvd,Pretko:2019omh}, and exhibit unusual phenomenology, such as gravitational physics and localization \cite{Xu:2006,Gu:2009jh,Xu:2010eg,Pretko:2017fbf}. Fractonic behaviours are to be connected with exotic global symmetries \cite{Seiberg:2020wsg,Vijay:2016phm,Gromov:2018nbv}. Systems with these symmetries challenge the common lore by which the low energy behaviour of every lattice system can be described by a continuum quantum field theory~: these lattice constructions are usually not Lorentz invariant, and they present an unusual ground state degeneracy, infinite in the continuum limit. Significant efforts have been made to better understand their non-standard behaviours and to extend the known theoretical frameworks to include them \cite{Pretko:2018jbi,Pretko:2020cko,Seiberg:2020wsg,Gromov:2018nbv,Gorantla:2022eem,Blasi:2022mbl}. Moreover, this inability to move is the main property shared by all fracton models, and ultimately can be of physical interest, for instance, in the development of quantum memories \cite{Haah:2011drr,Bravyi:quantum,Terhal,Ma:2017igk}. Lattice models describing fractons fall into two classes, depending on their particle content~:
``type I'', the most representative of which is the X-cube model \cite{Vijay:2016phm}, has both fractons and 1,2-dimensional particles, while
``type II'', of which the Haah's code \cite{Haah:2011drr} is the prototypical example, describes fractons only.
Fractons, like \acf{LG}, can be described by a gauge theory of a symmetric tensor field, which generalizes the ordinary Maxwell electromagnetism for a vector field \cite{Pretko:2016lgv}. This class of fracton theories, which can be related to the previous lattice models via a Higgs-like mechanism \cite{Ma:2018nhd,Bulmash:2018lid}, were first introduced to describe gravity-related phenomena \cite{Gu:2006vw,Xu:2006, Gu:2009jh,Xu:2010eg}, and later were developed into the actual fracton theory \cite{Pretko:2016lgv,Pretko:2016kxt,Pretko:2017xar,rasmussen}.
Written in terms of a rank-2 symmetric tensor field $A_{ij}(x)$ ($i,j...$ spatial indices), the typical starting point is a Maxwell-like Hamiltonian density
	\be
	H=E^{ij}E_{ij}+B^{ij}B_{ij}\sim E^2+B^2\ ,
	\ee
where the ``electric'' field $E^{ij}(x)$ is the conjugate momentum of $A_{ij}(x)$ as in standard electromagnetism, and the ``magnetic'' field $B_{ij}(x)$ is defined as the gauge invariant object depending on the lowest possible number of derivatives of $A_{ij}(x)$ \cite{Pretko:2017xar}. From these definitions, generalized Maxwell equations follow  \cite{Pretko:2016lgv}~: Faraday's equation is a relation between $E_{ij}(x)$ as conjugate momentum and the time derivative of $B_{ij}(x)$, while Amp\`ere law is a Hamilton's equation for $E_{ij}(x)$. Finally, the usual Gauss theorem in this picture is not really a theorem, but, rather, is imposed as a constraint. 
The gauge transformation of $A_{ij}(x)$    is crucial since, besides determining $B_{ij}(x)$, it is strictly related to the Gauss-like constraint, which has a key role in implementing the restricted mobility of fractons \cite{Nandkishore:2018sel,Pretko:2016lgv,Pretko:2016kxt,Pretko:2017xar,rasmussen,Pretko:2018jbi}. There are two possibilities \cite{Nandkishore:2018sel,Pretko:2020cko,Pretko:2016lgv,Pretko:2016kxt,Pretko:2017xar,Pretko:2018jbi}~:
\begin{itemize}

\item \textbf{  scalar charge theory :} the Gauss constraint and the gauge transformation of $A_{ij}(x)$ are
\bea
\partial_i\partial_jE^{ij} &=&0 \label{FMAX-dedeE=0}\\
\delta A_{ij} &=&\partial_i\partial_j\Lambda \label{FMAX-deA=dedelambda}\ .
\eea
In the presence of fractonic matter one can define a charge density operator $\rho(x)$ \cite{Pretko:2018jbi} and the Gauss constraint generalizes to \cite{Pretko:2016kxt}
\be
\partial_i\partial_jE^{ij}=\rho\ .
\label{FMAX-dedeE=rho}\ee
The restricted mobility becomes evident, since \eqref{FMAX-dedeE=rho} implies charge and dipole \mbox{($p^i(x)=x^i\rho(x)$)} neutrality when integrated on an infinite volume $V$ (or up to boundary terms) \cite{Nandkishore:2018sel,Pretko:2020cko,Pretko:2016lgv,Pretko:2016kxt,Pretko:2017fbf}
\bea
\int dV\partial_i\partial_jE^{ij}=\int dV\rho &=&0 \label{FMAX-constr1}\\
-\int dV\partial_iE^{ik}=\int dVx^k\rho=\int dVp^k &=&0\ .\label{FMAX-dipoleneutr}
\eea
Eq. \eqref{FMAX-dipoleneutr} states that
single charges cannot move (fractons) due to dipole conservation, while dipoles do.

\item \textbf{  vector charge theory :} in this case the Gauss constraint involves one derivative instead of two, and the gauge transformation of $A_{ij}(x)$ is a non-covariant version of the diffeomorphisms transformation
\bea
\partial_iE^{ij} &=&0\label{FMAX-}\\
\delta A_{ij} &=& \partial_i\Lambda_j+\partial_j\Lambda_i\ . \label{FMAX-}
\eea
The nature of the charge density operator thus changes, becoming a vector  $\rho^j(x)$  \cite{Pretko:2018jbi},  while the ``electric field'' remains a symmetric rank-2 tensor and the generalized Gauss constraint is \cite{Pretko:2016kxt}
\be
\partial_iE^{ij}=\rho^j\ ,
\label{FMAX-}\ee
which immediately implies \eqref{FMAX-dedeE=rho} together with the conservation laws \eqref{FMAX-constr1} and \eqref{FMAX-dipoleneutr}, but yields a further mobility constraint due to conservation of angular momentum           
			\begin{equation*}
			\int dV\epsilon_{0ijk}x^i\rho^j=-\int dV\epsilon_{0ijk}E^{ij}=0\ .
			\end{equation*}
Hence, the vector charges of the system can move only along a line, which is the spatial direction related to the charged vector itself, thus being 1-dimensional particles (a.k.a. lineons).
\end{itemize}
Mobility can be further restricted in both scalar and vector charge models by adding tracelessness $E^i_{\,i}=0$ of the ``electric field'' as an additional constraint. In these cases the models are referred to as ``traceless (scalar/vector) fracton models''. In the scalar case the elementary charges can still be identified as fractons, since their motion is already maximally constrained by means of the generalized Gauss law, however now tracelessness also implies the conservation of a component of the quadrupole momentum, due to which the dipoles of the system are bound to move only on a plane transverse to their direction, thus becoming 2-dimensional particles (also called ``planons''). On the contrary, in the vector charge model the quasiparticles can move in one dimension, and the tracelessness constraint on the electric-like field completely restricts their motion making them proper fractons. These models, properties and particle content can be summarized by the following table
\begin{table}[H]
	\resizebox{1\columnwidth}{!}{%
		\bgroup
		\setlength\tabcolsep{15pt}
		\def\arraystretch{1.5}{
			\begin{tabular}{|c|c|c|c|}
				\hline
				\textbf{The model}&\textbf{(Gauss) constraint}&\textbf{(integrated) conserved quantities}&\textbf{particle content}\\
				\hline
				\multirow{2}{*}{Scalar charge th.} & \multirow{2}{*}{$\partial_i\partial_jE^{ij}=\rho$} &$\rho$ charge&charge = fracton\\[-5px] 
				&&$\vec x\rho$ dipole &dipole = free\\
				\hline
				\multirow{2}{*}{Treceless scalar charge th.} & {$\partial_i\partial_jE^{ij}=\rho$} &$\rho\ ;\ \vec x\rho$
&charge = fracton\\[-5px] 
				&$E^i_{\ i}=0$&$x^2\rho$ $\sim$ quadrupole &dipole = planon\\
				\hline
				\multirow{2}{*}{Vector charge th.} & \multirow{2}{*}{$\partial_iE^{ij}=\rho^j$} &$\vec\rho$ charge ($\sim$dipole)&\multirow{2}{*}{charge = lineon}\\[-5px] 
				&&$\vec x\wedge\vec\rho$ angular momentum &\\
				\hline
				\multirow{2}{*}{Traceless vector charge th.} & {$\partial_iE^{ij}=\rho^j$} &$\vec\rho\ ;\ \vec x\wedge\vec\rho$&\multirow{2}{*}{charge = fracton}\\[-5px] 
				&$E^i_{\ i}=0$&$\vec x\cdot\vec\rho$ ($\sim$quadrupole)$\ ;\ (\vec x\cdot\vec\rho)\vec x+\frac 1 2 x^2\vec\rho$ &\\
				\hline

			\end{tabular}
		\egroup}
	}
	\label{}
	\caption[Fracton phenomenology]{\footnotesize{Fracton phenomenology.}}
\end{table}
\noindent
In Maxwell theory the $A_0(x)$ component of the gauge field $A_\mu(x)$ is a multiplier enforcing the standard Gauss constraint $\vec\nabla\cdot\vec E(x)=0$. Following this, in fracton models the Gauss constraint \eqref{FMAX-dedeE=0} is implemented by introducing a Lagrange multiplier, as done for instance in \cite{Pretko:2017xar}, sometimes called $A_0(x)$ to enhance the Maxwell analogy. This multiplier seems to have no relation with the $A_{ij}(x)$ tensor field and in addition, due to this ``by hand'' implementation, the Lagrangian acquires an inhomogeneous number of (spatial) derivatives \cite{Pretko:2017xar}. For these reasons, despite all the similarities we mentioned, while Maxwell theory has a natural covariant formulation, the above construction of fracton models appear to be intrinsically non-covariant. We shall see that the covariant formulation presented in the next Section will make all these ad hoc implementations come out naturally as \ac{QFT} principles, and the nature and properties of fracton quasiparticles will emerge completely.

\section{Maxwell Theory of Fractons} \label{sec MaxThFract}

In this Section we will show that the main results concerning fractons, in particular the existence of tensorial electric and magnetic fields, the Gauss constraints, the Maxwell-like Hamiltonian and the dipole response to ``electromagnetic'' fields through a ``Lorentz force'', to cite a few, are indeed  consequences of a $covariant$ action for a symmetric rank-2 tensor field $A_{\mu\nu}(x)$, invariant under the covariant extension of the fracton transformation \eqref{FMAX-deA=dedelambda}
\be
\delta_{fract}A_{\mu\nu}=\partial_\mu\partial_\nu\Lambda\ ,
\label{FMAX-fractonsymintro}\ee
which therefore plays, as usual in quantum field theory, a central role. We shall also show that from the gauge tensor field $A_{\mu\nu}(x)$ it is possible to construct a rank-3 tensor $F_{\mu\nu\rho}(x)$ which we may call the fracton field strength, invariant under \eqref{FMAX-fractonsymintro} and satisfying a kind of geometrical Bianchi identity. Quite surprisingly, the fracton action can be written in terms of the fracton field strength as $\int F^2$, as the ordinary Maxwell theory, and all the above mentioned equations characterizing fractons are nothing else than the ``Maxwell'' equations, without need of introducing any external constraint or particular request, and therefore are just consequences of the covariant symmetry \eqref{FMAX-fractonsymintro}. Moreover, electric and magnetic tensor fields emerge naturally, and
in terms of these the action and the energy density read, respectively, $\int (E^2 - B^2)$ and $(E^2 + B^2)$. Finally, the Lorentz force for fracton dipoles derived ``by intuition'' ($sic$) in \cite{Pretko:2016lgv} is here recovered as part of the conservation law for the stress-energy tensor. As a matter of fact, the covariant generalization described in this Section makes apparent that the fracton theory is, indeed, a direct extension of the standard electromagnetic theory which can be formulated covariantly according to the typical field theory chain
$$\begin{array}{ccccc}
\mbox{symmetry} &\rightarrow& \mbox{action} &\rightarrow& \mbox{equations of motion} \\
\delta_{fract}A_{\mu\nu}=\partial_\mu\partial_\nu\Lambda(x)  
&\rightarrow&
-\frac{1}{6}\int d^4x\;F^{\mu\nu\rho}F_{\mu\nu\rho}
&\rightarrow&
\partial_\mu F^{\alpha\beta\mu}=0\ ,
\end{array}
$$
which really appears as a higher rank extension of Maxwell theory.\\

The relation between fractons and gravitons has been already remarked \cite{Nandkishore:2018sel,Pretko:2020cko,Blasi:2022mbl,Pretko:2017fbf}. From the field theory point of view this is evident from the covariant extension \eqref{FMAX-fractonsymintro} of the fracton symmetry, which is a particular case of the stronger infinitesimal diffeomorphism transformation \cite{Blasi:2022mbl}
\be
\delta_{diff}A_{\mu\nu}=\partial_\mu\Lambda_\nu+\partial_\nu\Lambda_\mu\ .
\label{FMAX-diffintro}\ee
In practice, this means that, while the diff symmetry \eqref{FMAX-diffintro} uniquely defines the \ac{LG} action, the most general action invariant under \eqref{FMAX-fractonsymintro} is formed by two separately invariant terms~: the \ac{LG} action and the fracton action $\int F^2$, which is quite peculiar since, to our knowledge, this is the only case of a Lorentz invariant action which, although free and quadratic, shows a dimensionless constant which cannot be eliminated by a field redefinition, and which cannot be identified as a physical mass, like in topologically massive 3D gauge theory \cite{Deser:1981wh}. Coherently with this covariant picture, both fractons and \ac{LG} can be given an electromagnetic formulation (the first, as discussed, in terms of tensors, while the second, known as gravitoelectromagnetism \cite{Mashhoon:2003ax,Carroll:2004st,Chatzistavrakidis:2020wum}, involves vectors), but, as we shall see, they also share
 the ``Gauss'' constraint \eqref{FMAX-dedeE=0} (which is not an external constraint in our formalism) which underlies the fracton limited mobility property. As a first step towards the study of boundaries, the analogy with ordinary electromagnetism can be pushed further through the 
topological $\theta$-term that can be added to the Maxwell Lagrangian (see for instance \cite{tong} and references therein).
The role of an analogous boundary term has been studied in the case of fractons in \cite{Pretko:2017xar} and, as for dyons, the result is that  the ``electric'' charge gains an additional contribution related to a ``magnetic'' vector charge \cite{Nandkishore:2018sel,Pretko:2020cko,Pretko:2017xar}. As the standard Witten effect has consequences in condensed matter physics, this higher rank version of the fracton $\theta$-term might give interesting results in the context of higher order topological phases \cite{You:2019bvu}, which seem to be related to boundary effects in fractons, as we shall discuss in Section \ref{sec Frac+bd}. The case of a local, instead of constant, $\theta(x)$ is relevant in axion models \cite{Peccei:1977hh,Peccei:1977ur}, where Maxwell equations acquire additional contributions \cite{Sikivie:1983ip,Wilczek:1987mv}. In \cite{Chatzistavrakidis:2020wum} a local $\theta(x)$-term has been added to \ac{LG}, with mainly two consequences : 
a correction to the Newtonian gravitational field and a Witten-like effect for gravitational dyons, in which ``gravitipoles'' \cite{Zee:1985xqg} acquire mass. The results of \cite{Chatzistavrakidis:2020wum} for \ac{LG} suggest the possibility of generalizing what has been found in \cite{Pretko:2017xar} for fractons, since both \ac{LG} and fractons are described by a rank-2 symmetric tensor field.\\

There are of course, and fortunately, a few open questions, which deserve further efforts. The first is that we were not able to find a symmetry which separates fractons from gravitons. In other words~: while the diff symmetry \eqref{FMAX-diffintro} uniquely defines the \ac{LG} action, the fracton symmetry \eqref{FMAX-fractonsymintro} gives two invariant functionals. One must necessarily buy gravitons, together with fractons. The way out in field theory is to recover the fracton action for vanishing \ac{LG} constant, but, still, it would be more satisfying to pick up the fracton action by means of an additional symmetry. Moreover~: the fracton symmetry \eqref{FMAX-fractonsymintro} (but also the original \eqref{FMAX-deA=dedelambda}) is dimensionally problematic. In 4D the rank-2 tensor $A_{\mu\nu}(x)$ has mass dimension one, both in fracton and \ac{LG} theory. Hence, to be dimensionally homogeneous, the fracton gauge transformation would require a scalar gauge parameter $\Lambda(x)$ with negative dimension
	\begin{equation}
	d=4\quad\Rightarrow\quad[A]=1\quad;\quad\delta A_{\mu\nu}=\partial_\mu\partial_\nu\Lambda\quad\Rightarrow[\Lambda]=-1\ ,
	\end{equation}
but this would not be the case in 6D, which would be the most ``natural'' spacetime dimensions for fractons to live in
	\begin{equation}
	d=6\quad\Rightarrow\quad[A]=2\quad;\quad\delta A_{\mu\nu}=\partial_\mu\partial_\nu\Lambda\quad\Rightarrow[\Lambda]=0\ .
	\end{equation}
 As we shall see, this reflects also in the fact that, in 4D, the stress-energy tensor is not traceless, hence the theory is not scale invariant, differently from the classical Maxwell theory. Instead, tracelessness is recovered in 6D. \\[5px]
The above construction of a covariant theory for fractons will be presented in this Section as follows :
\etocsetnexttocdepth{2}

\begingroup
\parindent=0em
\etocsettocstyle{\rule{\linewidth}{\tocrulewidth}\vskip1.25\baselineskip}{\vskip-0.75\baselineskip\rule{\linewidth}{\tocrulewidth}\vskip1\baselineskip}
\makeatletter
  \edef\scr@tso@subsection@indent
    {\the\dimexpr\scr@tso@subsection@indent-\scr@tso@section@indent}
  \def\scr@tso@section@indent{0pt}
\makeatother
\localtableofcontents 
\endgroup
\noindent
In particular Section \ref{sec Fract+LG} we derive the theory defined by the fracton symmetry \eqref{FMAX-fractonsymintro}, composed by two terms, fracton and \ac{LG}. We then construct the rank-3 fracton field strength $F_{\mu\nu\rho}(x)$ and we show that both the fracton and the \ac{LG} actions can be written in terms of this tensor, which satisfies an identity analogous to the Bianchi one. We then compute the canonical momentum $\Pi^{\alpha\beta}(x)$ associated to $A_{\alpha\beta}(x)$ and derive the field equations of motion. In Section \ref{sec FracMax} we consider the fracton theory only, obtained by putting the \ac{LG} constant to zero. Without imposing any external constraint, we recover the main properties of the fractons simply from the equations of motion, which, written in terms of the electric and magnetic tensor fields,  impressively reminds the ordinary Maxwell equations.  In Section \ref{sec T+F} we derive the stress-energy tensor and physically identify its components, which are the higher rank extensions of their Maxwell counterparts. We then write, interpret and discuss the stress-energy tensor conservation laws. We then add matter to the theory, represented by a symmetric rank-2 tensor coupled to $A_{\alpha\beta}(x)$, and extend the previously found results in presence of matter. The most important achievement of this Section is the expression of the Lorentz force which describes how dipole fractons respond to the electromagnetic tensor fields. This Lorentz force coincides with the one conjectured in \cite{Pretko:2016lgv}. Finally, in Section \ref{sec theta} we add to the fracton action the generalized $\theta$-term, which, again, can be written both in terms of the electromagnetic tensor fields and of the fracton field strength, in complete analogy with Maxwell theory. We  recover and generalize previous results obtained in the context of \ac{LG} \cite{Chatzistavrakidis:2020wum} and of fractons \cite{Pretko:2017xar} giving to the $\theta$-parameter a local dependence. Some final remarks can be found in Section \ref{sec FinalFracton}.

\subsection{Fractons and linearized gravity}\label{sec Fract+LG}

\subsubsection*{The symmetry}

    We adopt the standard point of view of field theory, that is to consider the symmetry as the birth certificate of a theory. In our case, the symmetry,  hereinafter ``fracton'' symmetry,  is the covariant generalization of the extended electromagnetic transformation \eqref{FMAX-deA=dedelambda} invoked in \cite{Pretko:2016lgv,Pretko:2016kxt,Pretko:2017xar,rasmussen} for fractons, $i.e.$ 
\begin{equation}\label{FMAX-dA}
	\delta_{fract} A_{\mu\nu}=\partial_\mu\partial_\nu\Lambda\ ,
\end{equation}
where $A_{\mu\nu}(x)$ is a rank-2 symmetric tensor field and $\Lambda(x)$ a local scalar parameter. 
The fracton transformation \eqref{FMAX-dA} is obtained from the more general infinitesimal diffeomorphism transformation
\be\label{FMAX-diff}
\delta_{diff} A_{\mu\nu}=\partial_\mu\Lambda_\nu + \partial_\nu\Lambda_\mu  
\ee
for the particular choice of the gauge parameter 
$\Lambda_\mu(x)=\frac{1}{2}\partial_\mu\Lambda(x)$.
The most general 4D action invariant under the fracton symmetry \eqref{FMAX-dA} is a linear combination of two invariant actions
\be
S_{inv}=g_1S_{fract}+g_2S_{LG}\ ,
\label{FMAX-Sinvg1g2}\ee
where
\begin{align}
S_{fract} &=
\int d^4x 
\left(\partial_\rho A_{\mu\nu}\partial^\rho A^{\mu\nu}- 
\partial_\rho A_{\mu\nu} \partial^\mu A^{\nu\rho} \right)\label{FMAX-Sfract}\\
S_{LG} &= \int d^4x\left(\partial_\mu A\partial^\mu A-\partial_\rho A_{\mu\nu}\partial^\rho A^{\mu\nu}-2\partial_\mu A\partial_\nu A^{\mu\nu}+2\partial_\rho A_{\mu\nu} \partial^\mu A^{\nu\rho}\right)\ , \label{FMAX-SLG}
\end{align}
$g_1,g_2$ are dimensionless constants, and $A\equiv\eta^{\mu\nu}A_{\mu\nu}$ is the trace of the tensor field. 
The action $S_{LG}$ is readily recognized to be the linearized action for gravity \cite{Hinterbichler:2011tt}, while $S_{fract}$ is our candidate to be the covariant action for fractons, as we shall motivate. 
Hence, the space of 4D local integrated functionals invariant under the fracton symmetry \eqref{FMAX-dA} has dimension two, and one of the two constants $g_1$ and $g_2$ can be reabsorbed by a redefinition of $A_{\mu\nu}(x)$, so that we have the rather peculiar feature that the free quadratic theory defined by the action \eqref{FMAX-Sinvg1g2}, hence by the fracton transformation \eqref{FMAX-dA}, depends on one constant. To our knowledge, this is the only example of a free quadratic covariant theory depending on a constant which cannot be reabsorbed by a field redefinition, without being identified as a mass, like in 3D topologically massive gauge theories \cite{Deser:1981wh}.
In particular we have that $S_{fract}$ \eqref{FMAX-Sfract} and $S_{LG}$ \eqref{FMAX-SLG} are both invariant under the fracton transformation \eqref{FMAX-dA}
\be
\delta_{fract}S_{LG} = \delta_{fract}S_{fract} = 0\ ,
\label{FMAX-}\ee
but only the \ac{LG} action \eqref{FMAX-SLG} is invariant under the diff transformation \eqref{FMAX-diff}
\be
\delta_{diff}S_{LG}= 0\ ,
\label{FMAX-}\ee
while the fracton $S_{fract}$ \eqref{FMAX-Sfract}, hence the whole action $S_{inv}$ \eqref{FMAX-Sinvg1g2}, is not
\be
\delta_{diff}S_{fract} = g_1\int d^4x \left[
2\partial^\mu\partial^\nu A_{\mu\nu}\partial^\rho\Lambda_\rho
-\partial^2A_{\mu\nu}(\partial^\mu\Lambda^\nu+\partial^\nu\Lambda^\mu)
\right]\neq0\ .
\label{FMAX-notdiff}\ee

\subsubsection*{The fracton field strength}

The first step towards a Maxwell theory for fractons, which is the  main purpose here, is the construction of the ``building block'' of the theory, namely the extension of the electromagnetic field strength $F_{\mu\nu}(x)$
\be
\begin{split}
A_\mu &\rightarrow F_{\mu\nu}\ \ =\partial_\mu A_\nu - \partial_\nu A_\mu  \\
A_{\mu\nu} &\rightarrow F_{\mu\nu\rho} =\quad ?
\end{split}
\label{FMAX-}\ee
To this aim, we look for a rank-3 tensor built from the first derivative of the rank-2 tensor field $A_{\mu\nu}(x)$
\begin{equation}
F_{\mu\nu\rho}\equiv 
a_1\partial_\mu A_{\nu\rho}+a_2\partial_\rho A_{\mu\nu}+a_3\partial_\nu A_{\mu\rho}\ ,
	\label{FMAX-Fmunurhoin}\end{equation}
where $a_i$ are dimensionless constants. As the electromagnetic tensor $F_{\mu\nu}(x)$ is invariant under the ordinary gauge transformation $\delta_{gauge}A_{\mu}(x)=\partial_\mu\Lambda(x)$, in the same way we require that $F_{\mu\nu\rho}(x)$ is invariant under the fracton symmetry \eqref{FMAX-dA}, which gives a constraint on the coefficients $a_i$
\be
\delta_{fract} F_{\mu\nu\rho}=0\ \Rightarrow\ a_3=-(a_1+a_2)\ ,
\label{FMAX-deltaF=0}\ee
so that
\begin{equation}\label{FMAX-F}
F_{\mu\nu\rho}=
a_1\partial_\mu A_{\nu\rho}+a_2\partial_\rho A_{\mu\nu}-(a_1+a_2)\partial_\nu A_{\mu\rho}\ .
\end{equation}
As a consequence of its definition, the invariant tensor \eqref{FMAX-F} has the properties listed in Table \ref{FMAX-table1}, compared to those of the Maxwell field strength $F_{\mu\nu}(x)$.
\begin{table}[H]
\centering
  \begin{tabular}{ | l | c | c| }
    \hline
     & fractons & Maxwell \\ \hline
      invariance & $\delta_{fract}F_{\mu\nu\rho}=0$& $\delta_{gauge}F_{\mu\nu}=0$ \\ \hline
    cyclicity & $F_{\mu\nu\rho}+F_{\nu\rho\mu}+F_{\rho\mu\nu}=0$ & $F_{\mu\nu}+F_{\nu\mu}=0$ \\ \hline
    Bianchi & $\epsilon_{\alpha\mu\nu\rho}\partial^{\mu}F^{\beta\nu\rho}=0$ & 
   $ \epsilon_{\mu\nu\rho\sigma}\partial^\nu F^{\rho\sigma}=0$ \\
    \hline
  \end{tabular}
\caption[Properties of the fracton and Maxwell field strengths]{\label{FMAX-table1}\footnotesize{Properties of the fracton and Maxwell field strengths.}}
\end{table}
\noindent
We remark that the fracton invariance of $F_{\mu\nu\rho}(x)$ \eqref{FMAX-deltaF=0} and the property which we called ``cyclicity'' in Table \ref{FMAX-table1} are equivalent
\begin{equation}
	\delta_{fract} F_{\mu\nu\rho}=0\quad\Leftrightarrow\quad F_{\mu\nu\rho}+F_{\nu\rho\mu}+F_{\rho\mu\nu}=0\quad\Leftrightarrow\quad a_1+a_2+a_3=0\ .
	\end{equation}	
All the physically relevant quantities (like for instance the equations of motion and the conjugate momenta) are obtained by making functional derivatives with respect to $A_{\mu\nu}(x)$, which is a symmetric tensor field. With the aim of writing everything in terms of the tensor field strength $F_{\mu\nu\rho}(x)$, it is natural to ask that also this latter is symmetric by the change of two indices, for instance the first two
\be
F_{\mu\nu\rho}=F_{\nu\mu\rho}
\label{FMAX-Fmunu=Fnumu}\ ,\ee
which implies $a_2=-2a_1$\footnote{We checked that this is indeed the case, $i.e.$ $F_{\mu\nu\rho}(x)-F_{\nu\mu\rho}(x)$ is always ruled out.}.
Therefore, after a rescaling of $A_{\mu\nu}(x)$, our fracton field strength is
	\begin{equation}
	F_{\mu\nu\rho}=F_{\nu\mu\rho}=\partial_\mu A_{\nu\rho}+\partial_\nu A_{\mu\rho}-2\partial_\rho A_{\mu\nu}\ .
	\label{FMAX-Fmunurho}\end{equation}
Rather surprisingly, the same symmetric tensor \eqref{FMAX-Fmunurho} appears as an unnumbered comment in the final part of a 1988 paper by Y.S. Wu and A. Zee \cite{Wu:1988py} as a consequence of the covariant symmetry \eqref{FMAX-dA}, but in a completely different context, since fractons were not even conceived yet.\footnote{We thank Giandomenico Palumbo for this remark.} The actions \eqref{FMAX-Sfract} and \eqref{FMAX-SLG} can be written in terms of the fracton field strength $F_{\mu\nu\rho}(x)$ \eqref{FMAX-Fmunurho} as  
\bea
S_{fract} &=&
\frac{1}{6}\;\int d^4x\;F^{\mu\nu\rho}F_{\mu\nu\rho}
\label{FMAX-SfractF}\\
S_{LG} &=& 
\int d^4x\; \left(
\frac{1}{4}F^\mu_{\ \mu\nu} F_\rho^{\ \rho\nu}-\frac{1}{6}F^{\mu\nu\rho}F_{\mu\nu\rho}
\right)\ .
\label{FMAX-SLGF}
\eea
Notice that also the \ac{LG} action \eqref{FMAX-SLG} can be written in terms of the newly introduced tensor $F_{\mu\nu\rho}(x)$ \eqref{FMAX-Fmunurho}. 
The fact that the fractonic component of the total action $S_{inv}$ \eqref{FMAX-Sinvg1g2} turns out to be of the form $\int F^2$ tells us that we are on the right way to build a Maxwell theory of fractons, but the analogies are even more surprising in what follows.
\subsubsection*{The canonical momentum $\Pi^{\alpha\beta}(x)$}

In the theory of the fracton quasiparticles an important role is played by the momentum canonically conjugated to $A_{\mu\nu}(x)$ \cite{Nandkishore:2018sel,Pretko:2016lgv,Pretko:2016kxt,Pretko:2017xar,Du:2021pbc}. From \eqref{FMAX-Sinvg1g2} we have
\be
\Pi^{\alpha\beta}\equiv\frac{\partial\mathcal L_{inv}}{\partial(\partial_tA_{\alpha\beta})}
=
-g_1 F^{\alpha\beta0}
-g_2\left[
\eta^{\alpha\beta}F_{\lambda}^{\ \lambda0}-\frac{1}{2}\left(\eta^{0\alpha}F_{\lambda}^{\ \lambda\beta}+\eta^{0\beta}F_{\lambda}^{\ \lambda\alpha}\right) -F^{\alpha\beta0}\right]\ ,
\label{FMAX-E^ij inv}\ee
whose components are
\bea
	\Pi^{00}&=&0 \label{FMAX-Pi00=0}\\
	\Pi^{i0}&=&-g_1F^{i00}-\frac{1}{2}g_2 F_j^{\ ji}\label{FMAX-Pi0i}\\
	\Pi^{ij}&=&-g_1F^{ij0} +g_2(F^{ij0}-\eta^{ij}F_k^{\ k0}) \label{FMAX-Piij}\ .
\eea
From \eqref{FMAX-Pi00=0} we see that $A_{00}(x)$ is not a dynamical field for the whole theory (both fractons and \ac{LG}). For what concerns \ac{LG} alone, it is known \cite{Pretko:2017fbf,Hinterbichler:2011tt} that the components with a time index, $A_{00}(x)$ and $A_{0i}(x)$, have non-dynamical equations of motion, acting as Lagrange multipliers to enforce gauge
constraints, in the same way as $A_0(x)$ acts as a Lagrange multiplier enforcing Gauss law in Maxwell theory. The physical degrees of freedom are contained in the spatial symmetric tensor $A_{ij}(x)$. We shall see that this property concerning \ac{LG} holds for fracton theory too, which therefore remarkably shares close similarities with both \ac{LG} and Maxwell theory. We finally notice that for a particular combination of $g_1$ and $g_2$ the trace of $\Pi^{\alpha\beta}$ vanishes
\be
\eta_{\alpha\beta}\Pi^{\alpha\beta}
=\Pi^\alpha_{\ \alpha}=\Pi^i_{\ i}=-(g_1+2g_2)F_{\lambda}^{\ \lambda0}=0\quad \mbox{if $g_1+2g_2=0$}\ .
\label{FMAX-tracelessconstr}\ee
This corresponds to the fact that, as already remarked in \cite{Blasi:2022mbl} and as we shall see in the next Section \ref{sec gauge}, the theory defined by $S_{inv}$ \eqref{FMAX-Sinvg1g2} at \eqref{FMAX-tracelessconstr} does not depend on the trace of the tensor field $A_{\mu\nu}(x)$, further lowering the  number of degrees of freedom.

\subsubsection*{The field equations of motion}

As the fracton and \ac{LG} actions \eqref{FMAX-SfractF} and \eqref{FMAX-SLGF}, the field \ac{EoM} can be written in terms of the fracton field strength $F_{\mu\nu\rho}(x)$ as well
\begin{align}
\frac {\delta S_{inv}}{\delta A^{\alpha\beta}} &=
g_1\left[
\left(\partial^\mu\partial_\alpha A_{\mu\beta}+\partial^\mu\partial_\beta A_{\mu\alpha}\right)
-2\partial^2 A_{\alpha\beta}
\right]\label{FMAX-eom inv}\\
&+2g_2\left[
\eta_{\alpha\beta}\left(\partial_\mu\partial_\nu A^{\mu\nu}-\partial^2A\right)
+\partial_\alpha\partial_\beta A
+ \partial^2 A_{\alpha\beta}
-\partial^\mu\left(\partial_\alpha A_{\mu\beta}+\partial_\beta A_{\mu\alpha}\right)\right]\nonumber \\
	&=
g_1\partial^{\mu}F_{\alpha\beta\mu}
+g_2\left[\eta_{\alpha\beta}\partial_\mu F_\nu^{\ \nu\mu}-\frac{1}{2}\left(\partial_\alpha F^\mu_{\ \mu\beta}+\partial_\beta F^{\mu}_{\ \mu\alpha}\right)-\partial^{\mu}F_{\alpha\beta\mu}\right]=0\ ,
\nonumber
\end{align}
whose components are
\begin{itemize}
\item $\alpha=\beta=0$
\be
g_1\partial_iF^{00i} - g_2 \partial_i (F_\lambda^{\ \lambda i}+F^{00i}) =
2\partial_i (-g_1F^{i00}-\frac{1}{2}g_2F_j^{\ ji}) =
2\partial_i\Pi^{i0} = 0\ ,
\label{FMAX-eomtot00}\ee
where we used $F^{00i}=-2F^{i00}$ and the definition of the canonical momentum $\Pi^{i0}$ \eqref{FMAX-Pi0i}. 
\item $\alpha=0$, $\beta=i$
\be
	\begin{split}
	0&=g_1\partial_\lambda F^{0i\lambda}-\frac{g_2}{2}(\partial^0F_\lambda^{\ \lambda i} + \partial^iF_\lambda^{\ \lambda 0} + 2 \partial_\lambda F^{0i\lambda})\\
	&=-\partial_0\Pi^{i0}+ g_1 \partial_j F^{0ij}-\frac{g_2}{2} (\partial^i F_\lambda^{\ \lambda 0} + 2\partial_jF^{0ij})\ ;
	\end{split}
	\label{FMAX-eomtot0i}
\ee
\item $\alpha=i$, $\beta=j$
\be
g_1\partial^\mu F_{ij\mu} + g_2[
\eta_{ij}\partial_\mu F_\nu^{\ \nu\mu}-\frac{1}{2}(\partial_iF^\mu_{\ \mu j} +\partial_jF^\mu_{\ \mu i})-\partial^\mu F_{ij\mu}]=0\ .
\label{FMAX-eomtotij}\ee
\end{itemize}

\subsection{Maxwell theory for fractons}\label{sec FracMax}

In this Section we treat the case $g_2=0$, and we shall recover the main features generally attributed to the fracton quasiparticles \cite{Nandkishore:2018sel,Pretko:2016lgv,Pretko:2016kxt,Pretko:2017xar}, thus allowing us to justify the identification of $S_{fract}$ \eqref{FMAX-SfractF} as the action for fractons.

\subsubsection*{Electric/magnetic tensor fields and ``Maxwell'' equations}

As far as only fractons are considered, in \cite{Nandkishore:2018sel,Pretko:2016lgv,Pretko:2016kxt,Pretko:2017xar,Du:2021pbc} an electric tensor field $E^{ij}(x)$ is defined as spatial ``canonical momentum'' as follows
\begin{equation}
		E_{ij}\propto -\partial_tA_{ij}+\partial_i\partial_j A_0\ .
\label{FMAX-electricfielddefPretko}\end{equation}	
We would like to show here that $E_{ij}(x)$ \eqref{FMAX-electricfielddefPretko} can indeed be derived from the action \eqref{FMAX-SfractF} in a way which also clarifies which is the origin of the scalar field $A_0(x)$ appearing in \eqref{FMAX-electricfielddefPretko}. In fact, $A_0(x)$ cannot be directly part of a canonical momentum unless in the Lagrangian weird terms with three derivatives are admitted \cite{Pretko:2017xar}. The covariant extension \eqref{FMAX-dA} 
of the fracton symmetry has a central role in determining \eqref{FMAX-electricfielddefPretko}, without the need of any \textit{ad-hoc} introduction. In fact, starting from the fracton transformation \eqref{FMAX-dA} one gets the action $S_{inv}$ \eqref{FMAX-Sinvg1g2} from which the spatial canonical momentum \eqref{FMAX-Piij} is derived. In the case where only fractons are present, namely $g_2=0$, the canonical momentum $\left.\Pi^{ij}(x)\right|_{g_2=0}$ reads:
\begin{equation}\label{FMAX-Pi^ij}
\left.\Pi^{ij}\right|_{g_2=0}= -g_1 F^{ij0} = g_1\left(2\partial^0A^{ij}-\partial^jA^{0i}-\partial^iA^{0j}\right)\ ,
\ee
which differs from \eqref{FMAX-electricfielddefPretko}. Nevertheless, the  electric tensor field \eqref{FMAX-electricfielddefPretko} can be indeed obtained from the spatial canonical momentum  $\left.\Pi^{ij}(x)\right|_{g_2=0}$ using the \ac{EoM} \eqref{FMAX-eom inv} with $g_2=0$, $i.e.$ those derived from the fracton action \eqref{FMAX-SfractF} alone, which closely remind the usual Maxwell equations
\be
\partial^\mu F_{\alpha\beta\mu}=0\ .
\label{FMAX-eom2}\ee
In fact, taking \eqref{FMAX-eom2} at $\alpha=\beta=0$ (or, equivalently, \eqref{FMAX-eomtot00} at $g_2=0$) we have 
\be
\partial^iF_{00i}=
2\partial^i\left(\partial_0A_{0i}-\partial_iA_{00}\right)=0\ .
\label{FMAX-eom00F}\ee
A particular solution is given by 
\begin{equation}\label{FMAX-A0}
		A_{0\mu}=A_{\mu0}\equiv\partial_\mu A_0\ ,
\end{equation}
which introduces the missing scalar potential $A_0(x)$. 
What renders remarkable the solution \eqref{FMAX-A0}, which is a direct consequence of our covariant approach, is that it leads to recover, up to a constant,  the electric tensor field \eqref{FMAX-electricfielddefPretko}. In fact, using \eqref{FMAX-A0} in \eqref{FMAX-Pi^ij} we get
\begin{equation}
(\Pi^{ij}|_{g_2=0})|_\eqref{FMAX-A0}=2g_1\left(\partial^0A^{ij}-\partial^i\partial^j A^{0}\right)\equiv E^{ij}\ ,
\label{FMAX-electricfielddef}\end{equation}
which indeed coincides with the tensor electric field \eqref{FMAX-electricfielddefPretko}. Hence, finally, the answer to the question is the following~: the electric tensor field $E^{ij}(x)$ \eqref{FMAX-electricfielddefPretko} introduced in \cite{Pretko:2016lgv,Pretko:2016kxt,Pretko:2017xar,Du:2021pbc} is defined as the canonical momentum $\Pi_{ij}$ \eqref{FMAX-Pi^ij} of the fracton action $S_{fract}$ \eqref{FMAX-Sfract} evaluated on the \ac{EoM} \eqref{FMAX-eom00F}.
In addition to the properties listed in Table \ref{FMAX-table1}, which hold in general, the particular solution \eqref{FMAX-A0} implies also
	\begin{empheq}{align}
	&F^{i00}=F^{0i0}=F^{00i}=0\label{FMAX-F00i=0}\\
	&F^{ij0}=-2F^{0ij}=-2F^{i0j}\label{FMAX-Fij0=-2F0ij}\ ,
	\end{empheq}
which hold for fractons only. As anticipated, because of \eqref{FMAX-F00i=0}, for the fracton theory $g_2=0$ we have an additional Hamiltonian constraint, besides \eqref{FMAX-Pi00=0}
\be
(\Pi^{i0}|_{g_2=0})|_\eqref{FMAX-A0}=-g_1F^{i00}=0\ ,
\label{FMAX-Pi00=0v2}\ee
which corresponds to the fact that, like in \ac{LG}, also for fractons the degrees of freedom concern only the spatial components $A_{ij}(x)$.
Moreover, again in surprising analogy with Maxwell theory where the electric field and the field strength are related by $E^i(x)=-F^{0i}(x)$, we have that
\be
E^{ij}=-g_1 F^{ij0}=2g_1 F^{0ij}=2g_1 F^{i0j}\ ,
\label{FMAX-Eij}\ee
which makes the electric tensor field an invariant quantity at sight. Taking \eqref{FMAX-eom2} at $\alpha=0$ and $\beta=i$ (or, equivalently, \eqref{FMAX-eomtot0i} at $g_2=0$) we have 
\begin{equation}\label{FMAX-eom0i}
\partial_\mu F^{0i\mu} = \partial_jF^{0ij} = -\frac{1}{2}\partial_jF^{ij0} = 0\ ,
\ee
which, using \eqref{FMAX-Eij}, writes
\be
\partial_jE^{ij}=0\ ,
\label{FMAX-gauss-v}\ee
which is the vacuum Gauss law for the electric  tensor field \eqref{FMAX-electricfielddef}. It is the tensorial extension of
\begin{equation}
		\vec\nabla\cdot\vec E=0\ .
		\end{equation}
Eq. \eqref{FMAX-gauss-v} trivially implies
	\begin{equation}\label{FMAX-gauss}
	\partial_i\partial_jE^{ij}=0\ ,
	\end{equation}
 which, together with \eqref{FMAX-gauss-v}, is crucial for the property of limited mobility characterizing the fracton quasiparticles \cite{Nandkishore:2018sel,Pretko:2020cko,Pretko:2016lgv,Pretko:2016kxt}. As we shall show in a moment, while the Gauss-like equation \eqref{FMAX-gauss-v} holds only for the fracton action \eqref{FMAX-SfractF}, an equation formally identical to the limited mobility equation \eqref{FMAX-gauss} holds for the \ac{LG} action $S_{LG}$ \eqref{FMAX-SLGF}, too.  
In fact, taking the divergence $\partial_i$ of the whole \ac{EoM} \eqref{FMAX-eomtot0i} and using \eqref{FMAX-eomtot00} we have, at $g_1=0$, $i.e.$ for \ac{LG} only,
\be
\partial_i\partial^iF_\lambda^{\ \lambda0} + 2 \partial_i\partial_jF^{ij0}=
-\partial_i\partial_j\Pi^{ij}|_{g_1=0}=0\ ,
\label{FMAX-Piij=0LG}\ee
where we used the cyclicity property in Table \ref{FMAX-table1} of the tensor $F_{\mu\nu\rho}(x)$, which in particular implies
\be
\partial_i\partial_jF^{0ij}=-\frac{1}{2}\partial_i\partial_jF^{ij0}\ .
\label{FMAX-}\ee
The equation \eqref{FMAX-Piij=0LG} is formally identical to its fracton counterpart \eqref{FMAX-gauss}, and its possible consequences on the limited mobility of the gravitational waves are worth to investigate and to interpret.
Finally, taking \eqref{FMAX-eom2} at $\alpha=i$ and $\beta=j$ (or, equivalently, \eqref{FMAX-eomtotij} at $g_2=0$), we have
\begin{equation}\label{FMAX-eomfractij}
\partial_\mu F^{ij\mu}
=\partial_0F^{ij0}+\partial_kF^{ijk}
=-\frac{1}{g_1}\partial_0E^{ij}+\partial_kF^{ijk}=0\ ,
\ee
where we used the definition of the electric tensor field $E^{ij}(x)$ \eqref{FMAX-Eij}.
The fracton \ac{EoM} \eqref{FMAX-eomfractij} suggests to define the magnetic tensor field, in analogy  with the ordinary  vector magnetic field \mbox{$B_i(x)=\epsilon_{ijk}\partial^jA^k(x)=\frac{1}{2}\epsilon_{ijk}F^{jk}(x)$}, as
\begin{equation}\label{FMAX-Bij}
		B_{i}^{\ j}\equiv g\epsilon_{0ilk}\partial^lA^{jk} = \frac{g}{3}\epsilon_{0ikl}F^{jkl}\ ,
\end{equation}
where $g$ is a constant to be suitably tuned. Its inverse is
\begin{equation}\label{FMAX-B=F}
F^{ijk}\equiv-\frac{1}{g}\left(\epsilon^{0ikl}B^{\ j}_{l}+\epsilon^{0jkl}B^{\ i}_{l}\right)\ .
\end{equation}
Again, thanks to the definition of the fracton field strength $F_{\mu\nu\rho}(x)$ the definition of the magnetic tensor field, as for the electric one \eqref{FMAX-Eij}, is explicitely invariant under the defining covariant fracton symmetry. The \ac{EoM} \eqref{FMAX-eomfractij} then can be written
\be
-\frac{1}{g_1}\partial_0E^{ij}-\frac{1}{g}\left(\epsilon^{0ikl}\partial_kB^{\ j}_{l}+\epsilon^{0jkl}\partial_kB^{\ i}_{l}\right)=0\ ,
\label{FMAX-ampere}\end{equation}
which turns out to be completely analogous to the electromagnetic Amp\`ere law of electromagnetism in vacuum
		\begin{equation}
		-\partial_t\vec E+\vec\nabla\times\vec B=0\ ,
		\end{equation}
    of which \eqref{FMAX-ampere} is the tensorial extension. It coincides with Eq.(26) in \cite{Pretko:2016lgv}.
From the definition \eqref{FMAX-Bij} we find that the magnetic tensor field is traceless
\be
	B^{\;p}_p=0\label{FMAX-TrB=0}\ ,
\ee
	and satisfies
\be
	\partial^aB^{\;p}_a=0\ \label{FMAX-DivB=0}\ ,
\ee
which  is analogous to the standard Maxwell equation		
\begin{equation}
		\vec\nabla\cdot\vec B=0\ ,
		\end{equation}
and coincides with Pretko's second equation (38) in \cite{Pretko:2016lgv}. As in standard electromagnetism, the equation \eqref{FMAX-DivB=0} is a geometric property, consequence of the definition of the magnetic tensor field $B_i^{\ j}(x)$ \eqref{FMAX-Bij}.\\

Let us now study which information comes from the ``Bianchi'' identity in Table \ref{FMAX-table1}
\begin{itemize}
\item $\alpha=0,\beta=j$
\begin{equation}\label{FMAX-divB=0}
\epsilon_{0iab}\partial^{i}F^{jab}=\frac{3}{g}\partial^iB_{i}^{\;j}= 0\ ,
\end{equation}
we therefore recover the tensor magnetic Gauss law \eqref{FMAX-DivB=0}.
\item $\alpha=l,\beta=i$
\begin{equation}\label{FMAX-faraday}
		\begin{split}
		0&=\epsilon_{l\mu\nu\rho}\partial^{\mu}F^{i\nu\rho}\\
		&=\epsilon_{l\mu j0}\partial^\mu F^{ij0}+\epsilon_{l\mu jk}\partial^\mu F^{ijk}+\epsilon_{l\mu 0j}\partial^\mu F^{i0j}\\
		&=\frac{3}{2}\epsilon_{lmj0}\partial^m F^{ij0}+\epsilon_{l0 jk}\partial^0 F^{ijk}\\
		&=\frac{3}{2g_1}\epsilon_{0lmj}\partial^mE^{ij}+\frac{3}{g}\partial_0B^{\;i}_{l}\ ,
		\end{split}
	\end{equation}
where we used \eqref{FMAX-F00i=0}, \eqref{FMAX-Fij0=-2F0ij}, \eqref{FMAX-Eij} and the definition \eqref{FMAX-Bij}.
Eq.\eqref{FMAX-faraday} is new, and it is the tensorial extension of the Faraday equation of electromagnetism~: 
\begin{equation}
		\vec\nabla\times\vec E+\partial_t\vec B=0\ .
\end{equation}
It coincides with Pretko's Eq.(36) in \cite{Pretko:2016lgv}. 
\item
$\alpha=\beta=0$ and $\alpha=i,\ \beta=0$ 
are trivial identities.
\end{itemize}
Summarizing, from the \ac{EoM} \eqref{FMAX-eom2} and the ``Bianchi'' identity in Table \ref{FMAX-table1} we have the following strong analogy with classical electromagnetism :
	\begin{empheq}{align}
	\mbox{ \textbf{ Maxwell}} \qquad& \mbox{  \textbf{Fractons}} \nonumber \\
	\vec\nabla\cdot\vec E=0\qquad&\partial_jE^{ij}=0\label{FMAX-max1}\\
	\vec\nabla\cdot\vec B=0\qquad&\partial^aB^{\;p}_a=0\label{FMAX-max2}\\
	\vec\nabla\times\vec E-\partial_t\vec B=0\qquad&\epsilon_{0lmj}\partial^mE^{ij}-\frac{2g_1}{g}\partial^0B^{\;i}_{l}=0\label{FMAX-max3}\\
	\vec\nabla\times\vec B-\partial_t\vec E=0\qquad&-\partial_0E^{ij}-\frac{2g_1}{g}\frac{1}{2}\left(\epsilon^{0ikl}\partial_kB^{\ j}_{l}+\epsilon^{0jkl}\partial_kB^{\ i}_{l}\right)=0\ .\label{FMAX-max4}
	\end{empheq}
Setting 
	\begin{equation}\label{FMAX-galpha1}
	\frac{2g_1}{g}=-1
	\end{equation} 
    the last two equations  \eqref{FMAX-max3} and \eqref{FMAX-max4} are fully analogous to the corresponding ordinary Maxwell equations at the left hand side, and coincides with those introduced by Pretko in \cite{Pretko:2016lgv} from a completely different point of view, where actually it is written \eqref{FMAX-gauss} rather than the more fundamental \eqref{FMAX-gauss-v}.

\subsubsection*{Fracton action in terms of electric and magnetic tensor fields}

We have seen that the fracton action \eqref{FMAX-Sfract}, originally written in terms of the field $A_{\mu\nu}(x)$, can be written in terms of the tensor $F_{\mu\nu\rho}(x)$ as \eqref{FMAX-SfractF}. This makes apparent the strong analogy with the classical electromagnetic Maxwell theory, of which the fracton theory appears to be the higher rank generalization. This analogy is even more spectacular when the four equations \eqref{FMAX-max1}, \eqref{FMAX-max2}, \eqref{FMAX-max3} and \eqref{FMAX-max4} governing the theory are considered, which can be written in terms of the two electric and magnetic tensor fields $E^{ij}(x)$ \eqref{FMAX-electricfielddef} and $B_i^{\ j}(x)$ \eqref{FMAX-Bij}.
As in Maxwell theory, two equations, namely \eqref{FMAX-max1} and \eqref{FMAX-max4}, are the \ac{EoM} of the action \eqref{FMAX-SfractF}, while the other two, \eqref{FMAX-max2} and \eqref{FMAX-max3}, are consequences of the ``Bianchi'' identity written in Table \ref{FMAX-table1} for the tensor $F_{\mu\nu\rho}(x)$, hence have a geometrical nature. The analogy with electromagnetism can be pushed further by noting that the fracton action \eqref{FMAX-SfractF} can be written in terms of the electric and magnetic tensor fields as follows~:
\begin{equation}\label{FMAX-actionE,B}
		\begin{split}
		S_{fract}&
		= {\frac{g_1}{6}}\int d^4x\,F_{\mu\nu\rho}F^{\mu\nu\rho}\\
		&= {\frac{g_1}{6}}\int d^4x\left(F^{0ij}F_{0ij}+F^{ij0}F_{ij0}+F^{i0j}F_{i0j}+F^{ijk}F_{ijk}\right)\\
		&= {\frac{g_1}{6}}\int d^4x\left(\frac{3}{2}F^{ij0}F_{ij0}+F^{ijk}F_{ijk}\right)\\
		&= {\frac{g_1}{6}}\int d^4x\left[-\frac{3}{2} {\frac{1}{g_1^2}}E^{ij}E_{ij}- {\frac{2}{g^2}\epsilon^{0kmn}B^{\;l}_{n}\epsilon_{0kab}B_{\;c}^{b}\left(\delta^a_m\delta^c_l+\delta^a_l\delta^c_m\right)}\right]\\
		&=\frac{1}{2}\int d^4x\left(- {\frac{1}{2g_1}}E^{ij}E_{ij}+ {\frac{2g_1}{g^2}}B^{\;j}_{i}B_{\;j}^{i}\right)\ ,
		\end{split}
	\end{equation}
where, besides the definitions of the electric and magnetic tensor fields \eqref{FMAX-electricfielddef} and \eqref{FMAX-Bij}, we used the properties of $F_{\mu\nu\rho}(x)$ \eqref{FMAX-F00i=0} and \eqref{FMAX-Fij0=-2F0ij}, and the tracelessness of the tensor magnetic field \eqref{FMAX-TrB=0}. The result \eqref{FMAX-actionE,B} closely reminds the electromagnetic action, whose Lagrangian is proportional to $E^2-B^2$, provided that
\begin{equation}\label{FMAX-galpha}
	g^2=4g_1^2\ ,
	\end{equation}
which is compatible with the previously found constraint \eqref{FMAX-galpha1}, that we will assume from now on.

\subsection{Stress-energy tensor and fracton Lorentz force}\label{sec T+F}
\subsection*{The energy-momentum tensor and conservation laws}
The stress-energy tensor for the fracton action $S_{fract}$ \eqref{FMAX-SfractF} is
	\begin{equation}\label{FMAX-Tmunu}
		\begin{split}
		T_{\alpha\beta}&=\left.-\frac{2}{\sqrt{-g}}\frac{\delta S_{fract}}{\delta g^{\alpha\beta}}
		\right|_{g^{\alpha\beta}=\eta^{\alpha\beta}}\\
		&=\left.-\frac{ {g_1}}{3\sqrt{-g}}\frac{\delta }{\delta g^{\alpha\beta}}\int d^4x\sqrt{-g}g^{\mu\lambda}g^{\nu\gamma}g^{\rho\sigma}F_{\lambda\gamma\sigma}F_{\mu\nu\rho}
		\right|_{g^{\alpha\beta}=\eta^{\alpha\beta}}\\
		&= {\frac{g_1}{6}}\eta_{\alpha\beta}F^2-\frac{g_1}{3}\eta_{\alpha\gamma}\eta_{\beta\lambda}\left(2F^{\lambda\nu\rho}F_{\ \,\nu\rho}^\gamma+F^{\mu\nu\lambda}F_{\mu\nu}^{\ \ \gamma}\right)\ .
		\end{split}
	\end{equation}
Notice that taking the trace of $T_{\mu\nu}$, we have, in $d$-spacetime dimensions
\begin{equation}
	T=\eta^{\alpha\beta}T_{\alpha\beta}=\left.g_1\frac{(d-6)}{6}F^2\right|_{d=4}=-\frac{g_1}{3}F^2\ ,
	\label{FMAX-traceTmunu}\end{equation}
which does not vanish in $d=4$, differently from what happens in Maxwell theory. The tracelessness of the stress-energy tensor would be recovered in $d=6$, which, as already remarked in the introductory part at the beginning of Section \ref{sec MaxThFract}, seem to be the most natural, although unphysical, spacetime dimensions for fractons. The non-vanishing of the trace of the fracton stress-energy tensor is the sign that the theory, already at classical level, is not scale invariant. This suggests the existence of an energy scale. Now, since tracelessness is eventually related to the masslessness of the photon, the fact that the trace \eqref{FMAX-traceTmunu} does not vanish might suggest the existence of a mass (as the typical energy scale) for the fractons, which can be introduced in a similar way as in \ac{LG} \cite{Blasi:2017pkk,Blasi:2015lrg}.
The components of the stress-energy tensor are physically interpretable as follows~:
	\begin{itemize}
	\item $\alpha=\beta=0$ gives the energy density $T_{00}=u$  
		\begin{equation}\label{FMAX-T00}
			\begin{split}
T_{00}=u
&=
- \frac{g_1}{6}F^2+\frac{g_1}{3}\left(2F^{0\mu\nu}F_{0\mu\nu}+F^{\mu\nu0}F_{\mu\nu0}\right)
\\
&=
\frac{1}{4g_1}\left( E^{ij}E_{ij}- B^{\;j}_iB_{\;j}^i\right)+\frac{g_1}{2}\left(- {\frac{1}{g_1}}\right)\left( {\frac{1}{g_1}}\right)E^{ij}E_{ij}
\\
&=
-\frac{1}{4g_1}\left(E^{ij}E_{ij}+B^{\;j}_iB_{\;j}^i\right)\ ,
\end{split}
\end{equation}
where \eqref{FMAX-F00i=0}, \eqref{FMAX-Fij0=-2F0ij}, \eqref{FMAX-actionE,B}, \eqref{FMAX-Eij} and \eqref{FMAX-galpha} have been used. Again, this expression is formally identical to the corresponding electromagnetic result $u\propto E^2+B^2$.
From the positivity constraint of the energy density $u$ it  must be
		\begin{equation}
		g_1<0\ ,
		\end{equation}
and, from now on, we choose
\be
g_1=-1\ ;
\label{FMAX-g1=-1}\ee
\item $\alpha=0,\ \beta=i$ gives the Poynting vector $T_{0i}=S_i$ 	
\begin{equation}
\begin{split}
T_{0i}=S_i=
& 
\frac{1}{3}\eta_{i\lambda}\left(2F^{\lambda\nu\rho}F_{0\nu\rho}+F^{\mu\nu\lambda}F_{\mu\nu0}\right)
\\
=& 
\frac{1}{3}\eta_{ij}\left(2F^{jkl}F_{0kl}+F^{klj}F_{kl0}\right)
\\
=&
\frac{1}{6}\eta_{ij}E_{kl}\left(-2\epsilon^{0jlp}B_p^{\;k}-\cancel{\epsilon^{0klp}B_p^{\;j}}+\epsilon^{0kjp}B_p^{\;l}\right)
\\
=&\frac{1}{2}\epsilon_{0ilp}E^{kl}B^p_{\;k}\ ,
\end{split}
\end{equation}
which, as in Maxwell electromagnetism, is the vector product of the electric and magnetic tensor fields $\vec S\propto\vec E\times\vec B$\ ;
\item $\alpha=i,\ \beta=j$ gives the stress tensor $T_{ij}=\sigma_{ij}$ 
\begin{align}
T_{ij}&=
-\frac{1}{6}\eta_{ij}F^2+\frac{1}{3}\eta_{jk}\left(2F^{k\mu\nu}F_{i\mu\nu}+F^{\mu\nu k}F_{\mu\nu i}\right)\nonumber
\\
&=
-\frac{1}{6}\eta_{ij}F^2-\eta_{jk}E^{ka}E_{ia}+\frac{1}{3}\eta_{jk}\left(2F^{kab}F_{iab}+F^{abk}F_{abi}\right)
\\
&=
-\frac{1}{6}\eta_{ij}F^2-\eta_{jk}E^{ka}E_{ia}+\frac{1}{6}\eta_{jk}\left(2\delta^k_i B_a^{\;b}B^a_{\;b}-2B_i^{\;a}B^k_{\;a}+4B_a^{\;k}B^a_{\;i}-\epsilon^{0akp}\epsilon_{0biq}B_p^{\;b}B^q_{\;a}\right)\nonumber
\\	
&=
-\frac{1}{6}\eta_{ij}F^2-\eta_{jk}E^{ka}E_{ia}+\frac{1}{2}\eta_{jk}\left(\delta^k_i B_a^{\;b}B^a_{\;b}-B_i^{\;a}B^k_{\;a}+B_a^{\;k}B^a_{\;i}\right)\nonumber
\\	
&=
\eta_{ij}T_{00}-\eta_{jk}\eta_{il}E^{ka}E^l_{\;a}-\frac{1}{2}\eta_{jk}\eta_{il}\left(B^{la}B^k_{\;a}-B_a^{\;k}B^{al}\right)\ ,\nonumber
\end{align}
where we used
\begin{empheq}{align}
2F^{kab}F_{iab}
&=
-\frac{1}{2}\left(\epsilon^{0kbp}B_p^{\;a}+\epsilon^{0abp}B_p^{\;k}\right)\left(\epsilon_{0ibq}B^q_{\;a}+\epsilon_{0abq}B^q_{\;i}\right)
\\
&=
\frac{1}{2}\left(\delta^k_i B_a^{\;b}B^a_{\;b}-B_i^{\;a}B^k_{\;a}+4B_a^{\;k}B^a_{\;i}\right)\nonumber\\
F^{abk}F_{abi}
&=
-\frac{1}{4}\left(\epsilon^{0akp}B_p^{\;b}+\epsilon^{0bkp}B_p^{\;a}\right)\left(\epsilon_{0aiq}B^q_{\;b}+\epsilon_{0biq}B^q_{\;a}\right)
\\
&=
\frac{1}{2}\left(\delta^k_i B_a^{\;b}B^a_{\;b}-B_i^{\;a}B^k_{\;a}-\epsilon^{0akp}\epsilon_{0biq}B_p^{\;b}B^q_{\;a}\right)\ ,
\nonumber
\end{empheq}
and, from \cite{landau}, 
		\begin{equation}\label{FMAX-epsxeps}
		\epsilon^{0akp}\epsilon_{0biq}=-\delta^a_b(\delta^k_i\delta^p_q-\delta^p_i\delta^k_q)+\delta^a_i(\delta^k_b\delta^p_q-\delta^p_b\delta^k_q)-\delta^a_q(\delta^k_b\delta^p_i-\delta^p_b\delta^k_i)\ ,	
		\end{equation}
so that
	\begin{equation}
	\epsilon^{0akp}\epsilon_{0biq}B_p^{\;b}B^q_{\;a}=-\delta^k_iB_a^{\;b}B^a_{\;b}+B_i^{\;a}B_{\;a}^k+B_a^{\;k}B^a_{\;i}\ ,
	\end{equation}
because of the tracelessness of the magnetic tensor \eqref{FMAX-TrB=0}. 
Finally, we used also the fact that, because of 
\eqref{FMAX-actionE,B} and \eqref{FMAX-T00} we have  \mbox{$T_{00}=-\frac{1}{6}F^2+\frac{1}{2} B_a^{\;b}B^a_{\;b}$}.
As expected, the stress tensor is symmetric $i\leftrightarrow j$~:
\begin{equation}
\begin{split}
T_{ij}=
\frac{1}{4}\eta_{ij}
\left(
E^{ab}E_{ab}+B^{\;b}_aB_{\;b}^a
\right)
-\eta^{ab}E_{ia}E_{jb}-\frac{1}{2}\eta^{ab}\left(B_{ia}B_{jb}-B_{aj}B_{bi}\right)\ .
\end{split}
\end{equation}
Once again, the analogy with Maxwell theory, for which the stress tensor is
	\begin{equation}
	\sigma_{ij}=\frac{1}{2}\eta_{ij}\left(E^2+B^2\right)-E_iE_j-B_iB_j\ ,
	\end{equation}
is impressive.
\end{itemize}
Let us now discuss the (on-shell) conservation of the stress-energy tensor 
\be
\partial^\nu T_{\mu\nu} =0\ ,
\label{FMAX-partialTmunu=0}\ee
whose components are
\begin{itemize}
\item $\mu=0$ :
\begin{equation}\label{FMAX-F0}
\begin{split}
\partial^\nu T_{\nu0}
&=
\partial^0T_{00}+\partial^iT_{i0}
\\
&=
\partial^0u+\partial^iS_i
\\
&=
-\frac{1}{2}\left[ E_{ab}\partial_0E^{ab}+B^a_{\;b}\partial_0B_a^{\;b}
-\epsilon_{0ilp}\partial^i\left(E^{kl}B_{\;k}^p\right)\right]
\\ 
&=-\frac{1}{2}\left[
\epsilon^{0akl}E_{ab}\partial_kB_l^{\;b}+\epsilon_{0amn}B^a_{\;b}\partial^mE^{bn}
-\epsilon_{0ilp}\partial^i\left(E^{kl}B_{\;k}^p\right)\right]
\\
&=
\frac{1}{2}\epsilon_{0amn}\left[
E^{ab}\partial^mB^n_{\;b}+B^n_{\;b}\partial^mE^{ab}
-\partial^m\left(E^{ab}B^n_{\;b}\right)\right]
\\
&=0\ ,
\end{split}
\end{equation}
where we used the \ac{EoM} \eqref{FMAX-ampere} and \eqref{FMAX-faraday}. The continuity equation is therefore verified on-shell
\begin{equation}
	\partial^iS_i+\partial^0u=0\ .
\label{FMAX-continuity}\end{equation}

\item $\mu=i$ : 
\begin{align}
\partial^\nu T_{\nu i}&=\partial^0T_{0i}+\partial^jT_{ji}\nonumber\\
&=
\partial^0S_i+\partial^j\sigma_{ji}\nonumber
\\
&=
-\frac{1}{2}\epsilon_{0imn}\partial_0\left(E^{am}B^n_{\;a}\right)+\partial_iu-\eta_{il}\partial_k\left(E^{ka}E^l_{\;a}\right)-\eta_{il}\partial_k\left(B^{la}B^k_{\;a}-B_a^{\;k}B^{al}\right)\nonumber
\\
&=
-\frac{1}{2}\epsilon_{0imn}\left\{\left[\frac{1}{2}\left(\epsilon^{0akl}\partial_kB_l^{\;m}+\epsilon^{0mkl}\partial_kB_l^{\;a}\right)\right]B^n_{\;a}-\epsilon^{0nbc}\partial_bE_{ac}E^{am}\right\}+\nonumber
\\
&\quad
+\partial_iu-\eta_{il}\partial_k\left(E^{ka}E^l_{\;a}\right)-\eta_{il}\partial_k\left(B^{la}B^k_{\;a}-B_a^{\;k}B^{al}\right)\nonumber
\\
&=
-\frac{1}{4}\left[\epsilon_{0imn}\epsilon^{0akl}\partial_kB_l^{\;m}+\left(\delta^k_i\delta^l_n-\delta^k_n\delta^l_i\right)\partial_kB_l^{\;a}\right]B^n_{\;a}-\nonumber
\\
&\quad
-\frac{1}{2}\left(\delta^b_i\delta^c_m-\delta^b_m\delta^c_i\right)E^{am}\partial_bE_{ac}+\partial_iu-\eta_{il}\partial_k\left(E^{ka}E^l_{\;a}\right)\nonumber
\\
&\quad-\eta_{il}\partial_k\left(B^{la}B^k_{\;a}-B_a^{\;k}B^{al}\right)\nonumber
\\
&=
\frac{1}{4}\left[
\left(3B^n_{\;i}\partial_mB_n^{\;m}+2B_n^{\;m}\partial_mB^n_{\;i}\right)-2E^{ab}\partial_bE_{ai}\right]\label{FMAX-Fi}\\
&\neq 0\ ,\nonumber
\end{align}	
where \eqref{FMAX-ampere}, \eqref{FMAX-faraday}, \eqref{FMAX-epsxeps} and \eqref{FMAX-gauss-v} have been used. Differently from the continuity equation \eqref{FMAX-continuity}, the spatial components of the divergence of the stress-energy tensor do not vanish. Now, what should we expect actually ? If we think of the stress-energy tensor as the conserved current associated to the diffeomorphism symmetry, as the definition \eqref{FMAX-Tmunu} suggests, it should not be conserved in a theory like the fracton one, which is not diffeomorphism invariant \eqref{FMAX-notdiff}. On the other hand, we are facing here with a partial conservation of the stress-energy tensor, because its time component is indeed conserved, yielding the continuity equation \eqref{FMAX-continuity}, which relates the flux of the energy density to the divergence of the momentum density.  The partial conservation of the stress-energy tensor might be explained by observing that the fracton symmetry \eqref{FMAX-dA} is indeed a diff transformation \eqref{FMAX-diff} with a particular choice of the vector diff parameter.
\end{itemize}

\subsection*{Fracton Lorentz force}

It is interesting to study how the physics is modified if matter is introduced by means of a symmetric rank-2 tensor $J^{\mu\nu}(x)=J^{\nu\mu}(x)$ coupled to the fracton field $A_{\mu\nu}(x)$
\be
S_{fract}\rightarrow S_{tot}=S_{fract}+S_J\ ,
\label{FMAX-Stot}\ee
where $S_{fract}$ is the pure fractonic action \eqref{FMAX-Sfract} (or \eqref{FMAX-SfractF}), and $S_J$ is the matter action
\begin{equation}
	S_J\equiv-\int d^4x\,J^{\mu\nu}A_{\mu\nu}\ .
\label{FMAX-Smatter}\end{equation}
The \ac{EoM} \eqref{FMAX-eom2} modifies as
\begin{equation}\label{FMAX-eomJ}
	\partial_\mu F^{\alpha\beta\mu}=-J^{\alpha\beta}\ .
\end{equation}
We observe that, due to the cyclicity identity in Table \ref{FMAX-table1}, $J^{\alpha\beta}$ is conserved in the following sense
\begin{equation}\label{FMAX-divJ=0}
	\partial_\alpha\partial_\beta J^{\alpha\beta}=0\ .
	\end{equation}
The components of the \ac{EoM} \eqref{FMAX-eomJ} are
	\begin{itemize}
	\item $\alpha=\beta=0$ :
	\begin{equation}\label{FMAX-eomJ00}
	\partial_i F^{00i}=-J^{00}=0\ ,
	\end{equation}
which vanishes because of \eqref{FMAX-F00i=0}, consequence of \eqref{FMAX-A0}. Hence, there is no coupling with $A_{00}(x)$, as expected,  since it is not a dynamical degree of freedom of the theory, due to \eqref{FMAX-Pi00=0}.

\item $\alpha=0,\ \beta=i$ :
	\begin{equation}\label{FMAX-eomJ0i}
	\partial_j F^{0ij}=-J^{0i}\ ,
	\end{equation}
which, using \eqref{FMAX-Eij}, becomes
	\begin{equation}\label{FMAX-gauss-vett}
	\partial_jE^{ij}=2J^{i0}\ .
	\end{equation}
Taking the divergence of \eqref{FMAX-gauss-vett} we find the analogous of the Gauss law 	\begin{equation}\label{FMAX-gauss+mat}
	\partial_i\partial_jE^{ij}=\rho\ ,
	\end{equation}
where we defined the charge density
	\begin{equation}\label{FMAX-rho}
	\rho\equiv2\partial_iJ^{i0} \ .
	\end{equation}
This equation plays a central role in \cite{Nandkishore:2018sel,Pretko:2020cko,Pretko:2016lgv,Pretko:2016kxt,Pretko:2017xar}, since it yields not only the charge neutrality condition, but also the vanishing of the total dipole moment. In fact, integrating \eqref{FMAX-gauss+mat}, we get
	\begin{equation}
	\int dV\partial_i\partial_jE^{ij}=\int dV\,\rho=0\ ,
	\end{equation}
which states that the total charge inside an infinite volume is zero. Moreover, from \eqref{FMAX-gauss+mat} we also have
	\begin{equation}\label{FMAX-dipole cons}
	\int dVx^k\partial_i\partial_jE^{ij}=
	\int dV\,x^k\rho=\int dV\,p^k=0\ ,
	\end{equation}
according to which the dipole moment density, defined as 
\be
p^k=x^k\rho\ ,
\label{FMAX-dipoledensity}\ee 
of an infinite volume vanishes.

\item $\alpha=i,\ \beta=j$ :
	\begin{equation}\label{FMAX-eomJij}
	\partial_\mu F^{ij\mu}=-J^{ij}\ ,
	\end{equation}
which, using \eqref{FMAX-ampere}, becomes
	\begin{equation}\label{FMAX-eomJij2}
	-\partial_0E^{ij}+\frac{1}{2}\left(\epsilon^{0ikl}\partial_kB^{\ j}_{l}+\epsilon^{0jkl}\partial_kB^{\ i}_{l}\right)=J^{ij}\ .
	\end{equation}
Differentiating 
 with $\partial_i\partial_j$, we have
	\begin{equation}\label{FMAX-}
		\begin{split}
		0&=\partial_0\partial_i\partial_jE^{ij}+\partial_i\partial_jJ^{ij}\\
		&=\partial_0\rho+\partial_i\partial_jJ^{ij}\ ,
		\end{split}
	\end{equation}
    where \eqref{FMAX-gauss+mat} has been used. It is a kind of continuity equation \cite{Pretko:2016lgv,Du:2021pbc,Jain:2021ibh}, which can also be obtained from the conservation equation \eqref{FMAX-divJ=0}
	\begin{equation}
	0=\partial_\alpha\partial_\beta J^{\alpha\beta}=2\partial_0\partial_iJ^{0i}+\partial_i\partial_jJ^{ij}=\partial_0\rho+\partial_i\partial_jJ^{ij}\ ,
	\end{equation}
	where we have used the definition of the density charge $\rho(x)$ \eqref{FMAX-rho} and the fact that $J^{00}(x)=0$ \eqref{FMAX-eomJ00}.
	\end{itemize}
It is also interesting to see how the (partial) conservation of the stress-energy tensor is modified by the presence of matter. 
The continuity equation \eqref{FMAX-continuity} is modified as
\be
\partial^\nu T_{\nu0}
=
\partial^0T_{00}+\partial^iT_{i0}=
\partial^0u+\partial^iS_i= E_{ab}J^{ab}\ ,
\label{FMAX-continuitymatter}\ee
while the spatial components of \eqref{FMAX-Fi} acquire the term in the last row
\bea
\partial^\nu T_{\nu i}=\partial^0T_{0i}+\partial^jT_{ji}
=
\partial^0S_i+\partial^j\sigma_{ji} &=&
\frac{1}{4}\left[
\left(3B^n_{\;i}\partial_mB_n^{\;m}+2B_n^{\;m}\partial_mB^n_{\;i}\right)-2E^{ab}\partial_bE_{ai}\right]\nonumber\\
&&+
\frac{1}{2}\epsilon_{0imn}J^{am}B^n_{\ a}-2J^{a0}E_{ia}\ ,
\label{FMAX-Fimatter}\eea
The additional terms appearing in \eqref{FMAX-continuitymatter} and \eqref{FMAX-Fimatter} can be easily interpreted if, again, we think to the standard Maxwell theory of electromagnetism, where the divergence of the stress-energy tensor in presence of matter involves the 4D ``Lorentz force'' per unit volume on matter $f^\mu$~: 
\be
\partial_\nu T^{\mu\nu}+f^\mu=0\ .
\label{FMAX-divTmatter}\ee
At the right hand side of \eqref{FMAX-continuitymatter} appears
\be
f^0=E_{ab}J^{ab}\ ,
\label{FMAX-power}\ee
which is the analogous of the electromagnetic power $\vec{E}\cdot\vec{J}$. The last term at the right hand side of \eqref{FMAX-Fimatter}
\be
f^i=  2J_{a0}E^{ia}-\frac{1}{2}\epsilon^{0imn}J_{am}B_n^{\ a}
\label{FMAX-lorentzforce}\ee
can be traced back to the generalized Lorentz force on a dipole $p^i(x)$ moving with velocity $v^i$ proposed in \cite{Pretko:2016lgv} once we take
\be
J^{0i}\sim p^i\ ,
\label{FMAX-J0ipi}\ee
and
\be
J^{ij}\sim p^iv^j+p^jv^i\ .
\label{FMAX-Jijpivj}\ee
The first identification \eqref{FMAX-J0ipi} is compatible with \eqref{FMAX-dipole cons}, when \eqref{FMAX-gauss-vett} is taken into account, and the second relation \eqref{FMAX-Jijpivj} agrees with the microscopic lattice definition of the current of a dipole made in \cite{Pretko:2016lgv,Xu:2006}. What is remarkable is that we recover here as part of the conservation law of the stress-energy tensor the picture conjectured in \cite{Pretko:2016lgv}~: the isolated electric monopoles of the theory described by the action \eqref{FMAX-SfractF} are fractons, which do not respond to the electromagnetic fields and, hence, do not move, due to the dipole conservation constraint \eqref{FMAX-gauss+mat}. What we find here is that, instead, dipole motion, which preserves the global dipole moment, does respond to the electromagnetic field tensors $E^{ij}(x)$ and $B_i^{\ j}(x)$ according to \eqref{FMAX-lorentzforce}, like a conventional charge particle responds to an ordinary electromagnetic field. Hence, from \eqref{FMAX-lorentzforce}, we confirm the ``intuition'' proposed in \cite{Pretko:2016lgv} concerning the Lorentz force on a fracton dipole.
\normalcolor

\subsection{$\theta$-term}\label{sec theta}

In ordinary vector gauge field theory, it is known that a term can be added to the Maxwell action (or to its non-abelian extension, namely the Yang-Mills theory)~: the so called $\theta$-term, which has the form
\be
S_\theta\sim\theta\int d^4x\;\epsilon^{\mu\nu\rho\sigma}F_{\mu\nu}F_{\rho\sigma}\sim
\theta\int d^4x\;\vec{E}\cdot\vec{B}\ ,
\label{FMAX-Stheta}\ee
where $\theta$ is a constant parameter. The $\theta$-term represented by \eqref{FMAX-Stheta} is topological, since it does not depend on the spacetime metric, and it is a total derivative, hence it does not contribute to the \ac{EoM}. Nonetheless, the $\theta$-term is relevant in several contexts, like axion electrodynamics, the Witten effect and the strong CP problem (see for instance \cite{tong,Peccei:1977hh,Peccei:1977ur,Sikivie:1983ip,Wilczek:1987mv}). 
For what concerns the fracton theory,  in \cite{Pretko:2017xar} the $\theta$-term has been generalized as $E_{ij}B^{ij}$, in analogy with the fractonic Hamiltonian density, assumed to be proportional to $(E_{ij}E^{ij}+B_{ij}B^{ij})$.
Considering a compact tensor gauge field, which implies a ``magnetic'' monopole $(\partial_iB^{ij}=g^i\neq0)$, the introduction of the $\theta$-term gives to the Gauss constraint an additional contribution related to the ``magnetic'' field. As for dyons in the Witten effect \cite{Witten:1979ey}, the ``electric'' charge thus gains an additional contribution related to the ``magnetic'' vector charge \cite{Nandkishore:2018sel,Pretko:2020cko,Pretko:2017xar}. On the other hand  the possibility of a non-constant $\theta$-term has not yet been investigated in the context of fractons. 
The motivation for such a generalization comes from \ac{TI}, which are characterized by a step function $\theta$-term, which switches between $\theta=0$ outside the material and $\theta=\pi$ inside. Another example is given by
axion models \cite{Peccei:1977hh,Peccei:1977ur}, which 
describe a dynamical field coupled to photons via a local $\theta(x)$-term. This generates modified Maxwell equations \cite{Sikivie:1983ip,Wilczek:1987mv} as follows
	\begin{empheq}{align}
	\vec\nabla\cdot\vec E&=\rho-\vec\nabla\theta\cdot\vec B\label{FMAX-axion1}\\
	\vec\nabla\times\vec B-\partial_t\vec E&=\vec J+\partial_t\theta\,\vec B+\vec \nabla\theta\times\vec E\ ,\label{FMAX-axion2}
	\end{empheq}
where the additional terms contribute as an excess of charge ($\vec\nabla\theta\cdot\vec B$) and current ($\partial_t\theta\,\vec B+\vec \nabla\theta\times\vec E$) densities. Thus axion models are frequently used  in the context of condensed matter and \ac{TI} to mimic a non-constant $\theta$-term, like for instance in \cite{Rosenberg:2010ia}. As we are dealing with a rank-2 tensor theory, an interesting example is the case studied in \cite{Chatzistavrakidis:2020wum} in the context of LG, where modified gravitoelectromagnetic \cite{Mashhoon:2003ax} equations analogous to \eqref{FMAX-axion1} and \eqref{FMAX-axion2} are recovered.
According to the formalism presented in this Section, based on the Maxwell-like construction of a consistent theory for fractons, the analogous of the $\theta$-term should be the following
\begin{equation}
		S_\theta=\frac{1}{9}\int d^4x\,\theta\epsilon_{\mu\nu\rho\sigma}F^{\lambda\mu\nu} F_\lambda^{\;\rho\sigma}\\
		=\int d^4x\,\theta\epsilon_{\mu\nu\rho\sigma}\partial^\mu A^{\lambda\nu}\partial^\rho A_{\lambda}^\sigma\ .
\label{FMAX-Sthetafract}	\end{equation}
    We shall see that this is indeed the case by comparing the consequences of adding this term to the action $S_{fract}$ \eqref{FMAX-Sfract} to the known results concerning the $\theta$-term in the theory of fractons \cite{Pretko:2017xar,Nandkishore:2018sel,Pretko:2020cko}
    and of \ac{LG} \cite{Chatzistavrakidis:2020wum}. Notice that,  differently from the standard $\theta$-term introduced to solve the strong CP problem \cite{tong,Peccei:2006as}, $S_\theta$ is not topological, due to the contraction of the $\lambda$-indices in \eqref{FMAX-Sthetafract}. Moreover, here $\theta(x)$ is not constant, like in \cite{Chatzistavrakidis:2020wum}. For instance, $\theta(x)$ might be the Heaviside step function, which  would correspond to introducing a boundary at $x=0$, as we have seen in the previous Chapters \cite{Amoretti:2014iza,Bertolini:2020hgr,Bertolini:2021iku,Amoretti:2013xya,Bertolini:2022sao}. The contribution of $S_\theta$ to the \ac{EoM} is
\begin{equation}\label{FMAX-eomSt}
	\frac{\delta S_\theta}{\delta A^{\alpha\beta}}=-(\delta^\gamma_\alpha\delta^\sigma_\beta+\delta^\sigma_\alpha\delta^\gamma_\beta)\partial^\rho\theta\,\epsilon_{\mu\nu\rho\sigma}\eta_{\lambda\gamma}\partial^\mu A^{\lambda\nu}\ .
	\end{equation}
It is interesting to observe that this contribution is the same as the one that in \cite{Chatzistavrakidis:2020wum} gives the $\theta$-modified term of the gravitoelectromagnetic equations, where the electric and magnetic fields are vectors.
The \ac{EoM} \eqref{FMAX-eom2} acquire an additional term
\begin{equation}\label{FMAX-eomS+St}
	\frac{\delta S_{fract}}{\delta A^{\alpha\beta}}+\frac{\delta S_\theta}{\delta A^{\alpha\beta}}=0\ ,
	\end{equation}
whose components are
\begin{itemize}
\item $\alpha=\beta=0$
		\begin{equation}
		\partial_iF^{00i}-2\partial^k\theta\,\epsilon_{0ijk}\partial^iA^{0j}=0\ ,
		\end{equation}
which is still solved by $A^{0\mu}=\partial^\mu A^0$ \eqref{FMAX-A0}~;

\item $\alpha=0,\ \beta=i$ 
	\begin{equation}\label{FMAX-eomt0i}
	\partial_jE^{ij}-\partial^j\theta\, B_j^{\ i}=0\ ,
	\end{equation}
which is the tensorial extension of \eqref{FMAX-axion1}
	\begin{equation}
	\vec\nabla\cdot\vec E=-\vec\nabla\theta\cdot\vec B~;
	\end{equation}
	
\item $\alpha=i,\ \beta=j$ 
\begin{equation}
	\delta_{ab}^{ij}\left[\partial_0E^{ab}-\epsilon^{0akl}\partial_kB_l^{\;b}+\eta^{ac}\left(\epsilon^{0nm b}\partial_n\theta\, E_{mc}+\partial_0\theta\,B^b_{\;c}\right)\right]=0\ ,
	\end{equation}
where we defined the symmetrized delta 
\be
\delta^{ab}_{ij}\equiv\frac{1}{2}(\delta^a_i\delta^b_j+\delta^b_i\delta^a_j)\ ,
\label{FMAX-}\ee
which agrees with \eqref{FMAX-axion2} 	
\begin {equation}
	\vec\nabla\times\vec B-\partial_t\vec E=\partial_t\theta\,\vec B+\vec \nabla\theta\times\vec E\ .
	\end{equation}
\end{itemize}
Therefore Eqs. \eqref{FMAX-eomS+St} are generalized tensorial $\theta$-modified Maxwell equations. In particular, as in the standard modified Maxwell equations \cite{Wilczek:1987mv}, we can interpret the $\theta$-dependent terms as an excess of charge and current densities: 
	\begin{equation*}
	\partial_{j}E^{ij}=\tilde\rho^i\quad;\quad	-\partial_0E^{ij}+\frac{1}{2}\left(\epsilon^{0ikl}\partial_kB^{\ j}_{l}+\epsilon^{0jkl}\partial_kB^{\ i}_{l}\right)=\tilde J^{ij}\ ,
	\end{equation*}
with
	\begin{equation*}
	\tilde\rho^i\equiv\partial^j\theta\, B_j^{\ i}\quad;\quad\tilde J^{ij}\equiv\delta_{ab}^{ij}\eta^{ac}\left(\epsilon^{0nm b}\partial_n\theta\, E_{mc}+\partial_0\theta\,B^b_{\;c}\right)\ .
	\end{equation*}
\normalcolor
We have a further confirmation that $S_\theta$ \eqref{FMAX-Stheta} is indeed the correct $\theta$-term when we write it in terms of the electric and magnetic tensor fields \eqref{FMAX-electricfielddef} and \eqref{FMAX-Bij} 
\begin{equation}
S_\theta =
	-\frac{1}{3}\int d^4x\,\theta\eta_{lm}\epsilon_{0ijk}F^{il0}F^{mjk}
	=-\frac{1}{2}\int d^4x\,\theta E^{il}B_{il}\ ,
\label{FMAX-}\end{equation}
which  is the tensorial extension of the standard $\theta$-term $S_\theta\sim\theta\int \vec{E}\cdot\vec{B}$ for  constant $\theta$ \cite{tong}.

\subsection{Final remarks}\label{sec FinalFracton}

In Section \ref{sec literature} we have observed that in the Literature the usual approach to deal with a theory of fractons, is to treat space and time separately, hence non covariantly. The standard way to proceed is to take the {spatial} Gauss contraint \eqref{FMAX-dedeE=0}, written for a ``tensor electric field'' $E_{ij}(x)$,  as the tool to realize the defining property of fractons, $i.e.$ their limited mobility, by extending to fracton dipoles the usual conservation law holding for electric charges. This is usually achieved by introducing a ``tensor'', instead of a vector, potential $A_{ij}(x)$, obeying the generalized $spatial$ gauge transformation \eqref{FMAX-deA=dedelambda}. As we explained, the tensor field $E_{ij}(x)$ was, somehow, defined as the $spatial$ canonical momentum, as in \eqref{FMAX-electricfielddefPretko}. We say \textit{somehow} because in the definition of $E_{ij}(x)$ a scalar field $A_0(x)$ appears, as a multiplier introduced by hand in order to enforce the Gauss constraint.
Moreover, given the fracton limited mobility, it comes naturally to ask which is the generalization of the Lorentz force, and how the absence of motion could be compatible with the existence of an electromagnetic Lorentz force, and, above all, which is the elementary object on which such a force acts. Last but not least, in the Literature fractons are often seen in relation to gravity in a nontrivial way, starting from \cite{Pretko:2017fbf}. In this Section we adopted a covariant approach to the theory of fractons. This is not only a matter of formalism, but, rather, it allows to better understand the nature itself of these quasiparticles. The main contribution has thus been to show that, embedding the usual spatial and non-covariant theory of fractons in a more general \textit{covariant gauge field theory}, everything goes to the right place naturally, without introducing by hand any external ingredient. Our unique and starting point, as usual in field theory, is the covariant transformation \eqref{FMAX-fractonsymintro}, from which the most general \textit{covariant} invariant action \eqref{FMAX-Sinvg1g2} is derived. To cite a few new results following this approach, the relation with linearized gravity appears immediately, since the action \eqref{FMAX-Sinvg1g2} consists of two terms, one of which, namely \eqref{FMAX-SLG}, just describes linearized gravity. The theory of fractons as ``emergent electromagnetism'', as often has been called, is evident from the beginning as well, once we defined the rank-three ``electromagnetic'' field strength \eqref{FMAX-Fmunurho} by means of which the fracton Lagrangian writes as $F^2$, just like Maxwell theory (from which the title of the Section).  According to our field-theoretical point of view, the Gauss constraint is not an external constraint anymore, but turns out to be one of the equations of motion, the others formally coinciding with the Maxwell equations, which is mostly interesting, in our opinion. Finally, studying the conservation of the stress-energy tensor, we recovered the Lorentz force \eqref{FMAX-lorentzforce}, as it was correctly guessed in \cite{Pretko:2016lgv}, from which we see that, as one might expect, the force acts on fracton dipoles, and not on isolated charges, thus preserving the absence of mobility for isolated fractons. Here, we remark a subtle point which concerns the stress-energy tensor of the fracton action. In Section \ref{sec T+F} we computed the stress-energy tensor, defined as
\be
T_{\alpha\beta}=-\frac{2}{\sqrt{-g}}\frac{\delta S}{\delta g^{\alpha\beta}}\ ,
\label{FMAX-Tmunuintro}\ee
and we show that it has the correct components, which are exactly the higher rank generalizations of the Maxwell energy density ($T_{00}(x)$), of the Poynting vector ($T_{0i}(x)$) and of the stress tensor ($T_{ij}(x)$). The time component is conserved on shell, $i.e.\ \partial_\mu T^{\mu0}(x)=0,$ and gives the continuity equation relating the energy density to the Poynting vector. So far so good. But the space component of the stress-energy tensor conservation law is not exactly conserved. We find a breaking term which might be interpreted as follows. The stress-energy tensor \eqref{FMAX-Tmunuintro} is the conserved current associated to the infinitesimal diff invariance \eqref{FMAX-diffintro} \cite{Carroll:2004st}, { which is not} a symmetry of the theory defined by \eqref{FMAX-fractonsymintro}. Hence, the stress-energy tensor \eqref{FMAX-Tmunuintro} should not be conserved. Nonetheless, the fracton transformation \eqref{FMAX-fractonsymintro} is a particular case of the general diff transformation \eqref{FMAX-diffintro}. Hence, it is not that unexpected that the stress-energy tensor is $almost$ conserved. A further confirmation that $T_{\alpha\beta}(x)$ \eqref{FMAX-Tmunuintro} is the correct one is that, when matter is added, quite remarkably the conservation equation gives exactly the Lorentz force for fracton dipoles that has been conjectured in \cite{Pretko:2016lgv}.


\section{Gauge for Two} \label{sec gauge}
The aim here is to study the theory of fractons from the field theoretical point of view where, again, peculiar and unusual features appear. In our approach, presented in the previous Section, fractons are described by a gauge field theory involving a symmetric tensor field $A_{\mu\nu}(x)$ transforming as 
\be
\delta A_{\mu\nu} = \partial_\mu\partial_\nu \Lambda\ ,
\label{FGF-sym}\ee
where $\Lambda(x)$ is a scalar local gauge parameter. The transformation \eqref{FGF-sym} is a particular case of the infinitesimal diffeomorphisms
\be
\delta_{diff} A_{\mu\nu} = \partial_\mu\Lambda_\nu + \partial_\nu \Lambda_\mu\ ,
\label{FGF-diff}\ee
when the vector gauge parameter $\Lambda_{\mu}(x)$ is the derivative of a scalar field 
\be
\Lambda_{\mu}=\frac{1}{2}\partial_\mu\Lambda\ .
\label{FGF-difftofract}\ee
The most general action $S_{inv}$ \eqref{FMAX-Sinvg1g2} invariant under the fracton symmetry \eqref{FGF-sym} is the sum of two terms, one of which, \eqref{FMAX-SLG}, can be immediately identified as \ac{LG}  \cite{Carroll:2004st,Hinterbichler:2011tt}. This term is to be expected since the infinitesimal diffeomorphisms \eqref{FGF-diff}, which is the defining symmetry of \ac{LG}, embeds the fracton symmetry \eqref{FGF-sym} through \eqref{FGF-difftofract}. We have shown in the previous Section that the other term \eqref{FMAX-Sfract} describes a theory with pure fractonic features \cite{Bertolini:2022ijb}~: from this term one can recover all the properties characterizing the so called ``scalar charge theory'' of fractons \cite{Pretko:2016lgv}, including Maxwell-like equations of which the Gauss law is at the origin of the limited mobility property that defines fractons \cite{Prem:2017kxc,Pretko:2016kxt,Pretko:2016lgv,Nandkishore:2018sel,Pretko:2020cko}.  Since the transformation \eqref{FGF-sym} is a particular case of the more general infinitesimal diffeomorphism \eqref{FGF-diff}, which is a gauge transformation \cite{Carroll:2004st,Hinterbichler:2011tt,Blasi:2015lrg,Gambuti:2020onb}, 
the theory of the symmetric tensor field $A_{\mu\nu}(x)$ defined by the symmetry \eqref{FGF-sym} is a $gauge$ field theory as well. 
Evidence of this appears when trying to compute the propagator from the quadratic action $S_{inv}$ \eqref{FMAX-Sinvg1g2}, which, similarly to the electromagnetic Maxwell theory, leads to a non invertible matrix. Therefore a gauge fixing term must be added. For any other gauge field theory, this procedure goes smoothly thanks to the Faddeev-Popov procedure \cite{Faddeev:1967fc}, which focuses on the gauge parameter, and in particular on its tensorial character. In \eqref{FGF-sym} the gauge parameter is a scalar, and this would require a scalar gauge fixing condition. The most general covariant one is
\be
\partial^\mu\partial^\nu A_{\mu\nu} + \kappa\partial^2A=0\ ,
\label{FGF-scalargaugecond}\ee
where $\kappa$ is a constant gauge fixing parameter. This is analogous to the covariant Lorentz condition for the vector field $A_\mu(x)$
\be
\partial^\mu A_\mu=0\ .
\label{FGF-lorentzgaugecond}\ee
The standard situation in gauge field theory is that of having a gauge field, represented by a $p$-tensor field, and a $(p-1)$-tensorial gauge parameter. This is the case of all known gauge theories. We already mentioned Maxwell theory, or its non-abelian counterpart, the Yang-Mills theory, but this is also true for higher rank theories, like the topological $BF$ theories in any dimensions \cite{Birmingham:1991ty}. \acl{LG}, described by a symmetric tensor field and by the symmetry \eqref{FGF-diff}, does not escape this rule. In this case, the commonly used gauge fixing condition is vectorial \cite{Hinterbichler:2011tt,Blasi:2017pkk,Gambuti:2021meo}
\be
\partial^\nu A_{\mu\nu} + \kappa\partial_\mu A=0\ .
\label{FGF-vectorgaugecond}\ee
Due to the presence of the parameter $\kappa$, the gauge fixing conditions \eqref{FGF-scalargaugecond} and \eqref{FGF-vectorgaugecond}  represent a class of covariant gauges. For instance, in \eqref{FGF-vectorgaugecond} the particular case $\kappa=-\frac{1}{2}$ corresponds to the ``harmonic'' gauge fixing \cite{Carroll:2004st,Gambuti:2020onb}. The theory we are considering here is, again, quite peculiar. Not only because, as we have seen, it displays a ``non-coupling'' constant, but also because it is defined by the gauge transformation \eqref{FGF-sym}, which associates a scalar gauge parameter to a rank-2 symmetric tensor field. Concerning the gauge fixing procedure, one might therefore look at both sides of \eqref{FGF-sym}, each of which opens a different path. Looking at the right hand side of \eqref{FGF-sym}, one sees a scalar gauge parameter and this leads to adopt the scalar gauge condition \eqref{FGF-scalargaugecond}. This standard way has been followed in \cite{Blasi:2022mbl}, where the propagators have been derived and the degrees of freedom have been studied. The scalar gauge condition \eqref{FGF-scalargaugecond} has two important drawbacks. The first is that the Landau gauge $\xi=0$ turns out to be mandatory. The theory seems not to be defined outside this gauge. Now, it is true that physical results should not depend on the gauge choice, but, still, being forced to a unique choice is unpleasant, and it would be much preferable to find all the physical results in a generic gauge and to show that they do not, indeed, depend on a particular choice. The second reason is that the scalar gauge condition \eqref{FGF-scalargaugecond} does not allow to reach the limit of pure \ac{LG}, which necessarily needs the vector gauge condition \eqref{FGF-diff}. This results in a singularity both in the propagators of the theory and in the degrees of freedom, which indeed has been found in \cite{Blasi:2022mbl}~: the theory with the scalar gauge condition \eqref{FGF-scalargaugecond} is not defined in the limit of pure \ac{LG}. The alternative, which we consider in this Section, is to
focus on the left hand side of \eqref{FGF-sym}, where the same symmetric rank-2 tensor field of \ac{LG} appears, and decide to adopt the same vectorial gauge condition \eqref{FGF-vectorgaugecond} as \ac{LG}. In doing so, two questions should be answered~: 
\begin{enumerate}
\item is the vector gauge condition a good gauge fixing, or, equivalently, do the propagators exist in the pure fractonic limit, possibly without being forced to choose a particular gauge~? 
\item in gauge field theory the gauge fixing condition serves to eliminate the redundant degrees of freedom which render infinite the Green functions' generating functional $Z[J]$
\be
Z[J]=\int [dh_{\mu\nu}]e^{iS_{inv}+\int J^{\mu\nu}A_{\mu\nu}}\ .
\label{FGF-Z}\ee
Do the fact of imposing four (vector) gauge conditions instead of one (scalar) affect the number of physical degrees of freedom of the whole theory $S_{inv}$~\eqref{FMAX-Sinvg1g2}~? 
\end{enumerate}
The above are legitimate and well posed questions and, naively, one might answer positively to both. To the first simply because four conditions are more than one, and one expects that they are more than enough to invert the gauge fixed action to find the propagators; to the second for the same reason~: four conditions are more than one, and hence the degrees of freedom which are eliminated are too much and differ from those ``killed'' by the scalar gauge choice \eqref{FGF-scalargaugecond}. We shall see that the propagators are not singular in the limits of pure fractons or pure \ac{LG} and that the number of physical degrees of freedom is the same for the two gauge fixing choices \eqref{FGF-scalargaugecond} and \eqref{FGF-vectorgaugecond}, which therefore are equivalent, with the advantage that the vectorial choice \eqref{FGF-vectorgaugecond} does not constrain us to the Landau gauge and allows to easily recover \ac{LG}.\\

The analysis of the gauge structure of the covariant fracton theory presented in this Section is organized as follows :
\etocsetnexttocdepth{2}
\begingroup
\parindent=0em
\etocsettocstyle{\rule{\linewidth}{\tocrulewidth}\vskip1.25\baselineskip}{\vskip-0.75\baselineskip\rule{\linewidth}{\tocrulewidth}\vskip1\baselineskip}
\makeatletter
  \edef\scr@tso@subsection@indent
    {\the\dimexpr\scr@tso@subsection@indent-\scr@tso@section@indent}
  \def\scr@tso@section@indent{0pt}
\makeatother
\localtableofcontents 
\endgroup
\noindent
 In Section \ref{FGF-sec:the-model}, starting from the symmetry \eqref{FGF-sym}, the action of the theory is introduced, which consists of two terms: \ac{LG} and a fractonic term. The vector gauge condition \eqref{FGF-vectorgaugecond} is realized by adding a gauge fixing term to the action. In Section \ref{FGF-sec:propagators} the propagators are computed, and the singularities are studied, which correspond to particular phases of the theory. In Section \ref{FGF-sec:degrees-of-freedom} we study the degrees of freedom, and we verify that their counting coincides with the known one \cite{Blasi:2022mbl}, without the drawback of being confined to the Landau gauge, which reassures of the fact that the number of degrees of freedom does not depend on a particular choice and that the alternative vectorial gauge fixing condition \eqref{FGF-vectorgaugecond} is, indeed, a good one. In Section \ref{FGF-sec:summary-and-discussion} we discuss our results.

\subsection{The model}\label{FGF-sec:the-model}

Let us consider the 4D theory of a symmetric tensor field $A_{\mu\nu}(x)$ which transforms as \eqref{FGF-sym}.
The choice \eqref{FGF-difftofract} is motivated by the fact that the theory defined by the symmetry \eqref{FGF-sym} describes the so called ``fractons''. 
The most general action invariant under \eqref{FGF-sym} was found in Section \ref{sec MaxThFract} to be
\be
S_{inv}(g_1,g_2)=g_1 S_{fract}+g_2 S_{LG}\ ,
\label{FGF-Sinv}\ee
where
\bea
S_{fract} &=&
\int d^4x \left(
A^{\mu\nu}\partial^\rho \partial_\mu A_{\rho\nu} - A_{\mu\nu} \partial^2 A^{\mu\nu}\right)
\label{FGF-Sfract}\\
S_{LG} &=&
\int d^4x \left(
	- A \partial^2 A + A_{\mu\nu} \partial^2 A^{\mu\nu} 
	+2 A\partial_\mu \partial_\nu A^{\mu\nu}
	-2 A^{\mu\nu}\partial^\rho \partial_\mu A_{\nu\rho}
	\right)
\label{FGF-SLG}\ ,
\eea
and $A(x)$ is the Minkowskian trace of $A_{\mu\nu}(x)$
\be
A=\eta^{\mu\nu}A_{\mu\nu}\ .
\label{FGF-A}\ee
The action \eqref{FGF-Sinv} appears to be the linear combination of two terms,  which we recognize to be the action $S_{LG}$ of \ac{LG} \cite{Carroll:2004st,Hinterbichler:2011tt} and the pure covariant fractonic action previously described \cite{Blasi:2022mbl,Bertolini:2022ijb}. The actions $S_{LG}$ and $S_{fract}$ are separately invariant under \eqref{FGF-sym}
\be
\delta S_{LG} = \delta S_{fract} = 0\ ,
\label{FGF-}\ee
and $g_{1,2}$ are constants, on which we will come back in a moment. Notice that while the space of 4D local integrated functionals invariant under \eqref{FGF-sym} is the linear combination \eqref{FGF-Sinv} of two elements, the infinitesimal diffeomorphism symmetry \eqref{FGF-diff} uniquely determines one functional only~: the \ac{LG} action $S_{LG}$ \eqref{FGF-SLG}
\be
\delta_{diff}S_{LG} = 0\ ,
\label{FGF-}\ee
under which the fractonic action $S_{fract}$ \eqref{FGF-Sfract} is not invariant, as we have seen in \eqref{FMAX-notdiff}.  In other words, the ``fractonic'' symmetry \eqref{FGF-sym} is less constraining than the diffeomorphism transformation \eqref{FGF-diff}, of which it is a particular case. The action \eqref{FGF-Sinv} actually depends on one constant only, because of the possibility of redefining the gauge field by a multiplicative constant $c$  without affecting the physical content of the theory: $A_{\mu\nu}\rightarrow cA_{\mu\nu}$. Nevertheless, 
we will keep both $g_1$ and $g_2$, in order to track the contributions of the gravitational ($g_1\rightarrow 0$) and the fractonic ($g_2\rightarrow 0$) parts. 
 Concerning the gauge fixing, the standard way to realize the condition \eqref{FGF-vectorgaugecond} is to add to the invariant action \eqref{FGF-Sinv} the gauge fixing term
\be
S_{gf}(\xi,\kappa)= 
-\frac{1}{2\xi}\int d^4x \left(\partial^\nu A_{\mu\nu} +\kappa\partial_\mu A\right)^2\ ,
\label{FGF-SgfFP}\ee
as it has been done in \cite{Blasi:2015lrg,Blasi:2017pkk,Gambuti:2021meo,Gambuti:2020onb} for \ac{LG} alone. In a fully equivalent way, \eqref{FGF-SgfFP} can be linearized by means of 
a Lagrange multiplier $b^\mu(x)$, also known as Nakanishi-Lautrup field \cite{Nakanishi:1966zz,Lautrup:1967zz}~:
\begin{equation}
S_{gf}(\xi,\kappa)= 
\int d^4x \left[ b^\mu\left(\partial^\nu A_{\mu\nu} +\kappa\partial_\mu A\right)+ \frac{\xi}{2}b_\mu b^\mu\right]
\ .
\label{FGF-Sgf}\end{equation}
In $S_{gf}(\xi,\kappa)$ two gauge fixing parameters appear~: $\xi$ and $\kappa$. The first -$\xi$- governs the type of gauge fixing. For instance $\xi=0$ and $\xi=1$ are respectively the Landau and Feynman gauges. The second -$\kappa$- tunes the type of gauge fixing picked up by $\xi$. For instance, the Landau gauge in \ac{LG} corresponds to a class of gauge choices, and it is realized by $\xi=0$ and generic $\kappa$.

\subsection{Propagators}\label{FGF-sec:propagators}

To find the propagators of the theory, we write the gauge fixed action 
\be
S(g_1,g_2;\xi,\kappa) = S_{inv}(g_1,g_2) + S_{gf}(\xi,\kappa)
\label{FGF-gaugefixedaction}\ee
in momentum space~:
\be
\int d^4p
		\begin{pmatrix}
			\tilde{A}^{\mu\nu}(p) &
			\tilde{b}^\gamma(p)
		\end{pmatrix}
		\begin{pmatrix}
			{\Omega}_{\mu\nu,\alpha\beta}(p) & {\Lambda}^*_{\mu\nu,\lambda}(p)\\
			{\Lambda}_{\gamma,\alpha\beta}(p) & {H}_{\gamma\lambda}(p)
		\end{pmatrix}
		\begin{pmatrix}
			\tilde{A}^{\alpha\beta}(-p)\\
			\tilde{b}^\lambda(-p)
		\end{pmatrix},
\label{FGF-momentumaction}\ee
where $ {\Omega}_{\mu\nu,\alpha\beta}(p)$, $ {\Lambda}^*_{\mu\nu,\lambda}(p)$ and $ {H}_{\gamma\lambda}(p)$ are $p$-dependent tensor operators. The propagators of the theory are obtained by inverting the operator matrix appearing in \eqref{FGF-momentumaction}. In order to do this, it is useful to write $ {\Omega}_{\mu\nu,\alpha\beta}(p)$, $ {\Lambda}_{\alpha\beta,\mu}(p)$ and $ {H}_{\mu\alpha}(p)$ on the corresponding tensorial basis, as follows
\bea
 {\Omega}_{\mu\nu,\alpha\beta} &= &
 {t}\,A_{\mu\nu,\alpha\beta} +  {u}\,B_{\mu\nu,\alpha\beta}+ 
 {v}\,C_{\mu\nu,\alpha\beta}+  {z}\,D_{\mu\nu,\alpha\beta} + 
 {w}\,E_{\mu\nu,\alpha\beta}
\label{FGF-Omega}\\
 {\Lambda}_{\alpha\beta,\mu} &=&
 -\frac{i}{2}\left[ {f}\,(d_{\alpha\mu} p_\beta + d_{\beta\mu}p_\alpha) +  {g}\,d_{\alpha\beta} p_\mu +  {l}\,e_{\alpha\beta}p_\mu \right]
 \label{FGF-Lambda}\\
 {H}_{\mu\alpha} &=&  {r}\, d_{\mu\alpha} +  {s}\, e_{\mu\alpha}\ ,
\label{FGF-H}\eea
where $e_{\mu\nu}(p)$ and $d_{\mu \nu}(p)$ are transverse and longitudinal projectors, respectively,
\be
e_{\mu\nu}=\frac{p_\mu p_\nu}{p^2}\quad;\quad
d_{\mu\nu}=\eta_{\mu\nu}-e_{\mu\nu}\ ,
\label{FGF-projectors}\ee
and the rank-4 tensor ${\Omega}_{\mu\nu,\alpha\beta}(p)$ \eqref{FGF-Omega} is expanded on a basis of operators 
\be
X_{\mu\nu,\alpha\beta}\equiv (A,B,C,D,E)_{\mu\nu,\alpha\beta}
\label{FGF-baseX}\ee
which can be found in Appendix \ref{FGF-sec:basis-for-the-omega-tensors}, together with their properties. The coefficients appearing in \eqref{FGF-Omega}, \eqref{FGF-Lambda} and \eqref{FGF-H} are found to be
	\begin{align}
		 {t} &= (g_1+ 2g_2)p^2		&;&	&	 {u} &= 0	&;&&	 {v} &= (g_1-g_2) p^2	&;&&	 {z} &= \frac{1}{2}g_1 p^2	&;&&	 {w} &= 0\label{FGF-3.8}\\[5pt]
		 {f} &= \frac{1}{2}	&;&	&	 {g} &= \kappa	&;&&	 {l} &= 1+\kappa	&;&&	 {r} &= \frac{\xi}{2} &;&&	 {s} &= \frac{\xi}{2}\label{FGF-3.9}\ .
	\end{align}
The propagators of the theory are organized in a matrix of tensor operators as well
\begin{equation}\label{FGF-matricepropagatori}
		\begin{pmatrix}
			\hat{{G}}^{\alpha\beta,\rho\sigma}& \hat{{G}}^{\alpha\beta,\tau}\\
			\hat{{G}}^{*\lambda,\rho\sigma}& \hat{{G}}^{\lambda\tau}
		\end{pmatrix}\ ,
	\end{equation}
	where
\begin{align}
		\hat{G}_{\alpha\beta,\rho\sigma}   = \langle \tilde{A}_{\alpha\beta}\;  \tilde{A}_{\rho\sigma}  \rangle &=  \hat{t}\,A_{\alpha\beta,\rho\sigma} + 
		\hat{u}\,B_{\alpha\beta,\rho\sigma} + \hat{v}\,C_{\alpha\beta,\rho\sigma} + \hat{z}\,D_{\alpha\beta,\rho\sigma} + \hat{w}\,E_{\alpha\beta,\rho\sigma}
\label{FGF-hhprop}	\\
		\hat{G}_{\alpha\beta,\rho}   = \langle \tilde{A}_{\alpha\beta}\;  \tilde{b}_{\rho}  \rangle &=  i\left[\hat{f}\,(d_{\alpha\rho} p_\beta + d_{\beta\rho}p_\alpha) + \hat{g}\,d_{\alpha\beta} p_\rho +\hat{l}\,e_{\alpha\beta}p_\rho\right]\label{FGF-hbprop}\\
		\hat{G}_{\alpha\rho}   = \langle \tilde{b}_{\alpha}\;  \tilde{b}_{\rho}  \rangle &= \hat{r}\,d_{\alpha\rho} + \hat{s}\,e_{\alpha\rho}\ ,\label{FGF-bbprop}
	\end{align}
and the set of coefficients 
\be
\left\{\hat t\,,\, \hat u\,,\, \hat v\,,\, \hat z\,,\, \hat w\,,\, \hat f\,,\, \hat g\,,\, \hat l\,,\, \hat r\,,\, \hat s\right\}
\label{FGF-propcoeff}\ee
are determined by the request that the matrix of propagators \eqref{FGF-matricepropagatori} satisfies
\begin{equation}\label{FGF-eq:matriciale}
		\begin{pmatrix}
			  {\Omega}_{\mu\nu,\alpha\beta} &   {\Lambda}^*_{\mu\nu,\lambda}\\
			  {\Lambda}_{\gamma,\alpha\beta} &   {H}_{\gamma\lambda}
		\end{pmatrix}
		\begin{pmatrix}
			\hat{G}^{\alpha\beta,\rho\sigma} & \hat{G}^{\alpha\beta,\tau}\\
			\hat{G}^{*\lambda,\rho\sigma}& \hat{G}^{\lambda\tau}
		\end{pmatrix}
		=
		\begin{pmatrix}
			\mathcal{I}_{\mu\nu}^{\ \ \rho\sigma} &0 \\
			0 & \delta_{\gamma}^{\ \tau}
		\end{pmatrix}\ ,
	\end{equation}
where $\mathcal{I}_{\mu \nu, \alpha\beta}$ is the rank-4 tensor identity 
\be
\mathcal{I}_{\mu \nu, \rho \sigma} = \frac{1}{2} (\eta_{\mu \rho} \eta_{\nu \sigma} + \eta_{\mu \sigma} \eta_{\nu \rho})\ .
\label{FGF-identity}\ee
In Appendix \ref{FGF-sec:2g1g2neq0} we show that the four tensor equations \eqref{FGF-eq:matriciale} are solved by
	\begin{align}
		\hat{t} &= \frac{(4\kappa+1)}{(\kappa+1)(g_1+ 2g_2)p^2}&;&
		&\hat{u}& = \frac{\kappa(4\kappa+1)-2\xi(g_1+ 2g_2)}{(\kappa+1)^2(g_1+ 2g_2)p^2}\label{FGF-eq:uhat}\\
		\hat{v}& = \frac{1}{(g_1-g_2)p^2}&;&
		&\hat{z}& = \frac{4\xi}{(2\xi g_1 -1)p^2}\label{FGF-eq:zhat}\\
		\hat{w}& = \frac{-4\kappa}{(\kappa+1)(g_1+ 2g_2)p^2}&;&
		&\hat{f}& = \frac{-2}{(2\xi g_1-1)p^2}\label{FGF-eq:fhat}\\
		\hat{g} &= 0&;&
		&\hat{l} &= \frac{2}{(\kappa+1)p^2}\label{FGF-eq:lhat}\\
		\hat{r} &= \frac{4g_1}{(2\xi g_1-1)}&;&
		&\hat{s} &= 0\label{FGF-eq:shat}\ .
	\end{align}
Now we can answer the first of our questions, concerning the fractonic limit~: the vector gauge fixing condition \eqref{FGF-vectorgaugecond} is a good one not only for \ac{LG} ($g_1=0$), but also for the pure fractonic case ($g_2=0$), as expected. On the other hand, we see that the propagators are singular in four cases~:
\bea
g_1+ 2g_2 &=& 0 \label{FGF-2g1+g2}\\
g_1-g_2 &=& 0\label{FGF-g1-g2}\\
2\xi g_1-1 &=& 0 \label{FGF-2kappag2-g1}\\
\kappa+1 &=& 0 \label{FGF-kappa+1}
\ .
\eea
For what concerns the singularity \eqref{FGF-kappa+1}, it simply implies that the ``secondary'' gauge fixing parameter $\kappa$, which tunes the gauge fixing choice of the ``primary'' parameter $\xi$ in $S_{gf}$ \eqref{FGF-Sgf}, should be 
\be
\kappa\neq -1\ ,
\label{FGF-kneq-1}\ee
as it happens also in \ac{LG} \cite{Blasi:2015lrg,Gambuti:2020onb,Blasi:2017pkk,Gambuti:2021meo}. The remaining singularities involve the action parameters $g_1$ and $g_2$ and the gauge fixing parameter $\xi$. The poles in the propagators give us information on the structure of the theory. For instance, they might signal the presence of masses, possibly not standard, as in the topologically massive Maxwell-Chern-Simons theory \cite{Deser:1981wh}. Or they might indicate the presence of phase transitions in the theory, like in QCD \cite{Cabibbo:1975ig,Halasz:1998qr}, or in the sigma model \cite{Baym:1977qb,Grater:1994qx}, or in the fracton theory itself \cite{Blasi:2022mbl}.  Therefore, the singularities appearing in the propagators of the theory $S_{inv}$ should be treated separately and with care.
 \begin{itemize}
 \item${   g_1=g_2}$\hfill \\
In this case, after a field redefinition, the invariant action \eqref{FGF-Sinv}  is 
	\begin{equation}\label{FGF-g1=g2}
\left.S_{inv}(g_1,g_2)\right|_{g_1=g_2}
= \int d^4p\ (\tilde{A} p^2 \tilde{A} -2 \tilde{A} p_\mu p_\nu \tilde{A}^{\mu\nu} +\tilde{A}^{\mu\nu}p^\rho p_\mu \tilde{A}_{\nu\rho})\ .
	\end{equation}
Notice that defining	
	\begin{equation}
		\tilde{\mathcal{A}}^\rho \equiv p^\rho\tilde{A} - p_\nu \tilde{A}^{\rho\nu}\ ,
	\end{equation}
the action \eqref{FGF-g1=g2} trivializes into	
	\begin{equation}
	\left.S_{inv}(g_1,g_2)\right|_{g_1=g_2} = - \int d^4p\ \tilde{\mathcal{A}}^\rho \tilde{\mathcal{A}}_\rho\ ,
	\end{equation}
which does not contain any kinetic term. Hence, the singularity at $g_1=g_2$ is explained as a point where the theory trivializes and does not propagate, and this case will be excluded from now on.
\item${   g_1+ 2g_2=0}$\hfill \\
In this case the invariant action \eqref{FGF-Sinv}, after a field redefinition,  reads 
	\begin{equation}\label{FGF-2g1+g_2}
\left.S_{inv}(g_1,g_2)\right|_{g_1+ 2g_2=0}
= \int d^4p\ 
 \left(\tilde{A} p^2 \tilde{A} -3\tilde{A}_{\mu\nu} p^2 \tilde{A}^{\mu\nu} -2 \tilde{A} p_\mu p_\nu \tilde{A}^{\mu\nu} +4 \tilde{A}^{\mu\nu}p^\rho p_\mu \tilde{A}_{\nu\rho} \right)\ .
	\end{equation}
We see that with this choice the action does not depend on the trace $\tilde A (p)$. In fact, defining
\begin{equation}
		{\bar{A}}_{\mu\nu}(p)\equiv \tilde{A}_{\mu\nu}(p)-\frac{1}{4}\eta_{\mu\nu}\tilde{A}(p)\ ,
\label{FGF-tracelessh}	\end{equation}
with
	\begin{equation}
		{\bar{A}}(p) = 0\ ,
	\end{equation}
the action \eqref{FGF-2g1+g_2} can be written in terms of $\bar{A}_{\mu\nu}(p)$ only~:
\begin{equation}\label{FGF-eq:tracelessaction}
		\left.S_{inv}(g_1,g_2)\right|_{g_1+ 2g_2=0} = \int d^4p\ \left(-3{\bar{A}}_{\mu\nu}p^2{\bar{A}^{\mu\nu}}+4{\bar{A}^{\mu\nu}}p^\rho p_\mu {\bar{A}_{\nu\rho}}\right)\ .
	\end{equation}
Hence, in this case the theory is traceless, and the singularity at the point $g_1+ 2g_2=0$ indicates a change in the counting of the degrees of freedom, as we shall explicitly show. The gauge fixing term \eqref{FGF-Sgf} does not depend on the trace $A(x)$ anymore and, hence, on the gauge fixing parameter $\kappa$. In momentum space it reads
	\begin{equation}\label{FGF-Sgftraceless}
S_{gf}(\xi)=
 \int d^4p\ \left(-i\tilde{b}_\mu p_\nu \bar{A}^{\mu\nu} + \frac{\xi}{2}\tilde{b}_\mu \tilde{b}^\mu\right)\ .
	\end{equation}
The propagators of the traceless theory are well defined, and the coefficients, computed in Appendix \ref{FGF-sec:2g1g20}, are 	
	\begin{align}
		\hat{t} &= - \frac{1}{3p^2}	&\hat{u} &= \frac{2\xi}{(2\xi-1)p^2}\\
		\hat{v}&= - \frac{1}{3p^2}	&\hat{z} &= \frac{-4\xi}{(4\xi+1)p^2}\\
		\hat{w} &= 0	&\hat{f} &= \frac{2}{(4\xi+1)p^2}\\
		\hat{g} &= 0	& \hat{l} &= \frac{-2}{(2\xi-1)p^2}\\
		\hat{r} &= \frac{8}{(4\xi+1)}&	\hat{s} &= \frac{4}{(2\xi-1)}\ .
	\end{align}
From the above coefficients we see that the particular values of the primary gauge parameter $\xi=-1/4$ and $\xi=1/2$ should be excluded.
\item${   2\xi g_1-1=0}$\hfill \\	
We first notice that this singularity is not present  in the pure \ac{LG} case $g_1=0$, as it is readily seen from the coefficients $\hat z$ \eqref{FGF-eq:zhat}, $\hat f$ \eqref{FGF-eq:fhat} and $\hat r$ \eqref{FGF-eq:shat} of the propagators.
Then, we remark that $g_1$ is a physical parameter, hence cannot depend on $\xi$, which is a gauge, unphysical, parameter. This means that $2\xi g_1-1=0$ should be interpreted as a singularity in $\xi$ as a function of $g_1$, and not \textit{viceversa}. In other words, we shall not exclude values of $g_1$ in order to admit a particular gauge, but, rather, the singularity must be read as a condition on the gauge fixing parameter $\xi$~:
\be 
\xi\neq \frac{1}{2g_1}\ .
\label{FGF-kappacond}\ee 
We consider the gauge fixing term \eqref{FGF-SgfFP}, before the the introduction of the Lagrange multiplier $b^\mu(x)$. At $2\xi g_1-1=0$ the gauge fixed action is
	\begin{multline}
\left.S(g_1,g_2;\xi,\kappa)\right|_{2\xi g_1-1=0} =
\int d^4p\ 
\left[
		(g_2-g_1\kappa^2)\tilde{A}p^2\tilde{A}
		-(g_2-g_1)\tilde{A}_{\mu\nu}p^2\tilde{A}^{\mu\nu}+\right.\\\left.
		-2(g_2+g_1\kappa)\tilde{A}p_\mu p_\nu \tilde{A^{\mu\nu}}
		+2(g_2-g_1)\tilde{A}^{\mu\nu}p_\mu p^\rho \tilde{A}_{\nu\rho}\right]\ .
	\end{multline}
Using the general results of Appendix \ref{FGF-sec:2g1g2neq0} it is easy to verify that this theory does not have propagators. As a remark, if we choose also $\kappa+1=0$, we find a curious result
	\begin{equation}
\left.S(g_1,g_2;\xi,\kappa)\right|_{2\xi g_1-1=0,\kappa+1=0}
=(g_2-g_1)S_{LG}\ .
\label{FGF-nogf}	\end{equation}
This means that with the particular gauge choice which involves both the singularities in the two gauge parameters $\xi$ and $\kappa$,  the gauge fixing procedure fails in choosing one representative for each gauge orbit, which is what the gauge fixing is supposed to do. In fact, according to \eqref{FGF-nogf}, in this particular gauge, the fracton contribution disappears, and the gauge fixed action coincides with $S_{LG}$ alone, which still needs to be gauge fixed. It also appears the fact which we already know that for $g_1=g_2$, which is the trivial, non propagating case already considered, the action vanishes. We thus showed that for $\xi=1/2g_1$ and $\kappa=-1$ the gauge fixed action $S(g_1,g_2;\xi,\kappa)$ coincides with the invariant, not gauge fixed, action $S_{LG}$.

\end{itemize}

\subsection{Degrees of freedom}\label{FGF-sec:degrees-of-freedom}

The counting of the degrees of freedom of the theory described by the action $S_{inv}$ \eqref{FGF-Sinv} is a crucial point. This is true in general, but for this model it is even more true. What we already know from \cite{Blasi:2022mbl} is that, if we adopt the standard scalar gauge choice \eqref{FGF-scalargaugecond}, which is the natural one when dealing with a gauge transformation depending on a scalar parameter, the degrees of freedom turn out to be six, as in \ac{LG} alone (five in the traceless case), as if the fractonic contribution $S_{fract}$ \eqref{FGF-Sfract} was not present. However, in \cite{Blasi:2022mbl} the choice of the Landau gauge appears to be mandatory, which is rather unpleasant, although the physical results should not depend on the gauge choice. Still, it would be preferable to avoid such a restriction, which instead seems unavoidable in the scalar gauge. This, together with the fact that \ac{LG} cannot be reached as a limit, leads us to conclude that the scalar gauge choice is not that natural, as the direct application of the Faddeev-Popov procedure would suggest. The aim is therefore to see whether the alternative and, at first sight, exotic choice of the $vector$ gauge condition \eqref{FGF-vectorgaugecond} is an acceptable, and possibly better, one. In the previous Subsection we passed the first test: we have seen that the vector gauge condition leads to well defined propagators, with a pole which corresponds to the traceless theory, in accordance to the scalar case \cite{Blasi:2022mbl}. In this Subsection, instead, we face the trickier point of the counting of the degrees of freedom. Not only we should recover the known result, but, and more important, we should show that the number of degrees of freedom does not depend on the gauge choice, which was impossible with the scalar gauge condition \eqref{FGF-scalargaugecond} frozen in the Landau gauge. This fact is not obvious, since the role of the gauge fixing is to eliminate the redundant degrees of freedom, in order to render finite the path integral $Z[J]$  \eqref{FGF-Z}, and the justified fear is that the four conditions represented by the vector choice \eqref{FGF-vectorgaugecond} might lead us to underestimate the degrees of freedom with respect to the unique scalar condition \eqref{FGF-scalargaugecond}.
The usual way to count the degrees of freedom is to look for the constraints deriving from the equations of motion of the gauge fixed action
\be
S(g_1,g_2;\xi,\kappa)= 
S_{inv}(g_1,g_2) + S_{gf}(\xi,\kappa)\ ,
\label{FGF-Stot}\ee
where the invariant action $S_{inv}(g_1,g_2)$ and the gauge fixing term $S_{gf}(\xi,\kappa)$ are given by \eqref{FGF-Sinv} and \eqref{FGF-Sgf}, respectively. In momentum space, we get
\begin{align}
\frac{\delta S}{\delta \tilde{A}^{\mu\nu}} 
=&\;
2g_2\eta_{\mu\nu}p^2\tilde{A} + 2(g_1-g_2)p^2\tilde{A}_{\mu\nu}-2g_2\eta_{\mu\nu}p_\alpha p_{\beta}\tilde{A}^{\alpha\beta}-2g_2p_\mu p_\nu\tilde{A} 
 \label{FGF-eomh}\\
+&(2g_2-g_1)p^\alpha(p_\mu  \tilde{A}_{\alpha\nu} +p_\nu  \tilde{A}_{\alpha\mu})  +\frac{i}{2}(p_\nu\tilde{b}_\mu+p_\mu\tilde{b}_\nu)+i\kappa\eta_{\mu\nu}p_\alpha \tilde{b}^{\alpha} = 0\nonumber
\\
\frac{\delta S}{\delta \tilde{b}^{\mu}} =& -ip^\alpha \tilde{A}_{\alpha\mu} -i\kappa p_\mu\tilde{A} + \xi\tilde{b}_\mu = 0\ . \label{FGF-eomb}
\end{align}
If our task was just to count the degrees of freedom, given that they must not depend on the gauge choice, we would immediately find the result by choosing $\xi=\kappa=0$ in $S_{gf}$ \eqref{FGF-Sgf}, which belongs to the class of Landau gauges. The $\tilde b^\mu$-equation of motion \eqref{FGF-eomb} gives
\be
p^\alpha \tilde{A}_{\alpha\mu}=0\ ,
\label{FGF-xi=k=0}\ee
which are the four constraints needed to recover the six degrees of freedom (five in the traceless case) which we expect for the symmetric rank-2 tensor field $A_{\mu\nu}(x)$. But we want more, that is to show that this number does not depend on the gauge choice, which would render the vector gauge condition a good one under any respect. Achieved that, the vector choice would be even preferable to the scalar one, since the Landau gauge would not be the only possibility, and \ac{LG} could be obtained as a limit. Hence we proceed without choosing a particular gauge, and
we saturate \eqref{FGF-eomh} with $\eta^{\mu\nu}$, $e^{\mu\nu}(p)$ \eqref{FGF-projectors} and $p^\mu$~:
\begin{align}
		\eta^{\mu\nu}\frac{\delta S}{\delta \tilde{A}^{\mu\nu}} &= 2(g_1+ 2g_2)\left(p^2\tilde{A}-p_\alpha p_{\beta}\tilde{A}^{\alpha\beta}\right)+i(1+4\kappa)p_\alpha \tilde{b}^{\alpha} = 0\label{FGF-eq:satura1}\\
		e^{\mu\nu}\frac{\delta S}{\delta \tilde{A}^{\mu\nu}} &= i(1+\kappa)p_\alpha \tilde{b}^{\alpha} = 0\label{FGF-eq:satura2}\\
		p^\nu\frac{\delta S}{\delta \tilde{A}^{\mu\nu}} &= 2g_1p^\alpha \left(p^2 \tilde{A}_{\alpha\mu} -p_\mu  p^\beta\tilde{A}_{\alpha\beta}\right) + ip^2\tilde{b}_\mu +i(1+2\kappa)p_\mu p_\alpha \tilde{b}^{\alpha} = 0\ .\label{FGF-eq:satura3}
	\end{align}
Multiplying \eqref{FGF-eomb} by $p^\mu$, we get
	\begin{equation}
		p^\mu \frac{\delta S}{\delta \tilde{b}^{\mu}} = ip_\alpha p_\beta\tilde{A}^{\alpha\beta} +i\kappa p^2\tilde{A} - \xi p_\alpha \tilde{b}^{\alpha} = 0\ .\label{FGF-eq:satura4}
	\end{equation}
We separately study  the generic case $g_1+ 2g_2\neq0$ and the traceless case $g_1+ 2g_2=0$\ .

\subsubsection*{Case $g_1+ 2g_2\neq0$}\label{FGF-sec:case-2g1g2neq0}

From \eqref{FGF-eq:satura2}, remembering that $\kappa+1\neq0$, we get the condition
	\begin{equation}
		p_\alpha \tilde{b}^\alpha =0\ ,
	\end{equation}
which, plugged in \eqref{FGF-eomb}, \eqref{FGF-eq:satura1}, \eqref{FGF-eq:satura3} and \eqref{FGF-eq:satura4}, yields
	\begin{align}
		&\xi\tilde{b}_\mu = i\left(p^\alpha \tilde{A}_{\alpha\mu} +\kappa p_\mu\tilde{A} \right)\label{FGF-eq:Bl}\\
		&(g_1+ 2g_2)\left(p^2\tilde{A}-p_\alpha p_{\beta}\tilde{A}^{\alpha\beta}\right) = 0\label{FGF-eq:satura12}\\
		&2g_1p^\alpha \left(p^2 \tilde{A}_{\alpha\mu} -p_\mu p^\beta\tilde{A}_{\alpha\beta}\right) + ip^2\tilde{b}_\mu = 0\label{FGF-eq:satura32}\\
		&ip_\alpha p_\beta\tilde{A}^{\alpha\beta} +i\kappa p^2\tilde{A} = 0\ .\label{FGF-eq:satura42}
	\end{align}
Now, since we are outside the critical point $g_1+ 2g_2=0$ \eqref{FGF-2g1+g2}, from \eqref{FGF-eq:satura12} and \eqref{FGF-eq:satura42} we have
\bea
		p^2\tilde{A}&=&p_\alpha p_{\beta}\tilde{A}^{\alpha\beta}\label{FGF-eq:p2hpph}\\
		-\kappa p^2\tilde{A} &=& p_\alpha p_\beta\tilde{A}^{\alpha\beta}\ ,
\eea
that is
\begin{equation}
		(1+\kappa)p^2\tilde{A} =0\quad\Rightarrow\quad p^2\tilde{A} =0\ ,
\label{FGF-p2h=0}	\end{equation}
hence
\begin{equation}\label{FGF-eq:pcurrent}
		p_\alpha p_{\beta}\tilde{A}^{\alpha\beta} = 0\ .
\end{equation}
Notice that the conditions \eqref{FGF-p2h=0} and \eqref{FGF-eq:pcurrent} are the ones holding for \ac{LG} alone ($g_1=0$) \cite{Blasi:2017pkk,Gambuti:2021meo}. It appears, therefore, that the fracton contribution ($g_2=0$) \eqref{FGF-Sfract} to the total invariant action \eqref{FGF-Sinv} is irrelevant as far as the degrees of freedom are concerned, which is an unexpected result. Nonetheless we can directly check  this result. Substituting the conditions \eqref{FGF-p2h=0} and \eqref{FGF-eq:pcurrent} into \eqref{FGF-eq:satura32}, we find
	\begin{equation}\label{FGF-eq:misfatto}
		-ip^2\tilde{b}_\mu = 2g_1 p^2p^\alpha \tilde{A}_{\alpha\mu}\ .
	\end{equation}
Using \eqref{FGF-eq:Bl} and \eqref{FGF-p2h=0}, assuming $\xi\neq0$, $i.e.$ excluding for the moment the Landau gauge, equation \eqref{FGF-eq:misfatto} becomes
	\begin{equation}
		(2g_1\xi-1)p^2p^\alpha\tilde{A}_{\alpha\mu} =0\ .
	\end{equation}
We previously studied the case $2g_1\xi-1=0$, which we now exclude. This means that
	\begin{equation}\label{FGF-eq:ppph}
		p^2p^\alpha\tilde{A}_{\alpha\mu} = 0
	\end{equation}
	and, from \eqref{FGF-eq:misfatto},
	\begin{equation}
		p^2\tilde{b}_\mu = 0\ .
	\end{equation}
We now define
	\begin{equation}
		\tilde{J}_\alpha \equiv p^\beta\tilde{A}_{\alpha\beta}\ ,
\label{FGF-defJ}	\end{equation}
which, because of \eqref{FGF-eq:pcurrent}, is a conserved current
	\begin{equation}
		p^\alpha\tilde{J}_\alpha = 0\ .
\label{FGF-pJ}	\end{equation}
The solution of \eqref{FGF-pJ} is
\begin{equation}\label{FGF-eq:jB}
		\tilde{J}_\alpha = \epsilon_{\alpha\mu\nu\rho}p^\mu \tilde{B}^{\nu\rho}\ ,
	\end{equation}
where $\tilde{B}^{\nu\rho}$ is a generic antisymmetric tensor. Plugging \eqref{FGF-defJ} in \eqref{FGF-eq:ppph}, we have
\begin{equation}
p^2\tilde{J}_\alpha = p^2\epsilon_{\alpha\mu\nu\rho}p^\mu \tilde{B}^{\nu\rho} = 0\ ,
	\end{equation}
	and therefore
	\begin{equation}
		p^2\tilde{B}_{\rho\lambda} = p_\rho\tilde{B}_\lambda - p_\lambda \tilde{B}_\rho\ .
	\end{equation}
From \eqref{FGF-eq:jB} we then deduce that the current $J_\alpha$ vanishes 		\begin{equation}\label{FGF-eq:lastconstr}
		\tilde{J}_\alpha = p^\beta\tilde{A}_{\alpha\beta} = 0\ .
	\end{equation}
We now come back to the Landau gauge $\xi=0$, which has been excluded in achieving the above result. We now show that \eqref{FGF-eq:lastconstr} holds also in this case. The $\tilde{b}^\mu$-equation of motion in the Landau gauge is
	\begin{equation}
		p^\alpha \tilde{A}_{\alpha\mu} = -\kappa p_\mu\tilde{A}\ .
	\end{equation}
The degrees of freedom must not depend on the gauge choice. We therefore choose $\kappa=0$ and we get 
	\begin{equation}\label{FGF-eq:llgdof}
		{p^\alpha \tilde{A}_{\alpha\mu} = 0}\ ,
	\end{equation}
which represents four constraints on the 4D symmetric tensor field $\tilde A_{\mu\nu}(p)$. Hence the number of degrees of freedom are six, at least if $g_1+ 2g_2\neq0$. This coincides with the number of degrees of freedom of \ac{LG} alone \cite{Blasi:2017pkk,Gambuti:2021meo}. 
We have seen that $g_1+ 2g_2=0$ corresponds to the traceless case, which is interesting and will be treated separately. 

\subsubsection*{Case $g_1+ 2g_2=0$}\label{FGF-sec:case-2g1g20}

As we have seen in Section \ref{FGF-sec:propagators}, this case corresponds to the traceless theory. The gauge fixed action can be written in terms of the traceless field $\bar A_{\mu\nu}(x)$ \eqref{FGF-tracelessh}, with $\kappa=0$, and, in momentum space, it reads 
	\begin{equation}
	\left.S(g_1,g_2;\xi,\kappa)\right|_{g_1+ 2g_2=0,\kappa=0}= \int d^4p\ \left(
		-3\bar{A}_{\mu\nu}p^2\bar{A}^{\mu\nu}
		+4\bar{A}^{\mu\nu}p_\mu p^\rho \bar{A}_{\nu\rho} 
		-i\tilde{b}_\mu p_\nu \bar{A}^{\mu\nu}
		+\frac{\xi}{2}\tilde{b}_\mu\tilde{b}^\mu
		\right)\ ,
	\end{equation}
whose equations of motion are
\begin{align}
		\frac{\delta S}{\delta \bar{A}^{\mu\nu}}  &=
		-6p^2\bar{A}_{\mu\nu} 
		+4p^\alpha\left(p_\mu\bar{A}_{\alpha\nu}+p_\nu \bar{A}_{\alpha\mu}\right)-2\eta_{\mu\nu}p^\alpha p^\beta\bar{A}_{\alpha\beta}\label{FGF-eq:1}\\
		&+\tfrac{i}{2}\left(p_\nu\tilde{b}_\mu +p_\mu\tilde{b}_\nu\right)-\tfrac{i}{4}\eta_{\mu\nu}p_\alpha\tilde b^\alpha = 0\nonumber
	\\
		\frac{\delta S}{\delta \tilde{b}^\mu} &=  -ip^\nu \bar{A}_{\mu\nu}+\xi\tilde{b}_\mu =0\ ,
\label{FGF-eq:2}\end{align}
where we remind that, for a traceless, symmetric tensor field
	\be
	\frac{\delta\bar{A}^{\alpha\beta}}{\delta\bar{A}^{\mu\nu}}=\frac{1}{2}\left(\delta^\alpha_\mu\delta_\nu^\beta+\delta^\alpha_\nu\delta_\mu^\beta\right)-\frac{1}{4}\eta_{\mu\nu}\eta^{\alpha\beta}\ .
	\ee
Saturating \eqref{FGF-eq:1} with  $e^{\mu\nu}(p)$ and $p^\nu$, we get
\begin{align}
e^{\mu\nu}\frac{\delta S}{\delta \bar{A}^{\mu\nu}} &= ip_\alpha\tilde{b}^\alpha =0\label{FGF-eq:sat2}\\
p^\nu\frac{\delta S}{\delta \bar{A}^{\mu\nu}} &=-2p^2 p^\alpha \bar{A}_{\alpha\mu} + 2p_\mu p_\alpha p_\beta \bar{A}^{\alpha\beta}+\frac{i}{2}p^2\tilde{b}_\mu -\frac{i}{4}p_\mu p_\alpha \tilde{b}^\alpha =0\ .\label{FGF-eq:sat3}
\end{align}
From saturating \eqref{FGF-eq:2} with $p^\mu$, and \eqref{FGF-eq:sat2} we have
\be 
p_\alpha p_\beta \bar{A}^{\alpha\beta} = 0 \label{FGF-eq:key2}\ .
\ee
If $\xi\neq0$, the Lagrange multiplier $\tilde b_\mu(p)$ can be obtained from \eqref{FGF-eq:2} and, plugged in \eqref{FGF-eq:sat3}, using \eqref{FGF-eq:key2}, we get
	\begin{equation}
		(4\xi+1)p^2p^\alpha \bar{A}_{\alpha\mu} = 0\ .
	\end{equation}
We have already excluded the gauge choice $4\xi+1=0$, which is the propagator singularity $2\xi g_1-1=0$ at $g_1+ 2g_2=0$, hence
	\begin{equation}\label{FGF-eq:key1}
		p^2p^\alpha \bar{A}_{\alpha\mu} = 0\ .
	\end{equation}
Now, using the same argument of Section \ref{FGF-sec:case-2g1g2neq0}, involving $\tilde J_\alpha(p)$ \eqref{FGF-defJ} and $\tilde B^{\nu\rho}(p)$ \eqref{FGF-eq:jB}, Eqs. \eqref{FGF-eq:key2} and \eqref{FGF-eq:key1} imply the four constraints
	\begin{equation}\label{FGF-eq:key3}
		p^\alpha \bar{A}_{\alpha\mu} = 0\ .
	\end{equation}
On the other hand, when $\xi=0$, $i.e.$ in the Landau gauge, the equation of motion of the Lagrange multiplier \eqref{FGF-eq:2} directly gives the constraint \eqref{FGF-eq:key3}. Hence, in all cases we have four constraints on a traceless rank-2 symmetric tensor. Therefore, when $g_1+ 2g_2=0$, the degrees of freedom are five. \\

The results for the different values of the action constants $g_1$ and $g_2$ and of the gauge fixing parameters $\xi$ and $\kappa$ are summarized in Table \ref{FGF-table:dofsummary}.

\begin{table}[H]
	\resizebox{1\columnwidth}{!}{%
		\bgroup
		\setlength\tabcolsep{15pt}
		\def\arraystretch{1.5}{
			\begin{tabular}{|c|cccc|}
				\hline
				\multirow{2}{*}{$\mathbf{g_1}$, $\mathbf{g_2}$} & \multicolumn{2}{c|}{\textbf{vectorial gauge fixing}} & \multicolumn{2}{c|}{\textbf{scalar gauge fixing}}\\ \cline{2-5} 
				& \multicolumn{1}{c|}{\textbf{degrees of freedom}} & \multicolumn{1}{c|}{\textbf{forbidden gauges}} & \multicolumn{1}{c|}{\textbf{degrees of freedom}} & \textbf{forbidden gauges} \\ \hline
				$g_1\neq g_2\neq 0$, $g_1+ 2g_2\neq0$ & \multicolumn{1}{c|}{6} & \multicolumn{1}{c|}{$\xi = \frac{1}{2g_1}$, $\kappa=-1$} & \multicolumn{1}{c|}{6} & $\xi\neq0$ \\ \hline
				$g_1=0$ (LG) & \multicolumn{1}{c|}{6} & \multicolumn{1}{c|}{$\kappa=-1$} & \multicolumn{2}{c|}{not defined}    \\ \hline
				$g_1+ 2g_2=0$& \multicolumn{1}{c|}{5} & \multicolumn{1}{c|}{$\xi=\left\{\frac{1}{2}, -\frac{1}{4}\right\}$} & \multicolumn{1}{c|}{5} &  $\xi\neq0$  \\ \hline
				$g_1=g_2$& \multicolumn{4}{c|}{trivial}                                                    \\ \hline
			\end{tabular}
		\egroup}
	}
	\caption[Gauge-fixing : comparison with the scalar case]{\footnotesize{Summary of results and comparison with the scalar case}\label{FGF-table:dofsummary}}
\end{table}
\noindent

\subsection{Final remarks}\label{FGF-sec:summary-and-discussion}

The aim of this Section was to find a well defined gauge fixed theory, not constrained to a particular gauge choice, and where both limiting cases, fractonic and linearized gravity, could be reached smoothly. The idea is simply to look at the left hand side of \eqref{FGF-sym} and take the vectorial gauge fixing choice \eqref{FGF-vectorgaugecond},
which, again, is the most general covariant one and is the same of linearized gravity. It is not obvious at all that it could work. First, what does it mean that ``it works'' ? Besides the possibility of getting both fractons and gravitons as a limit, we asked two minimal requirements: that propagators are defined in a generic gauge, and that the counting of the degrees of freedom of the theory coincides with the one found in \cite{Blasi:2022mbl}, but without referring to a particular gauge choice. The result  is that both requests have been achieved. We have now a covariant gauge fixed theory of fractons and linearized gravity, which has six degrees of freedom, or five, since for a particular combination of fractons and gravitons the theory is traceless. All the results are gauge independent, therefore the vector gauge fixing \eqref{FGF-vectorgaugecond} seems to be a better choice than the standard scalar one \eqref{FGF-scalargaugecond}.


\section{Fractons with Boundary} \label{sec Frac+bd}
We now have a 4D covariant theory that describes fracton quasiparticles, whose gauge structure is  peculiar but well defined. We are thus able to apply the \ac{QFT} procedure of Chapter \ref{QFTapproach} to study the consequences of the introduction of a flat boundary, with the aim of investigating whether an induced 3D theory exists and, in that case, which is its physical meaning. Tightly related is the question of the existence of an algebraic structure on the boundary. The theory we are dealing with is not topological, and it was a common belief that only topological field theories show non-trivial boundary physics, until the study of the Maxwell case. Moreover, we have seen that when a boundary is introduced in a \ac{QFT}, the gauge symmetry plays a fundamental role, since it is the breaking of gauge invariance caused by the presence of the boundary that gives rise to an algebraic structure on the boundary which ``holographically'' induces a lower-dimensional gauge theory. The fracton symmetry \eqref{FMAX-dA} is unusual, due to the presence of a double derivative, and considering a boundary on such a model $a\ priori$ may have a non-trivial outcome. On the other hand it would not be the first case of a non-topological \ac{QFT} exhibiting an induced theory on the boundary. In fact, as we mentioned, this also happens in the case of Maxwell theory in 3D \cite{Maggiore:2019wie} and 4D \cite{Bertolini:2020hgr}, and we know from Section \ref{sec MaxThFract} that fracton models share many similarities with the electromagnetic theory \cite{Pretko:2016lgv,Bertolini:2022ijb}. Additionally, it has been shown that a fractonic $\theta$-term, which is a pure boundary term when $\theta$ is constant, gives rise to a 3D Chern-Simons-like  term and a  generalized Witten effect \cite{Pretko:2017xar}, with important consequences in condensed matter systems \cite{Pretko:2020cko,You:2019bvu}. 
A non-covariant Chern-Simons-like  term of that kind was also studied in \cite{Cappelli:2015ocj}, where the higher-spin formalism is associated to dipolar behaviours in the context of Hall systems. In order to find these answers, we proceed as follows
\etocsetnexttocdepth{2}
\begingroup
\parindent=0em
\etocsettocstyle{\rule{\linewidth}{\tocrulewidth}\vskip1.25\baselineskip}{\vskip-0.75\baselineskip\rule{\linewidth}{\tocrulewidth}\vskip1\baselineskip}
\makeatletter
  \edef\scr@tso@subsection@indent
    {\the\dimexpr\scr@tso@subsection@indent-\scr@tso@section@indent}
  \def\scr@tso@section@indent{0pt}
\makeatother
\localtableofcontents 
\endgroup
\noindent
In particular in Section \ref{FBD-sec-model+bd} the boundary is introduced in the invariant action \eqref{FMAX-Sinvg1g2} described in Section \ref{sec MaxThFract} together with the gauge fixing and the most general boundary term. From the total action the \ac{EoM} and the most general \ac{BC} are computed. Because of the presence of the boundary, the Ward identities of the theory are broken, and this allows to identify the boundary \ac{DoF}, 
represented by two traceless symmetric rank-2 tensor fields. Moreover, the broken Ward identities give rise to an algebraic structure, which can be identified as a generalized \ac{KM} algebra, which, in Section \ref{FBD-sec-3Dmodel}, we interpret as canonical commutators of a 3D action. In Section \ref{FBD-sec-hc} the bulk/boundary correspondence is obtained by requiring that the \ac{EoM} of the induced 3D action are compatible with the \ac{BC} of the 4D bulk theory. This can be achieved by suitably tuning the parameters appearing in the \ac{BC} and in the 3D action. To physically interpret the 3D theory we found, 
in Section \ref{FBD-sec-phys} we study its \ac{EoM}, which appear to be 
Gauss and Amp\`ere-like laws for the boundary tensor fields, exactly as in the ``ordinary'' fracton theory. In particular our boundary theory can be identified with a
traceless fracton model. In Section \ref{FBD-sec PT} we analyze the effect of taking into account discrete symmetries such as parity ($\Par$)  and time reversal ($\TR$). Finally in Section \ref{FBD-sec-summary} we discuss our results.

\subsection{The model with boundary}\label{FBD-sec-model+bd}

\subsubsection*{The action}
 We introduce the boundary in the invariant action $S_{inv}$ \eqref{FMAX-Sinvg1g2} by means of a Heaviside step function in the action :
	\be\label{FBD-Sbulk}
	S_{bulk} =
	\int d^4x\;\theta(x^3)\left\{\frac{g_1}{6}F^{\mu\nu\rho}F_{\mu\nu\rho}+g_2 \left(
	\frac{1}{4}F^\mu_{\ \mu\nu} F_\rho^{\ \rho\nu}-\frac{1}{6}F^{\mu\nu\rho}F_{\mu\nu\rho}
	\right)\right\}\ .
	\ee
Notice that in what follows we cannot just set $g_1=0$ and restrict our results to \ac{LG} alone, because 
 $S_{LG}$ \eqref{FMAX-SLG} is uniquely defined by the infinitesimal diffeomorphism transformation \eqref{FMAX-diff}, and not by its subset \eqref{FMAX-dA}.
The transformations \eqref{FMAX-diff} and \eqref{FMAX-dA} differs in two aspects~: the first, \eqref{FMAX-diff}, depends on a vector gauge parameter, while \eqref{FMAX-dA} has a scalar gauge parameter, hence the former is more restrictive. Secondly, \eqref{FMAX-diff} and \eqref{FMAX-dA} depend on one and two derivatives respectively. As we have seen, this results in a mismatch in the mass dimensions. In fact, since from the action \eqref{FBD-Sbulk} we have $[A_{\mu\nu}]=1$, due to the double derivatives in \eqref{FMAX-dA} it must be $[\Lambda]=-1$, which is an exotic dimension assignment for the scalar gauge parameter. Moreover, on the $x^3$-boundary, the field $A_{\mu\nu}(x)$ and its $\partial_3$-derivative must be treated as independent fields \cite{Maggiore:2019wie,Bertolini:2020hgr,Karabali:2015epa,Blasi:2019wpq}, like we have seen for the Maxwell case of Chapter \ref{ch nonTFT}. Hence, on the boundary we define
	\be
	\tilde A_{\mu\nu}\equiv\partial_3A_{\mu\nu}|_{x^3=0}\ ,
	\ee
with $[\tilde A_{\mu\nu}]=2$.
We add to the invariant action $S_{inv}$ \eqref{FMAX-Sinvg1g2} the gauge-fixing term 
	\be
	S_{gf}=\int d^4x\;\theta(x^3)b^\mu A_{\mu3}\ ,
	\ee
which is of a vectorial type, as discussed in the previous Section \ref{sec gauge}, where $b^\mu(x)$ is a Nakanishi-Lautrup Lagrange multiplier \cite{Nakanishi:1966zz,Lautrup:1967zz} implementing the axial gauge condition 
	\be\label{FBD-ax-gf}
	A_{\mu3}=0\ .
	\ee
As a consequence of the fact that the field and its $\partial_3$-derivative on the boundary are independent quantities, together with the usual external field $J^{ab}(x)$ coupled to $A_{ab}(x)$, it is necessary to couple a source $\tilde J^{ab}(x)$ also to the $\partial_3$-derivative of $A_{ab}(x)$ on the boundary. The external source term is then
	\be
	S_J=\int d^4x\left[\theta(x^3)J^{ab}A_{ab}+\delta(x^3)\tilde J^{ab}\tilde A_{ab}\right]\ .
	\ee
	The presence of a boundary in a \ac{QFT} naturally rises the question of which \ac{BC} should be assigned to the quantum fields and/or their derivatives. A possible way is to impose them by hand, but one should worry about the dependence of the results on the particular choice. We have seen that this arbitrariness affecting \acp{QFT} with boundary has been elegantly solved by Symanzik in his pioneering paper \cite{Symanzik:1981wd}, where a scalar \ac{QFT} with boundary was considered. According to Symanzik's approach, the \ac{BC} are not imposed by hand, but are determined by the theory itself. This is achieved by adding a boundary term to the action, as the most general one, satisfying the requests of locality, power counting and 3D Lorentz invariance. The \ac{BC} are then determined from the \ac{EoM}, modified by the boundary term
	\be\label{FBD-Sbd}
	S_{bd}=\int d^4x\delta(x^3)\left[\xi_0 A_{ab}A^{ab}+\xi_1\tilde A_{ab}A^{ab}+\xi_2\epsilon^{abc}A_{ai}\partial_bA_c^i+\xi_3A^2+\xi_4\tilde AA\right]\ ,
	\ee
where, due to the gauge condition \eqref{FBD-ax-gf},
	\be
	A\equiv\eta^{\mu\nu}A_{\mu\nu}=\eta^{ab}A_{ab}\quad;\quad\tilde A\equiv\eta^{\mu\nu}\tilde A_{\mu\nu}=\eta^{ab}\tilde A_{ab}\ ,
	\ee
and $\xi_i$ are constant parameters, whose mass dimensions are
	\be
	[\xi_0]=[\xi_3]=1\quad;\quad[\xi_1]=[\xi_2]=[\xi_4]=0\ .
	\ee
Notice that the general \ac{QFT} requirements which constrain $S_{bd}$ imply the presence of the Chern-Simons-like  $\xi_2$-term, which can be traced back to the covariant fractonic $\theta$-term \eqref{FMAX-Sthetafract}. In fact, this latter is 
\be
S_\theta=\int d^4x \theta(x^3) \epsilon^{\mu\nu\rho\sigma} \partial_\mu A_{\nu\lambda}\partial_\rho A_\sigma^\lambda\ ,
\label{FBD-thetaterm}\ee
which, integrating by parts, reduces to
\be
\int d^3X \epsilon^{abc} A_{a\lambda}\partial_b A_c^\lambda\ ,
\label{FBD-xi2term}\ee
which, on the gauge condition \eqref{FBD-ax-gf}, coincides with the $\xi_2$-term in \eqref{FBD-Sbd}.
The total action is then
	\be\label{FBD-Stot}
	S_{tot}=S_{bulk}+S_{gf}+S_{J}+S_{bd}\ .
	\ee
	
\subsubsection*{Equations of motion and boundary conditions}

The \ac{EoM} for $A_{\alpha\beta}(x)$ and $\tilde A_{\alpha\beta}(x)$ are
\begin{align}
\frac {\delta S_{tot}}{\delta A_{\alpha\beta}} =&\theta(x^3)\left\{(g_1-g_2)\partial_{\mu}F^{\alpha\beta\mu}
+g_2\left[\eta^{\alpha\beta}\partial_\mu F_\nu^{\ \nu\mu}-\tfrac{1}{2}\left(\partial^\alpha F_\mu^{\ \mu\beta}+\partial^\beta F_{\mu}^{\ \mu\alpha}\right)\right]+\right.\nonumber\\
&+\left.\delta^\alpha_a\delta^\beta_bJ^{ab}+\tfrac{1}{2}(b^\alpha\delta^\beta_3+b^\beta\delta^\alpha_3)\right\}+\label{FBD-eomA}\\
&+\delta(x^3)\left\{(g_1-g_2)F^{\alpha\beta3}+g_2\left[\eta^{\alpha\beta}F_\mu^{\ \mu3}-\tfrac{1}{2}\left(\eta^{\alpha3}F_\mu^{\ \mu\beta}+\eta^{\beta3}F_\mu^{\ \mu\alpha}\right)\right]+\right.\nonumber\\
&+\left.\delta^\alpha_a\delta^\beta_b\left[ 2\xi_0A^{ab}+\xi_1\tilde A^{ab}+\xi_2(\epsilon^{aij}\partial_iA_j^b+\epsilon^{bij}\partial_iA_j^a)+2\xi_3\eta^{ab}A+\xi_4\eta^{ab}\tilde A\right]\right\}=0\ ,\nonumber
\end{align}
and
\be
\begin{split}
\frac {\delta S_{tot}}{\delta \partial_3A_{\alpha\beta}} &=\theta(x^3)\left\{(g_2-g_1)F^{\alpha\beta3}
-g_2\left[\eta^{\alpha\beta}F_\mu^{\ \mu3}-\tfrac{1}{2}\left(\eta^{\alpha3}F_\mu^{\ \mu\beta}+\eta^{\beta3}F_{\mu}^{\ \mu\alpha}\right)\right]\right\}\\
&+\delta(x^3)\delta^\alpha_a\delta^\beta_b\left\{ \tilde J^{ab}+\xi_1A^{ab}+\xi_4\eta^{ab}A\right\}=0\ .\label{FBD-eomAt}
\end{split}
\ee	
The most general \ac{BC} are obtained by applying  $\lim_{\epsilon\to0}\int^\epsilon_0dx^3$ to the \ac{EoM}. From \eqref{FBD-eomA} we get
\begin{align}
&\left\{(g_1-g_2)F^{\alpha\beta3}+g_2\left[\eta^{\alpha\beta}F_\mu^{\ \mu3}-\tfrac{1}{2}\left(\eta^{\alpha3}F_\mu^{\ \mu\beta}+\eta^{\beta3}F_\mu^{\ \mu\alpha}\right)\right]+\right.\label{FBD-bcA}\\
&+\left.\delta^\alpha_a\delta^\beta_b\left[ 2\xi_0A^{ab}+\xi_1\tilde A^{ab}+\xi_2(\epsilon^{aij}\partial_iA_j^b+\epsilon^{bij}\partial_iA_j^a)+2\xi_3\eta^{ab}A+\xi_4\eta^{ab}\tilde A\right]\right\}_{x^3=0}=0\ .\nonumber
\end{align}
We observe that
\begin{itemize}
\item $\alpha=\beta=3$ is trivially realized ;
\item $\alpha=3,\ \beta=b$ :
	\be
	g_2\left(\partial^bA-\partial_aA^{ab}\right)_{x^3=0}=g_2F_\mu^{\ \mu b}|_{x^3=0}=0\ ;\label{FBD-bcA3i}
	\ee
\item $\alpha=a,\ \beta=b$ :
	\begin{align}
	\left[ 2\xi_0A^{ab}+2\left(g_2-g_1+\tfrac{\xi_1}{2}\right)\tilde A^{ab}+\right.&\xi_2\left(\epsilon^{aij}\partial_iA_j^b+\epsilon^{bij}\partial_iA_j^a\right)+\label{FBD-bcAij}\\
	+&\left.
	2\xi_3\eta^{ab}A+2\left(\tfrac{\xi_4}{2}-g_2\right)\eta^{ab}\tilde A\right]_{x^3=0}=0\ .\nonumber
	\end{align}
\end{itemize}
Going on-shell, $i.e.$ at vanishing external sources $\tilde J(x)=0$, taking the $\lim_{\epsilon\to0}\int^\epsilon_0dx^3$ of the \ac{EoM} \eqref{FBD-eomAt} we get
\be
\delta^\alpha_a\delta^\beta_b\left(\xi_1A^{ab}+\xi_4\eta^{ab}A\right)_{x^3=0}=0\ ,\label{FBD-bcAt}
\ee
and again we observe that
\begin{itemize}
\item $\alpha=3,$ $\beta$ free, is trivially realized ;
\item $\alpha=a,\ \beta=b$ :
\be
\left(\xi_1A^{ab}+\xi_4\eta^{ab}A\right)_{x^3=0}=0\ .\label{FBD-bcAtij}
\ee
\end{itemize}

\subsubsection*{\textcolor{Maroon}{Ward identities and boundary degrees of freedom}}

The \ac{EoM} \eqref{FBD-eomA} yields the following integrated Ward identity
	\be
	\begin{split}
	0&=\int dx^3\partial_a\partial_b\frac{\delta S_{tot}}{\delta A_{ab}}\\
	&=\int dx^3\theta(x^3)\left\{(g_1-g_2)\partial_a\partial_b\partial_{3}F^{ab3}+g_2\partial_a\partial^a \partial_3F_\mu^{\ \mu 3}+\partial_a\partial_bJ^{ab}\right\}\ ,
	\end{split}
	\ee
where we used the \ac{BC} \eqref{FBD-bcAij} and the cyclic property of $F_{\mu\nu\rho}(x)$ in Table \ref{FMAX-table1}. Integrating by parts we get
	\be\label{FBD-wi1}
	\int dx^3\theta(x^3)\partial_i\partial_jJ^{ij}=
	2(g_2-g_1)\partial_i\partial_j\tilde A^{ij}-2g_2\partial_i\partial^i\tilde A|_{x^3=0}\ .
	\ee
Analogously, from the \ac{EoM} \eqref{FBD-eomAt} we find
	\be
	\begin{split}
	0&=\int dx^3\partial_a\partial_b\frac{\delta S_{tot}}{\delta \partial_3A_{ab}}\\
	&=-\int dx^3\theta(x^3)\partial_a\partial_b\left[2(g_2-g_1)\partial^3A^{ab}-2g_2\eta^{ab}\partial^3A\right]+\partial_a\partial_b\tilde J^{ab}|_{x^3=0}\ ,
	\end{split}
	\ee
where we used the \ac{BC}  \eqref{FBD-bcAtij}. Integrating by parts
	\be\label{FBD-wi2}
	\partial_i\partial_j\tilde J^{ij}|_{x^3=0}=-2(g_2-g_1)\partial_i\partial_j A^{ij}+2g_2\partial_i\partial^iA|_{x^3=0}\ .
	\ee
	Notice that the second Ward identity \eqref{FBD-wi2}, associated to the $\tilde A_{ab}(x)$ field on the boundary, is local and not integrated as the first \eqref{FBD-wi1}. The two Ward identities \eqref{FBD-wi1} and \eqref{FBD-wi2} are analogous to those characterizing Maxwell theory with boundary both in 3D and 4D \cite{Maggiore:2019wie,Bertolini:2020hgr}, which we have seen in Chapter \ref{ch nonTFT} in Eqs. \eqref{eq:wi1} and \eqref{eq:wi2}. At vanishing external source $J^{ab}(x)=0$, the Ward identity \eqref{FBD-wi1} gives
	\be
	\partial_i\partial_j\left[(g_2-g_1)\tilde A^{ij}-g_2\eta^{ij}\tilde A\right]_{x^3=0}=0\ .\label{FBD-cc1}
	\ee
In \cite{Blasi:2022mbl, Bertolini:2023juh} it has been shown that, when $g_1=g_2$ in the invariant action $S_{inv} $ \eqref{FMAX-Sinvg1g2}, it is possible to redefine the components of $A_{\mu\nu}(x)$ in such a way that the theory has no kinetic term, hence it is not dynamical. Therefore in what follows we shall exclude the trivial case
\be
g_1=g_2\ .
\ee
For $g_1\neq g_2$, and $g_{1,2}\neq0$, \eqref{FBD-cc1} implies
	\begin{empheq}{align}
	\partial_i\partial_j\tilde A^{ij}(X)&=0\label{FBD-ddA=0}\\
	\Box \tilde A(X)&=0\ .\label{FBD-boxA=0}
	\end{empheq}
Eq.\eqref{FBD-ddA=0} is solved as follows \cite{Henneaux:2004jw,Bunster:2012km}
	\be
	\partial_i\left(\partial_j\tilde A^{ij}\right)=0\quad\Rightarrow\quad \partial_j\tilde A^{ij}=\epsilon^{imn}\partial_mC_n\ ,\label{FBD-dA=Cn}
	\ee
where $C_n(X)$ is a generic 3D vector field. Eq.\eqref{FBD-dA=Cn}, in turn, gives
	\be
	\partial_j\left(\tilde A^{ij}-\epsilon^{ijn}C_n\right)=0\quad\Rightarrow\quad \tilde A^{ij}-\epsilon^{ijn}C_n=2\epsilon^{jab}\partial_a\tilde a^{\ i}_{b}\ ,
	\ee
where $\tilde a_{ij}(X)$ is a generic rank-2 tensor field. On the other hand, $\tilde A^{ij}(x)$ is symmetric, hence $C_n=0$ and  we have
\normalcolor
	\be\label{FBD-sol1}
	\tilde A^{ij}(X)\equiv\epsilon^{iab}\partial_a\tilde a_b^{\ j}(X)+\epsilon^{jab}\partial_a\tilde a_b^{\ i}(X)\ .
	\ee
The tensor field $\tilde a_{ij}(X)$ represents the \ac{DoF} on the boundary, with $[\tilde a_{ij}]=1$. Moreover, since $\tilde A_{ij}(x)=\tilde A_{ji}(x)$ has six independent components, the boundary field $\tilde a_{ij}(X)$ must be symmetric as well
	\be
	\tilde a_{ij}=\tilde a_{ji}\ ,
	\ee
in order that the boundary \ac{DoF} does not exceed the number of components of its bulk ancestor $\tilde A_{ij}(x)$. The solution \eqref{FBD-sol1} is traceless 
	\be
	\tilde A(X)|_\eqref{FBD-sol1}=0\ ,
	\ee
so that the condition  \eqref{FBD-boxA=0} is automatically satisfied. 
Analogously, from the local Ward identity \eqref{FBD-wi2} we have, at vanishing external source $\tilde J^{ab}(x)=0$
	\be
	\partial_i\partial_j\left[(g_2-g_1) A^{ij}-g_2\eta^{ij} A\right]_{x^3=0}=0\ ,\label{FBD-cc2}
	\ee
whose solution is
	\be\label{FBD-sol2}
	 A^{ij}(X)\equiv\epsilon^{iab}\partial_a a_b^{\ j}(X)+\epsilon^{jab}\partial_a a_b^{\ i}(X)\ ,
	\ee
where $a_{ij}(X)=a_{ji}(X)$ are \ac{DoF} on the boundary with $[a_{ij}]=0$. Let us now consider the two broken Ward identities \eqref{FBD-wi1} and \eqref{FBD-wi2} and make functional derivatives with respect to $J^{mn}(X')$ and $\tilde J^{mn}(X')$. Referring to Appendix \ref{FBD-appAlgebra} for the details, we obtain the following equal time commutation relations
	\begin{empheq}{align}
	\left[\Delta\tilde A(X)\ ,\ A_{\textsc{mn}}(X')\right]_{x^0=x'^0}&=+i\partial_\textsc{m}\partial_\textsc{n}\{\delta(x^1-x'^1)\delta(x^2-x'^2)\}\label{FBD-[DAt,A]}\\
	\left[\Delta A(X)\ ,\ \tilde A_{\textsc{mn}}(X')\right]_{x^0=x'^0}&=-i\partial_\textsc{m}\partial_\textsc{n}\{\delta(x^1-x'^1)\delta(x^2-x'^2)\}\label{FBD-[DA,At]}\ ,
	\end{empheq}
where
	\begin{empheq}{align}
	\Delta \tilde A&\equiv2(g_1-g_2)\left(\partial_j\tilde A^{0j}+\partial_\textsc{a}\tilde A^{0\textsc{a}}\right)+2g_2\partial^0\tilde A\label{FBD-DAt}\\	
	\Delta A&\equiv2(g_1-g_2)\left(\partial_jA^{0j}+\partial_\textsc{a}A^{0\textsc{a}}\right)+2g_2\partial^0A\ .\label{FBD-DA}
	\end{empheq}
Importantly, the commutation relations \eqref{FBD-[DAt,A]} and \eqref{FBD-[DA,At]} resemble the generalized U(1) \ac{KM} algebra derived in \cite{You:2019bvu} for a 3D non-chiral bosonic theory that lives on the boundary of a 4D dipolar fracton theory. Hence, the theory described by the action $S_{inv}$ \eqref{FMAX-Sinvg1g2} has an algebraic structure on the boundary that is different from that of topological field theories \cite{Blasi:2008gt,Blasi:2011pf,Amoretti:2012hs,Amoretti:2013nv,Amoretti:2013xya,Amoretti:2014iza,Maggiore:2017vjf,Bertolini:2021iku,Bertolini:2022sao} and Maxwell theory \cite{Maggiore:2019wie,Bertolini:2020hgr,Hofman:2018lfz}. The reason for that lies in the structure of the fracton symmetry \eqref{FMAX-dA}, characterized by two derivatives, which prevents the presence, at the right hand side of \eqref{FBD-[DAt,A]} and \eqref{FBD-[DA,At]}, of the central charge term $\partial\delta$, typical of usual \ac{KM} algebras. This leads us to guess that a conserved current algebra might exist on the boundary of \ac{LG}, whose defining symmetry \eqref{FMAX-diff} depends on one derivative only. This would be in agreement with the conjecture concerning the existence of a \ac{KM} algebra in \ac{LG} mentioned in \cite{Hinterbichler:2022agn}. We shall discuss this situation (\ac{LG} with boundary) separately, in Chapter \ref{ch LG}.

\subsection{The induced 3D theory}\label{FBD-sec-3Dmodel}

\subsubsection*{Canonical variables}

We look for the transformations of the boundary fields $a_{ij}(X),\ \tilde a_{ij}(X)$ which preserve the solutions \eqref{FBD-sol1} and \eqref{FBD-sol2}. The most general ones are
	\begin{empheq}{align}
	\delta a_{mn}=&\eta_{mn}\phi+\partial_m\xi_n+\partial_n\xi_m+\partial_m\partial_n\lambda\quad;\quad\delta\tilde a_{mn}=0\label{FBD-damn}\\
	\tilde\delta\tilde a_{mn}=&\eta_{mn}\tilde\phi+\partial_m\tilde\xi_n+\partial_n\tilde\xi_m+\partial_m\partial_n\tilde\lambda\quad;\quad\tilde\delta a_{mn}=0\ ,\label{FBD-damnt}
	\end{empheq}
where $\lambda(X),\ \tilde\lambda(X),\ \phi(X),\ \tilde\phi(X),\ \xi(X)$ and $\tilde\xi(X)$ are generic local parameters.
The solutions $A_{ij}(X)$ \eqref{FBD-sol1}  and $\tilde A_{ij}(X)$ \eqref{FBD-sol2} remain unchanged, $i.e.$ $\delta A_{ij}=\tilde\delta\tilde A_{ij}=0$, if $\xi_m=\partial_m\xi',\ \tilde\xi_m=\partial_m\tilde\xi'$, so that \eqref{FBD-damn} and \eqref{FBD-damnt} reduce to
	\begin{empheq}{align}
	\delta a_{mn}=&\eta_{mn}\phi+\partial_m\partial_n\lambda\quad;\quad\delta\tilde a_{mn}=0\label{FBD-da1}\\
	\tilde\delta\tilde a_{mn}=&\eta_{mn}\tilde\phi+\partial_m\partial_n\tilde\lambda\quad;\quad\tilde\delta a_{mn}=0\ .\label{FBD-dat1}
	\end{empheq}
We decompose the boundary fields $a_{ij}(X)$ and $\tilde a_{ij}(X)$ in terms of their trace and traceless contributions, $i.e.$
	\begin{empheq}{align}
	a_{ij}&=\alpha_{ij}+\frac{1}{3}\eta_{ij}a\\
	\tilde a_{ij}&=\tilde\alpha_{ij}+\frac{1}{3}\eta_{ij}\tilde a\ ,
	\end{empheq}
where $a\equiv\eta^{ij}a_{ij},\ \tilde a\equiv\eta^{ij}\tilde a_{ij}$, and $\alpha_{ij}(X),\ \tilde\alpha_{ij}(X)$ are symmetric traceless fields
\be
\eta^{ij}\alpha_{ij}=\eta^{ij}\tilde\alpha_{ij}=0\ ,
\label{FBD-tracelessalpha}\ee
which transform as
	\begin{empheq}{align}
	\delta \alpha_{mn}=&\partial_m\partial_n\lambda-\frac{1}{3}\eta_{mn}\partial^2\lambda\quad;\quad\delta\tilde \alpha_{mn}=0\label{FBD-dalpha}\\
	\tilde\delta\tilde \alpha_{mn}=&\partial_m\partial_n\tilde\lambda-\frac{1}{3}\eta_{mn}\partial^2\tilde\lambda\quad;\quad\tilde\delta \alpha_{mn}=0\ .\label{FBD-dalphat}
	\end{empheq}
The solutions  \eqref{FBD-sol1} and \eqref{FBD-sol2} depend only on the traceless components
	\be\label{FBD-a,at sol}
	\begin{split}
	\tilde A^{ij}(X)&\equiv\epsilon^{iab}\partial_a \tilde \alpha_b^{\ j}(X)+\epsilon^{jab}\partial_a \tilde \alpha_b^{\ i}(X)\\
	A^{ij}(X)&\equiv\epsilon^{iab}\partial_a \alpha_b^{\ j}(X)+\epsilon^{jab}\partial_a \alpha_b^{\ i}(X)\ ,
	\end{split}
	\ee	
and the trace contributions disappear. The \ac{DoF} of the boundary theory are then described by the rank-2 traceless tensor fields $\alpha_{ij}(X)$ and $\tilde\alpha_{ij}(X)$. This is consistent with the fact that, as we showed,  on the boundary the bulk fields $A_{ij}(x)$ and $\tilde A_{ij}(x)$ have only five components, exactly as the 3D boundary fields $\alpha_{ij}(X)$ and $\tilde\alpha_{ij}(X)$, which are symmetric and traceless. The solutions \eqref{FBD-a,at sol} highly simplify the definitions of $\Delta A(X)$  \eqref{FBD-DA} and $\Delta\tilde A(X)$ \eqref{FBD-DAt} :
	\begin{empheq}{align}
	\Delta A|_\eqref{FBD-sol2}&=4(g_1-g_2)\epsilon^{0\textsc{mn}}\partial_\textsc{m}\partial^\textsc{a}\alpha_{\textsc{na}}\\
	\Delta \tilde A|_\eqref{FBD-sol1}&=4(g_1-g_2)\epsilon^{0\textsc{mn}}\partial_\textsc{m}\partial^\textsc{a}\tilde \alpha_{\textsc{na}}\ ,
	\end{empheq}
where we observe that only spatial derivatives appear. Considering the following combinations of the commutators \eqref{FBD-[DAt,A]} and \eqref{FBD-[DA,At]} and their traces 
	\begin{align}
	\left[\Delta\tilde A\ ,\ \!A_{\textsc{df}}'-\frac{1}{2}\eta_{\textsc{df}}\eta^{\textsc{mn}}A_{\textsc{mn}}'\right]=&\!\quad\frac{i}{2}\left(\delta^{\textsc{m}}_{\textsc{d}}\delta^{\textsc{n}}_{\textsc{f}}+\delta^{\textsc{n}}_{\textsc{d}}\delta^{\textsc{m}}_{\textsc{f}}-\eta^{\textsc{mn}}\eta_{\textsc{df}}\right)\partial_\textsc{m}\partial_\textsc{n}\delta^{(2)}(X-X')\label{FBD-DAt,A-TrA}\\
	\left[\Delta A\ ,\ \!\tilde A_{\textsc{df}}'-\frac{1}{2}\eta_{\textsc{df}}\eta^{\textsc{mn}}\tilde A_{\textsc{mn}}'\right]=&\!-\!\frac{i}{2}\left(\delta^{\textsc{m}}_{\textsc{d}}\delta^{\textsc{n}}_{\textsc{f}}+\delta^{\textsc{n}}_{\textsc{d}}\delta^{\textsc{m}}_{\textsc{f}}-\eta^{\textsc{mn}}\eta_{\textsc{df}}\right)\partial_\textsc{m}\partial_\textsc{n}\delta^{(2)}(X-X')\!\ ,
	\end{align}
and using the solutions \eqref{FBD-a,at sol}, we can  identify the following two canonical commutation relations at the boundary (the details can be found in Appendix \ref{FBD-appComm})
	\begin{empheq}{align}
	\left[q_{\textsc{ab}}\ ,\ p'^{\textsc{cd}}\right]=&\frac{i}{2}\left(\delta^{\textsc{c}}_{\textsc{a}}\delta^{\textsc{d}}_{\textsc{b}}+\delta^{\textsc{d}}_{\textsc{a}}\delta^{\textsc{c}}_{\textsc{b}}-\eta^{\textsc{cd}}\eta_{\textsc{ab}}\right)\delta^{(2)}(X-X')\label{FBD-al,ft}\\
	\left[\tilde q_{\textsc{ab}}\ ,\ \tilde p'^{\textsc{cd}}\right]=&\frac{i}{2}\left(\delta^{\textsc{c}}_{\textsc{a}}\delta^{\textsc{d}}_{\textsc{b}}+\delta^{\textsc{d}}_{\textsc{a}}\delta^{\textsc{c}}_{\textsc{b}}-\eta^{\textsc{cd}}\eta_{\textsc{ab}}\right)\delta^{(2)}(X-X')\ ,\label{FBD-alt,f}
	\end{empheq}
where
	\begin{empheq}{align}
	q_{\textsc{ab}}&\equiv \alpha _{\textsc{ab}}-\frac{1}{2}\eta_{\textsc{ab}} \alpha ^\textsc{m}_{\ \textsc{m}}\\
	p^{\textsc{cd}}&\equiv 2g_{12}\left(\tilde f^{\textsc{cd}0}-\frac{1}{2}\eta^{\textsc{cd}}\tilde f^{\ a0}_{a}\right)\\
	\tilde q_{\textsc{ab}}&\equiv\tilde \alpha _{\textsc{ab}}-\frac{1}{2}\eta_{\textsc{ab}}\tilde \alpha ^\textsc{m}_{\ \textsc{m}}\\
	\tilde p^{\textsc{cd}}&\equiv -2g_{12}\left(f^{\textsc{cd}0}-\frac{1}{2}\eta^{\textsc{cd}}f^{\ a0}_{a}\right)\ ,
	\end{empheq}
and
	\be
	g_{12}\equiv2(g_1-g_2)\ .
	\ee
In analogy to $F_{\mu\nu\rho}(x)$ \eqref{FMAX-Fmunurho}, $f_{abc}(X)$ and $\tilde f_{abc}(X)$ are defined as
	\begin{empheq}{align}
	\tilde f_{abc}&\equiv\partial_a\tilde \alpha_{bc}+\partial_b\tilde \alpha_{ac}-2\partial_c\tilde \alpha_{ab}\\
	f_{abc}&\equiv\partial_a\alpha_{bc}+\partial_b\alpha_{ac}-2\partial_c\alpha_{ab}\ .
	\end{empheq}	
It is interesting to notice that a canonical commutator  similar to those we found in \eqref{FBD-al,ft} and \eqref{FBD-alt,f}, appears in \cite{Du:2021pbc}, in the context of the traceless fracton models  \cite{Pretko:2016lgv,Pretko:2016kxt}. 
The aim of \cite{Du:2021pbc} is to build a non-abelian model for fractons in 2+1 dimensions. To do so the abelian traceless theory needs to be defined first. As for any fracton theory \cite{Nandkishore:2018sel,Pretko:2020cko,Pretko:2016lgv,Pretko:2016kxt}, the ``electric field'' $E_\textsc{ij}(x)$ is the conjugate momentum of $A_\textsc{ij}(x)$, from which the commutator holds
	\be
	[E^{\textsc{ij}},A_\textsc{mn}]=i(\delta^\textsc{i}_\textsc{m}\delta^\textsc{j}_\textsc{n}+\delta^\textsc{j}_\textsc{m}\delta^\textsc{i}_\textsc{n})\label{FBD-fract-comm}\ .
	\ee
After that, the scalar Gauss constraints is imposed, together with a tracelessness condition : 
	\be\label{FBD-Gauss-constr}
	\partial_\textsc{i}\partial_\textsc{j}E^\textsc{ij}=\rho\quad;\quad E^{\ \textsc{i}}_\textsc{i}=0\ ,
	\ee
 which imply  three conservation equations : 
	\be
	\int\rho=const\quad;\quad\int\vec x\rho=const\quad;\quad\int x^2\rho=const\ ,
	\ee
of charge, dipole and a component of the quadrupole, respectively. The main characteristic of fracton theories, $i.e.$ the limited mobility, is here translated into the fact that single charges cannot move, while dipole bound states can only move along their transverse direction. The constraints \eqref{FBD-Gauss-constr} imply that the tensor field $A_\textsc{ij}(x)$ transforms exactly as \eqref{FBD-da1} and \eqref{FBD-dat1}, which is a remarkable check of our reasoning. However, while in our case it is natural to identify the \ac{DoF} of the theory with the  traceless fields $\alpha_{ij}(X)$ and $\tilde\alpha_{ij}(X)$, in \cite{Du:2021pbc} the tracelessness condition is imposed as a kind of gauge fixing, while for us it comes from the solutions \eqref{FBD-a,at sol}. As a consequence, the definition \eqref{FBD-fract-comm} of the canonical commutator is no longer valid (since $A^{\ \textsc{i}}_\textsc{i}=E^{\ \textsc{i}}_\textsc{i}=0$ would not commute), and  the commutator for the traceless theory of fractons is defined as Dirac brackets \cite{Dirac:1950pj}, which turns out to be identical to ours \eqref{FBD-al,ft} and \eqref{FBD-alt,f}.
\normalcolor

\subsubsection*{The most general 3D action}\label{FBD-sec S3D}

The action of the 3D boundary theory is constructed as the most general local integrated functional of the traceless rank-2 symmetric tensor fields $\alpha_{ij}(X)$ and $\tilde\alpha_{ij}(X)$ compatible with
	\bi
	\item power-counting $[\alpha]=0,\ [\tilde\alpha]=1$ ;
	\item symmetry $\delta S=\tilde\delta S=0$, where $\delta$ and $\tilde\delta$ are defined in \eqref{FBD-dalpha} and \eqref{FBD-dalphat} ;
	\item canonical variables identified in \eqref{FBD-al,ft} and \eqref{FBD-alt,f} :  $\frac{\partial\mathcal L_{kin}}{\partial \dot q}=p$.
	\ei
In Appendix  \ref{FBD-appSgen} we show that the most general 3D action satisfying these three requests is
	\be \label{FBD-S3D}
	S_{3D}=\int d^3X\left(-\frac{2}{3}g_{12}\,\varphi_{abc}\tilde\varphi^{abc}+\omega_5\, \epsilon^{abc}\tilde\alpha^d_a\tilde\varphi_{dbc}\right)\ ,
	\ee
where we defined
	\begin{align}
	\varphi_{abc}=\varphi_{bac}&\equiv f_{abc}+\frac{1}{4}\left(-2\eta_{ab}f^d_{\ dc}+\eta_{bc}f^d_{\ da}+\eta_{ac}f^d_{\ db}\right)\label{FBD-phi}\\
	&=-2\partial_c\alpha_{ab}+\partial_a\alpha_{bc}+\partial_b\alpha_{ac}-\eta_{ab}\partial^d\alpha_{dc}+\frac{1}{2}\eta_{bc}\partial^d\alpha_{da}+\frac{1}{2}\eta_{ac}\partial^d\alpha_{db}\ ,\nonumber
	\end{align}
and, analogously, $\tilde\varphi_{abc}(X)$ in terms of $\tilde\alpha_{ab}(X)$, with the following properties of ciclicity and ``tracelessness''
	\begin{empheq}{align}
	&\varphi_{abc}+\varphi_{cab}+\varphi_{bca}=0=\tilde\varphi_{abc}+\tilde\varphi_{cab}+\tilde\varphi_{bca}\label{FBD-cicl}\\
	&\eta^{ab}\varphi_{abc}=\eta^{bc}\varphi_{abc}=\eta^{ab}\tilde\varphi_{abc}=\eta^{bc}\tilde\varphi_{abc}=0\ .\label{FBD-traceless}
	\end{empheq}
The $\omega_5$ term in \eqref{FBD-S3D} looks like a Chern-Simons term, and the similarity is even more evident if we explicit the $\tilde\alpha_{ab}(X)$ dependence, since $\epsilon^{abc} \tilde\alpha^d_a\tilde\varphi_{dbc}\propto\,\epsilon^{abc}\tilde\alpha^d_a\partial_b\tilde\alpha_{cd}$.
Intriguingly, this Chern-Simons-like  term resembles the massless limit of 3D self-dual massive gravity \cite{Dalmazi:2020xou,Aragone:1986hm}. This theory contains a 3D Fierz-Pauli mass term that breaks the gauge invariance \cite{Blasi:2017pkk,Blasi:2015lrg}. However, it has been shown \cite{Dalmazi:2008zh} that it is dual to linearized topologically massive gravity \cite{Deser:1990ay}, which is gauge invariant and contains the Chern-Simons-like  term together with the linearized 3D Einstein-Hilbert action. These two equivalent theories were originally proposed as a viable way to describe a single propagating massive graviton in 3D in contrast with the standard Einstein-Hilbert theory, which is topological in 3D and does not support any propagating spin-2 particle. In our case, the linearized Einstein-Hilbert term is replaced by the tensorial Maxwell-like term such that our boundary action still supports a propagating ``graviton''. Notice that a non-covariant version of our Chern-Simons-like  term has been also considered in \cite{Cappelli:2015ocj,Gromov:2017vir} in the context of fractional quantum Hall effect and chiral fractons. However, these non-covariant field theories, that can be seen as dual one to each other, do not take into account any tensorial Maxwell-like terms.
 Notice also that if $\omega_5=0$ it is possible to decouple the fields. In fact, by defining 
	\be
	\alpha_{ab}^\pm\equiv\sqrt{M}\,\alpha_{ab}\pm\frac{1}{\sqrt{M}}\tilde\alpha_{ab}\quad\Rightarrow\quad\varphi^\pm_{abc}=\sqrt{M}\,\varphi_{abc}\pm\frac{1}{\sqrt{M}}\tilde\varphi_{abc}\ ,
	\ee
where $M$ is a parameter with mass dimension $[M]=1$ and $[\alpha^\pm]=\frac{1}{2}$, the 3D action \eqref{FBD-S3D} becomes
	\be
		\begin{split}
	S_{3D}=\int d^3X&\left[\frac{g_{12}}{6}\left(\varphi^-_{abc}\varphi^{-\,abc}-\varphi^+_{abc}\varphi^{+\,abc}\right)+\right.\\
	&\left.\;+\frac{M\omega_5}{4}\left( \alpha^{+\,d}_{\,a}\epsilon^{abc}\varphi^+_{dbc}-2\alpha^{+\,d}_{\,a}\epsilon^{abc}\varphi^-_{dbc}+\alpha^{-\,d}_{\,a}\epsilon^{abc}\varphi^-_{dbc}\right)\right]\ ,
		\end{split}
	\ee
which for $\omega_5=0$ decouples:
	\be\label{FBD-S3Ddec}
	S_{3D}[\alpha^+,\alpha^-,\omega_5=0]=S^+_{3D}[\alpha^+]+S^-_{3D}[\alpha^-]\
	\ee
with
	\be
	S^\pm_{3D}\equiv\mp\frac{g_{12}}{6}\int d^3X\,\varphi^\pm_{abc}\varphi^{\pm\,abc}\ .
	\ee
As we shall show in Section \ref{FBD-sec PT}, this second case keeps the $\TR$-invariance of the boundary in agreement with the $\TR$-symmetry of the bulk action \eqref{FBD-Sbulk}.

\subsection{The bulk and the boundary : holographic contact}\label{FBD-sec-hc}

Once the most general 3D action \eqref{FBD-S3D} has been derived, we have to establish the ``holographic contact'' between this induced 3D theory and the 4D theory $S_{tot}$ \eqref{FBD-Stot}. This is accomplished by requiring that the \ac{EoM} of the 3D theory coincide with the \ac{BC} \eqref{FBD-bcA3i}, \eqref{FBD-bcAij}, \eqref{FBD-bcAtij} of the 4D theory. To do so we have at our disposal the $\xi_i$ parameters appearing in $S_{bd}$ \eqref{FBD-Sbd} and $\omega_5$ in $S_{3D}$ \eqref{FBD-S3D}. The \ac{EoM} of $S_{3D}$ are
	\be\label{FBD-eom1bd}
		\frac{\delta S_{3D}}{\delta\alpha_{mn}}  
=-2g_{12}\partial_a\tilde\varphi^{mna}=0\ ,
	\ee
where we used the cyclic property \eqref{FBD-cicl}, and
	\be\label{FBD-eom2bd}
		\frac{\delta S_{3D}}{\delta\tilde\alpha_{mn}}
		=-2g_{12}\partial_a\varphi^{mna}+\omega_5\left(\epsilon^{mab}\tilde\varphi^{n}_{\ ab}+\epsilon^{nab}\tilde\varphi^{m}_{\ ab}\right)=0\ .
	\ee
We now consider the \ac{BC} of the bulk theory \eqref{FBD-bcA3i}, \eqref{FBD-bcAij} and \eqref{FBD-bcAtij}, which we write in terms of the solutions \eqref{FBD-a,at sol} and of the definitions of $\varphi_{abc}(X),\ \tilde\varphi_{abc}(X)$ \eqref{FBD-phi} 
	\begin{align}
	&{\frac{1}{3}\partial^a\left(\epsilon_{aij}\varphi_b^{\ ij}+\epsilon_{bij}\varphi_a^{\ ij}\right)}=0
\label{FBD-bc1-bd}
\\[10px]
	&\frac{2}{3}\xi_0\left(\epsilon^{aij}\varphi^b_{\ ij}+\epsilon^{bij}\varphi^a_{\ ij}\right)+\frac{1}{3}\left(\xi_1-g_{12}\right)\left(\epsilon^{aij}\tilde\varphi^b_{\ ij}+\epsilon^{bij}\tilde\varphi^a_{\ ij}\right)-2\xi_2\partial_c\varphi^{abc}=0\label{FBD-bc2-bd}\\[10px]
	&
\frac{\xi_1}{3}\left(\epsilon^{aij}\varphi^b_{\ ij}+\epsilon^{bij}\varphi^a_{\ ij}\right)=0\ .\label{FBD-bc3-bd}
	\end{align}
The contact is governed by two coefficients : $\xi_1$, which appears in $S_{bd}$ \eqref{FBD-Sbd}, and $\omega_5$ in the action $S_{3D}$ \eqref{FBD-S3D}. The first - $\xi_1$ - is relevant because it determines the existence of the \ac{BC} \eqref{FBD-bc3-bd}, the second - $\omega_5$ - decouples the \ac{EoM} of the boundary fields $\alpha_{ab}(X),\ \tilde\alpha_{ab}(X)$, $i.e.$ eliminates the Chern-Simons-like  term from the action  \eqref{FBD-S3D}. Additionally, we remark that $\tilde\alpha_{ij}(X)$ appears only in the \ac{BC} \eqref{FBD-bc2-bd}, coupled to $(\xi_1-g_{12})$, which should not vanish, otherwise no contact is possible. To summarize, the constraints on the coefficients, up to now, are
	\be\label{FBD-vincoli}
	g_1\neq0\quad;\quad g_{12}\neq0\quad;\quad\xi_1\neq g_{12}\ .
	\ee
Therefore, depending on $\xi_1$ and $\omega_5$, we distinguish the following cases :	
\bi
\item
$\pmb{\xi_1\neq0,\ \omega_5\neq0}$ : using \eqref{FBD-bc3-bd} in \eqref{FBD-bc2-bd} (or setting $\xi_0=0$), we have
	\be\label{FBD-bc2-3}
	\frac{1}{3}\left(\xi_1-g_{12}\right)\left(\epsilon^{aij}\tilde\varphi^b_{\ ij}+\epsilon^{bij}\tilde\varphi^a_{\ ij}\right)-2\xi_2\partial_c\varphi^{abc}=0\ ,
	\ee
which coincides with the 3D \ac{EoM} \eqref{FBD-eom2bd}
	\be
	-2g_{12}\partial_a\varphi^{mna}+\omega_5\left(\epsilon^{mab}\tilde\varphi^{n}_{\ ab}+\epsilon^{nab}\tilde\varphi^{m}_{\ ab}\right)=0
	\ee	
if
	\be\label{FBD-hc1}
	\omega_5=\frac{1}{3}(\xi_1-g_{12})\neq0\quad;\quad \xi_2=g_{12}\neq0\quad;\quad\xi_1\neq0\ .
	\ee
Notice that setting $\omega_5=0$ would imply $\xi_1=g_{12}$, which we excluded in  \eqref{FBD-vincoli}. Up to a numerical coefficient, we now consider the following symmetric combination of the curl of the \ac{BC}  \eqref{FBD-bc3-bd}
	\be\label{FBD-curlBC3}
		\begin{split}
		0&=\epsilon_{mac}\partial^c\eqref{FBD-bc3-bd}^a_b+\epsilon_{bac}\partial^c\eqref{FBD-bc3-bd}^a_m\\
		&=6\partial^i\varphi_{bmi}\ ,
		\end{split}
	\ee
where we used the properties of tracelessness \eqref{FBD-traceless} and cyclicity  \eqref{FBD-cicl} of $\varphi_{abc}(X)$. We then use this result in the \ac{BC} \eqref{FBD-bc2-3}, which becomes
	\be\label{FBD-bc2-3-d3}
	\frac{1}{3}\left(\xi_1-g_{12}\right)\left(\epsilon^{aij}\tilde\varphi^b_{\ ij}+\epsilon^{bij}\tilde\varphi^a_{\ ij}\right)=0\ ,
	\ee
of which we compute again the curl
	\be
	0=\epsilon_{mac}\partial^c\eqref{FBD-bc2-3-d3}^a_b+\epsilon_{bac}\partial^c\eqref{FBD-bc2-3-d3}^a_m=2\left(\xi_1-g_{12}\right)\partial^i\tilde\varphi_{bmi}\ ,
	\ee
which finally coincides with the 3D \ac{EoM} \eqref{FBD-eom1bd}
	\be
	-2g_{12}\partial_a\tilde\varphi^{mna}=0\ .
	\ee
Notice that this second contact is obtained without the need of any additional constraint on the parameters, we just need \eqref{FBD-hc1}. Taking into account \eqref{FBD-hc1}, the 3D action  \eqref{FBD-S3D} becomes
	\be \label{FBD-S3Dhc}
	S_{3D}=\frac{1}{3}\int d^3X\left[-2g_{12}\,\varphi_{abc}\tilde\varphi^{abc}+(\xi_1-g_{12})\, \tilde\alpha^d_a\epsilon^{abc}\tilde\varphi_{dbc}\right]\ ,
	\ee
while the boundary term \eqref{FBD-Sbd} now is
	\be
	S_{bd}=\int d^4x\delta(x^3)\left[{\xi_0} A_{ab}A^{ab}+\xi_1\tilde A_{ab}A^{ab}+g_{12}\epsilon^{abc}A_{ai}\partial_bA_c^i+{\xi_3}(A^a_a)^2+{\xi_4}\tilde A^a_aA^b_b\right]\ ,
	\ee
where the coefficients $\xi_0,\ \xi_3$ and $\xi_4$ are free and can for instance be set to zero, while $\xi_1\neq\{0,g_{12}\}$.
\item
$\pmb{\xi_1\neq0,\ \omega_5=0}$ : the \ac{EoM} of the 3D theory \eqref{FBD-eom1bd} and \eqref{FBD-eom2bd} are 
	\be\label{FBD-eom-bd-ro=0}
	\partial_a\varphi^{mna}=0\quad;\quad	\partial_a\tilde\varphi^{mna}=0\ ,
	\ee
while, ignoring the first \ac{BC} \eqref{FBD-bc1-bd}, which is automatically solved by the third one \eqref{FBD-bc3-bd}, and using \eqref{FBD-bc3-bd} in \eqref{FBD-bc2-bd}, the remaining \ac{BC} are
	\begin{empheq}{align}
	\frac{1}{3}\left(\xi_1-g_{12}\right)\left(\epsilon^{aij}\tilde\varphi^b_{\ ij}+\epsilon^{bij}\tilde\varphi^a_{\ ij}\right)-2\xi_2\partial_c\varphi^{abc}\label{FBD-bc2-bd'}&=0\\[10px]
	\frac{\xi_1}{3}\left(\epsilon^{aij}\varphi^b_{\ ij}+\epsilon^{bij}\varphi^a_{\ ij}\right)&=0\ .\label{FBD-bc3-bd'}
	\end{empheq}
As in \eqref{FBD-curlBC3}, we can again compute 
	\be\label{FBD-rotBC2}
	0=\epsilon_{mac}\partial^c\eqref{FBD-bc3-bd'}^a_b+\epsilon_{bac}\partial^c\eqref{FBD-bc3-bd'}^a_m=2\xi_1\partial^i\varphi_{bmi}
	\ee
which coincides with the first \ac{EoM} of \eqref{FBD-eom-bd-ro=0}. If we use this result \eqref{FBD-rotBC2} in \eqref{FBD-bc2-bd'} (analogous to setting $\xi_2=0$) and consider the same combination as \eqref{FBD-rotBC2}, we obtain the second \ac{EoM} of \eqref{FBD-eom-bd-ro=0} and thus we get the second matching between bulk and boundary. In that case the 3D action \eqref{FBD-S3D} is
	\be \label{FBD-S3Dhc}
	S_{3D}=-\frac{2g_{12}}{3}\int d^3X\,\varphi_{abc}\tilde\varphi^{abc}\ ,
	\ee
while the boundary term $S_{bd}$ \eqref{FBD-Sbd} becomes 
	\be
	S_{bd}=\int d^4x\delta(x^3)\left[{\xi_0} A_{ab}A^{ab}+\xi_1\tilde A_{ab}A^{ab}+{\xi_2}\epsilon^{abc}A_{ai}\partial_bA_c^i+{\xi_3}(A^a_a)^2+{\xi_4}\tilde A^a_aA^b_b\right]\ ,
	\ee
where the coefficients $\xi_0,\ \xi_2,\ \xi_3$ and $\xi_4$ are free, $i.e.$ do not contribute to the contact between the bulk and the boundary and can be set to zero without loss of generality, provided that  $\xi_1\neq\{0,g_{12}\}$. Therefore a second holographic contact is possible if
	\be\label{FBD-hc2}
	\omega_5=0\quad;\quad\xi_1\neq\{0,g_{12}\}\ .
	\ee
This second result has a relevant consequence : $\omega_5=0$ allows to decouple the action, as seen  in \eqref{FBD-S3Ddec}.
\ei
We observe that the holographic contacts obtained in \eqref{FBD-hc1} and \eqref{FBD-hc2} affect the boundary action $S_{bd}$ \eqref{FBD-Sbd} in different ways, in particular in the first case \eqref{FBD-hc1} the number of free parameters from five reduces to three, while in the second case \eqref{FBD-hc2} it goes to four. Finally, if $\xi_1=0$, no complete matching between \ac{BC} and 3D \ac{EoM} is possible, in fact setting ${\xi_1=0}$, one of the \ac{BC} \eqref{FBD-bc3-bd} disappears. We are left with
	\begin{align}
	&{\frac{1}{3}\partial^a\left(\epsilon_{aij}\varphi_b^{\ ij}+\epsilon_{bij}\varphi_a^{\ ij}\right)}=0
\label{FBD-bc1-bd''}\\[10px]
	&\frac{2}{3}\xi_0\left(\epsilon^{aij}\varphi^b_{\ ij}+\epsilon^{bij}\varphi^a_{\ ij}\right)-\frac{g_{12}}{3}\left(\epsilon^{aij}\tilde\varphi^b_{\ ij}+\epsilon^{bij}\tilde\varphi^a_{\ ij}\right)-2\xi_2\partial_c\varphi^{abc}=0\label{FBD-bc2-bd''}\ .
	\end{align}
For what concerns the parameter $\omega_5$ in \eqref{FBD-S3D}, two cases are possible
\bi
\item
$\pmb{\omega_5\neq0}$ : the \ac{BC} \eqref{FBD-bc2-bd''} coincides with the 3D \ac{EoM} \eqref{FBD-eom2bd} 
	\be
	-2g_{12}\partial_a\varphi^{mna}+\omega_5\left(\epsilon^{mab}\tilde\varphi^{n}_{\ ab}+\epsilon^{nab}\tilde\varphi^{m}_{\ ab}\right)=0
	\ee	
if
	\be\label{FBD-hc3}
	\omega_5=-\frac{g_{12}}{3}\neq0\quad;\quad \xi_2=g_{12}\neq0\quad;\quad\xi_0=\xi_1=0\ ,
	\ee
however it is not possible to establish a link with the other \ac{EoM} \eqref{FBD-eom1bd}.
\item
$\pmb{\omega_5=0}$ : the \ac{EoM} of the 3D boundary theory are given by \eqref{FBD-eom-bd-ro=0}. A matching is possible with the \ac{BC} \eqref{FBD-bc2-bd''} only if $\xi_2=0$, and if we compute
	\be
	0=\epsilon_{mac}\partial^c\eqref{FBD-bc2-bd''}^a_b+\epsilon_{bac}\partial^c\eqref{FBD-bc2-bd''}^a_m=4\xi_0\partial^i\varphi_{bmi}-2g_{12}\partial^i\tilde\varphi_{bmi}\ ,
	\ee
which coincides with  a combination of the two \ac{EoM} of the boundary. We see that also in this case a complete holographic contact between 3D \ac{EoM} and bulk \ac{BC}, is not possible.
\ei
This enforces the fact that the $\xi_1$-term in $S_{bd}$ \eqref{FBD-Sbd} plays a key role in the holographic contact. To summarize
\begin{table}[H]
\centering
  \begin{tabular}{ | c| c |  }
    \hline
     & BC-EoM matching \\ \hline
      $\xi_1\neq0,\ \omega_5\neq0$ & $\omega_5=\frac{1}{3}(\xi_1-g_{12})\ ;\ \xi_2=g_{12}$\\ \hline
      $\xi_1\neq0,\ \omega_5=0$ & $\omega_5=0$  \\ \hline
      $\xi_1=0$ & no contact \\ \hline
  \end{tabular}
  \caption[Holographic contacts in fracton and \acs{LG} with boundary]{\footnotesize\label{FBD-table2}Holographic contacts.}
  \end{table}
  \noindent

\subsection{Physical interpretation of the 3D theory}\label{FBD-sec-phys}

To understand the physical content of the 3D theory described by the action $S_{3D}$ \eqref{FBD-S3D}, we study its \ac{EoM}. The first \ac{EoM} \eqref{FBD-eom1bd} for $m=n=0$ gives
	\be\label{FBD-eom1bd00}
	0=\partial_{\textsc{a}}\tilde\varphi^{00\textsc{a}}=-2\partial_{\textsc{a}}\partial^\textsc{a}\tilde\alpha^{00}+2\partial_\textsc{a}\partial^0\tilde\alpha^{\textsc{a}0}+\partial_\textsc{a}\partial_b\tilde\alpha^{\textsc{a}b}	\ .
	\ee
Taking the $\partial_\textsc{n}$-derivative of \eqref{FBD-eom1bd} for $m=0,\ n=\textsc n$, we get
	\be\label{FBD-Gauss}
		0=\partial_\textsc{n}\partial_a\tilde\varphi^{0\textsc{n}a}
		=-\frac{1}{4g_{12}}\partial_\textsc{a}\partial_\textsc{n}p^{\textsc{a}\textsc{n}}\ ,
	\ee
where we used \eqref{FBD-eom1bd00}, the cyclicity property \eqref{FBD-cicl} and the definition of conjugate momentum in terms of $\tilde\varphi_{abc}(X)$ \eqref{FBD-phi}
	\be
 	p^{\textsc{mn}}=2g_{12}\tilde\varphi^{\textsc{mn}0}\label{FBD-p}\ .
	\ee
We see that \eqref{FBD-Gauss} is a Gauss-like equation, analogous to the one related to the traceless scalar charge model of fractons in vacuum \cite{Pretko:2016lgv,Pretko:2016kxt}, which is at the base of the limited mobility property. We therefore realize that the induced 3D theory shows fractonic properties. To analyze the second \ac{EoM} \eqref{FBD-eom2bd}, we first compute the conjugate momentum of $\tilde\alpha_{ab}(X)$ :
	\be\label{FBD-pt}
	\tilde p^{\textsc{mn}}=\frac{\partial \mathcal L_{3D}}{\partial\dot{\tilde\alpha}_{\textsc{mn}}}
		=2g_{12}\varphi^{\textsc{mn}0}-\frac{3}{2}\omega_5\left(\epsilon^{0\textsc{am}}\tilde\alpha^\textsc{n}_{\ \textsc{a}}+\epsilon^{0\textsc{an}}\tilde\alpha^\textsc{m}_{\ \textsc{a}}\right)\ .
	\ee
The \ac{EoM} for $\tilde\alpha_{ab}(X)$ \eqref{FBD-eom2bd} at $m=n=0$ is
		\be\label{FBD-eom2bd00}
		\partial_\textsc{a}\varphi^{00\textsc{a}}=\frac{3\omega_5}{g_{12}}\epsilon^{0\textsc{ab}}\partial_\textsc{a}\tilde\alpha^0_{\ \textsc{b}}\ ,
		\ee
and, as in the previous case, taking the $\partial_\textsc{n}$-derivative of \eqref{FBD-eom2bd} at $m=0,\ n=\textsc n$ we have
	\be
		\begin{split}
			0=&-2g_{12}\partial_a\partial_\textsc{n}\varphi^{0\textsc{n}a}+3\omega_5\partial_\textsc{n}\left(\epsilon^{0\textsc{ab}}\partial_\textsc{a}\tilde\alpha^\textsc{n}_{\ \textsc{b}}+\epsilon^{\textsc{n}ab}\partial_a\tilde\alpha^0_{\ b}\right)\\
			=&\frac{1}{2}\partial_\textsc{m}\partial_\textsc{n}\tilde p^{\textsc{mn}}-\frac{3}{2}\omega_5\epsilon^{0\textsc{am}}\partial_\textsc{m}\partial_\textsc{n}\tilde\alpha^\textsc{n}_{\ \textsc{a}}\ ,
		\end{split}
	\ee
where we used the cyclic property of $\varphi_{abc}(X)$ \eqref{FBD-cicl} and \eqref{FBD-eom2bd00}. Here again we find a Gauss-like equation for the traceless scalar charge theory of fractons \cite{Pretko:2016lgv,Pretko:2016kxt}, but with a matter contribution at the right hand side
		\be\label{FBD-tGauss6.7}
		\partial_\textsc{m}\partial_\textsc{n}\tilde p^{\textsc{mn}}=\tilde\rho_5=\omega_5\tilde\rho\ ,
		\ee
where
		\be\label{FBD-rho}
		\tilde\rho\equiv{3}\epsilon^{0\textsc{am}}\partial_\textsc{m}\partial_\textsc{n}\tilde\alpha^\textsc{n}_{\ \textsc{a}}
		\ee
plays the role of charge. This gives an interesting interpretation of the Chern-Simons-like  term in the induced 3D action \eqref{FBD-S3D} as ``internal'' matter. Notice that this term coincides with the charge identified by Pretko in \cite{Pretko:2017xar}, where a non-covariant Chern-Simons-like  term is studied. In that case the Chern-Simons term comes from a non-covariant fractonic $\theta$-term in the bulk, and it is written in terms of a spatial traceless tensor. The charge $\tilde\rho(X)$ \eqref{FBD-rho} comes from a constraint generated by a Lagrange multiplier that is inherited by the Chern-Simons action from the definition of the $\theta$-term. We observe that this $\tilde\rho(X)$ charge implies, by definition, a dipole conservation. The 3D theory \eqref{FBD-S3D} depends on  two fields $\alpha_{ij}(X)$ and $\tilde\alpha_{ij}(X)$, hence the conjugate momenta are two as well  \eqref{FBD-p} and \eqref{FBD-pt}, which in fracton models play the role of ``electric'' fields :
	\begin{empheq}{align}
	 E^{\textsc{mn}}&\equiv p^{\textsc{mn}}=2g_{12}\tilde\varphi^{\textsc{mn}0}\label{FBD-E}\\
	\tilde E^{\textsc{mn}}&\equiv \tilde p^{\textsc{mn}}=2g_{12}\varphi^{\textsc{mn}0}-\frac{3}{2}\omega_5\left(\epsilon^{0\textsc{am}}\tilde\alpha^\textsc{n}_{\ \textsc{a}}+\epsilon^{0\textsc{an}}\tilde\alpha^\textsc{m}_{\ \textsc{a}}\right)\label{FBD-tE}\ ,
	\end{empheq}
which satisfy the Gauss equations \eqref{FBD-Gauss} and \eqref{FBD-tGauss6.7}, which we write
	\begin{empheq}{align}
	\partial_\textsc{m}\partial_\textsc{n}E^{\textsc{mn}}=&0\\
	\partial_\textsc{m}\partial_\textsc{n}\tilde E^{\textsc{mn}}=&\tilde\rho_5\ .\label{FBD-tGauss}
	\end{empheq}
Concerning the corresponding ``magnetic'' fields for this theory, inspired by the ordinary 4D electromagnetism, where $B_i\propto\epsilon_{0ijk}F^{jk}$, it is natural to define
	\begin{empheq}{align}
	&B^\textsc{m}\equiv g\epsilon_{0\textsc{ab}}\tilde\varphi^{\textsc{mab}}\quad;\quad	B_\textsc{m}\equiv-g\epsilon^{0\textsc{ab}}\tilde\varphi_{\textsc{mab}}\label{FBD-B1}\\
	&\tilde B^\textsc{m}\equiv \tilde g\epsilon_{0\textsc{ab}}\varphi^{\textsc{mab}}\quad;\quad\tilde	B_\textsc{m}\equiv-\tilde g\epsilon^{0\textsc{ab}}\varphi_{\textsc{mab}}\ .\label{FBD-Bt1}
	\end{empheq}
Notice that, while in 4D fracton theories both electric and magnetic fields are rank-2 tensors \cite{Bertolini:2022ijb}, in our 3D case, the electric field is still a tensor, while the magnetic field is a vector. Moreover, the definitions \eqref{FBD-B1} and \eqref{FBD-Bt1} are consistent with the fact that in ordinary 3D electromagnetism the electric field $\vec E(x)$ is a vector, while the magnetic field is a pseudo-scalar $B=\epsilon^{0ij}F_{ij}$ \cite{Boito:2018rdh}. As we shall see, our guess \eqref{FBD-B1} and \eqref{FBD-Bt1} will be confirmed by a consistent physical interpretation of a fractonic ``magnetic-like'' behaviour. In terms of $\alpha_{ab}(X),\ \tilde\alpha_{ab}(X)$ the magnetic fields read
	\be
	B^\textsc{m}=-3g\epsilon_{0\textsc{ab}}\left(\partial^\textsc{b}\tilde\alpha^\textsc{ma}+\frac{1}{2}\eta^{\textsc{ma}}\partial_d\tilde\alpha^{\textsc b d}\right)\quad;\quad\tilde	B^\textsc{m}=-3\tilde g\epsilon_{0\textsc{ab}}\left(\partial^\textsc{b}\alpha^\textsc{ma}+\frac{1}{2}\eta^{\textsc{ma}}\partial_d\alpha^{\textsc b d}\right)\ ,
	\ee
which imply
	\begin{empheq}{align}
	\tilde\varphi^{\textsc{abc}}&=-\frac{1}{3g}\left(\epsilon^{0\textsc{ac}}B^\textsc{b}+\epsilon^{0\textsc{bc}}B^\textsc{a}\right)\label{FBD-B}\\
	\varphi^{\textsc{abc}}&=-\frac{1}{3\tilde g}\left(\epsilon^{0\textsc{ac}}\tilde B^\textsc{b}+\epsilon^{0\textsc{bc}}\tilde B^\textsc{a}\right)\ \label{FBD-tB}\ .
	\end{empheq}
Due to tracelessness property \eqref{FBD-traceless}, we get 
\be\label{FBD-TrPhi=B}
\varphi_{\textsc a}^{\ \textsc{ab}}=\varphi^{00\textsc b}=-\frac{2}{3\tilde g}\epsilon^{0\textsc{ab}}\tilde B_\textsc{a}\quad;\quad \tilde \varphi_{\textsc a}^{\ \textsc{ab}}=\tilde \varphi^{00\textsc b}=-\frac{2}{3g}\epsilon^{0\textsc{ab}}B_\textsc{a}\ .
\ee
Notice that
	\begin{empheq}{align}
	\partial_\textsc{m}B^\textsc{m}=&-\frac{3}{2}g\,\epsilon_{0\textsc{ab}}\partial^\textsc{a}\left(\partial_0\tilde\alpha^{0\textsc b}-\partial_\textsc{c}\tilde\alpha^{\textsc{bc}}\right)\neq0\label{FBD-divBneq0}\\
	 \partial_\textsc{m}\tilde B^\textsc{m}=&-\frac{3}{2}\tilde g\,\epsilon_{0\textsc{ab}}\partial^\textsc{a}\left(\partial_0\alpha^{0\textsc b}-\partial_\textsc{c}\alpha^{\textsc{bc}}\right)\neq0\ ,\label{FBD-divtBneq0}
	\end{empheq}
which would suggest the presence, in the 3D theory \eqref{FBD-S3D}, of a fractonic ``magnetic''-like vortex. Consistently with the fact of having non-vanishing divergences of the magnetic vector fields, we find also a broken Bianchi identity, which also suggests the presence of a kind of magnetic fracton vortex. This would imply that a part of our fracton fields give rise to 2D fracton vortex defects that represent a lower-dimensional version of  the 3D fracton magnetic monopole proposed in \cite{Pretko:2017xar}. In fact, we have
	\be
	\epsilon^{mbc}\partial_m\varphi_{abc}=-\frac{3}{2}\epsilon_{mac}\partial^m\partial_d\alpha^{cd}\neq0\quad;\quad\epsilon^{mbc}\partial_m\tilde\varphi_{abc}=-\frac{3}{2}\epsilon_{mac}\partial^m\partial_d\tilde\alpha^{cd}\neq0 \ ,
	\ee
which for $a=0$ give the non-vanishing divergences \eqref{FBD-divBneq0} and \eqref{FBD-divtBneq0}. Setting instead $a=\textsc a$  we find
\begin{align}
\frac{1}{g}\partial_0B_\textsc{a}+\frac{1}{2g_{12}}\epsilon^{0\textsc{mb}}\partial_\textsc{m}E_{\textsc{ab}}&=\frac{3}{2}\epsilon_{m\textsc a b}\partial^m\partial_d\tilde\alpha^{bd}+\epsilon^{\textsc{m0b}}\partial_\textsc{m}\tilde\varphi_{\textsc{0ab}}\neq0\\
\frac{1}{\tilde g}\partial_0\tilde B_\textsc{a}+\frac{1}{2g_{12}}\epsilon^{0\textsc{mb}}\partial_\textsc{m}\tilde E_{\textsc{ab}}&=\frac{9}{4}\frac{\omega_5}{g_{12}}\partial^\textsc{m}\tilde\alpha_\textsc{ma}+\frac{3}{2}\epsilon_{m\textsc a b}\partial^m\partial_d\alpha^{bd}+\epsilon^{\textsc{m0b}}\partial_\textsc{m}\varphi_{\textsc{0ab}}\neq0\label{FBD-Bianchi2}
\end{align}
which have non-vanishing right hand sides. Nonetheless, we have two scalar identities for $\varphi_{abc}(X)$ and $\tilde\varphi_{abc}(X)$ :
	\begin{empheq}{align}
	\epsilon^{mbc}\partial_m\partial^a\varphi_{abc}=&-\frac{3}{2}\epsilon_{mac}\partial^m\partial^a\partial_d\alpha^{cd}=0 \\
	\epsilon^{mbc}\partial_m\partial^a\tilde\varphi_{abc}=&-\frac{3}{2}\epsilon_{mac}\partial^m\partial^a\partial_d\tilde\alpha^{cd}=0 \ .
	\end{empheq}
Going back to the 3D EoM, we consider  \eqref{FBD-eom1bd} at $m=\textsc m,\ n=\textsc n$
	\be
		\partial_0\tilde\varphi^{\textsc{mn}0}+\partial_\textsc{a}\tilde\varphi^{\textsc{mna}}
		=\frac{1}{2g_{12}}\partial_tE^{\textsc{mn}}-\frac{1}{3g}\partial_\textsc{a}\left(\epsilon^{0\textsc{ma}}B^{\textsc n}+\epsilon^{0\textsc{na}}B^{\textsc m}\right)\ ,
	\ee
where we used the definitions  \eqref{FBD-E}, \eqref{FBD-B}. We thus have
	\be\label{FBD-amp1}
	\partial_tE^{\textsc{mn}}-\frac{2g_{12}}{3g}\partial_\textsc{a}\left(\epsilon^{0\textsc{ma}}B^{\textsc n}+\epsilon^{0\textsc{na}}B^{\textsc m}\right)=0\ ,
	\ee
which, remarkably, coincides with the traceless analog of the Amp\`ere equation of 3D fractons identified in Eq.(16) of \cite{Pretko:2017kvd}, where our \eqref{FBD-amp1} is obtained as an \ac{EoM}  from a 3D Maxwell-like Hamiltonian defined \textit{ad hoc}. Differently from our case, the 3D fracton theory in \cite{Pretko:2017kvd} is  not traceless and, in particular,  $E^{\ \textsc{a}}_\textsc{a}\neq0$. The aim of \cite{Pretko:2017kvd} is to study the so called fracton-elasticity duality and, more specifically, the analog of our \eqref{FBD-amp1} is used to investigate the effect of creation of defects as consequence of longitudinal motion of dipoles, which in the traceless fracton theory is not present, since dipoles only move along their transverse direction. We keep considering the \ac{EoM} \eqref{FBD-eom2bd} at $m=\textsc m,\ n=\textsc n$ :
	\be
		\begin{split}
		0=&-2g_{12}\partial_a\varphi^{\textsc{mn}a}+\omega_5\left(\epsilon^{\textsc{m}ab}\tilde\varphi^{\textsc{n}}_{\ ab}+\epsilon^{\textsc{n}ab}\tilde\varphi^{\textsc{m}}_{\ ab}\right)\\
		=&-\partial_0\tilde E^{\textsc{mn}}+\frac{3}{2}\omega_5\partial_0\left(\epsilon^{0\textsc{bm}}\tilde\alpha^\textsc{n}_\textsc{b}+\epsilon^{0\textsc{bn}}\tilde\alpha^\textsc{m}_\textsc{b}\right)+\frac{2g_{12}}{3\tilde g}\partial_\textsc{a}\left(\epsilon^{0\textsc{ma}}\tilde B^{\textsc n}+\epsilon^{0\textsc{na}}\tilde B^{\textsc m}\right)+\\
&+3\omega_5\partial_{\textsc{b}}\left(\epsilon^{0\textsc{mb}}\tilde\alpha^\textsc{n}_0+\epsilon^{0\textsc{nb}}\tilde\alpha^\textsc{m}_0\right)\ ,
		\end{split}
	\ee
where we used the definitions  \eqref{FBD-tE}, \eqref{FBD-tB}. We have
	\be\label{FBD-amp2}
	\partial_t\tilde E^{\textsc{mn}}-\frac{2g_{12}}{3\tilde g}\partial_\textsc{a}\left(\epsilon^{0\textsc{ma}}\tilde B^{\textsc n}+\epsilon^{0\textsc{na}}\tilde B^{\textsc m}\right)=\mathcal{\tilde{J}}_5^{\textsc{mn}}\equiv\omega_5\mathcal{\tilde{J}}^{\textsc{mn}}\ ,
	\ee
which is analogous to the Amp\`ere equation, in presence of a tensorial current, again related to the Chern-Simons-like  term in \eqref{FBD-S3D}, which behaves as a matter term
	\be\label{FBD-J}
		\mathcal{\tilde{J}}^{\textsc{mn}}
		\equiv3\left[\frac{1}{2}\partial_0\left(\epsilon^{0\textsc{bm}}\tilde\alpha^\textsc{n}_\textsc{b}+\epsilon^{0\textsc{bn}}\tilde\alpha^\textsc{m}_\textsc{b}\right)+\partial_{\textsc{b}}\left(\epsilon^{0\textsc{mb}}\tilde\alpha^\textsc{n}_0+\epsilon^{0\textsc{nb}}\tilde\alpha^\textsc{m}_0\right)\right]\ .
	\ee
By computing  $\partial_\textsc{m}\partial_\textsc{n}$ of \eqref{FBD-amp2}, we also get
	\be\label{FBD-cont1}
		\partial_t\tilde\rho_5=\partial_\textsc{m}\partial_\textsc{n}\mathcal{\tilde{J}}_5^{\textsc{mn}}\ ,
	\ee
where we used the Gauss-like equation \eqref{FBD-tGauss}. Eq.\eqref{FBD-cont1} represents a continuity equation typical of scalar fracton theories \cite{Pretko:2016lgv,Pretko:2016kxt}  if $\omega_5\neq0$
	\be\label{FBD-cont}
	\partial_t\tilde\rho-\partial_\textsc{m}\partial_\textsc{n}\mathcal{\tilde{J}}^{\textsc{mn}}=0\ .
	\ee
In $\tilde{\mathcal{J}}^{\textsc{mn}}(X)$ \eqref{FBD-J}, the contribution associated to the time derivative coincides with the one defined by Pretko (Eq.(118) of \cite{Pretko:2017xar}) as ``generalized Hall response''. As in our case, it is derived from a Chern-Simons-like  term seen as a boundary contribution originated by a fractonic $\theta$-term in the bulk. In particular, it comes from the dynamical part of the action. From the action $S_{3D}$ \eqref{FBD-S3D} we can identify both the current and the Amp\`ere-like equation \eqref{FBD-amp2}, to which the current \eqref{FBD-J} contributes. Moreover, as already mentioned for \eqref{FBD-amp1}, this second equation \eqref{FBD-amp2} is compatible with the traceful version identified in \cite{Pretko:2017kvd} in the context of an analysis of 3D fracton-elasticity duality. Since $E^{\ \textsc{m}}_\textsc{m}=\tilde E^{\ \textsc{m}}_\textsc{m}=0$, by computing the trace of the Amp\`ere-like equations \eqref{FBD-amp1} and \eqref{FBD-amp2}, we find
	\begin{empheq}{align}
	\epsilon^{0\textsc{mn}}\partial_\textsc{m}B_\textsc{n}&=0\\
	\epsilon^{0\textsc{mn}}\partial_\textsc{m}\tilde B_\textsc{n}&=-\frac{3}{4}\frac{\tilde g}{g_{12}}\mathcal{\tilde J}^{\textsc{m}}_{5\ \textsc m}=-\frac{9}{2}\frac{g\omega_5}{g_{12}}\epsilon^{0\textsc{mn}}\partial_\textsc{m}\tilde\alpha_{0\textsc n}\ ,
	\end{empheq}
which consistently coincide with  the \ac{EoM} for $m=n=0$ \eqref{FBD-eom1bd00}, \eqref{FBD-eom2bd00} previously found, $i.e.$
	\be
	\partial_{\textsc{a}}\tilde\varphi^{00\textsc{a}}=0\quad;\quad\partial_\textsc{a}\varphi^{00\textsc{a}}=\frac{3\omega_5}{g_{12}}\epsilon^{0\textsc{ab}}\partial_\textsc{a}\tilde\alpha^0_{\ \textsc{b}}\ ,
	\ee
due to \eqref{FBD-TrPhi=B}. The \ac{EoM} of the 3D boundary theory may be interpreted as a traceless tensorial extension of the standard 3D Maxwell equations \cite{Boito:2018rdh}, as summarized in Table~\ref{3Dmax-frac}
\begin{table}[H]
\centering
\resizebox{1\columnwidth}{!}{
  \begin{tabular}{ | rl | c | c| }
    \hline
   &  &  Maxwell &  Boundary of \ac{LG} and fractons \\ \hline
fields& electric, magnetic & $\vec E\ ,\ B$& 	$E^{\textsc{ab}},\ B^\textsc{a}\ ;\ \tilde E^{\textsc{ab}},\ \tilde B^\textsc{a}$ \\ \hline
&in vacuum& $\vec\nabla\cdot\vec E=0$ & $\partial_\textsc{a}\partial_\textsc{b}E^\textsc{ab}=0$ \\[-5px]
  Gauss && $$ & $$ \\[-5px]

 &with matter& $\vec\nabla\cdot\vec E=\rho$ & $\partial_\textsc{a}\partial_\textsc{b}\tilde E^\textsc{ab}=\tilde\rho_5$ \\ \hline

 &in vacuum& $\partial_t\vec E-\vec\nabla_\bot B=0$ & $	\partial_tE^{\textsc{mn}}-\frac{2g_{12}}{3g}\partial_\textsc{a}\left(\epsilon^{0\textsc{ma}}B^{\textsc n}+\epsilon^{0\textsc{na}}B^{\textsc m}\right)=0$ \\[-5px] 

  Amp\`ere&&&\\[-5px]
&with matter&$\partial_t\vec E-\vec\nabla_\bot B=\vec J$ & $\partial_t\tilde E^{\textsc{mn}}-\frac{2g_{12}}{3\tilde g}\partial_\textsc{a}\left(\epsilon^{0\textsc{ma}}\tilde B^{\textsc n}+\epsilon^{0\textsc{na}}\tilde B^{\textsc m}\right)=\mathcal{\tilde{J}}_5^{\textsc{mn}}$ \\ \hline
  \end{tabular}
}
  \caption[Comparison between \acs{EoM} and 3D Maxwell]{\footnotesize\label{FBD-table3}Comparison between \acs{EoM} and 3D Maxwell.}
\label{3Dmax-frac}
\end{table}
\noindent
where $\vec\nabla_\bot B\equiv\epsilon^\textsc{0ij}\partial_\textsc{j}B$, and the results are consistent with what can be found in the fracton literature \cite{Pretko:2017kvd,Pretko:2016lgv,Pretko:2016kxt,Pretko:2017xar}. We thus recovered, as \ac{EoM}, the Gauss constraints related to the mobility  of the traceless fracton theory in 3D \cite{Pretko:2016lgv,Pretko:2016kxt}, where the Chern-Simons-like  term contributes as a matter term through $\tilde\rho(X)$ \eqref{FBD-rho}, also identified by Pretko in \cite{Pretko:2017xar}. This Chern-Simons-like  term plays the role of matter contribution also in the fractonic Amp\`ere equation \eqref{FBD-amp2}, as a current  $\tilde{\mathcal{J}}^{\textsc{mn}}(X)$ \eqref{FBD-J}. Here again the term is in accordance with the literature, and in particular with what has been defined as ``generalized Hall response'' in \cite{Pretko:2017xar}. The Amp\`ere equations  \eqref{FBD-amp1} and \eqref{FBD-amp2}, to which the current $\tilde{\mathcal{J}}^{\textsc{mn}}(X)$ belongs, can be traced back to fracton theories as well, and, more specifically, they have the same structure as the fractonic Amp\`ere equation used in  \cite{Pretko:2017kvd} to study a duality between the theory of fractons and the theory of elasticity. However the one considered in \cite{Pretko:2017kvd} refers to the traceful theory of fractons, whose aim is to study defects as consequence of longitudinal motion of dipoles, which in the traceless fracton theory is not present since dipoles only move along their transverse direction \cite{Pretko:2016lgv,Pretko:2016kxt}. Therefore it would be interesting  to understand if a fracton-elasticity duality also exists for the traceless model. We also notice that in  \cite{Gromov:2017vir} a charge $\tilde\rho(x)$, a current  $\tilde J_\textsc{ab}(x)$, and a continuity equation as \eqref{FBD-cont}  are identified  from a Chern-Simons-like  theory with torsion $T(x)$, $i.e. $  $\int d^3x\epsilon^{\mu\nu\rho} e_\mu^\textsc{a}T_{\nu\rho}^\textsc{a}$. In particular, the model coincides with the one proposed in \cite{Pretko:2017xar} for the non-covariant Chern-Simons-like  action for a specific choice of vielbein $e_\mu^\textsc{a}(x)$, and under the condition of ``area-preserving diffeomorphisms'', which seems to be strictly related to fracton models, as also studied in \cite{Du:2021pbc}. This intriguing role of torsion in 3D chiral fractons is extended to 4D fractons of Section \ref{sec MaxThFract}, where it has been shown that a linearized topological term with torsion \cite{Chatzistavrakidis:2020wum} gives rise to the fracton $\theta$-term \cite{Pretko:2017xar}.

\subsection{Discrete symmetries : parity and time reversal}\label{FBD-sec PT}

As extensively shown in the recent literature concerning non-perturbative aspects of quantum field theories, discrete symmetries play a central role in the identification of global anomalies and anomaly inflow, which are related to topological obstructions and impose strong constraints on the renormalization group flows, massive boundary states, quantum dualities and the vacua of quantum field theories \cite{Cordova:2017kue,Gaiotto:2017yup,Benini:2017dus,Delmastro:2021xox,Kapustin:2014dxa,Barkeshli:2016mew,Cordova:2019wpi}. Moreover, the anomaly inflow has been also extended to certain non-covariant fracton models \cite{Burnell:2021reh}. Thus, here we analyze some discrete symmetries, such as $\TR$ and $\Par$ in the context of the induced theory we just derived. In fact, we can further constrain the induced 3D action by requiring a matching between the discrete symmetries in the bulk and on the boundary. In particular, under $\Par$ and $\TR$ the bulk fields transform as follows
	\begin{align}
	\begin{split}
	\TR\Bigl\{A_{00}\,,\,A_{0\textsc{a}}\,,\,A_{\textsc{ab}}\Bigr\}&=\Bigl\{A_{00}\,,\,-A_{0\textsc{a}}\,,\,A_{\textsc{ab}}\Bigr\}\\
	\TR\Bigl\{\tilde A_{00}\,,\,\tilde A_{0\textsc{a}}\,,\,\tilde A_{\textsc{ab}}\Bigr\}&=\Bigl\{\tilde A_{00}\,,\,-\tilde A_{0\textsc{a}}\,,\,\tilde A_{\textsc{ab}}\Bigr\}
	\end{split}\\
	\begin{split}
	\Par \Bigl\{A_{00}\,,\,A_{0\textsc{a}}\,,\,A_{\textsc{ab}}\Bigr\}&=\Bigl\{A_{00}\,,\,-A_{0\textsc{a}}\,,\,A_{\textsc{ab}}\Bigr\}\\
	\Par \Bigl\{\tilde A_{00}\,,\,\tilde A_{0\textsc{a}}\,,\,\tilde A_{\textsc{ab}}\Bigr\}&=\Bigl\{\!-\tilde A_{00}\,,\tilde A_{0\textsc{a}}\,,-\tilde A_{\textsc{ab}}\Bigr\}
	\end{split}\\
	\TR A=A\quad;\quad\TR\tilde A=\tilde A\quad&\ \ ;\ \ \quad \Par A=A\quad;\quad \Par \tilde A=-\tilde A
	\ .
	\end{align}
The bulk action \eqref{FBD-Sbulk} is invariant under $\TR$. Instead, due to the presence of the boundary $x^3=0$, the action is no longer $\Par$-invariant. We now consider the boundary term \eqref{FBD-Sbd}, and distinguish between space and time indices
	\begin{align}
		S_{bd}=\int d^4x\delta(x^3)&\left[\xi_0 \left(A_{00}A^{00}+2A_{0\textsc{a}}A^{0\textsc{a}}+A_{\textsc{ab}}A^{\textsc{ab}}\right)+\xi_1\left(\tilde A_{00}A^{00}+2\tilde A_{0\textsc{a}}A^{0\textsc{a}}+\tilde A_{\textsc{ab}}A^{\textsc{ab}}\right)\right.+\nonumber\\
		+&\left.\xi_2\epsilon^{0\textsc{ab}}\left(A_{0i}\partial_\textsc{a}A_\textsc{b}^i-A_{\textsc{a}i}\partial_0A_\textsc{b}^i+A_{\textsc{a}i}\partial_\textsc{b}A_0^i\right)+\xi_3A^2+\xi_4\tilde AA\right]\ .\label{FBD-SbdPT}
	\end{align}
	We then observe that
	\bi
	\item $\Par S_{bd}=S_{bd}$ if $\xi_1=\xi_4=0$ ;
	\item $\TR S_{bd}=S_{bd}$ if $\xi_2=0$ ;
	\item $\TR\Par S_{bd}=S_{bd}$ if $\xi_1=\xi_2=\xi_4=0$ .
	\ei
Under these considerations, we can update Table \ref{FBD-table2} of holographic contacts with the discrete symmetries allowed on the boundary term $S_{bd}$
	\begin{table}[H]
\centering
  \begin{tabular}{ | l | c | c| }
    \hline
     &\ac{BC}-\ac{EoM} matching& Discrete symmetries of $S_{bd}$\\ \hline
      $\xi_1\neq0,\ \omega_5\neq0$ & $\omega_5=\frac{1}{3}(\xi_1-g_{12})\ ;\ \xi_2=g_{12}$& No\\ \hline
      $\xi_1\neq0,\ \omega_5=0$ & $\omega_5=0$ & $\TR$ \\ \hline
  \end{tabular}
  \caption[Holographic contacts : constraints and discrete symmetries]{\footnotesize\label{FBD-table4}Holographic contacts, constraints and possible symmetries on $S_{bd}$ \eqref{FBD-Sbd}.
}
\end{table}
\noindent
From Table \ref{FBD-table4} we see that imposing $\Par$ on $S_{bd}$ does not lead to a holographic contact, since the $\xi_1$ term, crucial for the existence of the induced 3D action, is not $\Par$-invariant. We also highlight a relation  between $\omega_5\leftrightarrow\xi_2\leftrightarrow\TR$, in fact $\TR$ symmetry is possible only when $\xi_2=0$, which is allowed only in the second holographic contact \eqref{FBD-hc2}, $i.e.$ when $\omega_5$ is set to zero as well (in the first case \eqref{FBD-hc1} the parameter $\xi_2$ is constrained by \eqref{FBD-vincoli}). Therefore the Chern-Simons-like  terms must be absent both in $S_{bd}$ \eqref{FBD-Sbd} and in $S_{3D}$ \eqref{FBD-S3D}, in order to have $\TR$ symmetry preserved on the boundary.
\normalcolor

\subsection{Summary and discussion}\label{FBD-sec-summary}

The questions that were raised at the beginning of this analysis in Section \ref{sec Frac+bd}, have thus found their answers. Indeed an algebraic structure on the boundary does exist, as a consequence of the breaking of the Ward identities, and it can be interpreted  as a generalization of the standard $U(1)$ \ac{KM} algebra, characterized by a double derivative, as it appears for instance also in \cite{You:2019bvu}, further stressing the suspected relations between fractons and higher order topological insulators. From the two broken Ward identities,
 the boundary \ac{DoF} of the induced theory are identified as two symmetric traceless rank-2 tensors $\alpha_{ij}(X)$ and $\tilde\alpha_{ij}(X)$. It is worth to remark that
on the boundary some \ac{DoF} disappear, since the boundary tensor fields turn out to be traceless. This might be due to the presence of a hidden symmetry, a guess that should be further investigated. 
The procedure to recover the induced theory leads to
the action $S_{3D}$ \eqref{FBD-S3D}, which is composed of a term similar to a higher-rank Maxwell contribution, written in terms of traceless rank-3 field strengths, which
mixes both fields $\alpha_{ij}(X)$ and $\tilde\alpha_{ij}(X)$, with a coefficient depending on the bulk constants $g_1$ and $g_2$, and a Chern-Simons-like  term for $\tilde\alpha_{ij}(X)$ with a free coefficient. 
Concerning the physical interpretation of our 3D induced theory $S_{3D}$ \eqref{FBD-S3D}, this can be identified with the ``traceless scalar charge'' model of fractons \cite{Pretko:2016lgv,Pretko:2016kxt,Du:2021pbc}. In fact the transformations of the boundary fields, the canonical commutators, the traceless conjugate momenta, $i.e.$ the ``electric fields'', coincide with what appears in the literature. 
This claim is confirmed also by the \ac{EoM} of the 3D induced theory, from which two Gauss-like laws are derived, which imply the defining property of the fracton quasiparticles, $i.e.$ their limited mobility. Thus, one of the main results of this Section is that a non-standard covariant 3D traceless fracton theory turns out to be holographically induced from a 4D ordinary traceful covariant fracton theory. This claim gets even stronger confirmation from other components of the \ac{EoM}, which can be identified with the Amp\`ere-like equations 
 of fractons \cite{Pretko:2017kvd}, further stressing the relation of fracton models with Maxwell theory. Concerning this analogy, we remark a close resemblance of our 3D action $S_{3D}$ \eqref{FBD-S3D} with Maxwell-Chern-Simons theory \cite{Deser:1981wh}, of which it appears to be a kind of spin-two generalization. A similar observation can also be found in \cite{Dalmazi:2008zh} in the context of self-dual massive gravity, where an identical covariant Chern-Simons term appears, and whose relation with our 3D model is worth to be further investigated. However, differently from the standard Maxwell-Chern-Simons theory, here all the coefficients are dimensionless, hence no topological mass can be identified. 
Therefore to better analyze this analogy, the study of the propagators would be helpful.
Notice also that the Chern-Simons coefficient is free, thus it can be switched off. The choice of keeping the Chern-Simons-like  term or not is relevant for the physical interpretation of the model: by switching it off,  the 3D action $S_{3D}$ \eqref{FBD-S3D} can be decoupled into two Maxwell-like terms, and the boundary theory is compatible with $\TR$-symmetry, which characterizes the phenomenology involved. For instance, we have seen in Chapters \ref{ch CSandBF} and \ref{ch CSandBFinCS} that the physics on the boundary of the topological BF models \cite{Cho:2010rk,Cirio:2013dxa,Blasi:2011pf,Amoretti:2012hs} is identified with the effective description of the edge states of \acl{TI}, where $\TR$ is preserved both on the bulk and on the boundary. On the other hand, keeping the Chern-Simons-like  term, $i.e.$ relaxing the $\TR$ constraint, the \ac{EoM} get a matter contribution. In particular the Chern-Simons-like  term plays the role of fractonic charge $\tilde\rho(X)$ \eqref{FBD-rho} and current $\tilde J_{\textsc{ij}}(X)$ \eqref{FBD-J} in two of the Maxwell-like equations, in accordance with \cite{Pretko:2017xar}.
Some final physical remarks are in order. Differently from the standard electromagnetic theory and the 4D traceful fracton model, here the magnetic-like vectors $B_\textsc{a}(X)$ and $\tilde B_\textsc{a}(X)$ do not have zero divergence, nor a Bianchi identity exists for the traceless rank-3 field strengths $\varphi_{abc}(X),\ \tilde\varphi_{abc}(X)$, which suggests the presence of fractonic 3D vortices. Additionally, 3D fracton models are known to be related to the elasticity theory of topological defects through a duality \cite{Pretko:2017kvd}. For instance the traceful Amp\`ere-like equation can be seen as describing the motion of these defects. Under this respect, it would be interesting to understand if and how our traceless boundary theory can be related to topological defects. Finally, there seems to be an interesting possible interpretation of the fractonic Chern-Simons-like  term as associated to torsion contributions, as in \cite{Gromov:2017vir}, which also would be worth to further analyze.

\chapter{Linearized Gravity} 

\label{ch LG}


In this Chapter, we consider the 4D theory of \ac{LG} with a planar boundary, motivated by the guessed relation between \ac{KM} algebras \cite{Kac:1967jr,Moody:1966gf} and \ac{LG} \cite{Chamseddine:1988tu,Houart:2005wh,Hinterbichler:2022agn}, providing a precise interpretation of the physical quantity characterizing \ac{KM} algebras -- its central charge. According to the ``general rule'', the central charge should be related to the \ac{LG} coupling constant, normally set to one by a redefinition of the field, which in \ac{LG} is a symmetric rank-2 tensor field. Therefore the positivity of the central charge could also give us a way to determine the sign of the \ac{LG} ``coupling'' constant, which in this case cannot be inferred from an energy constraint.
Apart from the \ac{KM} algebra of conserved current, the introduction of a boundary yields more. The presence of a boundary in the 3D Chern-Simons theory induces a 2D theory, the Tomonaga-Luttinger theory of a chiral scalar field \cite{Tomonaga:1950zz,Luttinger:1963zz,Haldane:1981zza,Maggiore:2017vjf}, which we may refer to as a ``holographic'' reduction of the 3D bulk theory. Similar dimensional reductions -- lower dimensional theories induced by the presence of a boundary in higher-dimensional ones -- can be performed in other theories, both topological and non-topological. For instance we have seen in the previous Chapter the case of the fractonic theory, whose symmetry is a particular case of the diffeomorphisms, from which we have observed that fractons always come together with \ac{LG}. Thus the theory with boundary studied in Section \ref{sec Frac+bd} comprises both fractonic contribution and \ac{LG}, but without allowing for the limit of $g_1=0$ $i.e.$ pure \ac{LG}, which is not contained in that analysis. The reason is that, as explained, \ac{LG} is a theory defined by a more general symmetry (the diff symmetry), which we know from Chapter \ref{QFTapproach} plays an important role in the search for the induced boundary theory. Indeed it is the breaking of the symmetry due to the presence of the boundary that leads to the algebra and the degrees of freedom on the lower dimensional theory. If the symmetry changes, the boundary theory may change as well. Therefore here we will apply the same mechanism of the previous Chapters to the case of pure \ac{LG} to find the holographically induced 3D theory of 4D \ac{LG} with a planar boundary.  In particular in Section \ref{sec GW} we will first briefly review the theory of \ac{LG} and its bulk physical content, and in Section \ref{sec LG+bd} we will proceed with the introduction of a planar boundary.

\section{The theory of gravitational waves}\label{sec GW}
We will therefore consider the theory of a symmetric tensor field $h_{\mu\nu}(x)$ which transforms under diffeomorphisms
	\be\label{LGdiff}
	\delta_{diff}h_{\mu\nu}=\partial_\mu\xi_\nu+\partial_\nu\xi_\mu\ .
	\ee
As we mentioned previously, this theory contains the theory of fractons of Chapter \ref{ch fractons} through a reduction of the diff transformation \eqref{LGdiff} into the longitudinal one \eqref{FMAX-fractonsymintro}, but the physical content is quite different, hence the difference in the notation~: the symmetric rank-2 tensor field $A_{\mu\nu}(x)$ of the previous Chapter was of fractonic nature, here we will refer to $h_{\mu\nu}(x)$ as the rank-2 symmetric tensor field of \ac{LG}. We are stepping from the microscopic world of condensed matter and immobile quasiparticles, to the linearized theory of gravity describing gravitational waves. Indeed this theory can be either derived from a pure \ac{QFT} point of view as the theory invariant under the diff symmetry \eqref{LGdiff}, as a perturbation theory in General Relativity, where the gauge field $h_{\mu\nu}(x)$ is a perturbation around the flat Minkowskian metric. In the first case, $i.e.$ \ac{QFT} perspective, the transformation \eqref{LGdiff} immediately gives the invariant \ac{LG} action we already mentioned in the previous Chapters
	\be
	S_{LG} = \int d^4x\left(\partial_\mu h\partial^\mu h-\partial_\rho h_{\mu\nu}\partial^\rho h^{\mu\nu}-2\partial_\mu h\partial_\nu h^{\mu\nu}+2\partial_\rho h_{\mu\nu} \partial^\mu h^{\nu\rho}\right)\ , \label{LG-SLG}
	\ee
which describes a theory of a massless symmetric rank-2 tensor field $h_{\mu\nu}(x)$ in a flat spacetime. It is relevant to mention here an important generalization in which this gauge field is given a mass by breaking gauge invariance. This is known as the Fierz-Pauli action \cite{ ,Hinterbichler:2011tt,Blasi:2015lrg,Blasi:2017pkk,Gambuti:2021meo}
	\be
	S_{FP} = \int d^4x\left[m^2(h^2-h_{\mu\nu}h^{\mu\nu})+\partial_\mu h\partial^\mu h-\partial_\rho h_{\mu\nu}\partial^\rho h^{\mu\nu}-2\partial_\mu h\partial_\nu h^{\mu\nu}+2\partial_\rho h_{\mu\nu} \partial^\mu h^{\nu\rho}\right]\ , \label{LG-SFP}
	\ee
which describes a single massive spin 2 degree of freedom of mass $m$. This model is of particular interest nowadays since it seems that observations of gravitational waves, which we will show to be the physics related to the \ac{LG} action \eqref{LG-SLG}, seem to indicate that gravitons are not massless indeed. Instead, starting from the the Einstein-Hilbert action of General Relativity
	\be
	S_{EH}=\int d^4x\sqrt{-g}\,R\ ,
	\ee
one can consider a weak field limit, in which the metric of the spacetime $g_{\mu\nu}(x)$ is a perturbation around flat spacetime
	\be\label{g->h}
	g_{\mu\nu}=\eta_{\mu\nu}+h_{\mu\nu}\ .
	\ee
Under this assumption translational invariance in curved spacetime, which is encoded in the Lie derivative $\mathcal L_\xi$ \cite{Carroll:2004st}, generates the diff transformation \eqref{LGdiff} through linearization~:
	\be
	\mathcal L_\xi g_{\mu\nu}=\nabla_\mu\xi_\nu+\nabla_\nu\xi_\mu\quad\xrightarrow{\eqref{g->h}}\quad\delta_{diff}h_{\mu\nu}=\partial_\mu\xi_\nu+\partial_\nu\xi_\mu\ .
	\ee
At the same time the Einstein's \ac{EoM}
	\be
	\frac{\delta S_{EH}}{\delta g^{\mu\nu}}=\mathcal G_{\mu\nu}=R_{\mu\nu}-\frac{1}{2}g_{\mu\nu}R=0\ ,
	\ee
matches the \ac{EoM} recoverd from the \ac{LG} action \eqref{LG-SLG} in the weak-field limit \eqref{g->h}, for which at first order in the perturbation the Ricci tensor and scalar are respectively
	\be
	R_{\mu\nu}
\sim\frac{1}{2}\left(\partial_\mu\partial^\lambda h_{\lambda\nu}+\partial_\nu\partial^\lambda h_{\lambda\mu}-\Box h_{\mu\nu}-\partial_\mu\partial_\nu h\right)\quad;\quad R
\sim\partial^\mu\partial^\nu h_{\mu\nu}-\Box h\ ,
	\ee
thus recovering
	\be
	\mathcal G_{\mu\nu}
\sim\partial_\mu\partial^\lambda h_{\nu\lambda}+\partial_\nu\partial^\lambda h_{\mu\lambda}-\Box h_{\mu\nu}-\partial_\mu\partial_\nu h-\eta_{\mu\nu}\partial^\alpha\partial^\beta h_{\alpha\beta}+\eta_{\mu\nu}\Box h=0\ .\label{EOMLG}
	\ee
In order to analyze the physical content of this equation the first step is to write the gauge field $h_{\mu\nu}(x)$ in terms of the irreducible representation of rotations \cite{Carroll:2004st}, as we have also done in Section \ref{sec Frac+bd} in \eqref{App al00}-\eqref{App altij}. Therefore we have
	\be\label{LG-rotations}
	h_{00}=-2\phi\quad;\quad h_{0i}=w_i\quad;\quad h^k_{\,k}=-6\psi\quad;\quad h_{ij}=2(s_{ij}-\eta_{ij}\psi)\ ,
	\ee
thus we have two scalars - $\phi(x),\psi(x)$ -, a vector - $w_i(x)$ -, and a traceless symmetric spatial tensor - $s_{ij}(x)$ -. One then has to fix a gauge which gives four conditions on the model, due to the nature of the gauge parameter of the symmetry. The typical (non-\ac{QFT}) way to proceed \cite{Carroll:2004st} is thus to impose the so called ``transverse gauge''
	\be\label{tgauge}
	\partial^iw_i=0\quad;\quad\partial^is_{ij}=0 \ .
	\ee
Clearly the same results can be obtained also by means of other choices, however this is the best one to quickly recover the physical content. Indeed, from this choice and the decomposition \eqref{LG-rotations}, the \ac{EoM} \eqref{EOMLG} can be analyzed in terms of its time and space components as follows
	\begin{align}
	\mathcal G_{00}&=2\nabla^2\psi=0\quad &&\Rightarrow\quad\psi=0\label{G00}\\
	\mathcal G_{0i}&=2\partial_0\partial_i\cancel{\psi}-\frac 1 2 \nabla^2w_i=0\quad&&\Rightarrow\quad w_i=0\label{G0i}\\
	\mathcal G_{ij}&=\left(\eta_{ij}\nabla^2-\partial_i\partial_j\right)(\phi-\cancel{\psi})+&&\hspace{-2em}2\eta_{ij}\partial_0^2\cancel{\psi}+\partial_0(\cancel{\partial_iw_j+\partial_jw_i})-\Box s_{ij}=0\label{Gij}
	\end{align}
where \eqref{G00} and \eqref{G0i} give Poisson's equations, whose smooth  solutions are only given by vanishing fields \cite{Carroll:2004st}. By computing the trace of the last equation and the fact that $\psi=w_i=0$, we get another Poisson's equation for $\phi(x)$, which thus must be zero as well. We therefore get
	\be
	\Box s_{ij}=0\ ,
	\ee
which is a wave equation for a traceless symmetric field. This is the core of gravitational waves. In terms of the gauge field $h_{\mu\nu}(x)$, the above conditions (Poisson's solutions and transverse gauge \eqref{tgauge}), are usually (misleadingly) translated into the so called "transverse-traceless" (\textsc{tt}) gauge. We say misleading because this does not only contain a gauge, but also solutions to the \ac{EoM} and, additionally, the results could be recovered also in different gauges. Anyway, the gauge field in the ``\textsc{tt}-gauge'' $h^{\textsc{tt}}_{\mu\nu}(x)$ satisfy the following relations
	\begin{align}
	h^{\textsc{tt}}_{0\mu}&=0\\
	\eta^{\mu\nu}h^{\textsc{tt}}_{\mu\nu}&=0\quad\mbox{(traceless)}\\	
	\partial^\mu h^{\textsc{tt}}_{\mu\nu}&=0\quad\mbox{(transverse)}\label{LG trans}\\
	\Box 	h^{\textsc{tt}}_{\mu\nu}&=0\ .\label{LG wave}
	\end{align}
From \eqref{LG wave} we can find plane waves solution
	\be
	h^{\textsc{tt}}_{\mu\nu}=c_{\mu\nu}e^{ik_\lambda x^\lambda}\ ,
	\ee
with $c_{\mu\nu}$ being purely spatial and traceless ($i.e.\ c_{0\mu}=0=\eta^{\mu\nu}c_{\mu\nu}$), and $k_\lambda$ is a lightlike vector, meaning that the gravitational waves move at the speed of light (unless some kind of mass term can be introduced in the theory, like for instance through the Fierz-Pauli term \eqref{LG-SFP}). By choosing an appropriate reference frame, for example by fixing an axis along the direction of propagation $k^\mu=(\omega,0,0,\omega)$, and considering the transversality equation \eqref{LG trans}, one finally get that $h^{\textsc{tt}}_{\mu\nu}(x)$ is represented by the following matrix
	\be
	h^{\textsc{tt}}_{\mu\nu}=	\left(\begin{array}{cccc}
		0&0&0&0\\
		0&h_+&h_\times&0\\
		0&h_\times&-h_+&0\\
		0&0&0&0
		\end{array}\right)\ , 
	\ee
which means that the gravitational wave has two polarizations which oscillates  transversely to its path \cite{Carroll:2004st}. These oscillations causes stretches of the space, which can be measured and are the bases around which experiments on detection of gravitational waves are built \cite{Carroll:2004st,}.

\section{Linearized gravity with boundary}\label{sec LG+bd}

The analysis of this Section on the effects of the presence of a boundary in the diff invariant theory of \ac{LG} is organized as follows. 
\etocsetnexttocdepth{2}
\begingroup
\parindent=0em
\etocsettocstyle{\rule{\linewidth}{\tocrulewidth}\vskip1.25\baselineskip}{\vskip-0.75\baselineskip\rule{\linewidth}{\tocrulewidth}\vskip1\baselineskip}
\makeatletter
  \edef\scr@tso@subsection@indent
    {\the\dimexpr\scr@tso@subsection@indent-\scr@tso@section@indent}
  \def\scr@tso@section@indent{0pt}
\makeatother
\localtableofcontents 
\endgroup
\noindent
Specifically in Section \ref{LG bulk}, we consider the 4D \ac{LG} theory with a planar boundary, implemented using a Heaviside theta distribution in the bulk action. We derive the most general \ac{BC} following a method introduced by Symanzik, without imposing them by hand. The presence of the boundary breaks the invariance under diffeomorphisms, resulting in a breaking of the diffeomorphism Ward identity. From this, conserved currents and their \ac{KM} algebra are derived using standard \ac{QFT} methods. The central charge appears to be proportional to the inverse of the \ac{LG} ``coupling'' constant, as in topological field theories. Section \ref{LG boundary} focuses on the identification of the holographically induced 3D theory. By first solving the current conservation equation, we find the 3D degrees of freedom and determine the most general transformations that preserve their invariance. Remarkably, we discover that these transformations are diffeomorphisms, which is not obvious. Thus, diffeomorphism invariance emerges as a consequence of our procedure rather than a mere requirement. Once we have the quantum fields with their transformations, we arrive at the most general 3D action, satisfying the additional \ac{QFT} requirements of locality and power counting. Using this, in Section \ref{LG HC} we establish the holographic connection, yielding two solutions for the induced 3D theory. Our results are summarized in Section \ref{LG ends}.

\subsection{The bulk model}\label{LG bulk}

\acl{LG} is the theory of a rank-2 symmetric tensor field $h_{\mu\nu}(x)$ on a flat Minkowskian background. We introduce a planar boundary at $x^3=0$ in the \ac{LG} action by means of a Heaviside step function, which confines the model on a half space with single-sided planar boundary
	\be\label{LG-LG-Sbulk}
	S_{bulk} =
	\lambda \int d^4x\;\theta(x^3) \left(
	\frac{1}{4}F^\mu_{\ \mu\nu} F_\rho^{\ \rho\nu}-\frac{1}{6}F^{\mu\nu\rho}F_{\mu\nu\rho}
	\right)\ ,
	\ee
where $F^{\mu\nu\rho}(x)$ is the rank-3 tensor field strength associated to the tensor field $h_{\mu\nu}(x)$ introduced in Section \ref{sec MaxThFract} \cite{Bertolini:2023sqa,Bertolini:2022ijb}
	\begin{equation}
	F_{\mu\nu\rho}=F_{\nu\mu\rho}=\partial_\mu h_{\nu\rho}+\partial_\nu h_{\mu\rho}-2\partial_\rho h_{\mu\nu}\ ,
	\label{LG-Fmunurho}\end{equation}
satisfying the following properties
	\begin{empheq}{align}
	F_{\mu\nu\rho}+F_{\nu\rho\mu}+F_{\rho\mu\nu}&=0\\
	\epsilon_{\alpha\mu\nu\rho}\partial^{\mu}F^{\beta\nu\rho}&=0\ .
	\end{empheq}
Notice that this field strength, defined in the context of Section \ref{sec MaxThFract}, is invariant unader the longitudinal diff (fracton) symmetry \eqref{FMAX-fractonsymintro}, but it is not invarint under the more general diff transformation \eqref{LGdiff}. The constant $\lambda$ in \eqref{LG-LG-Sbulk} could be reabsorbed through a redefinition of $h_{\mu\nu}(x)$, however we maintain it in order to keep track of the bulk contributions. Due to the presence of the boundary at $x^3=0$ the $x^3$-derivative of the gauge field at $x^3=0$ must be considered as independent from $h_{\mu\nu}(x)$ \cite{Karabali:2015epa}, thus we define
	\be\label{LG-ht}
	\tilde h_{\mu\nu}\equiv\partial_3h_{\mu\nu}|_{x^3=0}\ ,
	\ee
and the fields have the mass dimensions
	\be\label{LG-dimh}
	[h_{\mu\nu}]=1\quad;\quad[\tilde h_{\mu\nu}]=2\ .
	\ee
As already discussed, the \ac{LG} theory \eqref{LG-LG-Sbulk} in the absence of a boundary is invariant under the infinitesimal diffeomorphism transformation
	\be\label{LG-LG-diff}
	\delta_{diff} h_{\mu\nu}=\partial_\mu\Lambda_\nu+\partial_\nu\Lambda_\mu\ ,
	\ee
where $\Lambda_\mu(x)$ is a local vector parameter. \ac{LG} is a gauge field theory, for which a gauge should be fixed. We choose the axial gauge, as customary in presence of a boundary
	\be\label{LG-h3mu=0}
	h_{\mu3}=0\ .
	\ee
This can be realized by means of a vector Lagrange multiplier $b^\mu(x)$ through the gauge-fixing term
	\be
	S_{gf}=\int d^4x\;\theta(x^3)b^\mu h_{\mu3}\ .
	\ee
Moreover, the following source term is needed
	\be
	S_J=\int d^4x\left[\theta(x^3)J^{ab}h_{ab}+\delta(x^3)\tilde J^{ab}\tilde h_{ab}\right]\ ,
	\ee
	 where, together with the external field $J^{ab}(x)$ associated to the tensor gauge field $h_{ab}(x)$, on the boundary an additional external source $\tilde J^{ab}(x)$ is coupled to $\tilde h_{ab}(x)$ \eqref{LG-ht}. The presence of a boundary requires \ac{BC}. We  follow Symanzik's approach \cite{Symanzik:1981wd} previously discussed, by adding to the action a boundary term constrained only by power-counting and locality. This modifies the \ac{EoM} by a boundary contribution, and the \ac{BC} are then obtained by means of a simple variational principle. For our model the boundary term is
	\be\label{LG-LG-Sbd}
	S_{bd}=\int d^4x\delta(x^3)\left[\xi_0 h_{ab}h^{ab}+\xi_1\tilde h_{ab}h^{ab}+\xi_2\epsilon^{abc}h_{ai}\partial_bh_c^i+\xi_3h^2+\xi_4\tilde hh\right]\ ,
	\ee
where
	\be
	h\equiv\eta^{\mu\nu}h_{\mu\nu}\quad;\quad\tilde h\equiv\eta^{\mu\nu}\tilde h_{\mu\nu}\ ,
	\ee
and the $\xi_i,\ i=\{0,...,4\}$, are constant parameters with mass dimensions
	\be
	[\xi_0]=[\xi_3]=1\quad;\quad[\xi_1]=[\xi_2]=[\xi_4]=0\ .
	\ee
Notice that, due to the gauge-fixing condition \eqref{LG-h3mu=0}, the nontrivial part of the trace $h(x)$ is $\eta^{ab}h_{ab}(x)$. The full action of the model finally is
	\be\label{LG-Stot}
	S_{tot}=S_{bulk}+S_{gf}+S_{J}+S_{bd}\ .
	\ee

\subsection*{Equations of motion and boundary conditions}

Besides the \ac{EoM} of the Lagrange multiplier $b^\mu(x)$, which implements the axial gauge condition \eqref{LG-h3mu=0}
	\be
	\frac{\delta S_{tot}}{\delta b^\mu}=h_{\mu3}=0\ ,
	\ee
the \ac{EoM} of the gauge field $h_{\alpha\beta}(x)$ and its $\partial_3$-derivative $\tilde h_{\alpha\beta}(x)$ are, respectively
\begin{align}
\frac {\delta S_{tot}}{\delta h_{\alpha\beta}} &=\theta(x^3)\left\{\lambda \left[
\eta^{\alpha\beta}\partial_\mu F_\nu^{\ \nu\mu}-\partial_{\mu}F^{\alpha\beta\mu}-\tfrac{1}{2}\left(\partial^\alpha F_\mu^{\ \mu\beta}+\partial^\beta F_{\mu}^{\ \mu\alpha}\right)\right]+\delta^\alpha_a\delta^\beta_bJ^{ab}\right.\label{LG-LG-eomh}\\
&+\left.
\tfrac{1}{2}(b^\alpha\delta^\beta_3+b^\beta\delta^\alpha_3)\right\}
+\delta(x^3)\left\{\lambda \left[\eta^{\alpha\beta}F_\mu^{\ \mu3}-F^{\alpha\beta3}-\tfrac{1}{2}\left(\eta^{\alpha3}F_\mu^{\ \mu\beta}+\eta^{\beta3}F_\mu^{\ \mu\alpha}\right)\right]\right.\nonumber\\
&+\left.\delta^\alpha_a\delta^\beta_b\left[ 2\xi_0h^{ab}+\xi_1\tilde h^{ab}+\xi_2(\epsilon^{aij}\partial_ih_j^b+\epsilon^{bij}\partial_ih_j^a)+2\xi_3\eta^{ab}h+\xi_4\eta^{ab}\tilde h\right]\right\}=0\ ,\nonumber
\end{align}
and
\be
\begin{split}
\frac {\delta S_{tot}}{\delta \partial_3h_{\alpha\beta}} &=\lambda \theta(x^3)\left[F^{\alpha\beta3}
-\eta^{\alpha\beta}F_\mu^{\ \mu3}+\tfrac{1}{2}\left(\eta^{\alpha3}F_\mu^{\ \mu\beta}+\eta^{\beta3}F_{\mu}^{\ \mu\alpha}\right)\right]+\\
&+\delta(x^3)\delta^\alpha_a\delta^\beta_b\left\{ \tilde J^{ab}+\xi_1h^{ab}+\xi_4\eta^{ab}h\right\}=0\ .\label{LG-LG-eomht}
\end{split}
\ee
The \ac{BC} come from a variational principle applied on the \ac{EoM} as $\lim_{\epsilon\to0}\int_0^\epsilon dx^3(\mbox{\ac{EoM}})$, which corresponds to putting equal to zero the $\delta(x^3)$ contribution of the \ac{EoM} \eqref{LG-LG-eomh} and \eqref{LG-LG-eomht}. From $\lim_{\epsilon\to0}\int^\epsilon_0dx^3\eqref{LG-LG-eomh}$ we have
\begin{align}
&\left\{\lambda \left[-F^{\alpha\beta3}+\eta^{\alpha\beta}F_\mu^{\ \mu3}-\tfrac{1}{2}\left(\eta^{\alpha3}F_\mu^{\ \mu\beta}+\eta^{\beta3}F_\mu^{\ \mu\alpha}\right)\right]+\right.\label{LG-LG-BC}\\
&+\left.\delta^\alpha_a\delta^\beta_b\left[ 2\xi_0h^{ab}+\xi_1\tilde h^{ab}+\xi_2(\epsilon^{aij}\partial_ih_j^b+\epsilon^{bij}\partial_ih_j^a)+2\xi_3\eta^{ab}h+\xi_4\eta^{ab}\tilde h\right]\right\}_{x^3=0}=0\ .\nonumber
\end{align}
The nontrivial components are
\begin{itemize}
\item $\alpha=3,\ \beta=b$ :
	\be
	\lambda \left(\partial^bh-\partial_ah^{ab}\right)_{x^3=0}=\lambda F_\mu^{\ \mu b}|_{x^3=0}=0\ .\label{LG-LG-bch3i}
	\ee
\item $\alpha=a,\ \beta=b$ :
	\be
	\left[ 2\xi_0h^{ab}+(2\lambda +\xi_1)\tilde h^{ab}+\xi_2(\epsilon^{aij}\partial_ih_j^b+\epsilon^{bij}\partial_ih_j^a)+2\xi_3\eta^{ab}h+(\xi_4-2\lambda )\eta^{ab}\tilde h\right]_{x^3=0}\!\!\!=0\ .\label{LG-LG-bchij}
	\ee
\end{itemize}
Taking $\lim_{\epsilon\to0}\int^\epsilon_0dx^3\eqref{LG-LG-eomht}$ and going on-shell ($\tilde J=0$), we get 
\be
\delta^\alpha_a\delta^\beta_b\left(\xi_1h^{ab}+\xi_4\eta^{ab}h\right)_{x^3=0}=0\ ,\label{LG-LG-bcht}
\ee
whose nonvanishing components are $\alpha=a,\ \beta=b$, which give
\be
\left( \xi_1h^{ab}+\xi_4\eta^{ab}h\right)_{x^3=0}=0\label{LG-LG-bchtij}\ .
\ee
Notice that from \eqref{LG-LG-bch3i} and taking $\partial_a$-derivative of \eqref{LG-LG-bchtij} we have the following constraint on the boundary parameters
	\be
	\xi_1=-\xi_4\ .
	\ee
To summarize, the most general \ac{BC} on the planar boundary $x^3=0$ are the following
\begin{empheq}{align}
\partial^bh-\partial_ah^{ab}&=0\label{LG-BCh}\\
\xi_1\left( h^{ab}-\eta^{ab}h\right)&=0\label{LG-bc2}\\
 2\xi_0h^{ab}+(2\lambda +\xi_1)&(\tilde h^{ab}-\eta^{ab}\tilde h)+\xi_2(\epsilon^{aij}\partial_ih_j^b+\epsilon^{bij}\partial_ih_j^a)+2\xi_3\eta^{ab}h=0\ .\label{LG-bc3}
\end{empheq}
The \ac{BC} \eqref{LG-BCh} is universal, in the sense that it does not depend on $S_{bd}$ \eqref{LG-LG-Sbd}. It represents the conservation of a current on the boundary
\be
\partial_a K^{ab}=0\ ,
\label{LG-conscurrK}\ee
with
\be
K^{ab}\equiv h^{ab}-\eta^{ab}h\ .
\label{LG-K}\ee
On the other hand, we remark that if $\xi_1=0$, the \ac{BC} are given by \eqref{LG-BCh} and \eqref{LG-bc3}. If instead $\xi_1\neq0$, \eqref{LG-bc2} implies \eqref{LG-BCh} and the \ac{BC} are given by \eqref{LG-bc2} and \eqref{LG-bc3}.

\subsection*{Ward identities}

From the \ac{EoM} for $h_{\mu\nu}(x)$ \eqref{LG-LG-eomh} we get
	\be\label{LG-LG-ward-ab}
	0=\int dx^3\partial_a\frac{\delta S_{tot}}{\delta h_{ab}}=2\lambda \left(\partial^b\tilde h-\partial_a\tilde h^{ab}\right)_{x^3=0}+\int dx^3\theta(x^3)\partial_aJ^{ab}\ ,
	\ee
where we used the \ac{BC} \eqref{LG-LG-BC}. We thus obtain the following Ward identity
	\be\label{LG-LG-ward1}
	\int dx^3\theta(x^3)\partial_aJ^{ab}=-2\lambda \left(\partial^b\tilde h-\partial_a\tilde h^{ab}\right)_{x^3=0}\ ,
	\ee
 which is broken on the boundary $x^3=0$. 
 In the same way, from the \ac{EoM} of $\tilde h_{\mu\nu}(x)$ \eqref{LG-LG-eomht}, we find
	\be
		0=\int dx^3\partial_a\frac {\delta S_{tot}}{\delta \partial_3h_{ab}}=\partial_a\tilde J^{ab}|_{x^3=0}-2\lambda \left(-\partial_ah^{ab}+\partial^bh\right)_{x^3=0}\ ,
	\ee
which represents a local Ward identity, broken by the boundary
	\be\label{LG-ward2}
	\partial_a\tilde J^{ab}|_{x^3=0}=2\lambda \left(-\partial_ah^{ab}+\partial^bh\right)_{x^3=0}\ .
	\ee
Notice that the r.h.s. describes the conservation on the boundary  of the current $K^{ab}(X)$ \eqref{LG-K}, previously found as the \ac{BC} \eqref{LG-BCh}, hence we may write
	\be\label{LG-ward2=0}
	\partial_a\tilde J^{ab}|_{x^3=0}=0\ .
	\ee
Going on-shell ($J=\tilde J=0$), the broken Ward identity \eqref{LG-LG-ward1} yields
	\be
	\left(\partial^b\tilde h-\partial_a\tilde h^{ab}\right)_{x^3=0}=0\label{LG-cc1}\ ,
	\ee
which, again, is a current conservation equation
\be
\partial_a \tilde K^{ab}=0\ ,
\label{LG-conscurrKt}\ee
with
\be
\tilde{K}^{ab}\equiv \tilde{h}^{ab}-\eta^{ab}\tilde h\ .
\label{LG-Kt}\ee
Hence, the presence of a planar boundary in \ac{LG} theory has as a consequence the presence of conserved currents, which consist of the particular combinations \eqref{LG-K} and \eqref{LG-Kt}. 

\subsection*{Kac-Moody Algebra}

By computing the functional derivative with respect to $J^{mn}(x')$ of the broken Ward identity \eqref{LG-LG-ward1}, $i.e.$ $\frac{\delta}{\delta J^{mn}(x')}\eqref{LG-LG-ward1}$ :
	\be
	\int_0^\infty dx^3\partial_a\left(\frac{\delta^a_m\delta^b_n+\delta^a_n\delta^b_m}{2}\delta^{(4)}(x-x')\right)=-2\lambda \left(\eta_{ac}\partial^b-\delta^b_c\partial_a\right)\frac{\delta Z_c[J,\tilde J]}{\delta\tilde J_{ac}\delta J'^{mn}}\ ,
	\ee
we get the commutation relations
	\be
	\frac{1}{2}\left(\delta^a_m\delta^b_n+\delta^a_n\delta^b_m\right)\partial_a\delta^{(3)}(X-X')
=-2i\lambda \left(\eta_{ac}\eta^{b0}-\delta^b_c\delta^0_a\right)\left[\tilde h^{ac}\ ,\ h'_{mn}\right]\delta(x^0-x'^0)\ ,
	\ee
where we used the on-shell constraint \eqref{LG-cc1}. By setting
\bi
\item $b=0$ we have
	\be
	\left(\delta^a_m\delta^0_n+\delta^a_n\delta^0_m\right)\partial_a\delta^{(3)}(X-X')=4i\lambda \left[\tilde h^\textsc{d}_\textsc{d}\ ,\ h'_{mn}\right]\delta(x^0-x'^0)\ ,
	\ee
from which, integrating over time,
	\bi
	\item  $m=n=0$ gives
		\be
		\left[\tilde h^\textsc{d}_\textsc{d}\ ,\ h'_{00}\right]_{x^0=x'^0}=0\ .
		\ee
	\item $m=0,\ n=\textsc n$ we get
		\be
		\left[\tilde h^\textsc{d}_\textsc{d}\ ,\ h'_{0\textsc n}\right]_{x^0=x'^0}=-\frac{i}{4\lambda }\partial_\textsc{n}\delta^{(2)}(X-X')\ .
		\ee
This can be identified as a \ac{KM} algebraic structure \cite{Kac:1967jr,Moody:1966gf} with central charge
	\be\label{LG-cc}
	c=-\frac{1}{4\lambda}\ ,
	\ee
which implies 
	\be
	\lambda<0\ ,
	\ee
because of the positivity of central charge of \ac{KM} algebras \cite{Mack:1988nf,Becchi:1988nh}.
	\item $m=\textsc m,\ n=\textsc n$ gives
		\be
		\left[\tilde h^\textsc{d}_\textsc{d}\ ,\ h'_\textsc{mn}\right]_{x^0=x'^0}=0\ .
		\ee
	\ei
	\item $b=\textsc b$
	\be
	\left(\delta^a_m\delta^\textsc{b}_n+\delta^a_n\delta^\textsc{b}_m\right)\partial_a\delta^{(3)}(X-X')=4i\lambda \left[\tilde h^{0\textsc b}\ ,\ h'_{mn}\right]\delta(x^0-x'^0)\ .
	\ee
	\bi
	\item $m=n=0$
		\be
		\left[\tilde h^{0\textsc b}\ ,\ h'_{00}\right]_{x^0=x'^0}=0\ .
		\ee
	\item $m=0,\ n=\textsc n$
		\be
		\left[\tilde h^{0\textsc b}\ ,\ h'_{0\textsc n}\right]_{x^0=x'^0}=0\ .
		\ee
	\item $m=\textsc m,\ n=\textsc n$
		\be
		\left[\tilde h^{0\textsc b}\ ,\ h'_\textsc{mn}\right]_{x^0=x'^0}=-\frac{i}{4\lambda }\left(\delta^\textsc{a}_\textsc{m}\delta^\textsc{b}_\textsc{n}+\delta^\textsc{a}_\textsc{n}\delta^\textsc{b}_\textsc{m}\right)\partial_\textsc{a}\delta^{(2)}(X-X')\ .
		\ee
Here again a \ac{KM} algebraic structure is observed with the same central charge  $c$ \eqref{LG-cc}.
	\ei
\ei
We now compute the functional derivative of the broken Ward identity \eqref{LG-LG-ward1} with respect to $\tilde J^{mn}(x')$, $i.e.\ \frac{\delta}{\delta \tilde J^{mn}(x')}\eqref{LG-LG-ward1}$ :
	\be
	0=-2i\lambda \left(\eta_{ac}\eta^{b0}-\delta^b_c\delta^0_a\right)\left[\tilde h^{ac}\ ,\ \tilde h'_{mn}\right]_{x^0=x'^0}\ ,
	\ee
where we used the on-shell constraint \eqref{LG-cc1} and integrated over time. In particular we have at
	\bi
	\item $b=0$
		\be
		\left[\tilde h^\textsc{d}_\textsc{d}\ ,\ \tilde h'_{mn}\right]_{x^0=x'^0}=0\ ;
		\ee
	\item $b=\textsc b$
		\be
		\left[\tilde h^{0\textsc b}\ ,\ \tilde h'_{mn}\right]_{x^0=x'^0}=0\ .
		\ee
	\ei
Summarizing, from the integrated Ward identity \eqref{LG-LG-ward1} we get the semidirect sum of \ac{KM} algebras with the same central charge 
	\begin{empheq}{align}
	\left[\tilde h^\textsc{d}_\textsc{d}\ ,\ h'_{0\textsc n}\right]&=-\frac{i}{4\lambda }\partial_\textsc{n}\delta^{(2)}(X-X')\label{LG-[Trht,h0n]}\\
	\left[\tilde h^{0\textsc b}\ ,\ h'_\textsc{mn}\right]&=-\frac{i}{4\lambda }\left(\delta^\textsc{a}_\textsc{m}\delta^\textsc{b}_\textsc{n}+\delta^\textsc{a}_\textsc{n}\delta^\textsc{b}_\textsc{m}\right)\partial_\textsc{a}\delta^{(2)}(X-X')\label{LG-[ht0b,hmn]}\ .
	\end{empheq}
The above \ac{KM} algebraic structure has a physical meaning when expressed in terms of the conserved currents $K^{ab}(X)$ \eqref{LG-K} and $\tilde K^{ab}(X)$ \eqref{LG-Kt}, which are expressed in terms of the tensor fields $h^{ab}(X)$, $\tilde h^{ab}(X)$ and their traces. In fact, as a consequence of \eqref{LG-[Trht,h0n]} and \eqref{LG-[ht0b,hmn]} we find that $K^{ab}(X)$ and $\tilde K^{ab}(X)$ form a \ac{KM} algebra with central charge \eqref{LG-cc} whose non vanishing components are
	\begin{empheq}{align}
	\left[\tilde K^{00}\ ,\ K'_{0\textsc m}\right]&=-\frac{i}{4\lambda }\partial_\textsc{m}\delta^{(2)}(X-X')\label{LG-KK1}\\
	\left[\tilde K^{0\textsc b}\ ,\ K'_{mn}\right]&=-\frac{i}{4\lambda }\left(\delta^\textsc{a}_m\delta^\textsc{b}_n+\delta^\textsc{a}_n\delta^\textsc{b}_m-2\eta^{\textsc a\textsc b}\eta_{mn}\right)\partial_\textsc{a}\delta^{(2)}(X-X')\label{LG-KK2}\ .
	\end{empheq}
The existence of a \ac{KM} algebraic structure for conserved currents on the boundary of 4D \ac{LG} confirms the guess made in \cite{Hinterbichler:2022agn} as a particularly interesting possibility in connection with Weinberg's soft graviton theorems \cite{Weinberg:1965nx,He:2014laa,Kapec:2015vwa}.
	
\subsection{The boundary}\label{LG boundary}

\subsection*{The degrees of freedom}

The presence of a 3D boundary in the 4D theory described by the action $S_{tot}$ \eqref{LG-Stot} induces a 3D theory, whose field content is determined by the solution of the on-shell broken Ward identity \eqref{LG-LG-ward1}
	\be
	\partial_a\left(\tilde h^{ab}-\eta^{ab}\tilde h\right)_{x^3=0}=0\label{LG-cc1'}
	\ee
and of the \ac{BC} \eqref{LG-BCh}
	\be
	\partial_a\left(h^{ab}-\eta^{ab}h\right)_{x^3=0}=0\ .\label{LG-cc2'}
	\ee
Let us consider first \eqref{LG-cc1'}. Define 
	\be
	\tilde C^{ab}\equiv\tilde h^{ab}-\eta^{ab}\tilde h\ ,
	\ee
whose trace is
	\be
	\tilde C=\eta_{ab}\tilde C^{ab}=-2\tilde h\ .
	\ee
Eq. \eqref{LG-cc1'} then reads
	\be\label{LG-divC=0}
	\partial_a\tilde C^{ab}=0\ .
	\ee
In order to find the most general solution, let us parametrize the symmetric tensor $\tilde C^{ab}(X)$ as follows
	\be\label{LG-solC1}
	\tilde C^{ab}=\frac{1}{2}\left(\epsilon^{amn}\partial_m\tilde \Sigma_n^{\ b}+\epsilon^{bmn}\partial_m\tilde \Sigma_n^{\ a}\right)\ .
	\ee
Because of \eqref{LG-divC=0} it must be
	\be
	\epsilon^{bmn}\partial_m\partial_a\tilde \Sigma_n^{\ a}=0\ ,
	\ee
which is solved by
\be
\tilde \Sigma_n^{\ a}=\epsilon^{acd}\partial_c\tilde\sigma_{nd}+\partial_n\phi^a\ ,
\ee
but we observe that the $\phi^a(X)$ contribution trivializes $\tilde C^{ab}(X)$ \eqref{LG-solC1}. Hence
	\be
	\tilde\Sigma_n^{\ a}=\epsilon^{acd}\partial_c\tilde\sigma_{nd}\ .
	\ee
In terms of this result, $\tilde C^{ab}(X)$ \eqref{LG-solC1} solves \eqref{LG-divC=0}, and reads
	\be\label{LG-solC}
	\tilde C^{ab}=\epsilon^{bmn}\epsilon^{acd}\partial_m\partial_c\tilde\sigma_{nd}\ ,
	\ee
with $\tilde\sigma_{ab}(X)=\tilde\sigma_{ba}(X)$ as a consequence of the fact that $\tilde C^{ab}(X)$ is symmetric $\tilde C^{ab}(X)=\tilde C^{ba}(X)$, and with $[\tilde\sigma]=0$. Thus the general solution for $\tilde h^{ab}(X)$ is
		\begin{align}
		\tilde h^{ab}&=\tilde C^{ab}-\frac{1}{2}\eta^{ab}\tilde C\label{LG-h solved}\\
		&=-\frac{1}{2}\eta^{ab}(\partial_m\partial^m\tilde\sigma_{n}^{ \;n}-\partial^m\partial^n\tilde\sigma_{mn})+\partial_m\partial^m\tilde\sigma^{ab}+\partial^a\partial^b\tilde\sigma_{n}^{ \;n}-\partial_c(\partial^b\tilde\sigma^{am}+\partial^a\tilde\sigma^{bc})\ .\nonumber
		\end{align}
The Eq.\eqref{LG-cc2'} for $h^{ab}(x)$ has the same structure as \eqref{LG-cc1'}, therefore the solution has the same form \eqref{LG-h solved}. We finally get	
	\begin{align}
		\tilde h^{ab}&=\epsilon^{bmn}\epsilon^{acd}\partial_m\partial_c\tilde\sigma_{nd}+\frac{1}{2}\eta^{ab}(\partial_m\partial^m\tilde\sigma_{n}^{\;n}-\partial^m\partial^n\tilde\sigma_{mn})\label{LG-solht}\\
		&=-\frac{1}{2}\eta^{ab}(\partial_m\partial^m\tilde\sigma_{n}^{\;n}-\partial^m\partial^n\tilde\sigma_{mn})+\partial_m\partial^m\tilde\sigma^{ab}+\partial^a\partial^b\tilde\sigma_{n}^{\;n}-\partial_c(\partial^b\tilde\sigma^{ac}+\partial^a\tilde\sigma^{bc})\nonumber\\
		h^{ab}&=\epsilon^{bmn}\epsilon^{acd}\partial_m\partial_c \sigma_{nd}+\frac{1}{2}\eta^{ab}(\partial_m\partial^m \sigma_{n}^{\;n}-\partial^m\partial^n \sigma_{mn})\label{LG-solh}\\
		&=-\frac{1}{2}\eta^{ab}(\partial_m\partial^m \sigma_{n}^{\;n}-\partial^m\partial^n \sigma_{mn})+\partial_m\partial^m \sigma^{ab}+\partial^a\partial^b \sigma_{n}^{\;n}-\partial_c(\partial^b \sigma^{ac}+\partial^a \sigma^{bc})\ ,\nonumber
	\end{align}
which means that the fields of the induced 3D theory are identified as the rank-2 symmetric tensors $\sigma^{ab}(X)$ and $\tilde\sigma^{ab}(X)$. Moreover, these solutions are invariant under the following transformations of the boundary fields $\sigma_{ab}(X),\ \tilde\sigma_{ab}(X)$
	\begin{empheq}{align}
	\tilde\delta\tilde h_{ab}=0\quad&\Leftrightarrow\quad\tilde\delta\tilde\sigma_{mn}=\partial_m\tilde\xi_n+\partial_n\tilde\xi_m\label{LG-tdiff}\\
	\delta h_{ab}=0\quad&\Leftrightarrow\quad\delta\sigma_{mn}=\partial_m\xi_n+\partial_n\xi_m\ ,\label{LG-diff}
	\end{empheq}
which remarkably means that the induced boundary theory must be invariant under infinitesimal diffeomorphisms, which therefore is a consequence of the general method we followed to introduce a boundary in LG, without need of requiring it explicitly.

\subsection*{Most general 3D action}

As a consequence of the solutions $\tilde h_{ab}(x)$ \eqref{LG-solht} and $h_{ab}(x)$ \eqref{LG-solh} and of their mass dimensions \eqref{LG-dimh}, the boundary fields $\sigma_{ab}(X)$ and $\tilde\sigma_{ab}(X)$ should have mass dimensions $[\sigma]=-1$ and $[\tilde\sigma]=0$. However, in 3D the canonical choices for the mass dimensions of the tensor fields are two :
	\begin{enumerate}
	\item $[\sigma]=[\tilde\sigma]=1$, which can be realized by rescaling as follows
		\be
		\tilde\sigma\to \tilde M^{-1}\tilde\sigma\quad;\quad\sigma\to M^{-2}\sigma\ .
		\ee
However in this case power-counting and locality constrain the action to the following Chern-Simons/BF-like action \cite{Birmingham:1991ty}, 
	\be\label{LG-Scs}
	S=\int d^3x\epsilon^{abc}\left(a_1\sigma_{ad}\partial_b\sigma_c^{\ d}+a_2\tilde\sigma_{ad}\partial_b\sigma_c^{\ d}+a_3\tilde\sigma_{ad}\partial_b\tilde\sigma_c^{\ d}\right)
	\ee
which is not invariant under the diffeomorphism transformations $\delta$ \eqref{LG-diff} and $\tilde\delta$ \eqref{LG-tdiff} :
	\begin{empheq}{align}
	\delta S&=\int d^3x\epsilon^{abc}\left(2a_1\sigma_{ad}+a_2\tilde\sigma_{ad}\right)\partial_b\partial^d\xi_c\\
	\tilde\delta S&=\int d^3x\epsilon^{abc}\left(a_2\sigma_{ad}+2a_3\tilde\sigma_{ad}\right)\partial_b\partial^d\tilde\xi_c\ ,
	\end{empheq}
which indeed vanish only at the trivial case ($a_1=a_2=a_3=0$). Thus we must discard this possibility.
	\item $[\sigma]=[\tilde\sigma]=\frac{1}{2}$, achieved by rescaling
	\be\label{LG-rescaling}
	\tilde\sigma\to \tilde M^{-\frac{1}{2}}\tilde\sigma\quad;\quad\sigma\to M^{-\frac{3}{2}}\sigma\ ,
	\ee
which, instead, leads to a nontrivial solution, as we shall see in what follows.
	\end{enumerate}
The most general action invariant under the infinitesimal diffeomorphisms $\tilde\delta$ \eqref{LG-tdiff} and $\delta$ \eqref{LG-diff} has the following structure
	\be\label{LG-Sinv}
	S_{3D}[\sigma,\tilde\sigma]=\kappa S_{LG}[\sigma]+\tilde\kappa\tilde S_{LG}[\tilde\sigma]+\kappa_{m}S_{mix}[\sigma,\tilde\sigma]\ ,
	\ee
where $\kappa,\ \tilde\kappa,\ \kappa_m$ are dimensionless constants, and $S_{LG}[\sigma]$ and $\tilde S_{LG}[\tilde\sigma]$ are \ac{LG} contributions analogous to \eqref{LG-LG-Sbulk}, written in terms of the boundary tensor field $\sigma_{ab}(X)$ and $\tilde\sigma_{ab}(X)$, respectively
	\begin{empheq}{align}
	S_{LG} &= \int d^3x \left(\frac{1}{4}f^a_{\ ac} f_b^{\ bc}-\frac{1}{6}f^{abc}f_{abc}\right)\\
	\tilde S_{LG}&= \int d^3x \left(\frac{1}{4}\tilde f^a_{\ ac}\tilde f_b^{\ bc}-\frac{1}{6}\tilde f^{abc}\tilde f_{abc}\right)\ ,
	\end{empheq}
with
	\begin{empheq}{align}
	f_{abc}&=f_{bac}=\partial_a \sigma_{bc}+\partial_b \sigma_{ac}-2\partial_c \sigma_{ab}	\label{LG-fabc}\\
	\tilde f_{abc}&=\tilde f_{bac}=\partial_a \tilde\sigma_{bc}+\partial_b \tilde\sigma_{ac}-2\partial_c \tilde\sigma_{ab}\ ,
	\label{LG-tfabc}
	\end{empheq}
satisfying the ciclicity property
	\begin{empheq}{align}\label{LG-cicl f}
	f^{abc}+f^{bca}+f^{cab}&=0\\
	 \tilde f^{abc}+\tilde f^{bca}+\tilde f^{cab}&=0\ .\label{LG-cicl tf}
	\end{empheq}
Notice that no Chern-Simons or BF contributions like in \eqref{LG-Scs} are allowed as a consequence of the diffeomorphism invariances $\delta S_{3D} =\tilde\delta S_{3D} =0$. The $S_{mix} $ term in \eqref{LG-Sinv} is the most general one depending on both $\sigma_{ab}(X)$ and $\tilde\sigma_{ab}(X)$, compatible with power-counting and the invariances $\delta S_{mix}=\tilde \delta S_{mix}=0$. Excluding again Chern-Simons/BF-like contributions, which are not invariant under diffeomorphisms, we have
	\begin{equation}
		S_{mix} =\int d^3x\left\{a_0\partial_a\sigma\partial^a\tilde\sigma+a_1\partial_c\sigma_{ab}\partial^c\tilde\sigma^{ab}+a_2\partial_a\sigma\partial_b\tilde\sigma^{ab}+a_3\partial_a\tilde\sigma\partial_b\sigma^{ab}+a_4\partial_c\sigma_{ab}\partial^a\tilde\sigma^{bc}\right\}\ .
	\end{equation}
Under the diff transformation $\delta$ \eqref{LG-diff} we get
	\begin{equation}
		\begin{split}
		\delta S_{mix} 
		&=-\int d^3x\left\{\tilde\sigma^{ab}\left[\left(2a_1+a_4\right)\partial_a\partial^2\xi_b+\left(2a_2+a_4\right)\partial_a\partial_b\partial_m\xi^m\right]+
2\tilde\sigma\left(a_0+a_3\right)\partial_m\partial^2\xi^m\right\}\ ,
		\end{split}
	\end{equation}
which means that the action is invariant, $i.e.\ \delta S_{mix}=0$, if
	\be
		\begin{split}
		&a_3=-a_0\\
		&a_4=-2a_1\\
		&a_2=a_1	\ .
		\end{split}
	\ee
The $\delta$-invariant $S_{mix} $ action term is
	\begin{equation}\label{LG-Smix}
		S_{mix} =\int d^3x\left\{a_0\partial_a\sigma\partial^a\tilde\sigma+a_1\partial_c\sigma_{ab}\partial^c\tilde\sigma^{ab}+a_1\partial_a\sigma\partial_b\tilde\sigma^{ab}-a_0\partial_a\tilde\sigma\partial_b\sigma^{ab}-2a_1\partial_c\sigma_{ab}\partial^a\tilde\sigma^{bc}\right\}\ .
	\end{equation}
Requiring now invariance under $\tilde\delta$ \eqref{LG-tdiff}, we get
\begin{equation}
		\begin{split}
		\tilde\delta S_{mix} =0
		&=2\int d^3x\left\{\sigma_{ab}\left(a_0+a_1\right)\partial^a\partial^b\partial^m\tilde\xi_m
-\sigma\left(a_0+a_1\right)\partial_m\partial^2\tilde\xi^m\right\}\ ,
		\end{split}
	\end{equation}
hence it must be
	\be
	a_1=-a_0\ .
	\ee
Therefore the mixed action term \eqref{LG-Smix} invariant under both $\tilde\delta$ \eqref{LG-tdiff} and $\delta$ \eqref{LG-diff} is
	\begin{equation}\label{LG-Smix'}
		S_{mix} =a_0\int d^3x\left\{\partial_a\sigma\partial^a\tilde\sigma-\partial_c\sigma_{ab}\partial^c\tilde\sigma^{ab}-\partial_a\sigma\partial_b\tilde\sigma^{ab}-\partial_a\tilde\sigma\partial_b\sigma^{ab}+2\partial_c\sigma_{ab}\partial^a\tilde\sigma^{bc}\right\}\ .
	\end{equation}
After reabsorbing the $a_0$ parameter into $\kappa_m$ in \eqref{LG-Sinv}, we observe that using the definitions of $f^{abc}(X)$ \eqref{LG-fabc} and $\tilde f^{abc}(X)$ \eqref{LG-tfabc} $S_{mix} $ \eqref{LG-Smix'} can be written as
	\be\label{LG-Smix''}
	S_{mix} =\int d^3x \left(\frac{1}{4} f^a_{\ ac}\tilde f_b^{\ bc}-\frac{1}{6} f^{abc}\tilde f_{abc}\right)\ .
	\ee
The most general invariant action therefore is
	\be\label{LG-Sinv'}
		\begin{split}
		S_{3D} &=\kappa S_{LG} +\tilde\kappa \tilde S_{LG}+\kappa_{m}S_{mix} \\
		&=\int d^3x\left\{\kappa\left(\tfrac{1}{4}f^a_{\ ac} f_b^{\ bc}-\tfrac{1}{6}f^{abc}f_{abc}\right)+\tilde\kappa\left(\tfrac{1}{4}\tilde f^a_{\ ac}\tilde f_b^{\ bc}-\tfrac{1}{6}\tilde f^{abc}\tilde f_{abc}\right)+\right.\\
		&\qquad\qquad\!\!\!\!+\left.\kappa_m\left(\tfrac{1}{4} f^a_{\ ac}\tilde f_b^{\ bc}-\tfrac{1}{6} f^{abc}\tilde f_{abc}\right)\right\}\ .
		\end{split}
	\ee
We finally observe that $S_{mix} $ \eqref{LG-Smix''} can be written as
	\be
	S_{mix} =\int d^3x\epsilon^{abc}\epsilon^{def}\sigma_{ad}\partial_b\partial_e\tilde\sigma_{cf}\ ,
	\ee 
hence, replacing $\tilde\sigma_{ab}(X)$ with $\sigma_{ab}(X)$ we have an alternative way to write the 3D \ac{LG} action
	\be
	\begin{split}
	S_{LG} 
	&=\int d^3x(\epsilon^{abc}\partial_b\sigma_{am})(\epsilon^{pnm}\partial_n\sigma_{pc})\ ,
	\end{split}
	\ee 
whose \ac{EoM} are
	\be
	\epsilon^{ap_1p_2}\epsilon^{bp_3p_4}\partial_{p_1}\partial_{p_3}h_{p_2p_4}=0\ ,
	\ee
which are those of \ac{LG} written in an alternative and more compact way. A similar expression holds for 4D \ac{LG}, whose \ac{EoM} can be written as
	\be
	\epsilon^{\mu\alpha_1\alpha_2\alpha_3}\epsilon^{\nu\alpha_4\alpha_5\alpha_6}\eta_{\alpha_3\alpha_6}\partial_{\alpha_1}\partial_{\alpha_4}h_{\alpha_2\alpha_5}=0\ .
	\ee

\subsection*{Equations of motion of the 3D induced theory}

From the ciclicity property of $f^{abc}(X)$ and $\tilde f^{abc}(X)$ \eqref{LG-cicl f}, \eqref{LG-cicl tf}, we find the following \ac{EoM} for the boundary fields $\sigma_{ab}(X)$ and $\tilde\sigma_{ab}(X)$
	\begin{empheq}{align}
		\frac{\delta S_{3D} }{\delta\sigma_{mn}}=&\kappa\left[-\partial_af^{mna}+\eta^{mn}\partial_af_b^{\ ba}-\tfrac{1}{2}\left(\partial^mf_b^{\ bn}+\partial^nf_b^{\ bm}\right)\right]+\label{LG-eomSig}\\
&+\frac{\kappa_m}{2}\left[-\partial_a \tilde f^{mna}+\eta^{mn}\partial_a\tilde f_b^{\ ba}-\tfrac{1}{2}\left(\partial^m\tilde f_b^{\ bn}+\partial^n\tilde f_b^{\ bm}\right)\right]=0\nonumber\\[5px]
		\frac{\delta S_{3D} }{\delta\tilde\sigma_{mn}}=&\tilde\kappa\left[-\partial_a\tilde f^{mna}+\eta^{mn}\partial_a\tilde f_b^{\ ba}-\tfrac{1}{2}\left(\partial^m\tilde f_b^{\ bn}+\partial^n\tilde f_b^{\ bm}\right)\right]+\label{LG-eomtSig}\\
&+\frac{\kappa_m}{2}\left[-\partial_a  f^{mna}+\eta^{mn}\partial_a f_b^{\ ba}-\tfrac{1}{2}\left(\partial^m f_b^{\ bn}+\partial^n f_b^{\ bm}\right)\right]=0\ .\nonumber
	\end{empheq}	

\subsection{Contact between bulk and boundary}\label{LG HC}

It is possible to make a holographic contact between the 4D bulk theory described by the action $S_{tot}$ \eqref{LG-Stot} and the induced 3D theory whose action is $S_{3D} $ \eqref{LG-Sinv'} by requiring that the \ac{EoM} \eqref{LG-eomSig} and \eqref{LG-eomtSig} derived from $S_{3D} $ coincide with the \ac{BC} \eqref{LG-BCh}, \eqref{LG-bc2} and \eqref{LG-bc3} we found for the 4D bulk theory. This can be achieved by suitably fine tuning the $\xi_i$ parameters appearing in $S_{bd}$ \eqref{LG-LG-Sbd}, and $\kappa,\ \tilde\kappa,\ \kappa_m$ in $S_{3D} $ \eqref{LG-Sinv'}.
The first step is to write the \ac{BC} \eqref{LG-BCh}, \eqref{LG-bc2} and \eqref{LG-bc3} in terms of the boundary fields $\sigma_{ab}(X)$ and $\tilde\sigma_{ab}(X)$ through the solutions \eqref{LG-solht} and \eqref{LG-solh}. The \ac{BC} \eqref{LG-BCh} is the defining equation for $h_{ab}(X)$ on the boundary \eqref{LG-solh}, thus  the contact is automatically satisfied. Concerning  \eqref{LG-bc2}, using \eqref{LG-solh} we have, on $x^3=0$
	\be
	0= h^{ab}-\eta^{ab}h=M^{-\frac{3}{2}}\left[\partial^2\sigma^{ab}+\partial^a\partial^b\sigma-\partial_c\left(\partial^a\sigma^{bc}+\partial^b\sigma^{ac}\right)+\eta^{ab}\left(\partial^c\partial^d\sigma_{cd}-\partial^2\sigma\right)\right]
	\ee
where $M$ is the rescaling factor of $\sigma_{ab}(X)$ introduced in \eqref{LG-rescaling}. This can also be written as
	\be
	H^{mn}\equiv\left( h^{mn}-\eta^{mn}h\right)|_\eqref{LG-solh}=\frac{M^{-\frac{3}{2}}}{2}\left[-\partial_af^{mna}+\eta^{mn}\partial_af_b^{\ ba}-\tfrac{1}{2}\left(\partial^mf_b^{\ bn}+\partial^nf_b^{\ bm}\right)\right]\ .\label{LG-h-f}
	\ee
Analogously
	\be
	\tilde H^{mn}\equiv\left(\tilde h^{mn}-\eta^{mn}\tilde h\right)|_\eqref{LG-solh}=\frac{\tilde M^{-\frac{1}{2}}}{2}\left[-\partial_a\tilde f^{mna}+\eta^{mn}\partial_a\tilde f_b^{\ ba}-\tfrac{1}{2}\left(\partial^m\tilde f_b^{\ bn}+\partial^n\tilde f_b^{\ bm}\right)\right]\ ,\label{LG-th-tf}
	\ee
where $\tilde M$ is the rescaling factor of $\tilde\sigma_{ab}(X)$ introduced in \eqref{LG-rescaling}. We introduced $H^{ab}$ and $\tilde H^{ab}$ so that the contact between the bulk \ac{BC} and the boundary \ac{EoM} will be more evident, as we shall see. Indeed the 3D \ac{EoM} \eqref{LG-eomSig} and \eqref{LG-eomtSig} can be written as linear combination of \eqref{LG-h-f} and \eqref{LG-th-tf}
	\be\label{LG-H+tH}
	\alpha H^{mn}+\beta \tilde H^{mn}=0\ .
	\ee
Explicitly we have
	\begin{empheq}{align}
	\eqref{LG-eomSig}\ =\ 2{\kappa}M^{\frac{3}{2}} H^{ab}+\kappa_m \tilde M^{\frac{1}{2}}\tilde H^{ab}&=0\label{LG-eomH1''}\\
	\eqref{LG-eomtSig}\ =\ \kappa_mM^{\frac{3}{2}}H^{ab}+2\tilde\kappa\tilde M^{\frac{1}{2}}\tilde H^{ab}&=0\label{LG-eomH2''}\ .
	\end{empheq}
The \ac{BC} \eqref{LG-bc2} can be written as
	\be\label{LG-bc2H}
	\xi_1H^{ab}=0\ ,
	\ee
while the \ac{BC} \eqref{LG-bc3} cannot be written as \eqref{LG-H+tH}
	\be\label{LG-BCHh}
	 2\xi_0H^{ab}+(2\lambda +\xi_1)\tilde H^{ab}+\xi_2(\epsilon^{aij}\partial_ih_j^b+\epsilon^{bij}\partial_ih_j^a)+2(\xi_3+\xi_0)\eta^{ab}h=0\ ,
	\ee
unless
	\be
	\xi_3=-\xi_0\quad;\quad \xi_2=0\ ,
	\ee
in which case the \ac{BC} \eqref{LG-bc3} becomes
	\be
	2\xi_0H^{ab}+(2\lambda +\xi_1)\tilde H^{ab}=0\ ,\label{LG-bc2''}
	\ee
recalling that $\lambda$ is the coefficient of the bulk action $S_{bulk}$ \eqref{LG-LG-Sbulk}. Now that both \ac{EoM} and \ac{BC} have a similar structure, we can $holographycally$ match them by tuning their parameters so that (\ac{EoM})$\leftrightarrow$(\ac{BC}). Keeping in mind that the $\xi_1$ parameter defines two situations
	\bi
	\item $\xi_1=0$ : one \ac{BC} \eqref{LG-bc2''}
	\item $\xi_1\neq0$ : two \ac{BC} \eqref{LG-bc2H} and \eqref{LG-bc2''},
	\ei
let us look at the first case.
	\begin{enumerate}
	\item $\pmb{\xi_1=0}$ : the only \ac{BC} is, after a multiplication by $\frac{1}{2\lambda}$ (remember that $\lambda\neq0$, being the coupling constant of the bulk)
	\be\label{LG-bc1H}
	\frac{\xi_0}{\lambda}H^{ab}+\tilde H^{ab}=0\ .
	\ee
In the same way we have seen that the \ac{EoM} \eqref{LG-eomSig} and \eqref{LG-eomtSig} can be written as \eqref{LG-eomH1''}, \eqref{LG-eomH2''}
	\begin{empheq}{align}
	2\mu\frac{\kappa}{\kappa_m} H^{ab}+\tilde H^{ab}&=0\label{LG-eomH1}\\
	\frac{\mu}{2}\frac{\kappa_m}{\tilde\kappa}H^{ab}+\tilde H^{ab}&=0\label{LG-eomH2}\ ,
	\end{empheq}
where $\mu\equiv\sqrt{\tfrac{M^3}{\tilde M}}$ with $[\mu]=1$. They both match with the \ac{BC} \eqref{LG-bc1H} if
	\be\label{LG-hc xi1=0}
	\frac{\xi_0}{\lambda}=2\mu\frac{\kappa}{\kappa_m}=\frac{\mu}{2}\frac{\kappa_m}{\tilde\kappa}\quad\Rightarrow\quad\kappa_m^2=4\kappa\tilde\kappa\ ,\ \kappa\tilde\kappa>0\ .
	\ee
The implication on the 3D action \eqref{LG-Sinv'} is that the following redefinition of the fields is possible 
	\be\label{LG-rho}
	\rho_{ab}\equiv\sqrt\kappa\sigma_{ab}\pm\sqrt{\tilde\kappa}\tilde\sigma_{ab}\quad;\quad\Phi_{abc}\equiv\sqrt\kappa f_{abc}\pm\sqrt{\tilde\kappa}\tilde f_{abc}\ ,
	\ee
such that the action only depends on one field as 
	\be
		\begin{split}
		S_{3D}&=\frac{1}{6}\int d^3x\left(\frac{1}{\sqrt2}\eta^{ab}\Phi_m^{\ mc}-\Phi^{abc}\right)\left(\frac{1}{\sqrt2}\eta_{ab}\Phi^n_{\ nc}+\Phi_{abc}\right)\\
		&=\int d^3x\left(\frac{1}{4}\Phi_m^{\ mc}\Phi^n_{\ nc}-\frac{1}{6}\Phi^{abc}\Phi_{abc}\right)\\
		&=S_{LG}[\rho]\ ,\label{LG-S(rho)}
		\end{split}
	\ee
which is \ac{LG} in 3D. The sign $\pm$ in \eqref{LG-rho} depends on the sign of $\kappa_m$ as a consequence of the contact \eqref{LG-hc xi1=0} for which $\kappa_m=\pm2\sqrt{\kappa\tilde\kappa}$. Notice that if $\xi_1=0$, $S_{bd}$ \eqref{LG-LG-Sbd} becomes
	\be\label{LG-Sbdhc1}
	S_{bd}=\xi_0\int d^4x\delta(x^3)\left( h_{ab}h^{ab}-h^2\right)\ ,
	\ee
$i.e.$ the boundary action $S_{bd}$ \eqref{LG-LG-Sbd} does not depend on the $\partial_3$-derivative of the gauge field anymore. We recognize in $S_{bd}$ the Fierz-Pauli mass term \eqref{LG-SFP} \cite{Hinterbichler:2011tt,Blasi:2017pkk,Blasi:2015lrg,Gambuti:2020onb,Gambuti:2021meo}, 
which renders the relation with \ac{LG} even more remarkable. This allows to interpret $\xi_0$ as a Fierz-Pauli mass for the tensor field $h_{ab}(X)$ on the boundary $x^3=0$.
	\item $\pmb{\xi_1\neq 0}$ : the \ac{BC} are the following
	\begin{empheq}{align}
	H^{ab}&=0\label{LG-bcH}\\
	2\xi_0 \xcancel{H^{ab}}+(2\lambda+\xi_1)\tilde H^{ab}&=0\ .\label{LG-bcH+tH}
	\end{empheq}
We have to distinguish between two cases: $2\lambda+\xi_1=0$ and $2\lambda+\xi_1\neq0$. For $\xi_1=-2\lambda$ we are left with the \ac{BC} \eqref{LG-bcH} only, which depends on $h_{ab}(X)$, hence on $\sigma_{ab}(X)$ through the solution \eqref{LG-solh}. To have a contact, we have to switch off the $\tilde\sigma_{ab}(X)$ dependence in the 3D action \eqref{LG-Sinv'} (and in the \ac{EoM} \eqref{LG-eomH1''}, \eqref{LG-eomH2''}) by putting $\kappa_m=\tilde\kappa=0$. The induced theory in this case is \ac{LG} for the field $\sigma_{ab}(X)$
	\be
	S_{3D}=\kappa S_{LG}=\kappa\int d^3x\left(\tfrac{1}{4}f^a_{\ ac} f_b^{\ bc}-\tfrac{1}{6}f^{abc}f_{abc}\right)\ ,
	\ee
where we can reabsorb the $\kappa$ parameter through a redefinition of the field $\sigma_{ab}(X)$. The boundary action term $S_{bd}$ \eqref{LG-LG-Sbd} is 
	\be
	S_{bd}=\int d^4x\delta(x^3)\left[\xi_0 (h_{ab}h^{ab}-h^2)-2\lambda(\tilde h_{ab}h^{ab}-\tilde hh)\right]\ .
	\ee
Notice that also in this case the $\xi_0$ parameter plays the role of a Fierz-Pauli mass for $h_{ab}(X)$ on the boundary.
If instead $\xi_1\neq\{-2\lambda,0\}$ 	\begin{empheq}{align}
	H^{ab}&=0\label{LG-BCH}\\
	\tilde H^{ab}&=0\ .\label{LG-BCtH}
	\end{empheq}
Looking at the  3D boundary-side (\ac{EoM}) we can use the \ac{EoM} \eqref{LG-eomH1}
	\be
	\tilde H^{ab}=-2\mu\frac{\kappa}{\kappa_m}H^{ab}\label{LG-eomH1'}
	\ee
in the \ac{EoM} \eqref{LG-eomH2}, which becomes
	\be
	\frac{\mu}{2}\left(\frac{\kappa_m^2-4\kappa\tilde\kappa}{\tilde\kappa\kappa_m}\right)H^{ab}=0\quad;\quad\tilde\kappa,\ \kappa_m\neq0\ .\label{LG-eomH2'}
	\ee
Now we notice that if $\kappa^2_m-4\kappa\tilde\kappa=0$ the \ac{EoM} \eqref{LG-eomH2'} becomes trivial, and we only have one \ac{EoM}, which is \eqref{LG-eomH1'}, which can never match the two \ac{BC} \eqref{LG-bcH} and \eqref{LG-bcH+tH} at the same time. Indeed this case ($\kappa^2_m-4\kappa\tilde\kappa=0$) allows a contact only if we look at the \ac{BC} in the form \eqref{LG-bc2H} and \eqref{LG-bc2''} and set $\xi_1=0$, which coincide with Case 1 \eqref{LG-hc xi1=0}. Therefore $\kappa^2_m-4\kappa\tilde\kappa=0\ \Leftrightarrow\ \xi_1=0$. Considering $\kappa^2_m-4\kappa\tilde\kappa\neq0$ we can use the second \ac{EoM} \eqref{LG-eomH2'} back into the first one \eqref{LG-eomH1'} and get
	\begin{empheq}{align}
	H^{ab}&=0\\
	\tilde H^{ab}&=0\quad;\quad\kappa^2_m-4\kappa\tilde\kappa\neq0,\ \tilde\kappa,\ \kappa_m\neq0\ ,
	\end{empheq}
which matches exactly the \ac{BC} \eqref{LG-BCH} and \eqref{LG-BCtH}. Thus the holographic contact is possible for $\xi_1\neq\{-2\lambda,0\}$, $\kappa_m\neq\{0,2\sqrt{\kappa\tilde\kappa}\}$ and $\tilde\kappa\neq0$. Again $\xi_0$ does not affect the contact and can be interpreted as a Fierz-Pauli mass.
	\end{enumerate}
We summarize our results in the following Table \ref{LG-summary}
\begin{table}[H]
	\resizebox{1\columnwidth}{!}{
	\begin{tabular}{|c|c|c|c|}
	\hline
	$\pmb{S_{bd}}$ \textbf{parameters}&\textbf{Constraints}&$\pmb{S_{bd}=}$&$\pmb{S_{3D}=}$\\\hline
	$\xi_1=0,\ \xi_0\mbox{ free}$&$\kappa=\frac{\xi_0}{2\mu\lambda}\kappa_m\ ;\ \tilde\kappa=\frac{\mu\lambda}{2\xi_0}\kappa_m\ ;\ \kappa_m^2=4\kappa\tilde\kappa$ &$S_{bd}[h]$&$S_{LG}[\rho]$\\\hline
	$\xi_1=-2\lambda,\ \xi_0\mbox{ free}$&$\kappa\ \mbox{free}\ ;\ \tilde\kappa=0\ ;\ \kappa_m=0$&$S_{bd}[h,\tilde h]$&$S_{LG}[\sigma]$\\\hline
	$\xi_1\neq\{-2\lambda,0\},\ \xi_0\mbox{ free}$&$\kappa\ \mbox{free}\ ;\ \tilde\kappa\neq0\ ;\ \kappa_m\neq\{0,2\sqrt{\kappa\tilde\kappa}\}$&$S_{bd}[h,\tilde h]$&$\kappa S_{LG}+\tilde\kappa\tilde S_{LG}+\kappa_mS_{mix}$\\\hline
	\end{tabular}
}
\caption[Holographic contacts in \acs{LG} with boundary]{\footnotesize{Scheme of the contacts between bulk (\acs{BC}) and boundary (\acs{EoM}) with constraints on the parameters of $S_{bd}$ \eqref{LG-LG-Sbd} and of $S_{3D}$ \eqref{LG-Sinv'}.}}
\label{LG-summary}
\end{table}
\noindent
As we see from Table \ref{LG-summary}, depending on the value of the $\xi_1$ parameter of $S_{bd}$ \eqref{LG-LG-Sbd}, we found two possibilities for the 3D theory induced by the presence of a planar boundary on the 4D \ac{LG} theory
	\bi
	\item $\pmb{\xi_1=\{0,-2\lambda\}}$ the 3D induced theory is \ac{LG} for one symmetric rank-2 tensor field
		\be
		S_{3D}=S_{LG}\ .
		\label{LG-S3D1}\ee
The corresponding 4D $S_{bd}$ \eqref{LG-Sbdhc1} is a Fierz-Pauli mass term for the 4D tensor field $h_{ab}(X)$ whose mass parameter is $\xi_0$.
	\item $\pmb{\xi_1\neq\{0,-2\lambda\}}$ the 3D induced action depends on two rank-2 symmetric tensor fields $\sigma_{ab}(X)$ and $\tilde\sigma_{ab}(X)$. After a field redefinition, it reads
		\be
		S_{3D}[\sigma,\tilde\sigma]=S_{LG}[\sigma]+\tilde S_{LG}[\tilde\sigma]+kS_{mix}[\sigma,\tilde\sigma]\ ,
		\label{LG-S3D2}\ee
where $k$ is a constant which cannot be reabsorbed.
	\ei
In both cases, the 4D boundary term $S_{bd}$ \eqref{LG-LG-Sbd} contains a Fierz-Pauli mass term for the bulk tensor field $h_{ab}(X)$, whose mass parameter is $\xi_0$.

\subsection{Summary of results}\label{LG ends}

In this Section we studied the effect of the presence of a planar boundary on 4D \ac{LG}, realized by means of a Heaviside step function in the action \eqref{LG-LG-Sbulk}. Following the \ac{QFT} approach presented in Chapter \ref{QFTapproach} we observed that the presence of the boundary breaks the invariance under diffeomorphisms, which are the symmetry transformations of \ac{LG}. Correspondingly, the Ward identity which describes the invariance under diffeomorphisms \eqref{LG-LG-ward1} acquires a breaking, which is crucial, because from it the main information of the theory might be derived, namely the fields content, the symmetry transformations and the boundary algebra. We wrote ``might'' because it is not obvious that this can always be done. In fact, we stressed throughout this whole Thesis that this seemed to work for all \acp{TQFT}, where non trivial boundary dynamics has been first observed, and for a long time this property has been believed to be peculiar of these kind of theories. But we have seen that more recently, similar results have been found in non-topological field theories, like Maxwell theory \cite{Bertolini:2020hgr}, and this motivated the boundary investigations of this Thesis for more general theories, like we did in this Section for \ac{LG}. A first remarkable result is that on the boundary we found two conserved currents \eqref{LG-K} and \eqref{LG-Kt} which form the algebraic structure \eqref{LG-KK1} and \eqref{LG-KK2} of the \ac{KM} type, whose central charge is proportional to the inverse of the \ac{LG} ``coupling'' constant \eqref{LG-cc}. This confirms what has been guessed in \cite{Hinterbichler:2022agn}, where it was suspected the existence, in 4D \ac{LG}, of a \ac{KM} algebra as a particularly interesting possibility in connection with Weinberg's soft graviton theorems \cite{Weinberg:1965nx,He:2014laa,Kapec:2015vwa}.
Since the central charge of a \ac{KM} algebra must be positive, this is mostly useful to determine the sign of the overall \ac{LG} action, which otherwise should be determined by imposing that the energy density, that is the 00-component of the energy-momentum tensor, is positive, which in gravity is a known tricky issue \cite{Carroll:2004st,Misner:1973prb}. Moreover, we were able to solve the on-shell Ward identity \eqref{LG-cc1'} and the universal \ac{BC} \eqref{LG-cc2'} getting \eqref{LG-solht} and \eqref{LG-solh}, which allowed us to express, on the boundary, the 4D bulk fields $h^{ab}(X)$ and $\tilde h^{ab}(X)$ in terms of 3D fields which are the degrees of freedom of the induced 3D theory. We found that these latter, like their 4D ancestors, are rank-2 symmetric tensor fields~: $\sigma^{ab}(X)$ and $\tilde \sigma^{ab}(X)$. This, as \ac{LG} shows, seems to be peculiar of non topological \acp{QFT}. Indeed what is usually found in \acp{TQFT} is that the fields living on the $d-1$-dimensional boundary are tensors of lower rank with respect their $d$-dimensional counterpart~: from rank-2 tensors one finds vectors in the topological 4D BF theory \cite{Amoretti:2014iza} and the boundary reduction of the gauge field in Chern-Simons theory gives scalars, as we have seen in Chapters \ref{ch CSandBF} and \ref{ch CSandBFinCS}. Here, instead, the 3D boundary fields are rank-2 symmetric tensor fields as those of 4D \ac{LG}. And, quite interestingly, the transformation which keeps invariant the definition of the boundary fields turns out to be the diffeomorphisms \eqref{LG-tdiff} and \eqref{LG-diff}, which therefore are a consequence of the introduction of the boundary, rather than an \textit{a priori} request. Given the dynamical fields and the symmetry transformations, requiring locality and power counting allowed us to find the most general 3D action $S_{3D}$ \eqref{LG-Sinv'}, which consists of three terms. Each term being invariant by its own, $S_{3D}$ depends on three constants which we do not reduced by redefining the 3D fields as we could, but we fixed them by establishing a ``holographic'' contact as our last step. This has been realized by requiring that the \ac{EoM} of the 3D action $S_{3D}$ coincide with the \ac{BC} of the 4D theory. To do that, we had at our disposal the 4 parameters on which $S_{bd}$ \eqref{LG-LG-Sbd} depends and the three constants in $S_{3D}$ \eqref{LG-Sinv'}. As an outcome of this tuning, we found two possibilities, depending on the value of one particular parameter appearing in $S_{bd}$~: $S_{3D}$ describes either \ac{LG} for one single tensor field \eqref{LG-S3D1}, or the action \eqref{LG-S3D2}, containing two decoupled \ac{LG} terms for the boundary tensor fields $\sigma^{ab}(X)$ and $\tilde \sigma^{ab}(X)$ and one term which mixes them. As a last, but probably not least, fact, we remark that in any case the $S_{bd}$ action term which governs the holographic contact contains a mass term \eqref{LG-Sbdhc1} for the bulk tensor field $h_{ab}(x)$ of the particular Fierz-Pauli type \cite{Hinterbichler:2011tt,Blasi:2017pkk,Blasi:2015lrg,Gambuti:2020onb,Gambuti:2021meo}, with a free parameter $\xi_0$ which we can interpret as a mass.

\cleardoublepage 

\ctparttext{}
\part{Final Remarks}\label{partIV}

\chapter{Concluding Remarks} 

\label{ch end} 

We started with the idea of an analysis of boundary effects on theories both standard and not, within a formal \ac{QFT} approach. We ended up doing this and more. Indeed the formal requirements of \acp{QFT}, together with phenomenological inputs, also pushed us into the development of a new, well defined, \ac{QFT} for fracton quasiparticles \cite{Bertolini:2022ijb}. This was surely necessary in order to study boundary effects on fracton models in a precise and \ac{QFT}-oriented way, but the fact of having a proper \ac{QFT} for such models means even more, since one is now able to delve into the secrets of fractons without boundary as well, and without the necessity of invoking ``intuitions'' as it is explicitly admitted in the seminal papers \cite{Pretko:2016lgv,Pretko:2016kxt}, but just by applying straightforwardly \ac{QFT} first principles. All that would not have happened if it were not for phenomenological inputs, which made us look with interest into the theory of fractons, together with the desire to only trust first principles and that ``high flexibility and perfection of the \ac{QFT} formalism'' that motivated Symanzik. With this spirit we were thus able to build a new theory, but also to expand the dictionary of \acp{QFT} with boundary~: we observed the effect of a metric in \acp{TQFT} to have a non-trivial effect on the edge states, but we also showed that the bulk-boundary paradigm is much wider than expected, being extensible also to non-\acp{TQFT}. We have remarked many times that dealing with boundaries may have surprising outcomes, and this just by thinking about the standard, textbook example of \acp{TQFT} with boundary in flat spacetime, from which one starts with an unphysical theory and end up talking about experimental observation on the edge of Hall systems. But, as we shall see in more details in what follows, the results presented in this Thesis concerns an even wider variety of areas of physics : condensed matter (\acl{FQHE}, \acl{TI}, higher order \acl{TI}, \acl{QAH} Effect, fracton phases...), elasticity, gravity (massive, self dual, torsional effects...) all these are words that have appeared when discussing about the results we obtained on the edge of either \acp{TQFT}, fractons or \acl{LG} with boundary.

\subsection*{Final overview on the specific results}
The results achieved in this Thesis could be quickly summarized by saying that on the boundary of \acp{TQFT} on a generic manifold in curved spacetime Hall systems with accelerated edge modes are found, a 3D traceless fracton model can be recovered as induced by a traceful scalar charge theory of fractons in the bulk, and finally  a 3D \ac{LG}-like theory exists on the edge of 4D \ac{LG} itself. But there is much more to say about the results obtained from the works collected in this Thesis, both dealing with boundary effects and not. In particular
\paragraph{TQFTs in curved spacetimes : }even though the metric does not affect the bulk of the \ac{TQFT}, by definition of a topological theory, a memory of it is kept by the boundary, from which we observed that the metric does intervene in the local observables of the lower-dimensional theory, namely in the chiral velocities. The dictionary has thus been updated as follows
	\begin{align}
	&\mbox{\textbf{Flat} \ac{TQFT} with boundary}&&\Rightarrow\qquad\mbox{\textbf{constant} chiral edge velocities}\nonumber\\
	&\mbox{\textbf{Curved} \ac{TQFT} with boundary}&&\Rightarrow\qquad\mbox{\textbf{local} chiral edge velocities}\ .\nonumber
	\end{align}
This allows for a formal theoretical explanation of the observed accelerated edge modes of some condensed matter systems, which could not be justified in terms of the standard flat bulk-boundary approach, but only through an \textit{ad hoc} introduction coming from phenomenological requirements. Moreover, with our results we can make a contact between these phenomenological requirements (confining potential...) and the \ac{QFT} procedure : the induced metric $\gamma(x)$ play the role of the phenomenological potential $V(x)$ which makes the edge velocities acquire spacetime dependence $i.e.\ v=v(x)$. This same thing happen, also in the flat case, when thinking about the Chern-Simons coupling constant $\kappa$, which represents the filling factor $\nu$ of the phenomenological model on the boundary.  These results can be found in \cite{Bertolini:2021iku} and \cite{Bertolini:2022sao}.

\paragraph{New QFT for fractons : }built from a \ac{QFT} symmetry-based approach, this new \textit{covariant} theory for fractons allowed to recover all the results of the Literature from first principles of \ac{QFT}, and more. Indeed we observed
	\bi
	\item an immediate relation with \ac{LG}. This claim can be found in the fracton Literature \cite{Xu:2006,Gu:2009jh,Xu:2010eg,Pretko:2017fbf}, but it is immediately apparent from the field theoretical point of view. This relation appears as an additional term in the action with its own ``coupling''. The fact that a quadratic, Lorentz and gauge invariant theory depends on one unavoidable constant is quite uncommon, if not unique. It is not even clear how to call this constant, since it  cannot be a ``coupling'' constant, being the theory $S_{inv}$ \eqref{FMAX-Sinvg1g2} free and non interacting, nor a mass, being dimensionless. This peculiarity originates from the fact that the space of functionals invariant under the covariant fractonic transformation \eqref{FMAX-dA} has dimension two, instead of one as it commonly happens. To our knowledge, the only exception is given by the 3D Maxwell-Chern-Simons theory \cite{Deser:1981wh}, which depends on one ``true'' constant as well, but in that case the constant can be identified as a mass. We might say that gravitons may exist alone, while fractons necessarily come with gravitons, lacking, up to now, a symmetry which uniquely determines them.
	
	\item the presence of an energy momentum tensor which has strong analogies with the standard electromagnetic one. The possibility of computing it as a variation of the metric was allowed only by the fact that we had a covariant theory, and confirmed, through the computation of the ``Lorentz'' force, that the theory concerned fully mobile dipoles, as one would expect from a fracton model.  We however observe that such energy-momentum tensor is not conserved, which, again, could be a confirmation of the theory we are dealing with : standard non-covariant fracton models display broken translational symmetry, as a consequence of dipole conservation. In our covariant theory this can be encoded in the non-conservation of the energy-momentum tensor (its spatial components in particular). From a formal point of view one can also notice that this is in agreement with Weinberg-Witten theorem \cite{Weinberg:1980kq} when keeping in mind that here we have massless spin two objects
\begin{adjustwidth}{15px}{15px}
{\small 
Theorem 2 : A theory that allows the construction of a conserved Lorentz covariant energy-momentum tensor $T^{\mu\nu}$ for which $\int d^3x T^{\mu\nu}$ is the energy-momentum four-vector cannot contain massless particles of spin $j>1$.
}\end{adjustwidth}
	Additionally, another interesting observation on that regard is that the stress-energy tensor is not traceless, hence the theory is not scale invariant, differently from the classical Maxwell theory. Instead, tracelessness is recovered in 6D. Had we dealt with a field theory exercise, it would have been better to face the problem in 6D, where the gauge field has mass dimensions $[A_{\mu\nu}]=2$ and, consequently, the gauge parameter would have been given vanishing dimensions, as usual in gauge field theory. But the most general invariant action \eqref{FMAX-Sinvg1g2} consists of two terms~: $S_{LG}$ \eqref{FMAX-SLG} and $S_{fract}$ \eqref{FMAX-Sfract} which are respectively the actions for linearized gravity and for fractons, both physically relevant in 4D.
	\item every physical quantity is invariant at sight. The definition of the invariant rank-3 field strength $F_{\mu\nu\rho}(x)$ in terms of which the electric and magnetic tensor fields $E^{ij}(x)$ \eqref{FMAX-Eij} and $B^{ij}(x)$ \eqref{FMAX-Bij} of fractons can be written, makes them immediately invariant quantities under the fracton symmetry \eqref{FMAX-dA}, allowing for them to be identifiable as the physical objects of the theory.
	
	\item a nontrivial boundary physics. As a consequence of the peculiarity of the fractonic symmetry, a ``generalized'' \acl{KM} algebra is recovered on the boundary, which matches exactly the one observed in the context of higher order \acl{TI}, thus adding a checkmark on the always suspected relation between these materials and edge states of fracton models \cite{You:2019bvu,Pretko:2020cko}. The boundary theory display a curious phenomenon for which the edge \ac{DoF} turn out to be traceless, and the physical model appears to be related to the traceless scalar charge theory of fractons where a higher rank Chern-Simons contribution plays the role of matter, in particular as the charge in the Gauss constraint and of current in the Amp\`ere law. This means that the particle content of the boundary theory involves fractons as the immobile charges, and dipoles as one-dimensional particles (lineons). A broken Faraday equation also suggest the possible presence of fractonic vortices.	
	\ei
These results are collected in \cite{Bertolini:2022ijb}, \cite{Bertolini:2023juh} and \cite{Bertolini:2023sqa}.
\paragraph{Linearized Gravity : }we observe the emergence of a standard \ac{KM} algebra on the boundary which thus makes us able to give a confirmation on speculations about  its existence on the boundary of \ac{LG}. This example highlights the drastic consequences of considering a different  symmetry in the context of a theory with boundary. Indeed, in comparison with the results on the boundary of the theory of fractons and \ac{LG} of Section \ref{sec Frac+bd}, defined by invariance under longitudinal diffomorphisms (a.k.a. fracton symmetry),  the more general symmetry of diffeomorphisms, defining \ac{LG} alone, makes evident how results on the boundary are strictly related to the symmetry involved. This fact can be better expressed by the following Table, which compares the results in the two cases
\begin{table}[H]
\centering
  \begin{tabular}{ |r|c|c| }
\hline
   & \textbf{ Fractons and \ac{LG}} ($g_{1,2}\neq0$) & \textbf{ \ac{LG}} ($g_{1}=0$) \\\hline
{Defining symmetry} &Longitudinal diffs $\partial_\mu\partial_\nu\Lambda$& diffs { $\partial_\mu\Lambda_\nu+\partial_\nu\Lambda_\mu$ }\\
{Boundary algebra}& Generalized \ac{KM}&Standard \ac{KM}\\
{Boundary \ac{DoF}} & { $\alpha_{mn}$ symmetric, traceless }& $\sigma_{mn}$ symmetric\\
{ 3D  theory}  &Maxwell-Chern-Simons-like&3D \ac{LG}\\
{ Induced  physics}  &Traceless fracton phases&3D Linearized gravity-like\\
\hline
\end{tabular}
\caption[Fractons and \acs{LG} with boundary \textit{vs} \acs{LG} with boundary]{\footnotesize comparison between the theory of fractons and \acs{LG} with boundary and of \ac{LG} alone with boundary.}
\end{table}
\noindent
It is interesting to notice that it seems that the pure \ac{LG} case is the first situation where a theory with boundary does not involve results in the context of condensed matter. At least at first sight. These results were obtained in \cite{Bertolini:2023wie}.

\subsection*{Concerning the Future}
We conclude the Thesis with some remarks on future prospects which we think will be possible thanks to the results achieved here. From Chapter \ref{ch CSandBFinCS}, $i.e.$ \acp{TQFT} with boundary in curved spacetime, the first thing to mention is that as a consequence of the necessity of an update of the dictionary of \acp{TQFT} with boundary for explaining the observed local velocities of edge states of \ac{FQHE}, we are now also able to predict the existence of local velocities of edge modes in other Hall systems, like, in particular, \ac{TI}. Another interesting aspect to analyze from this new dictionary, is the relation between the induced metric on the boundary of \acp{TQFT}, and the confining/interaction potential responsible in the phenomenological models for such local velocities. For instance one could think of a perturbation expansion of the induced metric $\gamma_{ij}(x)=\eta_{ij}+h_{ij}(x)$, as one does in \ac{LG}, isolating the spacetime dependent part (which would be related to the metric perturbation $h_{ij}(x)$) from the standard, constant, contribution, and see how this non-constant, metric perturbation can encode the \textit{ad hoc} potential typically introduced in the models. That would complete the ``duality'' of the dictionary~: we knew the relation between the coupling and the filling factor $\kappa\leftrightarrow\nu$, and now we also add the relation between the metric and the potential $h_{ij}\leftrightarrow V$. Concerning Part \ref{partIII}, the fact of having found a non-trivial physics on the boundary of non-\acp{TQFT}, on its own, is already a striking result which should enhance the interest in investigating other non-\acp{TQFT} with boundary. Even more because of the richness in the physical content of those analyzed in the present work. Indeed form the study of both fracton and \ac{LG} theories with boundary, many questions and insights emerge. For instance on the boundary of the theory of fractons described by the action \eqref{FBD-Sbulk} a 3D traceless scalar charge theory arise, which is a new 3D fracton model. The tracelessness of the theory is an unexpected consequence of the \ac{DoF} induced by the boundary, thus a consequence of the fracton symmetry as a breaking from a Ward identity. We see in fact that a different, more general, symmetry such as the one defining  \ac{LG} ($i.e.$ diffs) gives traceful \ac{DoF}, meaning that maybe these disappearance of trace dependence is encoded in something hidden, which still has to be looked for. Concerning this 3D theory, it is known \cite{Pretko:2017kvd,} that fractons in 3D possess a duality with the theory of elasticity, as an analog higher-rank version of the particle-vortex duality \cite{Dunne:1998qy,Tong:2016kpv}, therefore the possibility that the 3D action recovered on the boundary shares a form of such higher spin duality is highly nontrivial and worth to be investigated. It will also be interesting to further analyze the relations which both the pure theory of \ac{LG} \eqref{LG-LG-Sbulk} and the fracton one \eqref{FBD-Sbulk} seem to have with a variety of models of gravity and massive gravity, on the boundary. Indeed we have seen that the Symanzik boundary term \eqref{LG-LG-Sbd}, necessary in \ac{LG}, needs to contain a Fierz-Pauli mass term \eqref{LG-SFP} in order to induce a lower-dimensional physics. Moreover, the Chern-Simons-like contribution contained in the 3D traceless theory \eqref{FBD-S3D} induced by fractons, strongly reminds a gravity model known as ``self-dual massive gravity'', and the full action appears as a higher rank Maxwell-Chern-Simons term. However it is interesting to notice that the coupling  seems to be dimensionless, which means that to better understand the theory and relations with the above-mentioned models the gauge structure and propagators of the theory will have to be computed. But in this Thesis results does not only involve boundaries. Indeed also the covariant \ac{QFT} of fractons is a novelty, and one of great interest. In fact, having now a new and well defined theory for these quasiparticles opens the doors at many other \ac{QFT}-oriented computations and physical analysis. The first that comes to mind is to apply the same approach of the four-dimensional bulk theory to a three-dimensional one, with the same spirit of building everything from first principles like symmetry, power-counting and covariance, and from this to look at the duality between fractons and elasticity from the \ac{QFT} side. In this context 3D fracton models also seem to share relations with theories with torsion \cite{Gromov:2017vir}, and since fractons and \ac{LG} are strongly intertwined (remind that no symmetry principle allow us to isolate fractons from gravity!), this fact makes very promizing the idea of investigating torsional effect from a theory of gravity, in order to see if fractons may come out of it and thus better understand their nature. This could also help us with the task of finding a way (symmetry principle,...) to isolate fractons from \ac{LG} without the need of switching off any coefficient by hand. Sharing so many similarities with Maxwell theory, a final remark concerns the possibility of promoting the fracton theory to a non-abelian one, and the possible quantization of the action \eqref{FMAX-Sinvg1g2}. As it is well known, the quantization of \ac{LG} is a long standing issue. As far as we know, a quantum field theory of fractons has not been achieved yet. In view of this, the covariant formulation adopted in this Thesis should be quite suitable, especially because the fractonic part \eqref{FMAX-Sfract} of the action \eqref{FMAX-Sinvg1g2} impressively reminds the electromagnetic Maxwell theory. Hence, one might think about a kind of ``fracton QED'', where matter is coupled to fractons. Under this respect, the vector gauge fixing studied in Section \ref{sec gauge} can be very useful, not being restricted to the Landau gauge.

\cleardoublepage 

\appendix
\ctparttext{}
\part*{\hspace{1.75em} \textls[80]{\scshape{appendix}}}


\chapter{Some technical remarks}
\section{Derivative of the Heaviside step distribution in curved spacetime} \label{appA}

 The scalar Dirac delta distribution $\delta^{(n)}(x-x')$ and the {corresponding} density $\tilde\delta^{(n)}(x-x')$ are related by \cite{Basler:1991st,Poisson:2011nh}
\begin{equation}
\delta^{(n)}(x-x') = \frac{\tilde\delta^{(n)}(x-x')}{\sqrt{-g}}\ ,
\label{1.5}\end{equation}
with $g\equiv g(x)$, acting on a test function $f(x)$ as follows
\begin{equation}
\int d^nx\,\sqrt{-g}\,\delta^{(n)}(x-x')f(x) = 
\int d^nx\,\tilde\delta^{(n)}(x-x')f(x)
=f(x')\ .
\label{}\end{equation}
Hence, the functional derivative of a generic dual vector field $V_\mu(x)$ is the (1,1) tensor
\begin{equation}
\frac{\delta V_\mu(x)}{\delta V_\nu(x')}=\delta^\nu_\mu\,\delta^{(n)}(x-x')\ .
\label{1.7}
\end{equation}
The Heaviside theta distribution is used to describe a boundary. Its general form is $\theta(f(x))$, where $f(x)=f(t,r,\theta)=0$ describes the equation of the hypersurface $\partial \mM$ of the manifold $\mathcal M$, $i.e.$
\begin{equation}\label{A.1}
\theta(f(x))= \begin{cases} 1\;, & \mbox{if } x\in \mathcal{M} \\ 0\;, & \mbox{if } x\not\in \mathcal{M}\ . \end{cases}
\end{equation}
For a generic manifold $\mathcal{M}$, Stokes' theorem states that
\begin{equation}\label{BF-1.6}
\int_\mathcal{M}d^nx\; \sqrt{-g}\, \nabla_\mu V^\mu 
=
\int_{\partial \mathcal{M}} d^{n-1}y\; \sqrt{-\gamma}\; e_\mu V^\mu\ ,
\end{equation}
where $e_\mu$ is the unit vector normal to the boundary $\partial\mathcal M$ described by the equation $f(x)=0$, $i.e.$
	\begin{equation}
	e_\mu=-\frac{\partial_\mu f}{\sqrt{g^{\mu\nu}\partial_\mu f\partial_\nu f}}\ ,
	\end{equation}
and $\gamma_{ij}$ is the induced metric on $\partial\mathcal M$. The presence of the boundary $f(x)=0$ in the l.h.s.  of \eqref{BF-1.6} can be implemented through the introduction of the step distribution \eqref{A.1}~:
	\begin{empheq}{align}\label{A.3}
	\int_\mathcal {M}d^nx\sqrt{|g|}\ \nabla_\mu V^\mu&=\int d^nx\sqrt{|g|}\ \theta(f(x))\nabla_\mu V^\mu\\
	&=-\int d^nx\sqrt{|g|}\ \nabla_\mu\theta(f(x)) V^\mu\ .\nonumber
	\end{empheq}
The integration on the r.h.s.  of \eqref{A.3} is performed over all spacetime ($i.e.$ boundary at infinity), then, when integrating by parts on the second line, the boundary term vanishes. We can also observe, from \eqref{A.3}, that the step function is a scalar quantity. Identifying the r.h.s.  of \eqref{BF-1.6} and \eqref{A.3} we get 	\begin{equation}\label{A.4}
	-\int d^nx\sqrt{|g|}\ \nabla_\mu\theta(f(x)) V^\mu=\int_{\partial \mathcal M}d^{n-1}x\sqrt{|\gamma|}\ e_\mu V^\mu\ .
	\end{equation}
In \eqref{A.4} we can identify $\nabla_\mu\theta(f(x))$ as a scalar Dirac delta $\delta_{\partial \mathcal M}$\footnote{We adopted the notation of \cite{Vassilevich:2004id}, p.13 Eq.(4.10). }
	\begin{equation}\label{A.5}
	-e_\mu\;\delta_{\partial \mathcal M}\equiv\nabla_\mu\theta(f(x))\ .
	\end{equation}
To see it explicitly and deduce how this distribution acts, we insert \eqref{A.5} back  in \eqref{A.4}
	\begin{equation}\label{A.6}
	\int d^nx\sqrt{|g|}\;\delta_{\partial \mathcal M}\;e_\mu V^\mu=\int_{\partial \mathcal M}d^{n-1}x\sqrt{|\gamma|}\ e_\mu V^\mu\ .
	\end{equation}
Considering a constant, radial boundary
	\begin{equation}\label{BF-n}
	f(x)=R-r\quad\Rightarrow\quad e_\mu=\frac{\delta_\mu^r}{\sqrt{g^{rr}}}=\delta_\mu^r\frac{\sqrt{|g|}}{\sqrt{|\gamma|}}\ ,
	\end{equation}
where we used \cite{Blau}~:
	\begin{equation}\label{BF-medamath}
	\sqrt{g^{rr}}=\frac{\sqrt{-\gamma}}{\sqrt{-g}}\ ,
	\end{equation}
we have
	\be
	\delta_{\partial \mathcal M}=\frac{\sqrt{|\gamma|}}{\sqrt{|g|}}\delta(f(x))=\frac{\sqrt{|\gamma|}}{\sqrt{|g|}}\delta(R-r)\ .
	\ee
Therefore, using \eqref{BF-n}, the derivative of the step function \eqref{A.5} simplifies to
	\begin{equation}\label{BF-1.9}
	\nabla_\mu\theta(R-r)
	=\partial_\mu\theta\left(f(x)\right)=-\delta_\mu^r\;\delta(r-R)\ .
	\end{equation}

\section{Basis for the $\Omega$-tensors}\label{FGF-sec:basis-for-the-omega-tensors}

In the momentum space gauge fixed action $S(g_1,g_2;\xi,\kappa)$ \eqref{FGF-gaugefixedaction} the kinetic operator $\Omega_{\mu\nu,\alpha\beta}(p)$ displays the following symmetries
\begin{equation}
		{\Omega}_{\mu\nu,\alpha\beta}(p) = {\Omega}_{\nu\mu,\alpha\beta}(p) = {\Omega}_{\mu\nu,\beta\alpha}(p) = {\Omega}_{\alpha\beta,\mu\nu}(p)\ .
\label{FGF-symmetries}	
\end{equation}
It can be expanded on a basis formed by a set of five rank-4 tensors, collectively denoted $X_{\mu\nu,\alpha\beta}(p)$  \cite{Blasi:2015lrg,Gambuti:2020onb,Blasi:2017pkk,Gambuti:2021meo,Blasi:2022mbl}
\be
X_{\mu\nu,\alpha\beta}\equiv (A,B,C,D,E)_{\mu\nu,\alpha\beta}
\label{FGF-baseX}\ee
with the same symmetry properties \eqref{FGF-symmetries}. Explicitly, the $X$-tensors read \cite{Blasi:2015lrg,Gambuti:2020onb,Blasi:2017pkk,Gambuti:2021meo,Blasi:2022mbl,Kugo:2014hja}
\begin{align}
    A_{\mu \nu, \alpha \beta} &= \frac{d_{\mu \nu} d_{\alpha \beta}}{3} \label{FGF-A}\\[10pt]
 B_{\mu \nu, \alpha \beta} &= e_{\mu \nu} e_{\alpha \beta} \label{FGF-B}\\[10pt]
  C_{\mu \nu, \alpha \beta} &= \frac{1}{2} \left(  d_{\mu \alpha} d_{\nu \beta} + d_{\mu \beta} d_{\nu \alpha} - \frac{2}{3} d_{\mu \nu} d_{\alpha \beta}  \right) \label{FGF-C}\\[10pt]
  D_{\mu \nu, \alpha \beta} &=  \frac{1}{2} \left(  d_{\mu \alpha} e_{\nu \beta} + d_{\mu \beta} e_{\nu \alpha} + e_{\mu \alpha} d_{\nu \beta} + e_{\mu \beta} d_{\nu \alpha}  \right)\label{FGF-D}\\[10pt]
  E_{\mu \nu, \alpha \beta} &= \frac{\eta_{\mu \nu} \eta_{\alpha \beta}}{4} \label{FGF-E} \; ,
\end{align}
and $e_{\mu\nu}(p)$ and $d_{\mu \nu}(p)$ are the transverse and longitudinal projectors \eqref{FGF-projectors}, 
which are idempotent and orthogonal
\begin{equation}
    e_{\mu \lambda} {e^\lambda}_\nu = e_{\mu \nu}, \ \mathrm{}\   d_{\mu \lambda} {d^\lambda}_\nu = d_{\mu \nu}, \ \mathrm{}\     e_{\mu \lambda} {d ^\lambda}_\nu  =0 \: .  
\label{FGF-defed}\end{equation}
The $X$-tensors have the following properties:
\begin{itemize}
    \item 
    decomposition of the rank-4 tensor identity $\mathcal{I}_{\mu \nu, \alpha\beta}$ ~:
    \be
    A_{\mu \nu , \alpha \beta} + B_{\mu \nu , \alpha \beta} + C_{\mu \nu , \alpha \beta} + D_{\mu \nu , \alpha \beta} =  \mathcal{I}_{\mu \nu , \alpha \beta}
    \label{FGF-idempotency} 
    \ee
    \be
            \mathcal{I}_{\mu \nu, \rho \sigma} = \frac{1}{2} (\eta_{\mu \rho} \eta_{\nu \sigma} + \eta_{\mu \sigma} \eta_{\nu \rho}) 
\label{FGF-identity}\ee
    \item idempotency~:
    \begin{equation}
    X_{\mu\nu}^{\ \ \rho\sigma}X_{\rho\sigma,\alpha\beta}=X_{\mu\nu,\alpha\beta}\ ;
    \label{FGF-idempotency}\end{equation}
    \item orthogonality of $A$, $B$, $C$ and $D$~:
    \begin{equation}
    X_{\mu \nu , \alpha \beta} {{X^\prime}^{\alpha \beta}}_{\rho \sigma} = 0\ \ \mbox{if}\
    (X,X^\prime)\neq E\ \mbox{and}\ X\neq X^\prime\ ;
    \label{FGF-orthogonality}\end{equation}
    \item contractions with $E$~:
    \begin{align} 
    A_{\mu \nu , \alpha \beta} {E^{\alpha \beta}}_{\rho \sigma} 
    &= \frac{d_{\mu \nu} \eta_{\rho \sigma}}{4}\label{FGF-AE}\\[10pt] 
    B_{\mu \nu , \alpha \beta} {E^{\alpha \beta}}_{\rho \sigma} 
    &= \frac{e_{\mu \nu} \eta_{\rho \sigma}}{4}\label{FGF-BE}\\[10pt]
    C_{\mu \nu , \alpha \beta} {E^{\alpha \beta}}_{\rho \sigma}
    &=D_{\mu \nu , \alpha \beta} {E^{\alpha \beta}}_{\rho \sigma}=0\label{FGF-CE-DE}\ .    \end{align}
    
\end{itemize} 


\chapter{Detailed calculations}



\section{Solutions of the boundary conditions for the BF model}\label{BF-appl}
The \ac{BC} for the 3D bulk theory are given by \eqref{BF-B.16}:
	\begin{equation}
	\left.\Lambda^{IJ}X_J\right|_{r=R}=0\ ,
	\end{equation}
where $I,J=\{i;j\}=\{0,2;0,2\}$,
\begin{equation}
\Lambda^{IJ}\equiv\left(\begin{array}{cc}
\alpha^{ij} & \zeta^{i j}-\kappa\epsilon^{i1 j} \\
\zeta^{j i} & \beta^{ i j} \\
\end{array}\right)=\left(
\begin{array}{cccc}
\alpha^{00} & \alpha^{02} & \zeta^{00} & \zeta^{02}-\hat\kappa\\
 \alpha^{20} & \alpha^{22} & \zeta^{20}+\hat\kappa & \zeta^{22}\\
 \zeta^{00} & \zeta^{20} & \beta^{00} & \beta^{02}\\
 \zeta^{02} & \zeta^{22}& \beta^{02} & \beta^{22}
\end{array}
\right)\ , 
\end{equation}
and
\begin{equation}
X_J\equiv\left(\begin{array}{c}
A_j\\
B_{j}
\end{array}\right)\ ,
\end{equation}
and $\hat\kappa$ is given by \eqref{BF-B.19}. Being the linear system of equations \eqref{BF-B.16} homogeneous, three of the four components $X_J$ can be written in terms of the fourth, provided that
	\begin{equation}
	\det\Lambda=0\ .
	\end{equation}
The choice we make is
	\begin{empheq}{align}
	B_\theta(X)&= -l_1B_t(X)\\
	A_\theta(X)&= -l_2B_t(X)\\
	A_t(X)&= -l_3B_t(X)\ ,
	\end{empheq}
with
\begin{align}
{l_1}&\equiv-\frac{-\alpha ^{00} (\beta ^{00} \zeta ^{22}- \beta ^{02} \zeta ^{20})+\alpha ^{02}( \beta ^{00} \zeta ^{02}- \beta ^{02} \zeta ^{00})+\zeta ^{00} \det\zeta}{\alpha ^{00} (\beta ^{02} \zeta ^{22}-\beta ^{22} \zeta ^{20})+\alpha ^{02} (\beta ^{22} \zeta ^{00}-\beta ^{02} \zeta ^{02})-(\zeta ^{02}-\hat\kappa ) \det\zeta}\label{BF-l1}\\
{l_2}&\equiv-\frac{\alpha ^{00}\det\beta-\beta ^{00}\zeta ^{02}(\zeta ^{02}-\hat\kappa)-\hat\kappa  \beta ^{02} \zeta ^{00}+2 \beta ^{02} \zeta ^{00} \zeta ^{02}-\beta ^{22}( \zeta ^{00})^2}{\alpha ^{00} (\beta ^{02} \zeta ^{22}-\beta ^{22} \zeta ^{20})+\alpha ^{02} (\beta ^{22} \zeta ^{00}-\beta ^{02} \zeta ^{02})-(\zeta ^{02}-\hat\kappa ) \det\zeta}\label{BF-l2}\\
{l_3}&\equiv-\frac{-\alpha ^{02} \det\beta+( \zeta ^{02} -\hat\kappa)(  \beta ^{00} \zeta ^{22}-\beta ^{02} \zeta ^{20})+\zeta ^{00} (\beta ^{22}\zeta ^{20}-\beta ^{02}\zeta ^{22})}{\alpha ^{00} (\beta ^{02} \zeta ^{22}-\beta ^{22} \zeta ^{20})+\alpha ^{02} (\beta ^{22} \zeta ^{00}-\beta ^{02} \zeta ^{02})-(\zeta ^{02}-\hat\kappa ) \det\zeta}\ ,\label{BF-l3}
\end{align}
where the request of non-Dirichelet solutions implies the requirements $l_i\neq0$ and $l_i^{-1}\neq0$.
We remind that these coefficients are local, depending on the induced metric determinant and/or components from the bulk parameters. In the case of $\TR$-invariant $S_{bd}$ \eqref{BF-xB.23'} discussed in Section \ref{sec TRBFinCS}, the coefficients \eqref{BF-vi'} are given by
\begin{align}
{l_1}|_{\eqref{BF-Thp}}&=-\gamma^{t\theta}\frac{\hat\beta  \alpha ^{00} \hat\zeta ^{20}+\hat\alpha  \beta ^{00} \hat\zeta ^{02}}{-\alpha ^{00} \beta ^{22} \hat\zeta ^{20}+\hat\zeta ^{02}\left[\hat\zeta ^{20} (\hat\zeta ^{02}-\hat\kappa )-\hat\alpha\hat\beta  \left(\gamma^{t\theta}\right)^2 \right]}\label{BF-l1'}\\
l_2|_{\eqref{BF-Thp}}&=-\frac{\alpha ^{00}\det\beta-\beta ^{00}\hat \zeta ^{02} (\hat\zeta ^{02}-\hat\kappa )}{-\alpha ^{00} \beta ^{22} \hat\zeta ^{20}+\hat\zeta ^{02}\left[\hat\zeta ^{20} (\hat\zeta ^{02}-\hat\kappa )-\hat\alpha\hat\beta  \left(\gamma^{t\theta}\right)^2 \right]}\label{BF-l2'}\\
l_3|_{\eqref{BF-Thp}}&=-\gamma^{t\theta}\frac{-\hat\alpha \det\beta-\hat\beta  \hat\zeta ^{20} (\hat\zeta ^{02}-\hat\kappa )}{-\alpha ^{00} \beta ^{22} \hat\zeta ^{20}+\hat\zeta ^{02}\left[\hat\zeta ^{20} (\hat\zeta ^{02}-\hat\kappa )-\hat\alpha\hat\beta  \left(\gamma^{t\theta}\right)^2 \right]}\ .\label{BF-l3'}
\end{align}

\section{Calculation of the fracton propagators}\label{FGF-sec:calculation-of-the-propagators}

\subsection{$g_1+2 g_2\neq0$}\label{FGF-sec:2g1g2neq0}

The matrix equation \eqref{FGF-eq:matriciale} yields 
\bea
			 \Omega_{\mu\nu,\alpha\beta}\hat G^{\alpha\beta,\rho\sigma} +  \Lambda^*_{\mu\nu,\lambda} \hat G^{*\lambda,\rho\sigma} &=& \mathcal{I}_{\mu\nu}^{\ \ \rho\sigma}\label{FGF-B1}\\
			 \Omega_{\mu\nu,\alpha\beta}\hat G^{\alpha\beta,\tau} +  \Lambda^*_{\mu\nu,\lambda}\hat G^{\lambda\tau} &=& 0\label{FGF-B2}\\
			 \Lambda_{\gamma,\alpha\beta}\hat G^{\alpha\beta,\rho\sigma} +   H_{\gamma\lambda}\hat G^{*\lambda,\rho\sigma} &=& 0\label{FGF-B3}\\
			 \Lambda_{\gamma,\alpha\beta}\hat G^{\alpha\beta,\tau} +   H_{\gamma\lambda}\hat G^{\lambda\tau} &=& \delta_{\gamma}^{\tau} \label{FGF-B4}\ .
\eea
The first equation \eqref{FGF-B1}, using the expansions \eqref{FGF-Omega}, \eqref{FGF-Lambda}, \eqref{FGF-hhprop} and the properties of the $X$-basis listed in Appendix \ref{FGF-sec:basis-for-the-omega-tensors}, gives a system of six equations (remember that our aim is to find the set of coefficients of the propagators \eqref{FGF-propcoeff})
\bea
			 4 {t}\hat{t} + 3 {t}\hat{w} - 4 {v}\hat{v}  + 6p^2  {g}\hat{g} &=& 0\\
			  {t}\hat{w} + 2p^2  {g}\hat{l} &=&0\\
			 p^2   l\hat{l} &=& 2\\
			 p^2   l\hat{g} &=&0\\
			  {v}\hat{v} &=& 1\\
		 {z}\hat{z}+p^2  {f}\hat{f} &=&1\ .
\eea
In the same way, from \eqref{FGF-B2}, \eqref{FGF-B3} and \eqref{FGF-B4} we get three more sets of equations
\allowdisplaybreaks
\bea
		2 {z}\hat{f}+ {f}\hat{r} &=&0\\
			 4 {t}\hat{g} + 2 {g}\hat{s} &=& 0\\
		 {l}\hat{s} &=& 0\\[10px]
		  {f}\hat{z} + 2 {r}\hat{f} &=&0\\
	4 {g}\hat{t}+ 3 {g}\hat{w}+ {l}\hat{w}+8 {s}\hat{g} &=&0\\
			4 {l}\hat{u} +3 {g}\hat{w}+ {l}\hat{w}+8 {s}\hat{l} &=&0\\[10px]
			p^2 {f}\hat{f} +  {r}\hat{r} &=&1\\
			3p^2  {g} \hat{g} + p^2  {l}\hat{l} +2 {s}\hat{s} &=&2\ .
\eea
The above systems of equations are easily solved and, using the coefficients of the kinetic term \eqref{FGF-3.8} and \eqref{FGF-3.9}, we finally get
	\begin{align}
		\hat{t} &= \frac{(4\kappa+1)}{(\kappa+1)(g_1+2g_2)p^2}&;&
		&\hat{u}& = \frac{\kappa(4\kappa+1)-2\xi(g_1+2g_2)}{(\kappa+1)^2(g_1+2g_2)p^2}\label{FGF-eq:uhatapp}\\
		\hat{v} &= \frac{1}{(g_1- g_2)p^2}&;&
		&\hat{z}& = \frac{4\xi}{(2\xi g_1 -1)p^2}\label{FGF-eq:zhatapp}\\
		\hat{w}& = \frac{-4\kappa}{(\kappa+1)(g_1+2g_2)p^2}&;&
		&\hat{f} &= \frac{-2}{(2\xi g_1-1)p^2}\label{FGF-eq:fhatapp}\\
		\hat{g} &= 0&;&
		&\hat{l} &= \frac{2}{(\kappa+1)p^2}\label{FGF-eq:lhatapp}\\
		\hat{r}& = \frac{4g_1}{(2\xi g_1-1)}&;&
		&\hat{s} &= 0\label{FGF-eq:shatapp}\ .
	\end{align}

\subsection{$g_1+2g_2=0$}\label{FGF-sec:2g1g20}

The action of the gauge fixed theory, after a field redefinition and setting $\kappa=0$ because this is the traceless case, is
\be
\left.S( g_1,g_2;\xi,\kappa)\right|_{g_1+2g_2=0;\kappa=0} = 
\left.S_{inv}( g_1,g_2)\right|_{g_1+2g_2=0} + S_{gf}(\xi)\ ,
\label{FGF-}\ee
where the invariant action $\left.S_{inv}(g_1,g_2)\right|_{g_1+2g_2=0}$ and  the gauge fixing term $S_{gf}(\xi)$ are given by \eqref{FGF-2g1+g_2} and \eqref{FGF-Sgftraceless}, respectively. It can be written in the form \eqref{FGF-momentumaction}, and the coefficients in $ \Omega_{\mu\nu,\alpha\beta}$ \eqref{FGF-Omega}, $ \Lambda_{\alpha\beta,\mu}$ \eqref{FGF-Lambda} and $  H_{\mu\alpha}$ \eqref{FGF-H} are 
	\begin{align}
		 {t} &= -3p^2 &;&&
		 {u} &= p^2 &;&&
		 {v} &= -3p^2&;&&
		 {z} &= -p^2&;&&
		 {w} &= 0 \label{FGF-B30}\\
		 {f} &= \frac{1}{2}&;&&
		 {g} &= 0&;&&
		 {l} &= 1&;&&
		 {r} &= \frac{\xi}{2}&;&&
		 {s} &= \frac{\xi}{2}\ .
	\end{align}

Following the same steps of the general case, we find the system of equations for the coefficients of the propagators
	\begin{align}
			 {v}\hat{v} &= 1 &
			 {z}\hat{z}+p^2  {f}\hat{f} &= 1\label{FGF-eq:systemea2}\\
			p^2 {f}\hat{f} +  {r}\hat{r} &= 1 &
			2 {z}\hat{f}+ {f}\hat{r} &= 0\label{FGF-eq:systemea4}\\
			 {f}\hat{z} + 2 {r}\hat{f} &= 0 &
			p^2 \hat{l} +2 {s}\hat{s} &= 2\label{FGF-eq:systemea6}\\
			4 {t}\hat{g} &= 0 &
			4 {u}\hat{l} + 2\hat{s} &= 0\label{FGF-eq:systemea8}\\
			4 {t}\hat{t} + 3 {t}\hat{w} &= 4 &
			\hat{w}+8 {s}\hat{g} &= 0\label{FGF-eq:systemea10}\\
			 {u}\hat{w} + 2p^2 \hat{g} &= 0 &
			4 {u}\hat{u} +  {u}\hat{w} +2p^2 \hat{l} &= 4\label{FGF-eq:systemea12}\\
			 {t}\hat{w} &= 0 &
			4\hat{u}+\hat{w}+8 {s}\hat{l} &= 0\ ,\label{FGF-eq:systemea14}
	\end{align}
which is simpler than the general case. In particular, since $t(p)\neq0$ in \eqref{FGF-B30}, from \eqref{FGF-eq:systemea8} and \eqref{FGF-eq:systemea10} we immediately get $\hat g = \hat w =0$. 
The solutions are therefore easily found 
	\begin{align}
			\hat{t} &= -\frac{1}{3p^2}&
			\hat{u} &= \frac{2\xi}{(2\xi-1)p^2}\\
			\hat{v} &= -\frac{1}{3p^2}&
			\hat{z} &= \frac{-4\xi}{(4\xi+1)p^2}\\
			\hat{w} &= 0&
			\hat{f} &= \frac{2}{(4\xi+1)p^2}\\
			\hat{g} &= 0&
			\hat{l} &= \frac{-2}{(2\xi-1)p^2}\\
			\hat{r} &= \frac{8}{(4\xi+1)}&
			\hat{s} &= \frac{4}{(2\xi-1)}\ . 
	\end{align}
	\allowdisplaybreaks[0]


\section{Fractons with boundary : detailed calculations}

\subsection{Commutators}

\subsubsection{The bulk : generalized Ka\c{c}-Moody algebra}\label{FBD-appAlgebra}

	Considering the first Ward identity \eqref{FBD-wi1} 
	\be\label{FBD-wi1A}
	\int dx^3\theta(x^3)\partial_i\partial_jJ^{ij}=2(g_2-g_1)\partial_i\partial_j\tilde A^{ij}-2g_2\partial_i\partial^i\tilde A|_{x^3=0}\ ,
	\ee
	we compute \\
	$\frac{\delta}{\delta J^{mn}(x')}\eqref{FBD-wi1A}$ :
	\begin{align}
	\partial_m\partial_n\delta^{(3)}(X-X')&=2(g_2-g_1)\partial_i\partial_j\frac{\delta^2Z_c[J,\tilde J]}{\delta J^{mn}(X')\tilde J_{ij}(X)}-2g_2\eta_{ij}\partial_a\partial^a\frac{\delta^2Z_c[J,\tilde J]}{\delta J^{mn}(X')\tilde J_{ij}(X)}\nonumber\\
	&=2i\left[(g_2-g_1)\delta^k_i\delta^l_j-g_2\eta^{kl}\eta_{ij}\right]\partial_k\partial_l\langle T(A_{mn}(X')\tilde A^{ij}(X))\rangle\nonumber\\
	&=2i\left[(g_2-g_1)\delta^k_i\delta^l_j-g_2\eta^{kl}\eta_{ij}\right]\cancel{\langle T(A_{mn}(X')\partial_k\partial_l\tilde A^{ij})\rangle}+\nonumber\\
&+2i\left[(g_2-g_1)\delta^k_i\delta^l_j-g_2\eta^{kl}\eta_{ij}\right]\left\{\left[\partial_l\tilde A^{ij}(X),A_{mn}(X')\right]\delta^0_k\delta(x^0-x'^0)+\right.\nonumber\\
&+\left.\partial_k\left(\left[\tilde A^{ij}(X),A_{mn}(X')\right]\delta^0_l\delta(x^0-x'^0)\right)\right\}\nonumber\\
	&=2i\left[(g_2-g_1)\left(\partial_j\tilde A^{0j}+\partial_\textsc{a}\tilde A^{0\textsc{a}}\right)-g_2\partial^0\tilde A\ ,\ A'_{mn}\right]\delta(x^0-x'^0)+\nonumber\\
	&+2i\partial_0\left\{\left[(g_2-g_1)\tilde A^{00}+g_2\tilde A\ ,\ A'_{mn}\right]\delta(x^0-x'^0)\right\}\ ,
	\end{align}
where we used the conserved current equation \eqref{FBD-cc1}. Integrating over $dx^0$, we finally get to the following equal time commutators
	\begin{empheq}{align}
	\left[\Delta\tilde A(X)\ ,\ A_{0n}(X')\right]_{x^0=x'^0}&=0\label{FBD-app[DAt,A0n]}\\
	\left[\Delta\tilde A(X)\ ,\ A_{\textsc{mn}}(X')\right]_{x^0=x'^0}&=i\partial_\textsc{m}\partial_\textsc{n}\{\delta(x^1-x'^1)\delta(x^2-x'^2)\}\ ,\label{FBD-comm1}
	\end{empheq}
where we defined
	\be\label{FBD-DA't}
	\Delta \tilde A\equiv2(g_1-g_2)\left(\partial_j\tilde A^{0j}+\partial_\textsc{a}\tilde A^{0\textsc{a}}\right)+2g_2\partial^0\tilde A\ .
	\ee
In the same way, we now compute
	$\frac{\delta}{\delta \tilde J^{mn}(x')}\eqref{FBD-wi1A}$ :
	\be
	\begin{split}
	0&=2(g_2-g_1)\partial_i\partial_j\frac{\delta^2Z_c[J,\tilde J]}{\delta\tilde J^{mn}(X')\tilde J_{ij}(X)}-2g_2\eta_{ij}\partial_a\partial^a\frac{\delta^2Z_c[J,\tilde J]}{\delta\tilde J^{mn}(X')\tilde J_{ij}(X)}\\
	&=2i\left[(g_2-g_1)\left(\partial_j\tilde A^{0j}+\partial_\textsc{a}\tilde A^{0\textsc{a}}\right)-g_2\partial^0\tilde A\ ,\ \tilde A'_{mn}\right]\delta(x^0-x'^0)+\\
	&\quad+2i\partial_0\left\{\left[(g_2-g_1)\tilde A^{00}+g_2\tilde A\ ,\ \tilde A'_{mn}\right]\delta(x^0-x'^0)\right\}\ ,
	\end{split}
	\ee
where we used again the conserved current equation \eqref{FBD-cc1}. By integrating over time and using the definition \eqref{FBD-DA't}, we find the following equal time commutator
	\be
	\left[\Delta\tilde A(X)\ ,\ \tilde A_{mn}(X')\right]_{x^0=x'^0}=0\ .
	\ee
Taking the second broken Ward identity \eqref{FBD-wi2}
	\be\label{FBD-wi2A}
	\partial_i\partial_j\tilde J^{ij}|_{x^3=0}=-2(g_2-g_1)\partial_i\partial_j A^{ij}+2g_2\partial_i\partial^iA|_{x^3=0}\ .
	\ee
we compute 	$\frac{\delta}{\delta J^{mn}(x')}\eqref{FBD-wi2A}$ :
	\be
	\begin{split}
	0&=-2(g_2-g_1)\partial_i\partial_j\frac{\delta^2Z_c[J,\tilde J]}{\delta J^{mn}(X') J_{ij}(X)}+2g_2\eta_{ij}\partial_a\partial^a\frac{\delta^2Z_c[J,\tilde J]}{\delta J^{mn}(X') J_{ij}(X)}\\
	&=-2i\left[(g_2-g_1)\left(\partial_j A^{0j}+\partial_\textsc{a} A^{0\textsc{a}}\right)-g_2\partial^0 A\ ,\  A'_{mn}\right]\delta(x^0-x'^0)+\\
	&\quad-2i\partial_0\left\{\left[(g_2-g_1)  A^{00}+g_2 A\ ,\  A'_{mn}\right]\delta(x^0-x'^0)\right\}\ ,
	\end{split}
	\ee
where we used \eqref{FBD-cc2}. Integrating over $dx^0$ we find the equal time commutator
	\be
	\left[\Delta A(X)\ ,\  A_{mn}(X')\right]_{x^0=x'^0}=0\ ,
	\ee
where $\Delta A(X)$ is defined as  \eqref{FBD-DA't}
	\be\label{FBD-DA'}
	\Delta A\equiv2(g_1-g_2)\left(\partial_jA^{0j}+\partial_\textsc{a}A^{0\textsc{a}}\right)+2g_2\partial^0A\ .
	\ee
We finally compute $\frac{\delta}{\delta\tilde J^{mn}(x')}\eqref{FBD-wi2A}$ :
	\begin{align}
	\partial_m\partial_n\delta^{(3)}(X-X')&=-2(g_2-g_1)\partial_i\partial_j\frac{\delta^2Z_c[J,\tilde J]}{\delta\tilde J^{mn}(X') J_{ij}(X)}+2g_2\eta_{ij}\partial_a\partial^a\frac{\delta^2Z_c[J,\tilde J]}{\delta\tilde J^{mn}(X') J_{ij}(X)}\nonumber\\
	&=-2i\left[(g_2-g_1)\left(\partial_j A^{0j}+\partial_\textsc{a} A^{0\textsc{a}}\right)-g_2\partial^0 A\ ,\ \tilde A'_{mn}\right]\delta(x^0-x'^0)+\nonumber\\
	&\quad-2i\partial_0\left\{\left[(g_2-g_1) A^{00}+g_2 A\ ,\ \tilde A'_{mn}\right]\delta(x^0-x'^0)\right\}\ ,
	\end{align}
where we used \eqref{FBD-cc2} and from which, integrating over $dx^0$, we find
	\begin{empheq}{align}
	\left[\Delta A(X)\ ,\ \tilde A_{0n}(X')\right]_{x^0=x'^0}&=0\\
    	\left[-\Delta A(X)\ ,\ \tilde A_{\textsc{mn}}(X')\right]_{x^0=x'^0}&=i\partial_\textsc{m}\partial_\textsc{n}\{\delta(x^1-x'^1)\delta(x^2-x'^2)\}\ .\label{FBD-comm2}
	\end{empheq}
	
\subsubsection{The boundary : canonical commutators}\label{FBD-appComm}

We take the commutator \eqref{FBD-[DAt,A]}  and its trace, in the following equal time combination
	\be\label{FBD-DAt,A-TrA appB}
	\left[\Delta\tilde A\ ,\ A_{\textsc{df}}'-\frac{1}{2}\eta_{\textsc{df}}\eta^{\textsc{mn}}A_{\textsc{mn}}'\right]=\frac{i}{2}\left(\delta^{\textsc{m}}_{\textsc{d}}\delta^{\textsc{n}}_{\textsc{f}}+\delta^{\textsc{n}}_{\textsc{d}}\delta^{\textsc{m}}_{\textsc{f}}-\eta^{\textsc{mn}}\eta_{\textsc{df}}\right)\partial_\textsc{m}\partial_\textsc{n}\delta^{(2)}(X-X')\ .
	\ee
In terms of the solutions on the boundary \eqref{FBD-a,at sol} we have
	\begin{empheq}{align}
	 A_{\textsc{df}}(X)|_\eqref{FBD-sol2}&=\epsilon_{\textsc{d}ab}\partial^a \alpha^b_{\ \textsc{f}}(X)+\epsilon_{\textsc{f}ab}\partial^a \alpha^b_{\ \textsc{d}}(X)\label{FBD-sol2'}\\
	 \eta^{\textsc{mn}}A_{\textsc{mn}}|_\eqref{FBD-sol2}&=2\epsilon_{0\textsc{bc}}\partial^\textsc{b}\alpha^{0\textsc{c}}=A_{00}(X)|_\eqref{FBD-sol2}\label{FBD-TrA=A00}\\
	\Delta \tilde A|_\eqref{FBD-sol1}&=\partial_\textsc{m}\partial_\textsc{n}\left[g_{12}\left(\epsilon^{0\textsc{mb}}\tilde \alpha_{\textsc{b}}^{\ \textsc{n}}+\epsilon^{0\textsc{nb}}\tilde \alpha_{\textsc{b}}^{\ \textsc{m}}\right)\right]\ ,\label{FBD-DAt-sol}
	\end{empheq}
where $g_{12}\equiv2(g_1-g_2)$. As a consequence of the tracelessness of $A_{ij}(X)$, we can use \eqref{FBD-TrA=A00} and write the commutator \eqref{FBD-app[DAt,A0n]} for $n=0$ as follows
	\be\label{FBD-DA,TrA=0}
	\left[\Delta\tilde A(X)|_\eqref{FBD-sol1}\ ,\ A_{00}(X')|_\eqref{FBD-sol2}\right]=\left[\Delta\tilde A(X)|_\eqref{FBD-sol1}\ ,\ \eta^{\textsc{mn}}A_{\textsc{mn}}(X')|_\eqref{FBD-sol2}\right]=0\ ,
	\ee
then, using \eqref{FBD-sol2'},  \eqref{FBD-DAt-sol} and \eqref{FBD-DA,TrA=0}, the commutator \eqref{FBD-DAt,A-TrA appB} becomes
	\be
	\begin{split}
	\frac{i}{2}\partial_\textsc{m}\partial_\textsc{n}\left\{...
	\right\}^{\textsc{mn}}_{\textsc{df}}&=\left[\Delta\tilde A(X)|_\eqref{FBD-sol1}\ ,\ A_{\textsc{df}}(X')|_\eqref{FBD-sol2}\right]\\
	&=\partial_\textsc{m}\partial_\textsc{n}\left[g_{12}\left(\epsilon^{0\textsc{mb}}\tilde \alpha_{\textsc{b}}^{\ \textsc{n}}(X)+\epsilon^{0\textsc{nb}}\tilde \alpha_{\textsc{b}}^{\ \textsc{m}}(X)\right)\ ,\ \epsilon_{\textsc{d}ab}\partial^a \alpha^b_{\ \textsc{f}}(X')+\epsilon_{\textsc{f}ab}\partial^a \alpha^b_{\ \textsc{d}}(X')\right]\ ,
	\end{split}
	\ee
from which we can identify the following canonical commutation relation
	
	\be\label{FBD-P,Q}
	\left[Q^{\textsc{mn}}\ , \ P'_{\textsc{df}}\right]=i\left(\frac{\delta^{\textsc{m}}_{\textsc{d}}\delta^{\textsc{n}}_{\textsc{f}}+\delta^{\textsc{n}}_{\textsc{d}}\delta^{\textsc{m}}_{\textsc{f}}}{2}-\frac{1}{2}\eta^{\textsc{mn}}\eta_{\textsc{df}}\right)\delta^{(2)}(X-X')\ ,
	\ee
with
	\begin{empheq}{align}
	Q^{\textsc{mn}}=Q^{\textsc{nm}}&\equiv\epsilon^{0\textsc{mb}}\tilde \alpha_{\textsc{b}}^{\ \textsc{n}}+\epsilon^{0\textsc{nb}}\tilde \alpha_{\textsc{b}}^{\ \textsc{m}} \\
	P_{\textsc{df}}=P_{\textsc{fd}}\ &\equiv g_{12}\left(\epsilon_{\textsc{d}ab}\partial^a \alpha^b_{\ \textsc{f}}+\epsilon_{\textsc{f}ab}\partial^a \alpha^b_{\ \textsc{d}}\right)
	\end{empheq}
and $Q^\textsc{m}_{\ \textsc{m}}=P^\textsc{m}_{\ \textsc{m}}=0$. We can go further, multiplying both right and left hand sides of \eqref{FBD-P,Q} by \mbox{$\epsilon_{0\textsc{ma}}\epsilon^{0\textsc{fk}}=\delta_\textsc{m}^\textsc{k}\delta_\textsc{a}^\textsc{f}-\delta_\textsc{m}^\textsc{f}\delta_\textsc{a}^\textsc{k}$}
	\be
	\left[\epsilon_{0\textsc{ma}}Q^{\textsc{mn}}\ , \ \epsilon^{0\textsc{fk}}P'_{\textsc{df}}\right]=i(\delta_\textsc{m}^\textsc{k}\delta_\textsc{a}^\textsc{f}-\delta_\textsc{m}^\textsc{f}\delta_\textsc{a}^\textsc{k})\left(\frac{\delta^{\textsc{m}}_{\textsc{d}}\delta^{\textsc{n}}_{\textsc{f}}+\delta^{\textsc{n}}_{\textsc{d}}\delta^{\textsc{m}}_{\textsc{f}}}{2}-\frac{1}{2}\eta^{\textsc{mn}}\eta_{\textsc{df}}\right)\delta^{(2)}(X-X')\ ,
	\ee
for which
	\be
	\begin{split}
	\epsilon_{0\textsc{ma}}Q^{\textsc{mn}}&=\epsilon_{0\textsc{ma}}\left(\epsilon^{0\textsc{mb}}\tilde \alpha_{\textsc{b}}^{\ \textsc{n}}+\epsilon^{0\textsc{nb}}\tilde \alpha_{\textsc{b}}^{\ \textsc{m}}\right)\\
	&=-2\tilde \alpha_\textsc{a}^{\ \textsc{n}}+\delta^\textsc{n}_\textsc{a}\tilde \alpha^\textsc{m}_{\ \textsc{m}}
	\end{split}
	\ee
which is the traceless spatial part of $\tilde \alpha_{{mn}}(X) $. Then
	\be
	\begin{split}
	\epsilon^{0\textsc{fk}}P_{\textsc{df}}&=g_{12}\left[(\delta^0_a\delta^\textsc{k}_b-\delta^0_b\delta^\textsc{k}_a)\partial^a\alpha^b_{\ \textsc{d}}+\epsilon^{0\textsc{fk}}\epsilon_{0\textsc{da}}\partial^\textsc{a}\alpha^0_{\ \textsc{f}}-\epsilon^{0\textsc{fk}}\epsilon_{0\textsc{db}}\partial^0\alpha^\textsc{b}_{\ \textsc{f}}\right]\\
	&=g_{12}\left[ 2(\partial^0\alpha^\textsc{k}_{\ \textsc{d}}-\partial^\textsc{k}\alpha^0_{\ \textsc{d}})+\delta^\textsc{k}_\textsc{d}(\partial^\textsc{a}\alpha^0_{\ \textsc{a}}-\partial^0\alpha^\textsc{b}_{\ \textsc{b}})\right]\ .
	\end{split}
	\ee
Finally, at the right hand side we have
	\be
	(\delta_\textsc{m}^\textsc{k}\delta_\textsc{a}^\textsc{f}-\delta_\textsc{m}^\textsc{f}\delta_\textsc{a}^\textsc{k})\left(\frac{\delta^{\textsc{m}}_{\textsc{d}}\delta^{\textsc{n}}_{\textsc{f}}+\delta^{\textsc{n}}_{\textsc{d}}\delta^{\textsc{m}}_{\textsc{f}}}{2}-\frac{1}{2}\eta^{\textsc{mn}}\eta_{\textsc{df}}\right)=\frac{1}{2}\left(\delta^\textsc{n}_\textsc{a}\delta^\textsc{k}_\textsc{d}-\delta^\textsc{n}_\textsc{d}\delta^\textsc{k}_\textsc{a}-\eta_{\textsc{ad}}\eta^{\textsc{kn}}\right)\ .
	\ee
By properly raising and lowering the indices with {$\eta^{\textsc{dl}}\eta_{\textsc{nb}}$}, we finally get
	\be
	4g_{12}\biggl[\tilde \alpha_\textsc{ab}-\eta_\textsc{ab}\tilde \alpha^\textsc{m}_{\ \textsc{m}}\ ,\  (\partial^0\alpha^\textsc{kl}-\partial^\textsc{k}\alpha^{0\textsc{l}})'+\eta^\textsc{kl}(\partial_\textsc{f}\alpha^{0\textsc{f}}-\partial^0\alpha^\textsc{f}_{\ \textsc{f}})'\biggr]=\tfrac{i}{2}\left(\delta^{\textsc{k}}_{\textsc{a}}\delta^{\textsc{l}}_{\textsc{b}}+\delta^{\textsc{l}}_{\textsc{a}}\delta^{\textsc{k}}_{\textsc{b}}-\eta^{\textsc{kl}}\eta_{\textsc{ab}}\right)\delta^{(2)}(X-X')\ ,
	\ee
where the primed quantities  depend on $X'$. At the right hand side we have the index symmetry $\textsc{a}\leftrightarrow \textsc{b}$ and $\textsc{c}\leftrightarrow \textsc{d}$, while at the left hand side the symmetry is only for  $\textsc{a}\leftrightarrow \textsc{b}$. We thus symmetrize the result as follows
	\be
	\frac{1}{2}\left(\biggl[..._{\textsc{ab}}\ ,\ ...^\textsc{cd}\biggr]+\biggl[..._{\textsc{ab}}\ ,\ ...^\textsc{dc}\biggr]\right)=\frac{i}{2}\left(\delta^{\textsc{c}}_{\textsc{a}}\delta^{\textsc{d}}_{\textsc{b}}+\delta^{\textsc{d}}_{\textsc{a}}\delta^{\textsc{c}}_{\textsc{b}}-\eta^{\textsc{cd}}\eta_{\textsc{ab}}\right)\delta^{(2)}(X-X')\ ,
	\ee
obtaining
	\be
	\left[\tilde \alpha _{\textsc{ab}}-\frac{1}{2}\eta_{\textsc{ab}}\tilde \alpha ^\textsc{m}_{\ \textsc{m}}\ ,\ -g_{12}\left(2f'^{\textsc{cd}0}-\eta^{\textsc{cd}}f'^{{\ a}0}_{a}\right)\right]=\frac{i}{2}\left(\delta^{\textsc{c}}_{\textsc{a}}\delta^{\textsc{d}}_{\textsc{b}}+\delta^{\textsc{d}}_{\textsc{a}}\delta^{\textsc{c}}_{\textsc{b}}-\eta^{\textsc{cd}}\eta_{\textsc{ab}}\right)\delta^{(2)}(X-X')\ ,\label{FBD-qt,p}
	\ee
where $f_{abc}(X)$ is analogous to $F_{\mu\nu\rho}(x)$ \eqref{FMAX-Fmunurho}, but referred to $\alpha_{ab}(X)$, $i.e.$
	\be\label{FBD-fabc}
	f_{abc}\equiv\partial_a\alpha_{bc}+\partial_b\alpha_{ac}-2\partial_c\alpha_{ab}\ .
	\ee
From \eqref{FBD-qt,p} we can identify the new canonical variables as
	\begin{align}
	\tilde q_{\textsc{ab}}&\equiv\tilde \alpha _{\textsc{ab}}-\frac{1}{2}\eta_{\textsc{ab}}\tilde \alpha ^\textsc{m}_{\ \textsc{m}}\\
	\tilde p^{\textsc{cd}}&\equiv -2g_{12}\left(f^{\textsc{cd}0}-\frac{1}{2}\eta^{\textsc{cd}}f^{\ a0}_{a}\right)=-2g_{12}\left[(\partial^\textsc{c}\alpha^{0\textsc{d}}+\partial^\textsc{d}\alpha^{0\textsc{c}}-2\partial^0\alpha^{\textsc{cd}})-\eta^{\textsc{cd}}\partial_a\alpha^{0a}\right]\ .
	\end{align} 
Not surprisingly, starting from the commutator \eqref{FBD-[DA,At]}, and proceeding as we just did, we land on an analogous result (up to a sign) with $\alpha\leftrightarrow\tilde \alpha$ switched, $i.e.$ we get
	\be
	\left[\alpha _{\textsc{ab}}-\frac{1}{2}\eta_{\textsc{ab}}\alpha ^\textsc{m}_{\ \textsc{m}}\ ,\ g_{12}\left(2\tilde f'^{\textsc{cd}0}-\eta^{\textsc{cd}}\tilde f'^{{\ a}0}_{a}\right)\right]=\frac{i}{2}\left(\delta^{\textsc{c}}_{\textsc{a}}\delta^{\textsc{d}}_{\textsc{b}}+\delta^{\textsc{d}}_{\textsc{a}}\delta^{\textsc{c}}_{\textsc{b}}-\eta^{\textsc{cd}}\eta_{\textsc{ab}}\right)\delta^{(2)}(X-X')\ ,\label{FBD-q,pt}
	\ee
where $\tilde f_{abc}(X)$ refers to $\tilde\alpha_{ab}(X)$
	\be
	\tilde f_{abc}\equiv\partial_a\tilde \alpha_{bc}+\partial_b\tilde \alpha_{ac}-2\partial_c\tilde \alpha_{ab}\ .
	\ee
From \eqref{FBD-q,pt} we can identify another set of canonical variables
	\begin{align}
	q_{\textsc{ab}}&\equiv \alpha _{\textsc{ab}}-\frac{1}{2}\eta_{\textsc{ab}} \alpha ^\textsc{m}_{\ \textsc{m}}\\
	p^{\textsc{cd}}&\equiv 2g_{12}\left(\tilde f^{\textsc{cd}0}-\frac{1}{2}\eta^{\textsc{cd}}\tilde f^{\ a0}_{a}\right)=2g_{12}\left[(\partial^\textsc{c}\tilde\alpha^{0\textsc{d}}+\partial^\textsc{d}\tilde\alpha^{0\textsc{c}}-2\partial^0\tilde\alpha^{\textsc{cd}})-\eta^{\textsc{cd}}\partial_a\tilde\alpha^{0a}\right]\ .
	\end{align}
We observe that the canonical variables $\tilde q_{\textsc{ab}}(X),\ q_{\textsc{ab}}(X)$ in \eqref{FBD-qt,p} and \eqref{FBD-q,pt}, depend on the traceless spatial part of the fields $\tilde\alpha_{ij}(X)$ and $\alpha_{ij}(X)$.

\subsection{The most general action}\label{FBD-appSgen}

The most general action of the 3D boundary theory must be compatible with
	\bi
	\item power-counting $[\alpha]=0,\ [\tilde\alpha]=1$ ;
	\item symmetry $\delta S=\tilde\delta S=0$, where $\delta$ and $\tilde\delta$ are defined in \eqref{FBD-dalpha} and \eqref{FBD-dalphat} ;
	\item canonical variables, $i.e.\ \frac{\partial\mathcal L_{kin}}{\partial \dot q}=p$, identified in \eqref{FBD-al,ft} and \eqref{FBD-alt,f}.
	\ei
From the first two requests we have that the most general action must be
		\begin{align}
		S_{3D}&=\int d^3X\left\{\omega_1\left(\partial_c\alpha_{ab}\partial^c\alpha^{ab}-\tfrac{3}{2}\partial_c\alpha_{ab}\partial^a\alpha^{bc}\right)+\omega_2\left(\partial_c\alpha_{ab}\partial^c\tilde\alpha^{ab}-\tfrac{3}{2}\partial_c\alpha_{ab}\partial^a\tilde\alpha^{bc}\right)-\right.\nonumber\\
		&\quad\left.-3\omega_3\,\alpha^d_a\epsilon^{abc}\partial_b\alpha_{cd}-3\omega_4\,\tilde\alpha^d_a\epsilon^{abc}\partial_b\alpha_{cd}-3\omega_5\,\tilde\alpha^d_a\epsilon^{abc}\partial_b\tilde\alpha_{cd}\right\}\label{FBD-appS3D-gen}\\
		&=\int d^3X\left\{\tfrac{\omega_1}{6}\,\varphi_{abc}\varphi^{abc}+\tfrac{\omega_2}{6}\,\varphi_{abc}\tilde\varphi^{abc}+\omega_3\, \alpha^d_a\epsilon^{abc}\varphi_{dbc}+\omega_4\, \alpha^d_a\epsilon^{abc}\tilde\varphi_{dbc}+\omega_5\, \tilde\alpha^d_a\epsilon^{abc}\tilde\varphi_{dbc}\right\}\nonumber
		\end{align}
where $\omega_i$ are constants, $[\omega_2]=[\omega_5]=0,\ [\omega_1]=[\omega_4]=1,\ [\omega_3]=2$, and we defined the tensor
	\be\label{FBD-appphi}
		\begin{split}
		\varphi_{abc}&\equiv f_{abc}+\frac{1}{4}\left(-2\eta_{ab}f^d_{\ dc}+\eta_{bc}f^d_{\ da}+\eta_{ac}f^d_{\ db}\right)\\
		&=-2\partial_c\alpha_{ab}+\partial_a\alpha_{bc}+\partial_b\alpha_{ac}-\eta_{ab}\partial^d\alpha_{dc}+\frac{1}{2}\eta_{bc}\partial^d\alpha_{da}+\frac{1}{2}\eta_{ac}\partial^d\alpha_{db}\ ,
		\end{split}
	\ee
and its analog $\tilde\varphi_{abc}(X)$ with respect to $\tilde\alpha_{ab}(X)$, with the following properties
	\begin{empheq}{align}
	&\varphi_{abc}=\varphi_{bac}\quad;\quad	\tilde\varphi_{abc}=\tilde\varphi_{bac}\\
	&\varphi_{abc}+\varphi_{cab}+\varphi_{bca}=0=\tilde\varphi_{abc}+\tilde\varphi_{cab}+\tilde\varphi_{bca}\label{FBD-appcicl}\\
	&\delta\varphi_{abc}=\tilde\delta\tilde\varphi_{abc}=0\\
	&\eta^{ab}\varphi_{abc}=\eta^{bc}\varphi_{abc}=\eta^{ab}\tilde\varphi_{abc}=\eta^{bc}\tilde\varphi_{abc}=0\ .\label{FBD-apptraceless}
	\end{empheq}
Notice that 
	 \begin{empheq}{align}
 	\tilde\varphi^{\textsc{mn}0}&=\tilde f^{\textsc{mn}0}-\frac{1}{2}\eta^{\textsc{mn}}\tilde f_a^{\ a0}=\frac{1}{2g_{12}}p^{\textsc{mn}}\label{FBD-appp}\\
 	\varphi^{\textsc{mn}0}&=f^{\textsc{mn}0}-\frac{1}{2}\eta^{\textsc{mn}} f_a^{\ a0}=-\frac{1}{2g_{12}}\tilde p^{\textsc{mn}}\ .
	\end{empheq}
We rewrite the fields $\alpha_{ab}(X)$ and $\tilde\alpha_{ab}(X)$ according to the representation of the rotation group, as follows
	\begin{empheq}{align}
	\alpha_{00}&=-4\psi=\alpha^\textsc{a}_\textsc{a}\label{FBD-apppsi}\\
	\alpha_{0\textsc{a}}&=v_{\textsc{a}}\label{FBD-appv}\\
	\alpha_{\textsc{ab}}&=2s_{\textsc{ab}}-2\eta_{\textsc{ab}}\psi\quad;\quad s_{\textsc{ab}}\equiv\frac{1}{2}\left(\alpha_{\textsc{ab}}-\frac{1}{2}\eta_{\textsc{ab}}\alpha^\textsc{d}_\textsc{d}\right)=\frac{1}{2}q_{\textsc{ab}}\label{FBD-apps}
	\end{empheq}
and
	\begin{empheq}{align}
	\tilde\alpha_{00}&=-4\tilde\psi=\tilde\alpha^\textsc{a}_\textsc{a}\label{App al00}\\
	\tilde\alpha_{0\textsc{a}}&=\tilde v_{\textsc{a}}\\
	\tilde\alpha_{\textsc{ab}}&=2\tilde s_{\textsc{ab}}-2\eta_{\textsc{ab}}\tilde\psi\quad;\quad \tilde s_{\textsc{ab}}\equiv\frac{1}{2}\left(\tilde\alpha_{\textsc{ab}}-\frac{1}{2}\eta_{\textsc{ab}}\tilde\alpha^\textsc{d}_\textsc{d}\right)=\frac{1}{2}\tilde q_{\textsc{ab}}\label{App altij}\ ,
	\end{empheq}
in terms of which the commutators \eqref{FBD-al,ft} and \eqref{FBD-alt,f} become
	\begin{align}
	\left[2s_{\textsc{ab}}\ ,\ 2g_{12}\tilde\varphi'^{\textsc{cd}0}\right]=&\frac{i}{2}\left(\delta^{\textsc{c}}_{\textsc{a}}\delta^{\textsc{d}}_{\textsc{b}}+\delta^{\textsc{d}}_{\textsc{a}}\delta^{\textsc{c}}_{\textsc{b}}-\eta_{\textsc{ab}}\eta^{\textsc{cd}}\right)\delta^{(2)}(X-X')\label{FBD-appq,p}\\
	\left[2\tilde s_{\textsc{ab}}\ ,\ -2g_{12}\varphi'^{\textsc{cd}0}\right]=&\frac{i}{2}\left(\delta^{\textsc{c}}_{\textsc{a}}\delta^{\textsc{d}}_{\textsc{b}}+\delta^{\textsc{d}}_{\textsc{a}}\delta^{\textsc{c}}_{\textsc{b}}-\eta_{\textsc{ab}}\eta^{\textsc{cd}}\right)\delta^{(2)}(X-X')\ .\label{FBD-appqt,pt}
	\end{align}
Compatibility of the action $S_{3D}$ \eqref{FBD-appS3D-gen} with the first commutator \eqref{FBD-appq,p}, $i.e.$
	\be\label{FBD-appLkin}
	\frac{\partial\mathcal L_{3D}}{\partial \dot q_{\textsc{ab}}}=p^{\textsc{ab}}\ ,
	\ee
requires
	\be
	\omega_1=\omega_3=\omega_4=0\ ,
	\ee
and $\omega_5$ is left free. Finally, distinguishing between time and space indices in the term of \eqref{FBD-appS3D-gen} which has $\omega_2$ as coefficient, we have
	\be
	\varphi_{abc}\tilde\varphi^{abc}=\frac{3}{2}\varphi_{00\textsc{a}}\tilde\varphi^{00\textsc{a}}+2\varphi_{0\textsc{ab}}\tilde\varphi^{0\textsc{ab}}+\varphi_{\textsc{ab}0}\tilde\varphi^{\textsc{ab}0}+\varphi_{\textsc{abc}}\tilde\varphi^{\textsc{abc}}\ ,
	\ee
where we observed that $\varphi_{\textsc{a}00}=-\frac{1}{2}\varphi_{00\textsc{a}}$ (and the same for $\tilde\varphi_{\textsc{a}00}$). Additionally, by using \eqref{FBD-apppsi}, \eqref{FBD-appv} and \eqref{FBD-apps}, we have
	\begin{align}
	\varphi_{00\textsc{a}}&=6\partial_\textsc{a}\psi+\partial_0v_\textsc{a}+\partial^\textsc{b}q_{\textsc{ab}}\\
	\varphi_{\textsc{ab}0}&=-2\partial_0q_{\textsc{ab}}+\partial_\textsc{a}v_\textsc{b}+\partial_\textsc{b}v_\textsc{a}-\eta_{\textsc{ab}}\partial^\textsc{d}v_\textsc{d}\\
	\varphi_{0\textsc{ab}}&=-2\partial_\textsc{b}v_\textsc{a}+\partial_\textsc{a}v_\textsc{b}+\partial_0q_{\textsc{ab}}+\tfrac{1}{2}\eta_{\textsc{ab}}\partial^\textsc{d}v_\textsc{d}\\
	\varphi_{\textsc{abc}}&=-2\partial_\textsc{c}q_{\textsc{ab}}+\partial_\textsc{b}q_{\textsc{ac}}+\partial_\textsc{a}q_{\textsc{bc}}+6\eta_{\textsc{ab}}\partial_\textsc{c}\psi-3\eta_{\textsc{bc}}\partial_\textsc{a}\psi-3\eta_{\textsc{ac}}\partial_\textsc{b}\psi-\nonumber\\
	&\quad-\partial^0(\eta_{\textsc{ab}}v_\textsc{c}-\tfrac{1}{2}\eta_{\textsc{ac}}v_\textsc{b}-\tfrac{1}{2}\eta_{\textsc{bc}}v_\textsc{a})-\partial^\textsc{d}(\eta_{\textsc{ab}}q_{\textsc{dc}}-\tfrac{1}{2}\eta_{\textsc{ac}}q_{\textsc{db}}-\tfrac{1}{2}\eta_{\textsc{bc}}q_{\textsc{da}})\ .
	\end{align}
The only terms containing $\dot q$ contributions are related to $\varphi_{\textsc{ab}0}$ and $\varphi_{0\textsc{ab}}$, $i.e.$ in the 3D Lagrangian they only appear in $\frac{\omega_2}{6}(2\varphi_{0\textsc{ab}}\tilde\varphi^{0\textsc{ab}}+\varphi_{\textsc{ab}0}\tilde\varphi^{\textsc{ab}0})$. Keeping in mind this, consistently with its definition, for a traceless tensor in $d$ dimensions we have
	\be\label{FBD-apptraceless-der}
	\frac{\partial s_{\mu\nu}}{\partial s_{\alpha\beta}}=\frac{\delta^\alpha_\mu\delta^\beta_\nu+\delta^\alpha_\nu\delta^\beta_\mu}{2}-\frac{1}{d}\eta^{\alpha\beta}\eta_{\mu\nu}\ ,
	\ee
the compatibility condition \eqref{FBD-appLkin} implies
	\be
		\begin{split}
		2g_{12}\tilde\varphi^{\textsc{ab}0}&=\frac{\partial\mathcal L_{3D}}{\partial \dot q_{\textsc{ab}}}\\
		&=\frac{\omega_2}{6}\left(2\frac{\partial\varphi_{0\textsc{mn}}}{\partial \dot q_{\textsc{ab}}}\tilde\varphi^{0\textsc{mn}}+\frac{\partial\varphi_{\textsc{mn}0}}{\partial \dot q_{\textsc{ab}}}\tilde\varphi^{\textsc{mn}0}\right)\\
		&=\frac{\omega_2}{6}\left(-2\tilde\varphi^{\textsc{ab}0}+\tilde\varphi^{0\textsc{ab}}+\tilde\varphi^{0\textsc{ba}}\right)\\
		&=-\frac{\omega_2}{2}\tilde\varphi^{\textsc{ab}0}\ ,
		\end{split}
	\ee
due to the cyclicity of $\tilde\varphi_{abc}(X)$ \eqref{FBD-appcicl}. We thus find
	\be
	\omega_2=-4g_{12}\ ,
	\ee
from which the 3D action \eqref{FBD-appS3D-gen} becomes
	\be \label{FBD-appS3D}
		\begin{split}
		S_{3D}&=\int d^3X\left[-4g_{12}\left(\partial_c\alpha_{ab}\partial^c\tilde\alpha^{ab}-\frac{3}{2}\partial_c\alpha_{ab}\partial^a\tilde\alpha^{bc}\right)+3\omega_5\,\tilde\alpha^d_a\epsilon^{abc}\partial_b\tilde\alpha_{cd}\right]\\
		&=\int d^3X\left[-g_{12}\left(\frac{2}{3}f_{abc}\tilde f^{abc}-\frac{1}{2}f_a^{\ ab}\tilde f^c_{\ cb}\right)+\omega_5\, \tilde\alpha^d_a\epsilon^{abc}\tilde f_{dbc}\right]\\
		&=\int d^3X\left(-\frac{2}{3}g_{12}\,\varphi_{abc}\tilde\varphi^{abc}+\omega_5\, \tilde\alpha^d_a\epsilon^{abc}\tilde\varphi_{dbc}\right)\ .
		\end{split}
	\ee
The second commutator \eqref{FBD-appqt,pt} yields the same result, with $\alpha\leftrightarrow\tilde\alpha$. The two choices are alternative and equivalent, in fact by choosing the first \eqref{FBD-appq,p} we have $q\sim\alpha$, $p\sim\tilde\alpha$, while the second \eqref{FBD-appqt,pt} corresponds to $\tilde q\sim\tilde\alpha$, $\tilde p\sim\alpha$. Something similar also happens in Maxwell theory \cite{Bertolini:2020hgr}.


\cleardoublepage

\label{app:bibliography} 

\manualmark 
\markboth{\spacedlowsmallcaps{\bibname}}{\spacedlowsmallcaps{\bibname}} 
\refstepcounter{dummy}
\sloppy
\addtocontents{toc}{\protect\vspace{\beforebibskip}} 
\addcontentsline{toc}{chapter}{\tocEntry{\bibname}}

\printbibliography 

\end{document}